\documentclass[10pt,journal]{IEEEtran}
\usepackage{research5IEEE}
\usepackage{psfrag} 
\usepackage{graphicx}
\usepackage{fancyhdr}
\usepackage{amsthm}
\usepackage{verbatim}

\usepackage{amsmath, amssymb}
\usepackage{mydef}

\allowdisplaybreaks[3]

\newtheorem{theorem}{Theorem}[section]
\newtheorem{lemma}[theorem]{Lemma}

\newtheorem{corollary}[theorem]{Corollary}

\newtheorem{proposition}[theorem]{Proposition}
\newtheorem{definition}[theorem]{Definition}

\newtheorem{remark}[theorem]{Remark}

\begin{document}


\title{On the AWGN MAC with Imperfect Feedback} \author{Amos~Lapidoth
  and Mich\`ele~Wigger \thanks{A.~Lapidoth is with the Departement of
    Information Technology and Electrical Engineering, ETH Zurich,
    Switzerland. Email: lapidoth@isi.ee.ethz.ch. M.~Wigger was with
    the Departement of Information Technology and Electrical
    Engineering, ETH Zurich, Switzerland.  She is now with the
    Communications and Electronics Departement, Telecom ParisTech,
    France. Email: michele.wigger@telecom-paristech.fr. Part of the
    results in this paper were presented at the Kailath colloquium on
    feedback communication 2006, Stanford University, June 2006; and
    at the Israel IEEE convention for electrical engineers 2006,
    Eilat, Israel, Nov.  2006. The research was partly supported by
    the Swiss National Science Foundation under Grant 200021-111863/1.}}

\date{}
\maketitle

\begin{abstract}
 New achievable rate regions are derived for the two-user additive
  white Gaussian multiple-access channel with noisy feedback. The
  regions exhibit the following two properties. Irrespective of the
  (finite) Gaussian feedback-noise variances, the regions include rate
  points that lie  outside the no-feedback capacity region,
  and when the feedback-noise variances tend to 0 the regions converge
  to the perfect-feedback capacity region.
  
  The new achievable regions also apply to the partial-feedback setting
  where one of the transmitters has a noisy feedback link and the
  other transmitter has no feedback at all. Again, irrespective of the
  (finite) noise variance on the feedback link, the regions
  include rate points that lie  outside the no-feedback
  capacity region. Moreover, in the case of perfect partial
  feedback, i.e., where the only feedback link is noise-free, for
  certain channel parameters the new regions include rate points
  that lie  outside the Cover-Leung region. This answers in
  the negative the question posed by van der Meulen as to whether the
  Cover-Leung region equals the capacity region of the Gaussian
  multiple-access channel with perfect partial feedback.
  
  Finally, we propose new achievable regions also for a setting where
  the receiver is cognizant of the realizations of the noise sequences
  on the feedback links.

\end{abstract}

\section{Introduction}
\label{sec-1}

In \cite{gaarderwolf75} Gaarder and Wolf showed that perfect feedback
from the receiver to the transmitters increases the capacity of some
memoryless multiple-access channels (MACs). That this also holds for
the two-user additive white Gaussian noise (AWGN) MAC was shown by
Ozarow in \cite{ozarow85}, where he also determined the capacity
region of this channel with perfect feedback. Here, we study the
capacity region of the two-user AWGN MAC when the feedback is
imperfect. We consider the following settings:
\begin{itemize}
\item \emph{noisy feedback} where the feedback links are corrupted by
  AWGN;
\item \emph{noisy partial feedback} where one of the two transmitters has
  a noisy feedback link whereas the other transmitter has  no
  feedback at all;
\item \emph{perfect partial feedback} where one of the two
  transmitters has a perfect (noise-free) feedback link whereas the
  other transmitter has no feedback at all; and
\item \emph{noisy feedback with  receiver side-information} where 
  both transmitters have noisy feedback links and the  receiver (but not the transmitters)
  is cognizant of the feedback-noise sequences.
\end{itemize}
The last setting arises, for example, when the receiver
actively feeds back a quantized version of the channel output over perfect
feedback links, and the feedback noises model the quantization
noises, which are known to the receiver. (The MAC with quantized
feedback has also been considered in \cite{shavivsteinberg08} but
under the assumption of a rate limitation on the feedback links and
for the discrete memoryless case.) We show that in all these settings the
capacity region is strictly larger than the no-feedback capacity
region.  Moreover, we show that for noisy feedback the capacity region
tends to Ozarow's perfect-feedback capacity region \cite{ozarow85} as
the feedback-noise variances tend to zero.  Finally, in the case of
perfect partial feedback we show that for certain channel parameters
the capacity region strictly contains the Cover-Leung region
\cite{coverleung81}, a region that was originally derived for the
perfect-feedback setting and that was later shown by Carleial
\cite{carleial82} and (for the discrete memoryless case) by Willems
and van der Meulen \cite{willemsmeulen83} to be achievable also in the
perfect partial-feedback setting. This answers in the negative the
question posed by van der Meulen in \cite{vandermeulen87} as to
whether the Cover-Leung region equals the capacity region of the
AWGN MAC with perfect partial feedback.

To derive these results we propose coding schemes for the described
settings and analyze the rates that they achieve. The idea behind our
schemes is to generalize Ozarow's capacity-achieving perfect-feedback
scheme to imperfect feedback. Ozarow's scheme is based on the
following strategy.  The transmitters first map their messages onto
message points in the interval $[-\frac{1}{2},\frac{1}{2}]$. They then
successively refine the receiver's estimates of these message points by
sending scaled versions of the receiver's linear minimum mean-squared
errors (LMMSE) of the message points.  Besides achieving capacity,
Ozarow's scheme has the advantage of a double-exponential decay of the
probability of error.  However, a drawback of the scheme is that it is
extremely sensitive to noise on the feedback links: it does not
achieve any positive rate if the feedback links are not noise-free \cite{kimlapidothweissman06}. To
overcome this weakness, we propose to apply an outer code around a
modified version of Ozarow's scheme where the transmitters---rather than
refining the message points---successively refine the input symbols
from the outer code.  We further modify Ozarow's scheme by allowing
the transmitters to refine the input symbols by sending
arbitrary linear updates (i.e., not necessarily LMMSE-updates) and
by allowing the number of refinements of each input symbol to be a
constant, which can be optimized and which does not grow with the
blocklength.  These modifications yield a scheme which achieves high
rates also for channels with imperfect feedback. In particular, for
noisy feedback and for noisy partial feedback our scheme exhibits the
following key properties:
\begin{itemize}
\item for all finite feedback-noise variances, our scheme achieves
  rate points that lie outside the capacity region without
  feedback, and
\end{itemize}
for noisy feedback
\begin{itemize}
\item the  scheme  achieves rate regions that converge to
  Ozarow's perfect-feedback capacity region when the feedback-noise
  variances tend to zero.
\end{itemize}

Previous achievable regions for the AWGN MAC with imperfect feedback
were given by Carleial \cite{carleial82}, by Willems et al.
\cite{willemsvandermeulenschalkwijk83-2}\footnote{The result in
  \cite{willemsvandermeulenschalkwijk83-2} is for the discrete
  memoryless case, but it easily extends to the Gaussian case.}, and
by Gastpar \cite{gastpar05}.  Carleial \cite{carleial82} and Willems
et al. \cite{willemsvandermeulenschalkwijk83-2} generalized the
Cover-Leung coding scheme \cite{coverleung81}. Gastpar's result is
also based on Ozarow's scheme and on the idea of modifying it to use
only a finite number of refinements which does not grow with the
blocklength.\footnote{The idea of using a finite number of refinements
  was already mentioned in \cite{schalkwijkkailath66}.  However, only
  in combination with zero rate or nonvanishing probability of error.}
All these regions collapse to the no-feedback capacity region when the
feedback-noise variances exceed a certain threshold.  Moreover, as the
feedback-noise variances tend to zero the regions in \cite{carleial82}
and \cite{willemsvandermeulenschalkwijk83-2} converge to the
Cover-Leung region, which is a strict subset of Ozarow's region
\cite{brosslapidothwigger08}.\footnote{It can be shown that the
  achievable rate region in \cite{gastpar05} converges to Ozarow's
  region when the feedback-noise variances tend to 0.} 

Kramer studied the discrete memoryless MAC with imperfect feedback,
and presented a coding scheme for this setup that is based on code
trees~\cite{kramer03, kramer98}.

Outer bounds on the capacity region of the AWGN MAC with noisy
feedback were derived by Gastpar and Kramer \cite{gastparkramer06_1}
and Tandon and Ulukus \cite{tandonulukus08} based on the idea of
dependence-balance \cite{hekstrawillems89}. These outer bounds do not
in general coincide with any known achievable regions.

The rest of the paper is outlined as follows. This section is
concluded with remarks on notation; Section~\ref{sec:channelmodel}
describes the channel models in more detail;
Section~\ref{sec:previous} discusses some previous achievability
results; Section~\ref{part:noisyfb} describes our results and the new
coding schemes for the setting with noisy feedback;
Section~\ref{sec:par} for the setting with noisy or perfect partial
feedback; and Section~\ref{sec:noisysi} for the setting with noisy
feedback where the receiver has side-information;
Section~\ref{sec:summary} finally summarizes the paper.

In the following $A^{\ell}$ denotes the $\ell$-tuple $(A_1,\ldots,A_\ell)$, i.e.,
$A^\ell=\trans{ ( A_1, A_2, \ldots, A_{\ell})}$;
$\diag{a_1,\ldots,a_\ell}$ denotes the diagonal matrix with diagonal
entries $a_1, \ldots, a_\ell$; $\mat{I}_{\ell}$ denotes the $\ell \times
\ell$ identity matrix; $\trans{\mat{A}}$ denotes the transpose of a
matrix $\mat{A}$, $|\mat{A}|$ its determinant, and $\tr{\mat{A}}$ its
trace.  Also, for zero-mean random vectors $\vect{S}$ and $\vect{T}$
we define the covariance matrices
$\mat{K}_{\vect{S},\vect{T}}\triangleq \E{ \vect{S}
  \trans{\vect{T}}}$ and $\mat{K}_{\vect{S}}\triangleq
\E{\vect{S}\trans{\vect{S}}}$.  For a two-dimensional rate region
$\set{R}$ we denote by $\cl{\set{R}}$ its closure and by
$\mathring{\set{R}}$ its interior.

\section{Channel Model}\label{sec:channelmodel}

This paper focuses on the AWGN MAC with two transmitters that wish to
transmit messages $M_1$ and $M_2$ to a single receiver.  The two
messages are assumed to be independent and uniformly
distributed over the discrete finite sets $\set{M}_1$ and $\set{M}_2$.

 \begin{figure}[tbp]
  \centering 
\psfrag{Encoder 1}[lc][lc]{\footnotesize Trans.1}
\psfrag{Encoder 2}[lc][lc]{\footnotesize Trans.2}
\psfrag{InEnc1}[lc][lc]{\footnotesize In.Enc.1}
\psfrag{InEnc2}[lc][lc]{\footnotesize In.Enc.2}
\psfrag{OutEnc1}[lc][lc]{\footnotesize Out.Enc.1}
\psfrag{OutEnc2}[lc][lc]{\footnotesize Out.Enc.2}
\psfrag{Decoder}[lc][lc]{\footnotesize  Receiver}
\psfrag{InDec}[lc][lc]{\footnotesize In.Dec.}
\psfrag{OutDec}[lc][lc]{\footnotesize Out.Dec.}
\psfrag{X1}[cc][cc]{\footnotesize{ $X_{1,t}$}}
\psfrag{X2}[cc][cc]{\footnotesize{ $X_{2,t}$}}
\psfrag{Y}[cc][cc]{\footnotesize{$Y_{t}$}}
\psfrag{Z}[cc][cc]{\footnotesize{$Z_{t}$}}
\psfrag{*}[cc][cc]{\footnotesize{$+$}}
\psfrag{+}[cc][cc]{\footnotesize{$+$}}
\psfrag{W1}[cc][cc]{\footnotesize{$W_{1,t}$}}
\psfrag{W2}[cc][cc]{\footnotesize{$W_{2,t}$}}
\psfrag{Y1}[cc][cc]{\footnotesize{$V_{1,t}$}}
\psfrag{Y2}[cc][cc]{\footnotesize{$V_{2,t}$}}
\psfrag{M1}[cc][cc]{\footnotesize{$M_1$}}
\psfrag{M2}[cc][cc]{\footnotesize{$M_2$}}
\psfrag{Mhat}[lc][lc]{\footnotesize{$\begin{pmatrix}\hat{M}_1\\\hat{M}_2\end{pmatrix} $}}
  \includegraphics[width=0.48\textwidth]{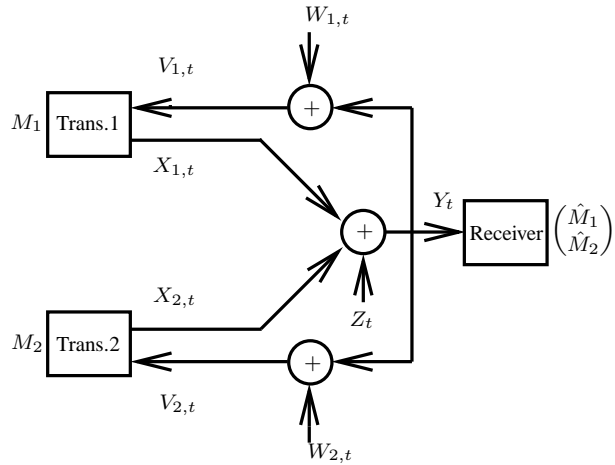}
    \caption{AWGN MAC with noisy feedback.}
  \label{fig:noisyfb} 
\end{figure}

To describe the channel model (see Figure~\ref{fig:noisyfb}), we introduce the sequence $\{Z_{t}\}$ of
independent and identically distributed (IID) zero-mean variance-$\N$
Gaussian random variables that will be used to model the additive
noise at the receiver. Using this sequence we can describe the
time-$t$ channel output $Y_{t}$ corresponding to the time-$t$ channel
inputs $x_{1,t}$ and $x_{2,t}$ by
\begin{equation*}
Y_{t}= x_{1,t} +  x_{2,t} +Z_{t}.
\end{equation*} 
The sequence $\{Z_{t}\}$ is assumed to be independent of the messages
$(M_{1}, M_{2})$.  Also, we introduce the IID sequence of
bivariate zero-mean Gaussians $\left\{ \bigl( W_{1,t}, W_{2,t} \bigr) \right\}$
of covariance matrix
\begin{IEEEeqnarray}{rCl}
  \KW&  \triangleq & 
  \begin{pmatrix}
    \E{W_{1,t}^{2}} & \E{W_{1,t}W_{2,t}} \\
    \E{W_{1,t}W_{2,t}} & \E{W_{2,t}^{2}}
  \end{pmatrix}  \nonumber \\ &=&   
  \begin{pmatrix}
    \sigma_1^2 & \sigma_1 \sigma_2 \varrho \\
    \sigma_1 \sigma_2 \varrho & \sigma_2^2 ,
  \end{pmatrix}\label{eq:KW}
\end{IEEEeqnarray}
where $\sigma_1,\sigma_2\geq 0$ and $\varrho\in[-1,1]$. The sequence
$\left\{ \bigl( W_{1,t}, W_{2,t} \bigr) \right\}$ is used to model the
additive noise corrupting the feedback links. The time-$t$ feedback
output $V_{\nu,t}$ at Transmitter~$\nu$ can then be modeled as
\begin{equation*}%
  V_{\nu,t} =  Y_{t} + W_{\nu,t}, \qquad \nu \in \{1,2\}.
\end{equation*}
The sequence $\{(W_{1,t},W_{2,t})\}$ is assumed to be
independent of $(M_{1}, M_{2}, \{Z_{t}\})$.

The transmitters observe the feedback outputs in a causal fashion, 
i.e., they compute their time-$t$ channel
inputs $X_{1,t}$ and $X_{2,t}$ after observing all prior
feedback outputs $V_{1,1},\ldots,V_{1,t-1}$ and
$V_{2,1},\ldots,V_{2,t-1}$.  Thus, for $\nu\in\{1,2\}$, Transmitter~$\nu$ computes its channel
inputs by mapping the Message $M_\nu$ and the previous feedback
outputs $V_{\nu,1}, \ldots, V_{\nu,t-1}$ into the time-$t$ channel
input $X_{\nu,t}$,
\begin{equation}\label{eq:encoding1}
  X_{\nu, t} = \varphi_{\nu,t}^{(n)}\left(M_{\nu}, V_{\nu,1}, \ldots,
  V_{\nu, t-1} \right) ,\qquad t \in \{1,\ldots, n\},
\end{equation}
for some sequences of encoding functions
\begin{equation}\label{eq:encoding}
\varphi_{\nu,t}^{(n)}\colon \; \set{M}_\nu \times \Reals^{t-1} \rightarrow \Reals 
,\qquad t \in \{1,\ldots, n\},
\end{equation} 
where $n$ denotes the blocklength of the scheme.  We only allow
encoding functions that satisfy the power constraints
\begin{equation}\label{eq:power}
  \frac{1}{n} \sum_{t=1}^{n} \E{\left(\varphi_{\nu,t}^{(n)}\left(M_{\nu}, V_{\nu,1},
  \ldots, V_{\nu, t-1}\right)\right)^{2}} \leq P_{\nu},
\end{equation}
where the expectation is over the messages and the realizations of the
channel, i.e., the noise sequences $\{Z_t\}$, $\{W_{1,t}\}$, and
$\{W_{2,t}\}$.\footnote{The achievability results in this paper
  remain valid also when the expected average block-power constraints
  \eqref{eq:power} are replaced by
  average block-power constraints that hold with probability 1.}

A \emph{ blocklength-$n$ powers-$(P_1,P_2)$ feedback-code} of rate
pair $\left(\frac{1}{n} \log(|\set{M}_1|),\frac{1}{n}
  \log(|\set{M}_2|)\right)$ is a triple
\begin{equation*}
\left( \left\{\varphi_{1,t}^{(n)}\right\}_{t=1}^n,
  \left\{\varphi_{2,t}^{(n)}\right\}_{t=1}^n, \phi^{(n)}\right)
\end{equation*}
where 
\begin{equation*}%
\phi^{(n)} \colon \; \Reals^n \rightarrow \set{M}_{1} \times
\set{M}_{2}
\end{equation*}
and where $\left\{\varphi_{1,t}^{(n)}\right\}$ and
$\left\{\varphi_{2,t}^{(n)}\right\}$ are of the form \eqref{eq:encoding} and
satisfy \eqref{eq:power}.  In the following we say that a rate pair
$(R_1,R_2)$ is achievable if for every $\delta>0$ and every sufficiently
large $n$ there exists a blocklength-$n$ powers-$(P_1,P_2)$
feedback code of rates exceeding $R_1-\delta$ and
$R_2-\delta$ such that the average probability of a decoding error,
\begin{IEEEeqnarray*}{rCl}
 \Prv{ \phi^{(n)}\left(Y_1,
 \ldots, Y_n\right) \neq (M_{1}, M_{2})}
\end{IEEEeqnarray*}
tends to 0 as the blocklength $n \rightarrow \infty$.  The set of all
achievable rate pairs for this setting is called the capacity region
and is denoted $\capa_{\textnormal{NoisyFB}}(\pow_1,\pow_2,\N,
\KW)$.

The case $\sigma_1^2=\sigma_2^2=0$ corresponds to the special case
when the feedback links are noise-free. We refer to this
setting as the ``perfect-feedback'' setting and denote the capacity
region by $\capa_{\textnormal{PerfectFB}}(P_1,P_2,N)$, i.e.,
\begin{equation*}
\capa_{\textnormal{PerfectFB}}(P_1,P_2,N)\triangleq
\capa_{\textnormal{NoisyFB}}\left(P_1,P_2,N,\mat{0} \right)
\end{equation*}
where $\mat{0}$ is the $2\times 2$ all-zero matrix.

 \begin{figure}[tbp]
  \centering 
\psfrag{Encoder 1}[lc][lc]{\footnotesize Trans.1}
\psfrag{Encoder 2}[lc][lc]{\footnotesize Trans.2}
\psfrag{InEnc1}[lc][lc]{\footnotesize In.Enc.1}
\psfrag{InEnc2}[lc][lc]{\footnotesize In.Enc.2}
\psfrag{OutEnc1}[lc][lc]{\footnotesize Out.Enc.1}
\psfrag{OutEnc2}[lc][lc]{\footnotesize Out.Enc.2}
\psfrag{Decoder}[lc][lc]{\footnotesize  Receiver}
\psfrag{InDec}[lc][lc]{\footnotesize In.Dec.}
\psfrag{OutDec}[lc][lc]{\footnotesize Out.Dec.}
\psfrag{X1}[cc][cc]{\footnotesize{ $X_{1,t}$}}
\psfrag{X2}[cc][cc]{\footnotesize{ $X_{2,t}$}}
\psfrag{Y}[cc][cc]{\footnotesize{$Y_{t}$}}
\psfrag{Z}[cc][cc]{\footnotesize{$Z_{t}$}}
\psfrag{*}[cc][cc]{\footnotesize{$+$}}
\psfrag{+}[cc][cc]{\footnotesize{$+$}}
\psfrag{W1}[cc][cc]{\footnotesize{$W_{1,t}$}}
\psfrag{W2}[cc][cc]{\footnotesize{$W_{2,t}$}}
\psfrag{Y1}[cc][cc]{\footnotesize{$V_{1,t}$}}
\psfrag{Y2}[cc][cc]{\footnotesize{$V_{2,t}$}}
\psfrag{M1}[cc][cc]{\footnotesize{$M_1$}}
\psfrag{M2}[cc][cc]{\footnotesize{$M_2$}}
\psfrag{Mhat}[lc][lc]{\footnotesize{$\begin{pmatrix}\hat{M}_1\\ \hat{M}_2\end{pmatrix}$}}
  \includegraphics[width=0.48\textwidth]{%
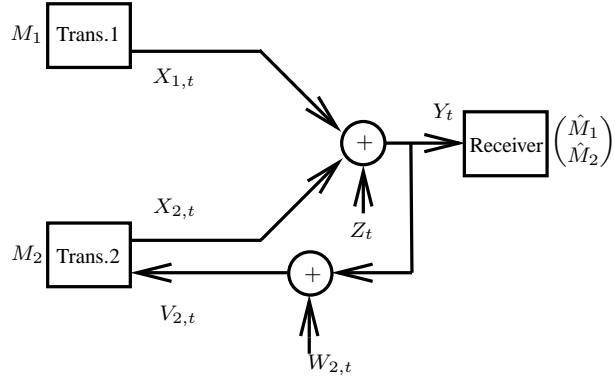}
    \caption{AWGN MAC with noisy partial feedback.}
  \label{fig:noisypartialfb} 
\end{figure}

\begin{figure}[tbp]
  \centering 
\psfrag{Encoder 1}[lc][lc]{\footnotesize Trans.1}
\psfrag{Encoder 2}[lc][lc]{\footnotesize Trans.2}
\psfrag{InEnc1}[lc][lc]{\footnotesize In.Enc.1}
\psfrag{InEnc2}[lc][lc]{\footnotesize In.Enc.2}
\psfrag{OutEnc1}[lc][lc]{\footnotesize Out.Enc.1}
\psfrag{OutEnc2}[lc][lc]{\footnotesize Out.Enc.2}
\psfrag{Decoder}[lc][lc]{\footnotesize  Receiver}
\psfrag{InDec}[lc][lc]{\footnotesize In.Dec.}
\psfrag{OutDec}[lc][lc]{\footnotesize Out.Dec.}
\psfrag{X1}[cc][cc]{\footnotesize{ $X_{1,t}$}}
\psfrag{X2}[cc][cc]{\footnotesize{ $X_{2,t}$}}
\psfrag{Y}[cc][cc]{\footnotesize{$Y_{t}$}}
\psfrag{Z}[cc][cc]{\footnotesize{$Z_{t}$}}
\psfrag{*}[cc][cc]{\footnotesize{$+$}}
\psfrag{+}[cc][cc]{\footnotesize{$+$}}
\psfrag{W1}[cc][cc]{\footnotesize{$W_{1,t}$}}
\psfrag{W2}[cc][cc]{\footnotesize{$W_{2,t}$}}
\psfrag{Y1}[cc][cc]{\footnotesize{$V_{1,t}$}}
\psfrag{Y2}[cc][cc]{}%
\psfrag{M1}[cc][cc]{\footnotesize{$M_1$}}
\psfrag{M2}[cc][cc]{\footnotesize{$M_2$}}
\psfrag{Mhat}[lc][lc]{\footnotesize{$\begin{pmatrix}\hat{M}_1\\ \hat{M}_2\end{pmatrix}$}}
  \includegraphics[width=0.48\textwidth]{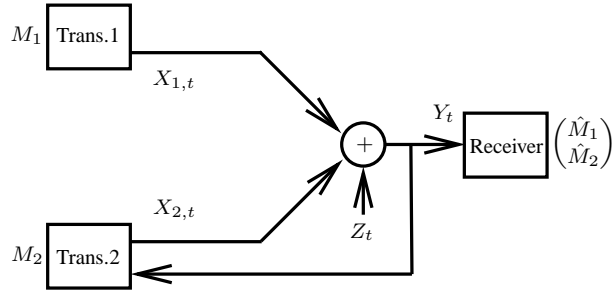}
    \caption{AWGN MAC with perfect partial feedback.}
  \label{fig:perfectpartialfb} 
\end{figure}

In addition to the noisy-feedback setting we also consider the
``partial-feedback'' setting (see 
Figure~\ref{fig:noisypartialfb}) where only one of the two
transmitters has feedback. We assume that the transmitter with
feedback is Transmitter~2.  For the partial-feedback setting
\eqref{eq:encoding1} and \eqref{eq:encoding} are modified by requiring
that the
sequence $\{X_{1,1},\ldots, X_{1,n}\}$ be a function of
Message~$M_1$ only. Since the sole feedback link can be noisy we shall
refer to this setting also as ``noisy partial feedback'' and denote
its capacity region by
$\capa_{\textnormal{NoisyPartialFB}}(P_1,P_2,N,\sigma_2^2)$, where
$\sigma_2^2\geq 0$ denotes the noise variance on the feedback link to
Transmitter~2. In the special case of $\sigma_2^2=0$, i.e., when the
sole feedback link is noise-free, we refer to the setting as ``perfect
partial feedback'' (see Figure~\ref{fig:perfectpartialfb}) and
denote the capacity region by
$\capa_{\textnormal{PerfectPartialFB}}(\pow_1,\pow_2, \N)$.

By the ``no-feedback'' setting we refer to the classical MAC  where neither
transmitter has a feedback link. In this case~\eqref{eq:encoding1} and
\eqref{eq:encoding} have
to be modified so both sequences $\{X_{1,1},\ldots,X_{1,n}\}$ and
$\{X_{2,1},\ldots, X_{2,n}\}$ are functions of the respective messages
only. We denote the capacity region of this MAC by
$\capaMAC(P_1,P_2,N)$.

\begin{figure}[tbp]
  \centering 
\psfrag{Encoder 1}[lc][lc]{\footnotesize Trans.1  }
\psfrag{Encoder 2}[lc][lc]{\footnotesize Trans.2  }
\psfrag{InEnc1}[lc][lc]{\footnotesize In.Enc.1}
\psfrag{InEnc2}[lc][lc]{\footnotesize In.Enc.2}
\psfrag{OutEnc1}[lc][lc]{\footnotesize Out.Enc.1}
\psfrag{OutEnc2}[lc][lc]{\footnotesize Out.Enc.2}
\psfrag{Decoder}[lc][lc]{\footnotesize  Receiver}
\psfrag{InDec}[lc][lc]{\footnotesize In.Dec.}
\psfrag{OutDec}[lc][lc]{\footnotesize Out.Dec.}
\psfrag{X1}[cc][cc]{\footnotesize{ $X_{1,t}$}}
\psfrag{X2}[cc][cc]{\footnotesize{ $X_{2,t}$}}
\psfrag{Y}[cc][cc]{\footnotesize{$Y_{t}$}}
\psfrag{Z}[cc][cc]{\footnotesize{$Z_{t}$}}
\psfrag{*}[cc][cc]{\footnotesize{$+$}}
\psfrag{+}[cc][cc]{\footnotesize{$+$}}
\psfrag{W1}[cc][cc]{\footnotesize{$W_{1,t}$}}
\psfrag{W2}[cc][cc]{\footnotesize{$W_{2,t}$}}
\psfrag{Y1}[cc][cc]{\footnotesize{$V_{1,t}$}}
\psfrag{Y2}[cc][cc]{\footnotesize{$V_{2,t}$}}
\psfrag{M1}[cc][cc]{\footnotesize{$M_1$}}
\psfrag{M2}[cc][cc]{\footnotesize{$M_2$}}
\psfrag{Mhat}[lc][lc]{\footnotesize{$\begin{pmatrix}\hat{M}_1\\ \hat{M}_2\end{pmatrix}$}}
  \includegraphics[width=0.48\textwidth]{%
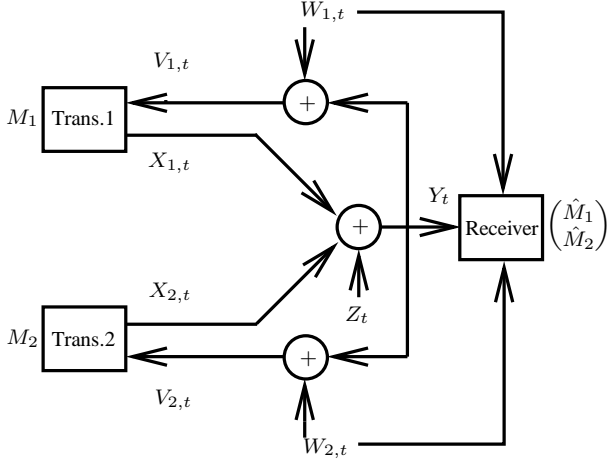}
    \caption{AWGN MAC with noisy feedback and receiver side-information.}
  \label{fig:noisyfbSI} 
\end{figure}

Finally, we also consider a noisy-feedback setting where the receiver
perfectly knows the realizations of the Gaussian noise sequences
$\{W_{1,t}\}$ and $\{W_{2,t}\}$ corrupting the feedback signals (see
also Figure~\ref{fig:noisyfbSI}).\footnote{Since we do not consider
  any delay constraints and the receiver cannot actively feed back a
  signal, it does not matter whether the receiver learns the
  feedback-noise sequences $\{W_{1,t}\}$ and $\{W_{2,t}\}$ causally or
  acausally.} We refer to this setting as the ``noisy feedback with
receiver side-information'' setting. For this setting the
formal description of the communication scenario is the same as in the
noisy-feedback setting, except for the decoder
$\phi^{(n)}_{\textnormal{SI}}$ which is of the form
\begin{IEEEeqnarray*}{rCl}
\phi^{(n)}_{\textnormal{SI}}  :\; \qquad\Reals^{n}\times \Reals^{n}\times
\Reals^{n}\;&\longrightarrow& \;\set{M}_1\times \set{M}_2, \nonumber\\
 (Y_1^{n},
W_{1}^{n},W_{2}^n)\;& \longmapsto &\; (\hat{M}_1, \hat{M}_2).
\end{IEEEeqnarray*}
We denote the capacity region of the MAC with noisy feedback and
perfect receiver side-information by
$\capa_{\textnormal{NoisyFBSI}}(\pow_1,\pow_2, \N, \KW)$.

\section{Previous Results}\label{sec:previous}
We survey some previous results that are needed in
the sequel.

The capacity region of the classical AWGN MAC without feedback
$\capaMAC(P_1,P_2,N)$ was independently determined by Cover
\cite{cover75} and Wyner
\cite{wyner74} 
and is given by the set of all rate pairs $(R_1,R_2)$ satisfying
\begin{subequations}\label{eq:C_NOFB}
\begin{IEEEeqnarray}{rCl}
 R_1 & \leq & \frac{1}{2}\log\left( 1+ \frac{\pow_1}{\N}\right),
\label{eq:C1_NOFB} \\
R_2 & \leq & \frac{1}{2}\log\left( 1+ \frac{\pow_2}{\N}\right),
\label{eq:C2_NOFB} \\
 R_1+R_2 & \leq & \frac{1}{2}\log\left( 1+ \frac{\pow_1+\pow_2}{\N}\right).\label{eq:C12_NOFB}
\end{IEEEeqnarray}
\end{subequations}

 The capacity region of the AWGN MAC with perfect feedback 
$\capa_{\textnormal{PerfectFB}}(P_1,P_2,N)$ was determined by Ozarow \cite{ozarow85}:
\begin{equation}\label{eq:unionOzarow}
\capa_{\textnormal{PerfectFB}}(P_1,P_2,N)=\bigcup_{\rho \in [0,1]}
\mathcal{R}_{\textnormal{Oz}}^\rho(P_1,P_2,N),
\end{equation}
where
$\mathcal{R}_{\textnormal{Oz}}^\rho(P_1,P_2,N)$ is the set of all
 rate pairs $(R_1,R_2)$ satisfying
\begin{subequations}\label{eq:OZ}
\begin{IEEEeqnarray}{rCl}
 R_1 & \leq & \frac{1}{2}\log\left( 1+ \frac{\pow_1\left(1-\rho^2\right)}{\N}\right),
\label{eq:OZ1} \\
R_2 & \leq & \frac{1}{2}\log\left( 1+ \frac{\pow_2\left(1-\rho^2\right)}{\N}\right),
\label{eq:OZ2}\\
 R_1+R_2 &  \leq & \frac{1}{2}\log\left( 1+ \frac{\pow_1+\pow_2 + 2
    \sqrt{\pow_1 \pow_2} \rho}{\N}\right). \IEEEeqnarraynumspace \label{eq:C_PerfectFB}
\end{IEEEeqnarray}
\end{subequations}
We next describe some properties of the regions
$\mathcal{R}_{\textnormal{Oz}}^\rho(P_1,P_2,N)$ and 
$\capa_{\textnormal{PerfectFB}}(P_1,P_2,N)$ that will be needed in
subsequent sections. Some of the properties, Remarks \ref{rem:zero}--\ref{rem:shape} and Remark~\ref{rem:maxsumpoint}, were reported
in \cite{ozarow85}.
\begin{definition}\label{def:rho}
  The parameter $\rho^*(P_1,P_2,N)$ (for short $\rho^*$) is defined as
  the unique solution in the interval $[0,1]$ of the following quartic
  equation in $\rho$
\begin{IEEEeqnarray}{lCl}\label{eq:rhostar}
\lefteqn{\N(\N+\pow_1+\pow_2+2 \sqrt{\pow_1 \pow_2}\rho) }\qquad\nonumber \\
&=&  
(\N+ \pow_1(1-\rho^2))(\N+\pow_2(1-\rho^2)).
\end{IEEEeqnarray}
\end{definition}
\begin{remark}\label{rem:zero}
Equation~\eqref{eq:rhostar} is equivalent to the right-hand side (RHS) of
\eqref{eq:C_PerfectFB} being equal to the sum of the RHSs
of \eqref{eq:OZ1} and \eqref{eq:OZ2}.
\end{remark}
That \eqref{eq:rhostar} has a unique solution in the interval $[0,1]$
can be seen as follows. At $\rho=0$ the left-hand side (LHS) of
\eqref{eq:rhostar} is smaller than its RHS, whereas for
$\rho=1$ the LHS is larger. Since the expressions on both
sides of \eqref{eq:rhostar} are continuous, by the Intermediate Value
Theorem there must exist at least one solution to \eqref{eq:rhostar}
in $[0,1]$. The uniqueness of the solution follows by noting that the
LHS of \eqref{eq:rhostar} is strictly increasing in $\rho$
whereas the RHS is strictly decreasing in $\rho\in[0,1]$.

Next, we discuss the region
$\set{R}_{\textnormal{Oz}}^{\rho}(P_1,P_2,N)$ and examine the rate
constraints \eqref{eq:OZ} defining the
region. The RHS of single-rate constraint
\eqref{eq:OZ1} and the RHS of \eqref{eq:OZ2} are both strictly
decreasing in $\rho\in[0,1]$, whereas the RHS of the sum-rate
constraint \eqref{eq:C_PerfectFB} is strictly increasing in $\rho$. By
these properties, by Definition~\ref{def:rho}, and by
Remark~\ref{rem:zero} we have:
\begin{remark}\label{rem:incdec}
  For $\rho=\rho^*$ the sum of the RHSs of the single-rate constraints
  \eqref{eq:OZ1} and \eqref{eq:OZ2} equals the RHS of the sum-rate
  constraint \eqref{eq:C_PerfectFB}; for $\rho\in[0,\rho^*)$ the sum
  of the RHSs of \eqref{eq:OZ1} and \eqref{eq:OZ2} is strictly larger
  than the RHS of \eqref{eq:C_PerfectFB}; and for $\rho \in(\rho^*,1]$ the
  sum of the RHSs of \eqref{eq:OZ1} and \eqref{eq:OZ2} is strictly
  smaller than the RHS of \eqref{eq:C_PerfectFB}.
\end{remark}
\begin{remark}\label{rem:shape}
  For every $\rho\in[0,\rho^*)$ the rate region
  $\set{R}_{\textnormal{Oz}}^{\rho}(P_1,P_2,N)$ has the shape of a pentagon
  and for every $\rho\in[\rho^*,1]$ the rate region
  $\set{R}_{\textnormal{Oz}}^{\rho}(P_1,P_2,N)$ has the shape of a
  rectangle. Furthermore, all rectangles
  $\set{R}_{\textnormal{Oz}}^{\rho}(P_1,P_2,N)$ for $\rho\in(\rho^*,1]$ are
  strictly contained in the rectangle
  $\set{R}_{\textnormal{Oz}}^{\rho^*}(P_1,P_2,N)$, and thus in
  \eqref{eq:unionOzarow} it is enough to take the union over all
  $\rho\in[0,\rho^*]$.
\end{remark}
For the next two observations we introduce the notation of a dominant
corner point as in \cite{rimoldiurbanke96}. A corner point of a given
rate region is called \emph{dominant} if it is of maximum sum-rate in
the considered region.

\begin{figure}[tbp]
 
\psfrag{R1}{ $R_1$}
\psfrag{R2}{ $R_2$}
\psfrag{R1Oz}{\footnotesize $\set{R}_{1,\textnormal{Oz}}^{\rho}(P_1,P_2,N)$}
\psfrag{R2Oz}{\footnotesize $\set{R}_{2,\textnormal{Oz}}^{\rho}(P_1,P_2,N)$}
\psfrag{ROz}{\footnotesize $\set{R}_{\textnormal{Oz}}^{\rho}(P_1,P_2,N)$}
\psfrag{CPerfectFB}{\footnotesize $\capa_{\textnormal{PerfectFB}}(P_1,P_2,N)$}
 \psfrag{boundary points}[lt][lt]{\large
   $\substack{\textnormal{Boundary point of \hspace{0.5cm}} \\ \textnormal{ sum-rate} \geq\frac{1}{2}\log\left(1+\frac{P_1+P_2}{N}\right)}$}
  \includegraphics[width=0.3\textwidth]{%
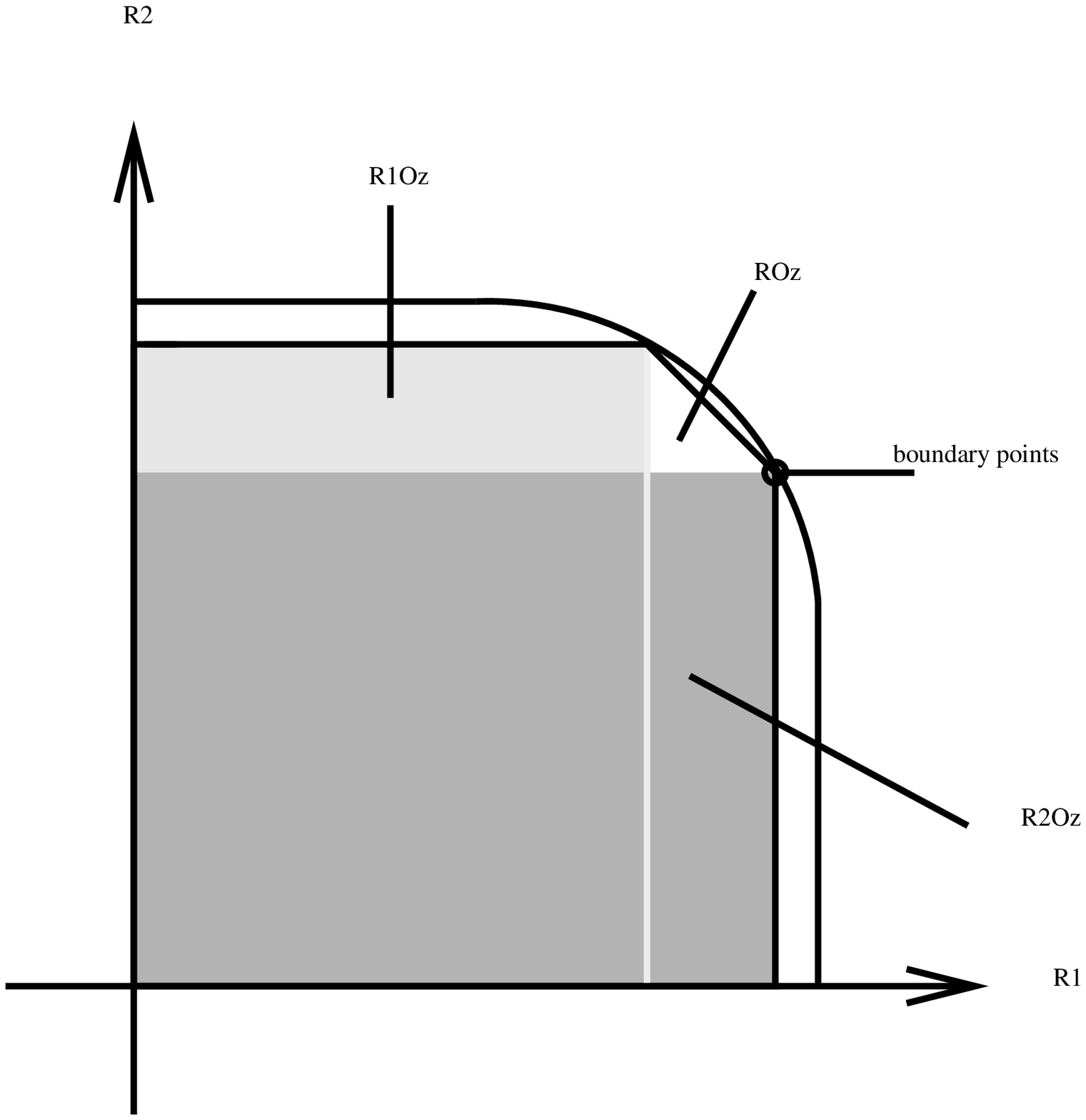}
    \caption{Perfect-feedback capacity region with an example of
      $\set{R}_{\textnormal{Oz}}^{\rho},
    \set{R}_{1,\textnormal{Oz}}^{\rho}$, and
    $\set{R}_{2,\textnormal{Oz}}^{\rho}$ for $0< \rho < \rho^*$.}
  \label{fig1} 
\end{figure}

 \begin{remark}\label{rem:boundary}
   To every boundary point of $\capa_{\textnormal{PerfectFB}}(P_1,P_2,N)$
   that has sum-rate larger or equal to
   $\frac{1}{2}\log\left(1+\frac{P_1+P_2}{N}\right)$ there exists a $\rho \in
   [0,\rho^*]$ such that this point is a dominant corner point  of the
   region $\set{R}_{\textnormal{Oz}}^{\rho}(P_1,P_2,N)$ (see
   Fig.\ref{fig1}).
\end{remark}
Remark~\ref{rem:boundary} follows by Remark~\ref{rem:shape}, by
continuity considerations, and by the monotonicities of the
constraints~\eqref{eq:OZ1}--\eqref{eq:C_PerfectFB}, see
Remark~\ref{rem:incdec}. To state the next observation we define:
\begin{definition}\label{def:Roz12}
For each $\rho\in[0,\rho^*]$, we define
$\set{R}_{1,\textnormal{Oz}}^{\rho}(P_1,P_2,N)$ as the set of all rate pairs
$(R_1,R_2)$ satisfying
\begin{IEEEeqnarray*}{rCl}
R_{1}&\leq& \frac{1}{2}\log \left( 1+ \frac{P_1(1-\rho^2)}{N}\right),\\ 
R_2 & \leq & \frac{1}{2} \log \left( \frac{P_1+P_2 +2 \sqrt{P_1 P_2
    }\rho +N}{P_1(1-\rho^2)+N}\right).
\end{IEEEeqnarray*}
Similarly, $\set{R}_{2,\textnormal{Oz}}^{\rho}(P_1,P_2,N)$ as the set
of all rate pairs $(R_1,R_2)$ satisfying
\begin{IEEEeqnarray*}{rCl}
  R_{1}&\leq& \frac{1}{2}\log\left( \frac{P_1+P_2 +2 \sqrt{P_1 P_2
      }\rho +N}{P_2(1-\rho^2)+N}\right),\\
  R_2 & \leq & \frac{1}{2} \log \left( 1+
    \frac{P_2(1-\rho^2)}{N}\right).
\end{IEEEeqnarray*}
\end{definition} 
Notice that by Remark~\ref{rem:shape},
$\set{R}_{1,\textnormal{Oz}}^{\rho^*}(P_1,P_2,N) =
\set{R}_{2,\textnormal{Oz}}^{\rho^*}(P_1,P_2,N)=
\set{R}_{\textnormal{Oz}}^{\rho^*}(P_1,P_2,N)$.
Also, for every $\rho\in[0,\rho^*]$ the regions
$\set{R}_{1,\textnormal{Oz}}^{\rho}(P_1,P_2,N)$ and
$\set{R}_{2,\textnormal{Oz}}^{\rho}(P_1,P_2,N)$ are rectangles
with dominant corner point equal to one of the dominant corner points
of $\set{R}_{\textnormal{Oz}}^{\rho}(P_1,P_2,N)$, see Figure~\ref{fig1}. By
these observations and by Remark~\ref{rem:boundary} we obtain:
\begin{remark}\label{rem:alternPFBCap}
The perfect-feedback capacity region can be expressed as 
\begin{IEEEeqnarray}{rCl}\label{eq:PFBCap}
\lefteqn{\capa_{\textnormal{PerfectFB}}(P_1,P_2,N)} \; \nonumber \\
& =&
\bigcup_{\rho\in[0,\rho^*]} \left( \set{R}_{1,\textnormal{Oz}}^{\rho}
  (P_1,P_2,N) \cup  \set{R}_{2,\textnormal{Oz}}^{\rho}
  (P_1,P_2,N) \right).\IEEEeqnarraynumspace
\end{IEEEeqnarray}
\end{remark} 
  
The final remark follows from Remark~\ref{rem:boundary} and from the
strict monotonicity in $\rho$ of the RHS of the sum-rate constraint
\eqref{eq:C_PerfectFB}.
\begin{remark}\label{rem:maxsumpoint}
  The dominant corner point of the rectangle
  $\set{R}_{\textnormal{Oz}}^{\rho^*}(P_1,P_2,N)$ is the only rate point of
  maximum sum-rate in $\capa_{\textnormal{PerfectFB}}(P_1,P_2,N)$.
\end{remark}

We next first present an achievability result for general discrete
memoryless MACs and AWGN MACs with perfect feedback due to Cover and
Leung \cite{coverleung81}.  The scheme is known to achieve capacity
for a specific class of discrete memoryless MACs with perfect feedback
\cite{willems82}.  However, for general channels it can be suboptimal,
e.g., for Gaussian channels.  For Gaussian channels the optimization
problem defining the Cover-Leung region is solved by jointly Gaussian
inputs, see \cite{vinodh07,brosslapidothwigger08}, and therefore the
Cover-Leung region is given by
\begin{equation*}%
\set{R}_{\textnormal{CL}}(P_1,P_2,N) = \bigcup_{\rho_1,\rho_2 \in [0,1]}
\set{R}_{\textnormal{CL}}^{(\rho_1,\rho_2)}(P_1,P_2,N),
\end{equation*} 
where $\set{R}_{\textnormal{CL}}^{(\rho_1,\rho_2)}(P_1,P_2,N)$ comprises
all rate pairs $(R_{1}, R_{2})$ satisfying
\begin{subequations}\label{eq:CL}
\begin{IEEEeqnarray}{rCl}
 R_1 & \leq & \frac{1}{2}\log\left( 1+ \frac{\pow_1\left(1-\rho_1^2\right)}{\N}\right),
\label{eq:R1CL} 
\\
 R_2 & \leq & \frac{1}{2}\log\left( 1+ \frac{\pow_2\left(1-\rho_2^2\right)}{\N}\right),
\label{eq:R2CL}
\\
 R_1+R_2 & \leq & \frac{1}{2}\log\left( 1+ \frac{\pow_1+\pow_2 + 2
    \sqrt{\pow_1 \pow_2}  \rho_1 \rho_2}{\N}\right).\nonumber \\\IEEEeqnarraynumspace\label{eq:R12CL}
\end{IEEEeqnarray}
\end{subequations}
Carleial \cite{carleial82} and Willems \cite{willemsmeulen83}
independently proved that to achieve the Cover-Leung region
$\set{R}_{\textnormal{CL}}(P_1,P_2,N)$ it suffices that only one of the two
transmitters have a perfect feedback link, i.e., they proved that the
Cover-Leung region is achievable also in a perfect partial-feedback
setting. Thereupon, van der Meulen in a survey paper on
multiple-access channels with feedback \cite{vandermeulen87} posed the
question whether the Cover-Leung region equals the capacity region for
discrete memoryless MACs or AWGN MACs with perfect partial
feedback. We will answer this question in the negative for Gaussian
channels by proving that for certain channel parameters $(P_1,P_2,N)$
there exist rate pairs that lie outside the Cover-Leung
region $\set{R}_{\textnormal{CL}}(P_1,P_2,N)$ but that are achievable in the
perfect partial-feedback setting.

For the AWGN MAC with perfect partial feedback Willems, van
der Meulen, and Schalkwijk proposed an encoding scheme
\cite{willemsvandermeulenschalkwijk83} which is based on the scheme by
Schalkwijk and Kailath \cite{schalkwijkkailath66}. Unfortunately, the
achievable rate region can only be stated in an implicit form and is
difficult to evaluate analytically and to compare to the Cover-Leung
region.

In \cite{carleial82} Carleial proposed a coding scheme for the
discrete memoryless MAC and the AWGN MAC with ``generalized''
feedback. In the Gaussian case, ``generalized'' feedback includes as
special cases noisy feedback, noisy partial feedback, and perfect
partial feedback.  We present Carleial's region for the AWGN MAC with
noisy feedback in Appendix~\ref{sec:Carleial}, where we also prove
that if the feedback noise variances $\sigma_1^2$ and $\sigma_2^2$
exceed a certain threshold depending on the channel parameters
$P_1,P_2$, and $N$, then Carleial's region collapses to the
no-feedback capacity region in \eqref{eq:C_NOFB}
(Proposition~\ref{th:usecar} in Appendix~\ref{sec:Carleial}). For
perfect partial feedback and for perfect feedback Carleial's scheme
equals the Cover-Leung region $\set{R}_{\textnormal{CL}}(P_1,P_2,N)$.
Hence, in the case of perfect feedback Carleial's scheme is known to
be strictly suboptimal for the two-user AWGN MAC.

Another coding scheme for the MAC with imperfect feedback was
proposed by Willems et al. in
\cite{willemsvandermeulenschalkwijk83-2}.
Although proposed for discrete memoryless channels, the
modifications to treat the Gaussian case are straightforward, and we
state their achievable rate region for the AWGN MAC with noisy
feedback in 
Appendix~\ref{sec:Willems}. 
Like Carleial's scheme, Willems et al.'s scheme collapses to
the no-feedback capacity region when the feedback-noise variances
$\sigma_1^2$ and $\sigma_2^2$ exceed a certain threshold
(Proposition~\ref{eq:prop_useful} in 
Section~\ref{sec:Willems}), and for
perfect feedback or perfect partial feedback the region equals the
Cover-Leung region. Thus, for very noisy feedback, for perfect
feedback, and for perfect partial feedback Carleial's region and
Willems et al.'s region coincide.

\section{Noisy Feedback}\label{part:noisyfb}

In this section we focus on the setup with noisy feedback. For this
setup we present new achievable regions, and based on these new regions
we derive new qualitative
properties of the capacity region (Section~\ref{sec:main}). We also
present the coding schemes corresponding to our new achievable regions
(Sections~\ref{sec:simple}--\ref{sec:ext}).

\subsection{Results}\label{sec:main}


In this section we present our results for noisy feedback. We
begin with some definitions. For given positive integer $\eta$;
$\eta$-dimensional column-vectors $\vect{a}_1, \vect{a}_2$; and
$\eta\times\eta$-matrices $\mat{B}_1, \mat{B}_2$, we define the $\eta
\times 2 $ matrix
\begin{IEEEeqnarray}{rCl}\label{eq:Ab}
\Ab
\eqdef &\blkr{\vect{a}_1}{\vect{a}_{2}},
\end{IEEEeqnarray}
the $2\eta \times 2$ matrix
\begin{IEEEeqnarray}{rCl}\label{eq:Ad}
\Ad&
\eqdef &\blkd{\vect{a}_1}{\vect{0}}{\vect{0}}{\vect{a}_{2}},
\end{IEEEeqnarray}
the $\eta \times 2 \eta$ matrix 
\begin{IEEEeqnarray}{rCl}\label{eq:barB}
\Bb& \eqdef& \blkr{
  \B_1}{\B_2},
\end{IEEEeqnarray}
the $2\eta \times \eta$ matrix 
\begin{IEEEeqnarray}{rCl}\label{eq:Bt}
\Bt& \eqdef& \begin{pmatrix}
  \B_1\\\B_2\end{pmatrix},
\end{IEEEeqnarray}
the $2 \eta \times 2\eta$ block-diagonal matrix 
\begin{IEEEeqnarray}{rCl}\label{eq:Bd}
\Bd& \eqdef& \blkd{
  \B_1}{\mat{0}}{\mat{0}}{\B_2},
\end{IEEEeqnarray}
and the $2\eta \times 2\eta$ matrix
\begin{IEEEeqnarray}{rCl}\label{eq:BB}
\BB& \eqdef& \blkd{
  \B_1}{\B_1}{ \B_2}{\B_2}.
\end{IEEEeqnarray}


Our first achievable region for noisy feedback is obtained by
evaluating the rates that are achieved by the concatenated scheme in
Section~\ref{sec:gen} ahead.  (An alternative formulation of
this achievable region is presented in Section~\ref{sec:achaltnoisy}.)
\begin{definition}\label{def:Reta}
   Let $\eta$ be a positive
  integer, let  $\vect{a}_1,\vect{a}_2$ be $\eta$-dimensional vectors,
  let $\B_1,\B_2$ be $\eta\times \eta$ strictly lower-triangular matrices,
  and let  $\mat{C}$ be a $2\times \eta$ matrix. Depending on the
  matrix $\mat{C}$ the rate region $\RateP{\nch}{\eta,
    \vect{a}_1,\vect{a}_2,\B_1,\B_2,\C}$ is defined as follows. 
\begin{itemize}
\item If the product $\mat{C}\trans{\mat{C}}$ is
  nonsingular,\footnote{Whenever $\eta \in \Nat$ is larger than 1,
    there is no loss in optimality in restricting attention to
    matrices $\C$ so that $\C\trans{\C}$ is nonsingular. { However,
      for completeness, we consider all possible choices of the matrix
      $\mat{C}$.}} then $\RateP{\nch}{\eta,
    \vect{a}_1,\vect{a}_2,\B_1,\B_2,\C}$ is defined as the set of all
  rate-pairs $(R_1,R_2)$ satisfying the three rate
  constraints~\eqref{eq:Rgen1} on top of the next page, where $\Ab$
  and $\Bb$ are defined in \eqref{eq:Ab} and \eqref{eq:barB} and where
  $\otimes$ denotes the Kronecker product.
\item If the product $\mat{C}\trans{\C}$ is singular but $\mat{C}\neq \mat{0}$, then $\RateP{\nch}{\eta,
    \vect{a}_1,\vect{a}_2,\B_1,\B_2,\C}$ is defined as the set of all
  rate pairs $(R_1,R_2)$ satisfying
  \eqref{eq:Rgen1} when the $2\times \eta$ matrix
  $\mat{C}$ is replaced by the $\eta$-dimensional row-vector obtained
  by choosing one of the non-zero rows of $\C$.\footnote{When
    $\C\trans{\C}$ is singular then the two rows of $\C$ are linearly
    dependent and it does not matter which non-zero row is chosen.}
\item If  $\mat{C}=\mat{0}$, then  $\RateP{\nch}{\eta,
    \vect{a}_1,\vect{a}_2,\B_1,\B_2,\C}$ is defined as the set
  containing only the origin.
\end{itemize}
\end{definition}
\begin{definition}\label{def:R}
   Define the rate region
  $\Rate{P_1,P_2,N,\KW}$ (or for short $\set{R}$) as
 \begin{IEEEeqnarray}{lCl}
\Rate{P_1,P_2,N,\KW}\nonumber  \\ \eqdef \cl{ \bigcup_{\eta,\vect{a}_1,\vect{a}_2, \B_1,\B_2,\C}
 \hspace{-0.2cm} \RateP{\nch}{\eta,\vect{a}_1,\vect{a}_2,\B_1,\B_2,\C} } \nonumber \\\label{eq:un}
\end{IEEEeqnarray}
where the union is over all tuples 
$(\eta,\vect{a}_1,\vect{a}_2, \B_1,\B_2,\C)$ satisfying the trace
constraints \eqref{eq:powergen} on top of the next page, and where the matrices $\Ad, \Bt,\Bd$, and $\BB$ are defined in \eqref{eq:Ad},
\eqref{eq:Bt}, \eqref{eq:Bd}, and \eqref{eq:BB}.\footnote{Since $\mat{B}_1$ and $\mat{B}_2$ are strictly lower-triangular, the
matrix $\left(\I_{2\eta} - \Bbb \right)$ is nonsingular and its
inverse exists.} 
\begin{figure*}[!t]
  \normalsize
\allowdisplaybreaks[4]
\begin{subequations}\label{eq:Rgen1}
\begin{IEEEeqnarray}{rCl}
R_1 &\leq & \frac{1}{2\eta}\log \frac{ \left|\mat{C} \left( \vect{a}_1
        \trans{\vect{a}}_1 +N \I_{\eta} + \Bb (\KW \otimes \I_{\eta})
        \trans{\Bb}\right)\trans{\mat{C}}
\right|  }{ \left|\C    \left( N \I_{\eta} + \Bb (\KW \otimes \I_{\eta})
        \trans{\Bb}\right) \trans{\C}\right|}
\label{eq:R1gen1}
\\
R_2&\leq &\frac{1}{2\eta}\log \frac{\left| \mat{C} \left( \vect{a}_2
      \trans{\vect{a}_2} +N \I_{\eta} + \Bb (\KW \otimes \I_{\eta})
      \trans{\Bb} \right)\trans{\mat{C}}\right|}{\left| \mat{C} \left( N \I_{\eta} + \Bb (\KW
        \otimes \I_{\eta}) \trans{\Bb}\right)
      \trans{\mat{C}}\right|}
  \label{eq:R2gen1}\\
 R_1+R_2& \leq&  \frac{1}{2\eta}\log \frac{\left| \mat{C} \left(
      \Ab \trans{\Ab}
+ N \I_{\eta} + \Bb
        (\KW \otimes \I_{\eta})\trans{\Bb}
 \right)\trans{\C}\right|}{\left| \mat{C} \left( N \I_{\eta} + \Bb
        (\KW \otimes \I_{\eta})\trans{\Bb}\right) \trans{\mat{C}}
    \right|} \label{eq:R12gen1}
\end{IEEEeqnarray}
\end{subequations}
\begin{subequations}\label{eq:powergen}
\begin{IEEEeqnarray}{lCl}
\textnormal{tr}\left( \blkr{\I_\eta}{0}   \left(\I_{2\eta} - \Bbb
    \right)^{-1}  \left( \Ad\trans{\Ad}+ N  \Bt\trans{\Bt}
 + \Bd(\KW \otimes \I_{\eta}) \trans{\Bd}\right)  \left(\I_{2\eta} - \Bbb
    \right)^{-\T} \blkc{\I_\eta}{0} \right)&\leq &\eta P_1
    \label{eq:powergen2}\\
\textnormal{tr}\left( \blkr{0}{\I_\eta}   \left(\I_{2\eta} - \Bbb
    \right)^{-1} \left( \Ad\trans{\Ad}+ N  \Bt\trans{\Bt}+  \Bd(\KW \otimes \I_{\eta}) \trans{\Bd}\right)   \left(\I_{2\eta} - \Bbb
    \right)^{-\T} \blkc{0}{\I_\eta} \right)&\leq& \eta P_2 \ \label{eq:powergen1}
\end{IEEEeqnarray}
\end{subequations}
  \hrulefill
\end{figure*}
\end{definition}
\begin{theorem}[Noisy Feedback]\label{th:general}
  The capacity region $\capa_{\textnormal{NoisyFB}}(P_1,P_2,N, \KW)$  of the two-user AWGN MAC with noisy feedback contains the rate
  region $\Rate{P_1,P_2,N,\KW}$, i.e.,
\begin{equation*}
\capa_{\textnormal{NoisyFB}}(P_1,P_2,N, \KW) \supseteq
\Rate{P_1,P_2,N,\KW}.
\end{equation*} 
\end{theorem}
\begin{proof} 
  The proof is based on the concatenated scheme in
  Section~\ref{sec:gen}. As will be described ahead,
  for each choice of parameters $\para$ our concatenated scheme
  achieves the capacity region of the AWGN MAC $\xi_1,\xi_2 \mapsto
  \left( \hat{\Xi}_1,\hat{\Xi}_2\right)$ in \eqref{eq:newdec} scaled
  by a factor $\eta^{-1}$, i.e., it achieves the region $\RateP{\nch}{\para}$. The
  details of the proof are omitted.
\end{proof}

\begin{remark}
  Evaluating the achievable region $\Rate{\ch}$ seems to be difficult
  even numerically. More easily computable (but possibly smaller)
  achievable regions are obtained by taking the union on the RHS of
  \eqref{eq:un} only over a subset of the parameters $\para$
  satisfying \eqref{eq:powergen}. In
  Appendices~\ref{sec:achproof} and \ref{sec:alterna} we present two
  such subsets and their corresponding achievable regions
  (Corollaries~\ref{cor:ach} and~\ref{cor:alte}). In
  Section~\ref{sec:choicenoisy}, we present more general guidelines on how
  to choose the parameters $\para$. 
\end{remark}

\begin{proposition}[Monotonicity and Convergence of the region $\Rateshort$]\label{eq:monotonicity}
  The achievable region
  $\Rate{P_1,P_2,N,\KW}$ satisfies the following three properties:
\begin{enumerate}
\item\label{pa} Given $P_1,P_2,N>0$, it is monotonically decreasing in $\KW$
  with respect to the Loewner order, i.e., for positive semidefinite
  matrices $\KW$ and $\KW'$:
\begin{IEEEeqnarray*}{rCl}
\lefteqn{\Big(\KW \succeq \KW'\Big) \Longrightarrow}\qquad
\\ & &   \Big(
\set{R}(P_1,P_2,N, \KW) \\ & & \hspace{2cm}\subseteq \Rate{P_1,P_2,N,\KW'}\Big).
\end{IEEEeqnarray*}
\item\label{pb} Given $\KW\succeq 0$ and $N>0$, it is continuous in $P_1$ and
  $P_2$, i.e., for all $P_1,P_2>0$:
\begin{IEEEeqnarray*}{rCl}\lefteqn{
\cl{ \bigcup_{\delta>0} \Rate{P_1-\delta,P_2-\delta,N,\KW}}} \qquad \\ &=& \Rate{P_1,P_2,N,\KW}.\hspace{3cm}
\end{IEEEeqnarray*}
\item\label{pc} Given $P_1,P_2,N>0$, it converges to the perfect-feedback
  achievable region $\Rate{P_1,P_2,N, \mat{0}}$ as the feedback-noise
  variances tend to 0 irrespective of the feedback-noise correlations,
  i.e.,
\begin{IEEEeqnarray}{rCl}\label{eq:Part2}
\lefteqn{\cl{\Ue \IK \Rate{P_1,P_2,N,\mat{K}}}}\nonumber\qquad  \\
&=&\Rate{P_1,P_2,N,\mat{0}}. \hspace{3cm}
\end{IEEEeqnarray}
\end{enumerate}
\end{proposition}
\begin{proof}
See Appendix~\ref{sec:monoto}.
\end{proof}

Specializing Theorem~\ref{th:general} to symmetric channels, i.e., to
$P_1=P_2=P$ and $\sigma_1^2=\sigma_2^2=\sigma^2$, and to $\eta=2$ and
the choice of parameters $\vect{a}_1$, $\vect{a}_2$, $\B_1, \B_2,$ and
$\mat{C}$ presented in Section~\ref{sec:enc} yields the following
Corollary~\ref{th:co1}. 
\begin{corollary}[Symmetric Noisy Feedback Channels---Sub-Optimal
  Choice of Parameters]\label{th:co1}
  The capacity region $\capa_{\textnormal{NoisyFB}}(P,P,N,\KW)$ of the
  \emph{symmetric} two-user AWGN MAC
  with noisy feedback, i.e.,  where
  \begin{equation*} \KW=\begin{pmatrix} \sigma^2 & \sigma^2
      \varrho\\\sigma^2 \varrho & \sigma^2 \end{pmatrix},
\end{equation*}
contains all rate pairs $(R_1,R_2)$ satisfying the rate
constraints~\ref{eq:ratesub} on top of the next page.
\begin{figure*}[!t]
  \normalsize
\allowdisplaybreaks[4]
\begin{subequations}\label{eq:ratesub}
\begin{IEEEeqnarray}{rCl}
R_1,R_2& \leq& \frac{1}{4} \log \left(1+\frac{2P}{N} \right)  +\frac{1}{4}
\log\left(1+
  \frac{P^2}{(2P+N)\left(P+N+\sigma^2+\frac{2P}{N}(\sigma^2-\varrho
      \sigma^2 )\right)} \right)\\
R_1+R_2 & \leq & \frac{1}{2} \log \left(1+\frac{2P}{N} \right) +\frac{1}{4}
\log\left(1+
  \frac{2P^2}{(2P+N)\left(P+N+\sigma^2+\frac{2P}{N}(\sigma^2-\varrho
      \sigma^2) \right)} \right)
\end{IEEEeqnarray}
\end{subequations}
\begin{IEEEeqnarray}{rCl}\label{eq:equalrate}
 R &\leq& \frac{1}{4} \log \left( 1 + \frac{2 \pow}{\N} \right)
+\frac{1}{8} \log \left( 1+\frac{2\pow^2}{(2 \pow +\N)\left(\pow+\N+\sigma^2+ \frac{2 \pow}{\N}\sigma^2(1-\varrho)\right) } \right) 
\end{IEEEeqnarray}
  \hrulefill
\end{figure*}
In particular, it contains the equal-rate point $(R,R)$ whenever it
satisfies \eqref{eq:equalrate} on top of the next page. 
\end{corollary}
From Corollary~\ref{th:co1} it is easily seen that the capacity of the
symmetric noisy-feedback setup  is larger than the no-feedback capacity, no matter how
large (but finite) the feedback-noise variance $\sigma^2$ is. The
following stronger result holds:
\begin{theorem}[Noisy Feedback is Always Beneficial]\label{th:C_unsym}
 For every feedback-noise covariance matrix $\KW$
  \begin{IEEEeqnarray*}{rCl}
    \capaMAC(\pow_1,\pow_2,\N)\;& \subset&  \;
    \capa_{\textnormal{NoisyFB}}(\pow_1,\pow_2,\N,\KW),
\end{IEEEeqnarray*}
where the inclusion is strict.
\end{theorem}
\begin{proof}
  Follows by Theorem~\ref{th:C_unsympar} ahead, which establishes that
  noisy partial feedback always increases capacity, and by observing
  that---since Transmitter~1 can always ignore its feedback---noisy
  feedback cannot be worse than noisy partial feedback, i.e., for all
  covariance matrices $ \KW= \begin{pmatrix} \sigma_1^2 & \sigma_1
    \sigma_2 \varrho\\ \sigma_1\sigma_2 \varrho & \sigma_2^2
  \end{pmatrix}$:
\begin{IEEEeqnarray*}{rCl}
\lefteqn{\capa_{\textnormal{NoisyFB}}\left(\pow_1,\pow_2,\N,
    \KW\right)}\qquad \\ & \supseteq& \capa_{\textnormal{NoisyPartialFB}}(\pow_1,\pow_2,\N, \sigma_2^2).
\end{IEEEeqnarray*} 
\end{proof}

Specializing Theorem~\ref{th:general} to perfect feedback, i.e.,
$\KW=\mat{0}$, and to the choice of parameters presented in
Section~\ref{sec:powerstar} yields the following remark.
\begin{remark}[Perfect Feedback]\label{th:perfectfbeta}
  For the two-user AWGN MAC with perfect
  feedback our concatenated scheme achieves all rate pairs inside the
  region $\set{R}_{\textnormal{Oz}}^{\rho^*}(P_1,P_2,N)$, i.e.,
\begin{equation*}
  \Rate{P_1,P_2,N, \mat{0}} \supseteq
\set{R}_{\textnormal{Oz}}^{\rho^*}(P_1,P_2,N).
\end{equation*}
 \end{remark}
\begin{proof} Is based on the specific choice of  parameters in
  Section~\ref{sec:powerstar}, i.e., on the regions
  $\set{\tilde{R}}_{\eta}(P_1,P_2,N,\mat{0})$ in Remark~\ref{def:remarkperf}.
For details, see Section~\ref{sec:proofremperf}.
\end{proof}

We next consider the noisy-feedback setting in the asymptotic regime
where the noise variances on both feedback links vanish.
Proposition~\ref{prop:maxsumratesigma} ahead shows that our achievable
regions in Theorem~\ref{th:general} converge to the point of maximum
sum-rate in $\capa_{\textnormal{PerfectFB}}$ when the feedback-noise
variances tend to 0, irrespective of the feedback-noise correlation.

\begin{proposition}[Convergence to Maximum Sum-Rate of
  $\capa_{\textnormal{PerfectFB}}$] \label{prop:maxsumratesigma}
Our achievable region satisfies
\begin{IEEEeqnarray}{rCl}\label{eq:maxsum}
\lefteqn{\cl{\Ue \IK 
\Rate{P_1,P_2,N, \mat{K}} }} \qquad \nonumber \\ &\supseteq &\set{R}_{\textnormal{Oz}}^{\rho^*}(P_1,P_2,N).\hspace{3cm}\IEEEeqnarraynumspace
\end{IEEEeqnarray}
Thus, by Remark \ref{rem:maxsumpoint} our achievable regions in
Theorem~\ref{th:general} asymptotically approach the point of maximum
sum-rate in the perfect-feedback capacity region.
\end{proposition}
\begin{proof}
Follows directly by Proposition~\ref{eq:monotonicity}, Part~\ref{pc}),
and by Remark~\ref{th:perfectfbeta}.
\end{proof}
\begin{remark}
  We can strengthen Proposition~\ref{prop:maxsumratesigma} as follows:
  Inclusion \eqref{eq:maxsum} remains valid if the region
  $\Rate{P_1,P_2,N,\mat{K}}$ is replaced by the union
  $\left(\bigcup_{\eta\in\Nat}\set{\tilde{R}}_{\eta}(P_1,P_2,N,\mat{K})\right)$,
  where the regions $\set{\tilde{R}}_{\eta}(P_1,P_2,N,\mat{K})$ are
  defined in Definition~\ref{def:Rtilde},
  and represent the regions achieved by our concatenated scheme for
  the specific choice of parameters presented in
  Section~\ref{sec:powerstar}.
\end{remark}

Our last achievability result for noisy feedback is based
on the rate-splitting scheme in Section~\ref{sec:entireregion}. Before
stating the result in Proposition~\ref{prop:0}, we define:
\begin{definition}\label{def:rsgen}
For fixed
  $\eta\in \Nat$; fixed $\eta$-dimensional vectors
  $\vect{a}_1,\vect{a}_2$; $\eta\times \eta$ strictly lower-triangular
  matrices $\B_1,\B_2$; and $2\times \eta$ matrix $\mat{C}$ define
  the region $\RateoneP{P_1',N,\KW}{\eta,
    \vect{a}_1,\vect{a}_2,\B_1,\B_2,\C}$ as the set of all rate pairs
  $(R_1,R_2)$ that for some nonnegative
  $R_{1,\textnormal{CS}},R_{1,\textnormal{NF}}$ summing to $R_1$
  satisfy the following two conditions:
\begin{IEEEeqnarray*}{rCl}
(R_{1,\textnormal{CS}},R_{2}) & \in &
\RateP{N+P_1',\KW}{\eta,\vect{a}_1,\vect{a}_2,\B_1,\B_2,\C}
\end{IEEEeqnarray*}
and 
\begin{IEEEeqnarray}{rCl}
  R_{1,\textnormal{NF}} & \leq & \frac{1}{2\eta} \log \left( \frac{\left| P_1'
        \I_{\eta} + N\I_{\eta}+ \Bb ( \KW \otimes
        \I_{\eta}) \trans{\Bb}\right|}{\left|
        N\I_{\eta}+ \Bb ( \KW \otimes \I_{\eta})
        \trans{\Bb}\right|} \right)\nonumber \\\label{eq:77}
\end{IEEEeqnarray}
where $\Bb$ is defined in \eqref{eq:barB}.

Similarly, 
define the region
$\RatetwoP{P_2',N,\KW}{\eta,\vect{a}_1,\vect{a}_2,\B_1,\B_2,\C}$ 
analogously to the region
$\RateoneP{P_1',N,\KW}{\eta,\vect{a}_1,\vect{a}_2,\B_1,\B_2,\C}$, but with
exchanged indices 1 and 2.
\end{definition}
\begin{definition}
Define the rate region $\Rateone{P_1',P_1'',P_2,N,\KW}$
(or for short $\set{R}_{\textnormal{RS,1}}$) as 
\begin{IEEEeqnarray*}{lCl}
\Rateone{P_1',P_1'',P_2,N,\KW}\quad \\  \eqdef \textnormal{cl}\left(\;\bigcup \RateoneP{P_1',N,\KW}{\eta,\vect{a}_1,\vect{a}_2,\B_1,\B_2,\C}\right)
\end{IEEEeqnarray*}
where the union is over all tuples
$(\eta,\vect{a}_1,\vect{a}_2,\B_1,\B_2,\C)$ satisfying the trace constraints
\eqref{eq:powergen} for powers $P_1''$ and
$P_2$, noise variance $(N+P_1')$, and feedback-noise covariance matrix
$\KW$.
Similarly, 
define the region $\Ratetwo{P_1,P_2',P_2'',N,\KW}$ (or for short
$\set{R}_{\textnormal{RS},2}$) as
\begin{IEEEeqnarray*}{lCl}
\Ratetwo{P_1,P_2',P_2'',N,\KW}\quad \nonumber \\    \eqdef \textnormal{cl}\left(
\; \bigcup\RatetwoP{P_2',N,\KW}{\eta,\vect{a}_1,\vect{a}_2,\B_1,\B_2,\C}\right)
\end{IEEEeqnarray*}
where the union is over all tuples
$(\eta,\vect{a}_1,\vect{a}_2,\B_1,\B_2,\C)$ satisfying the trace
constraints \eqref{eq:powergen} for powers
$P_1$ and $P_2''$, noise variance $(N+P_2')$, and feedback-noise
covariance matrix $\KW$.
\end{definition}
\begin{proposition}[Rate-Splitting for Noisy Feedback]\label{prop:0}
The capacity region $\capa_{\textnormal{NoisyFB}}(P_1,P_2,N,\KW)$
contains the region
$\Rateone{P_1',(P_1-P_1'),P_2,N,\KW}$ for
any $P_1'\in[0,P_1]$, and it contains the region
$\Ratetwo{P_1,P_2',(P_2-P_2'),N,\KW}$ for
any $P_2'\in[0,P_2]$:
\begin{IEEEeqnarray*}{rCl}
\lefteqn{\capa_{\textnormal{NoisyFB}}(P_1,P_2,N, \KW)} \nonumber  \quad\\ &  \supseteq&
 \hspace{-0.1cm} \bigcup_{\substack{ P_1'\in[0,P_1]}}
  \Rateone{P_1',(P_1-P_1'), P_2,N,\KW}
\end{IEEEeqnarray*}
and 
\begin{IEEEeqnarray*}{rCl}
 \lefteqn{   \capa_{\textnormal{NoisyFB}}(P_1,P_2,N,
  \KW)}\nonumber\qquad  \\&  \supseteq & \hspace{-0.1cm}
  \bigcup_{\substack{ P_2'\in
      [0,P_2]}}
 \Ratetwo{P_1,P_2',(P_2-
    P_2'),N,\KW}.
\end{IEEEeqnarray*}
\end{proposition}
\begin{proof} The rate region is achieved by the rate-splitting scheme
  in Section~\ref{sec:entireregion}. The
  analysis is based on Theorem~\ref{th:general}, on the capacity of a
  Gaussian multi-input antenna/multi-output antenna channel with
  noise sequences that are temporally-white but correlated across the antennas, and on a genie-aided
  argument as in \cite{rimoldiurbanke96} and
  \cite[p.~419]{wozencraftjacobs65}. The details are omitted.
\end{proof}
\begin{proposition}[Monotonicity and Convergence of
  Regions $\Rateoneshort$ and
  $\Ratetwoshort$]\label{prop:monotonicity2} The
  achievable region $\Rateone{P_1',P_1'',N, \KW}$ satisfies the
  following three properties:
\begin{enumerate}
\item Given $P_1',P_1'',P_2,N>0$, it is monotonically decreasing in $\KW$ with respect to the
  Loewner order, i.e., for positive semidefinite matrices $\KW$ and $\KW'$:
\begin{IEEEeqnarray*}{lCl}\lefteqn{
\Big(\KW \succeq \KW' \Big) \Longrightarrow }\hspace{1cm} \\
& & \Big(\Rateone{P_1',P_1'',P_2,N, \KW} \\ & & \hspace{1cm} \subseteq
\Rateone{P_1',P_1'',P_2,N,\KW'} \Big).
\end{IEEEeqnarray*}
\item Given $\KW \succeq 0$ and $N>0$, it is continuous in $P_1', P_1'',$ and $P_2$,
  i.e., for all $P_1',P_1'',P_2>0$:
\begin{IEEEeqnarray*}{rCl}
\lefteqn{\cl{\bigcup_{\delta>0}
  \Rateone{P_1'\!-\!\delta,P_1''\!-\!\delta,P_2\!-\!\delta,N,\KW}}}\qquad \\
&   = &\Rateone{P_1',P_1'',P_2,N,\KW}.\hspace{3cm}
\end{IEEEeqnarray*}
\item Given $P_1',P_1'',P_2,N>0$, it converges to the perfect-feedback achievable region
  $\Rateone{P_1',P_1'', P_2,N,\mat{0}}$ as
  the feedback-noise variances tend to 0 irrespective of the
  feedback-noise correlations, i.e.,
\begin{IEEEeqnarray*}{rCl}
\lefteqn{\cl{\Ue \IK \Rateone{P_1',P_1'',P_2,N,
  \mat{K}}}}\qquad \\ &  = &\Rateone{P_1',P_1'',P_2,N,
  \mat{0}}.\hspace{4cm}
\end{IEEEeqnarray*}
\end{enumerate}
Similarly, for $\Ratetwo{P_1,P_2',P_2'',N, \KW}$.
\end{proposition}
\begin{proof}
  Follows from Proposition~\ref{eq:monotonicity} and because for fixed
  $\vect{a}_1,\vect{a}_2,\B_1,\B_2$, and $\C$ the RHS of
  \eqref{eq:77} satisfies the following three properties. It is
  monotonically decreasing in $\KW$ with respect to the Loewner order,
  it is continuous in $P_1'$, and it converges to
  $\frac{1}{2}\log\left(1+\frac{P_1'}{N}\right)$ as the feedback-noise
  variances tend to 0 irrespective of the feedback-noise
  correlations. The details are omitted.
\end{proof}

With the rate-splitting extension in Section~\ref{sec:entireregion}
and Propositions~\ref{prop:0} and \ref{prop:monotonicity2},
Remark~\ref{th:perfectfbeta} and
Proposition~\ref{prop:maxsumratesigma} can be generalized to all the
boundary points of the capacity region $\capa_{\textnormal{PerfectFB}}$.
\begin{remark}[Perfect Feedback]\label{rem:Cperfectentire}
  For the two-user AWGN MAC with perfect
  feedback our rate-splitting scheme in Section~\ref{sec:entireregion}
  achieves all rate pairs in Ozarow's perfect-feedback capacity region
  $\capa_{\textnormal{PerfectFB}}(P_1,P_2,N)$:
\begin{IEEEeqnarray}{rCl}\label{eq:Cperfectinc}
\lefteqn{    \capa_{\textnormal{PerfectFB}}(P_1,P_2,N)}
\nonumber \\&  = & 
\left(
  \bigcup_{\substack{ P_1'\in[0,P_1]}}
  \Rateone{P_1',(P_1-P_1'), P_2,N,\mat{0}} \right)  \nonumber \\ & & \quad\cup \left(
  \bigcup_{\substack{ P_2'\in[0,P_2]}}
  \Ratetwo{P_1,P_2', (P_2-P_2'),N,\mat{0}} \right). \nonumber \\
\end{IEEEeqnarray}
In fact, for each $\rho\in[0,\rho^*]$ there exists a
$P_1'(\rho)\in[0,P_1]$ so that 
\begin{subequations}\label{eq:Rrho}
 \begin{IEEEeqnarray}{rCl} \label{eq:Rrho1}
\lefteqn{\Rateone{P_1'(\rho),(P_1-P_1'(\rho)), P_2,N,\mat{0}}} \qquad
\nonumber 
\\ & \supseteq& \set{R}_{1,\textnormal{Oz}}^\rho(P_1,P_2,N),\hspace{3cm}
\end{IEEEeqnarray}
and a $P_2'(\rho)\in[0,P_2]$ so that
 \begin{IEEEeqnarray}{rCl}\label{eq:Rrho2}
\lefteqn{\Ratetwo{P_1,P_2'(\rho),(P_2-P_2'(\rho)), N,\mat{0}}} \qquad\nonumber
\\ & \supseteq &\set{R}_{2,\textnormal{Oz}}^\rho(P_1,P_2,N).\hspace{3cm}
 \end{IEEEeqnarray}
\end{subequations}
\end{remark}
\begin{proof}
  By Remark~\ref{rem:alternPFBCap}, Equality~\eqref{eq:Cperfectinc}
  follows directly from \eqref{eq:Rrho}. For a proof of
  \eqref{eq:Rrho}, see
  Section~\ref{sec:Cperfectentire}.
\end{proof}

\begin{proposition}[Convergence to Boundary of $\capa_{\textnormal{PerfectFB}}$]\label{prop:splitmaxsumratesigma}
   For every
  $\rho\in[0,\rho^*(P_1,P_2,N)]$ we can find some $P_1'(\rho)\in[0,P_1]$
  so that
\begin{subequations}\label{eq:largeR}
\begin{IEEEeqnarray}{rCl}\label{eq:largeR1}
\lefteqn{\cl{\Ue \IKtwo \!\! \Rateone{ P_1'(\rho), (P_1-P_1'(\rho)),
  P_2, N, \mat{K}}} } \qquad\nonumber  \\
&\supseteq &\set{R}_{1,\textnormal{Oz}}^{\rho}(P_1,P_2,N).\hspace{4.2cm}
\end{IEEEeqnarray}
Similarly, for every $\rho\in[0,\rho^*(P_1,P_2,N)]$ we can find some
$P_2'(\rho)\in [0,P_2]$ so that 
 \begin{IEEEeqnarray}{rCl}\label{eq:largeR2}
\lefteqn{\cl{ \Ue \IKtwo \!\!\Ratetwo{ P_1, P_2'(\rho),(P_2-P_2'(\rho)),
  N, \mat{K}}}}\qquad \nonumber\\
&\supseteq &\set{R}_{2,\textnormal{Oz}}^{\rho}(P_1,P_2,N).\hspace{4.2cm}
\end{IEEEeqnarray}
\end{subequations}
Thus, by Remark~\ref{rem:boundary} and Definition~\ref{def:Roz12} our achievable regions in
Proposition~\ref{prop:0} asymptotically approach all boundary points
of the perfect-feedback capacity region.
\end{proposition}
\begin{proof}
See Section~\ref{sec:ProofProp2}.
\end{proof}

Propositions~\ref{prop:0} and
\ref{prop:splitmaxsumratesigma} combined with
Remark~\ref{rem:alternPFBCap} yield the following continuity result.
\begin{theorem}[Continuity of Noisy-Feedback Capacity Region]\label{th:C_sym}
  For all $P_1, P_2,N>0$:
\begin{IEEEeqnarray*}{rCl}
\lefteqn{\cl{\Ue \IK \capa_{\textnormal{NoisyFB}} (\pow_1, \pow_2, \N,
  \K)}}\qquad \\ & =& \capa_{\textnormal{PerfectFB}}(\pow_1, \pow_2, \N).\hspace{3cm}
\end{IEEEeqnarray*}
\end{theorem}
\begin{proof}
See Section~\ref{sec:conv} for details.
\end{proof}

\subsection{Simple Scheme}\label{sec:simple}
We present a simple coding scheme for the
noisy-feedback setting. It is a special case of the 
\emph{concatenated scheme} in
Section~\ref{sec:gen} ahead: the simple scheme  with parameters 
$a_{1,1}, a_{1,2}, a_{2,1}, a_{2,2}, b_1, b_2$ coincides with the
concatenated scheme for noisy feedback with parameters $\eta=2,
\vect{a}_{1}=\trans{\begin{pmatrix} a_{1,1} &a_{2,1}\end{pmatrix}},
\vect{a}_{2} = \trans{\begin{pmatrix} a_{2,1} &a_{2,2}\end{pmatrix}}$,
$\mat{B}_{1} =
\begin{pmatrix} 0 & 0 \\ b_1 & 0\end{pmatrix}$, $\mat{B}_{2} =
\begin{pmatrix} 0 & 0 \\ b_2 & 0\end{pmatrix}$, $\mat{C}=\mat{I}_2$.
We present the simple scheme here separately,
because it is easier and yet powerful enough to establish 
Corollary~\ref{th:co1} and Theorem~\ref{th:C_unsym}.

Prior to communication a blocklength-$n$, rate-$R_1$ codebook
$\set{C}_1$ and a blocklength-$n$, rate-$R_2$ codebook $\set{C}_2$ are
generated and revealed to both transmitters and to the receiver. The
codewords of codebook $\set{C}_1$ are chosen independently with the
$n$ components $\Xi_{1,1}(m_1), \ldots, \Xi_{1,n}(m_1)$ of the
$m_1$-th codeword chosen IID zero-mean unit-variance Gaussian.  The
codebook $\set{C}_2$ is drawn similarly.  Messages $M_1$ and $M_2$ are
then transmitted over $2n$ channel uses by sending each symbol of the
$n$-length codewords $\Xi_1^n(M_1)$ and $\Xi_2^n(M_2)$ over two
consecutive channel uses.  More precisely, at odd time steps
$t=2(k-1)+1$, for $k\in\{1,\ldots, n\}$, Transmitter~1 sends
\begin{IEEEeqnarray}{rCl}\label{eq:X1odd}
X_{1,2(k-1)+1} = a_{1,1} \Xi_{1,k}, 
\end{IEEEeqnarray}
and Transmitter~2 sends 
\begin{IEEEeqnarray}{rCl}\label{eq:X2odd}
X_{2,2(k-1)+1} = a_{2,1} \Xi_{2,k}.
\end{IEEEeqnarray}
 At even
time steps $t=2k$, for $k\in\{1,\ldots, n\}$, Transmitter~1 sends
\begin{IEEEeqnarray}{rCl}\label{eq:X1even}
X_{1,2k} = a_{1,2} \Xi_{1,k}-b_1V_{1,2(k-1)+1}, 
\end{IEEEeqnarray}
and Transmitter~2 sends 
\begin{IEEEeqnarray}{rCl}\label{eq:X2even}
X_{2,2k} = a_{2,2} \Xi_{2,k}-b_2 V_{2,2(k-1)+1}. 
\end{IEEEeqnarray}
To ensure that the two input sequences $\{X_{1,t}\}_{t=1}^{2n}$ and
$\{X_{2,t}\}_{t=1}^{2n}$  satisfy the power
constraints~\eqref{eq:power}, the parameters $a_{1,1}, a_{1,2},
a_{2,1}, a_{2,2}, b_1,$ and $b_2$ are chosen as to simultaneously satisfy
\begin{subequations}\label{eq:sp}
\begin{equation}
a_{1,1}^2+ (a_{1,2}- b_1 a_{1,1})^2+ b_1^2(a_{2,1}^2 + N+ \sigma_1^2 ) \leq 2 P_1 \label{eq:sp1}
\end{equation}
and 
\begin{equation}
a_{2,1}^2+ (a_{2,2}- b_2 a_{2,1})^2 + b_2^2(a_{1,1}^2+ N+  \sigma_2^2) \leq 2 P_2.\label{eq:sp2}
\end{equation}
\end{subequations}
The receiver uses an optimal decoding rule to decode Messages $M_{1}$
and $M_2$ based on the observed sequence of channel outputs ${Y}_{1},
\ldots, Y_{2n}$.

To describe the performance of the scheme, let $\Xi_1, \Xi_2,
Z_{\textnormal{odd}},$ and $Z_{\textnormal{even}}$ be 
independent zero-mean Gaussian random variables, where $\Xi_1$ and
$\Xi_2$ are of variance~$1$ and $Z_{\textnormal{odd}}$ and
$Z_{\textnormal{even}}$ of variance~$N$. Independent thereof, let the pair $(W_1,W_2)$ be
a zero-mean bivariate Gaussian of covariance matrix $\KW$
as defined in \eqref{eq:KW}. Also, let $Y_{\textnormal{odd}}$ and
$Y_{\textnormal{even}}$ be defined as
\begin{IEEEeqnarray*}{rCl}
Y_{\textnormal{odd}} & \triangleq  &  a_{1,1} \Xi_1 + a_{2,1} \Xi_2 + Z_{\textnormal{odd}}, \\
Y_{\textnormal{even}} & \triangleq  & a_{1,2} \Xi_1+ a_{2,2} \Xi_2 - b_1 V_{1,\textnormal{odd}}- b_2 V_{2,\textnormal{odd}} + Z_{\textnormal{even}};
\end{IEEEeqnarray*}
 and $V_{1,\textnormal{odd}}$ and $V_{2,\textnormal{odd}}$ be defined as
\[V_{\nu, \textnormal{odd}}\triangleq Y_{\textnormal{odd}}+ W_{\nu}, \qquad \nu\in\{1,2\}.\]
The performance of the simple scheme is then described as follows.
 The  scheme achieves all nonnegative rate
pairs $(R_1,R_2)$ that simultaneously satisfy
\begin{subequations}\label{eq:I}
\begin{IEEEeqnarray}{rCl}\label{eq:I1}
R_1 & \leq & \frac{1}{2} I(\Xi_1; Y_{\textnormal{odd}}, Y_{\textnormal{even}}|\Xi_2), \\ 
R_2 & \leq & \frac{1}{2} I(\Xi_2; Y_{\textnormal{odd}}, Y_{\textnormal{even}}|\Xi_1), \label{eq:I2}\\ 
R_1+R_2 & \leq & \frac{1}{2} I(\Xi_1, \Xi_2; Y_{\textnormal{odd}}, Y_{\textnormal{even}}), \label{eq:I3}
\end{IEEEeqnarray}
\end{subequations}
or equivalently---as obtained by evaluating the mutual information expressions on the
RHSs of \eqref{eq:I}---it achieves all nonnegative rate pairs
$(R_1,R_2)$ that simultaneously satisfy
\begin{IEEEeqnarray*}{lCl}
R_1 & \leq & \frac{1}{4}\log \Bigg( 1+ \frac{a_{1,1}^2}{N} \\ & & \hspace{1.4cm}+
  \frac{a_{1,2}^2}{b_1^2 \sigma_1^2 + b_2^2 \sigma_2^2+ 2b_1b_2\varrho \sigma_1\sigma_2+ N} \Bigg) \\
 R_2 & \leq & \frac{1}{4}\log \Bigg( 1+ \frac{a_{2,1}^2}{N} \\ &&
  \hspace{1.4cm} + \frac{a_{2,2}^2}{b_1^2 \sigma_1^2 +b_2^2
  \sigma_2^2+ 2b_1b_2\varrho \sigma_1\sigma_2+ N} \Bigg)
\end{IEEEeqnarray*}
and 
\begin{IEEEeqnarray*}{rCl}
\lefteqn{ R_1+R_2} \;\;\\ & \leq & \frac{1}{4}\log\Bigg(1+
    \frac{a_{1,1}^2+a_{2,1}^2}{N} \\
& & \hspace{1.4cm} + \frac{a_{1,2}^2+a_{2,2}^2}{b_1^2 \sigma_1^2+  b_2^2 \sigma_2^2+ 2b_1b_2\varrho \sigma_1\sigma_2+ N} \nonumber
  \\ & &
    \hspace{1.4cm} +\frac{(a_{1,1} a_{2,2}-a_{2,1}a_{1,2})^2}{N( b_1^2 \sigma_1^2+   b_2^2 \sigma_2^2+ 2b_1b_2\varrho \sigma_1\sigma_2+ N)}\Bigg)
\end{IEEEeqnarray*}
 for some choice of the parameters $a_{1,1}, a_{1,2}, a_{2,1}, a_{2,2}, b_1, b_2$ satisfying \eqref{eq:sp}.

\subsection{Concatenated Scheme}\label{sec:gense}
We first present our concatenated coding scheme with general
parameters in Section~\ref{sec:gen}; in Section~\ref{sec:choicenoisy}
we then give guidelines on how to choose the parameters of this
concatenated scheme.

\subsubsection{Scheme}\label{sec:gen}
 \begin{figure*}[tbp]
  \centering 
\psfrag{Encoder 1}[lc][lc]{\footnotesize Transmitter 1}
\psfrag{Encoder 2}[lc][lc]{\footnotesize Transmitter 2}
\psfrag{InEnc1}[lc][lc]{\footnotesize Inner Enc.1}
\psfrag{InEnc2}[lc][lc]{\footnotesize Inner Enc.2}
\psfrag{OutEnc1}[lc][lc]{\footnotesize Outer Enc.1}
\psfrag{OutEnc2}[lc][lc]{\footnotesize Outer Enc.2}
\psfrag{Decoder}[lc][lc]{\footnotesize  Receiver}
\psfrag{InDec}[lc][lc]{\footnotesize Inner Dec.}
\psfrag{OutDec}[lc][lc]{\footnotesize Outer Dec.}
\psfrag{InnerChannel}[lc][lc]{ ``new'' MAC}
\psfrag{X1}[cc][cc]{\footnotesize{ $X_{1,t}$}}
\psfrag{X2}[cc][cc]{\footnotesize{ $X_{2,t}$}}
\psfrag{Y}[cc][cc]{\footnotesize{$Y_{t}$}}
\psfrag{Z}[cc][cc]{\footnotesize{$Z_{t}$}}
\psfrag{*}[cc][cc]{\footnotesize{$+$}}
\psfrag{+}[cc][cc]{\footnotesize{$+$}}
\psfrag{W1}[cc][cc]{\footnotesize{$W_{1,t}$}}
\psfrag{W2}[cc][cc]{\footnotesize{$W_{2,t}$}}
\psfrag{Y1}[cc][cc]{\footnotesize{$V_{1,t}$}}
\psfrag{Y2}[cc][cc]{\footnotesize{$V_{2,t}$}}
\psfrag{M1}[cc][cc]{\footnotesize{$M_1$}}
\psfrag{M2}[cc][cc]{\footnotesize{$M_2$}}
\psfrag{Mhat}[c][c]{\footnotesize{$\begin{pmatrix}\hat{M}_1\\ \hat{M}_2\end{pmatrix}$}}
  \includegraphics[width=0.95\textwidth]{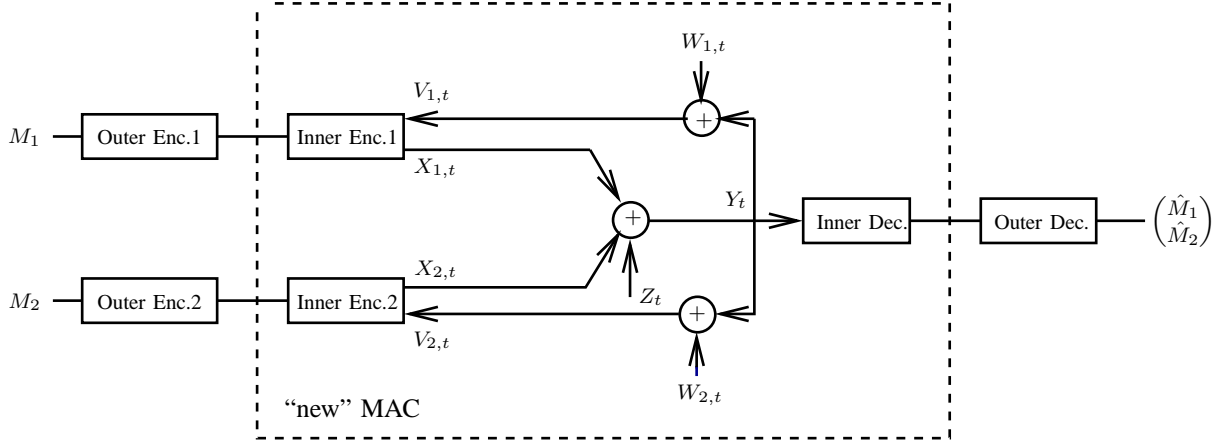}
    \caption{Structure of concatenated scheme.}
  \label{fig:concatenated} 
\end{figure*}


We propose an encoding scheme with a concatenated structure where each
of the encoders and the decoder consists of an outer part and an inner
part. (Here the inner parts are the parts that are closer to the
physical channel, see Figure~\ref{fig:concatenated}.) In our scheme the various parts fulfill the
following tasks.  The outer encoders map the messages into codewords
(without using the feedback) and feed these codewords to their
corresponding inner encoders. The inner encoders produce for every fed
symbol a sequence of $\eta$ channel inputs to the MAC with feedback,
for some positive integer $\eta$. In particular, when fed the
symbol $\xi_1\in \Reals$, Inner Encoder 1 produces $\eta$ inputs which
depend on $\xi_1$ and on the observed feedback outputs; all
symbols fed to the inner encoder are treated in the same way. Inner
Encoder 2 is analogously defined.  The $\eta$ symbols which the MAC
outputs for every pair of input symbols $(\xi_1,\xi_2)$ are then
linearly mapped by the inner decoder to a pair of estimates
$(\hat{\Xi}_1,\hat{\Xi}_2)$, and the estimates are fed to the
outer decoder.  Thus, the outer decoder is fed with a vector in
$\Reals^2$ every $\eta$ channel uses.  Based on the sequence of
vectors produced by the inner decoder, the outer decoder then decodes
the transmitted messages.

Consequently, the inner encoders and the inner decoder transform each
subblock of $\eta$ channel uses of the original MAC into a single
channel use of a ``new'' time-invariant and memoryless MAC which for
given inputs $\xi_1\in\Reals$ and $\xi_2\in \Reals$ produces the channel output
$\trans{(\hat{\Xi}_1,\hat{\Xi}_2)}\in\Reals^2$. We denote the new MAC by
$\xi_1,\xi_2\mapsto (\hat{\Xi}_1,\hat{\Xi}_2)$. We can then think of
the overall scheme as a \emph{no-feedback scheme} over the new MAC
$\xi_1,\xi_2\mapsto (\hat{\Xi}_1,\hat{\Xi}_2)$. As a consequence, the
capacity of the original MAC with feedback, which we denote by
$x_1,x_2 \mapsto Y$, is inner bounded by the capacity of the new MAC
$\xi_1,\xi_2\mapsto (\hat{\Xi}_1,\hat{\Xi}_2)$ without feedback but
scaled by $\eta^{-1}$ to account for the fact that to send the symbols
$\xi_1,\xi_2$ over the new MAC the original channel is used $\eta$
times.

We first sketch some of the properties of the inner encoders and the
inner decoder and postpone their detailed description to after the
description of the outer encoders and decoder. We
choose the inner encoders and the inner decoder so that the MAC
$\xi_1,\xi_2\mapsto (\hat{\Xi}_1,\hat{\Xi}_2)$ can be described by
\begin{IEEEeqnarray}{rCl}\label{eq:newdec0}
\begin{pmatrix} \hat{\Xi}_1 \\ \hat{\Xi}_2 \end{pmatrix} = \mat{A}
\begin{pmatrix} \xi_1 \\\xi_2 \end{pmatrix} + \vect{T},
\end{IEEEeqnarray} 
where $\mat{A}$ is a deterministic $2\times 2$ matrix and where
$\vect{T}$ is a bivariate Gaussian whose law does not depend on the
pair of inputs $(\xi_1,\xi_2)$. Also, the inner encoders are designed
so that if both outer encoders satisfy a unit average block-power
constraint (over time and messages) and if at every epoch the symbols
produced by the outer encoders are zero-mean (when averaged over the
messages), then the channel inputs to the original MAC $x_1,x_2\mapsto
Y$ satisfy the average power constraints \eqref{eq:power}.

For the outer code (encoders and decoder) we choose a capacity
achieving zero-mean code for the MAC $\xi_1,\xi_2 \mapsto
(\hat{\Xi}_1,\hat{\Xi}_2)$ under an average block-power constraint of
1. Note that there is no loss in optimality in restricting ourselves
to zero-mean codes because subtracting the mean of the
code can only reduce its average power (averaged over time and
messages) and does not change the performance on an additive noise MAC
such as \eqref{eq:newdec0}.  We shall need the property that the outer
encoders produce zero-mean symbols in the power-analysis of the input
sequences to the original channel $x_1,x_2\mapsto Y$.

For the inner encoders and the inner decoder we choose linear
mappings. To obtain a compact description of the linear mappings we
stack the $\eta$ channel inputs,
$X_{\nu,1},\ldots,X_{\nu,\eta}$, produced by Inner Encoder~$\nu$ in an
$\eta$-dimensional column vector 
\begin{equation*}
\vect{X}_{\nu} \eqdef \trans{(X_{\nu,1},\ldots, X_{\nu,\eta})}, \qquad
\nu \in \{1,2\}, 
\end{equation*} 
and similarly we stack the $\eta$ feedback outputs, $V_{\nu,1},\ldots,
V_{\nu,\eta}$, observed by
Inner Encoder~$\nu$ in the
$\eta$-dimensional vector
\begin{equation*}
\vect{V}_{\nu}=\trans{(V_{\nu,1},\ldots, V_{\nu,\eta})}, \qquad
\nu\in\{1,2\}.
\end{equation*}
We can then describe our choice of the inner encoders as follows.
When fed the input symbol $\xi_\nu\in\Reals$, Inner Encoder~$\nu$ produces
\begin{equation} \label{eq:encgenn}
\vect{X}_{\nu} =  \vect{a}_\nu \xi_\nu+ \mat{B}_{\nu}   \vect{V}_{\nu}, \qquad \nu\in\{1,2\},
\end{equation}
where $\vect{a}_{\nu}$ are $\eta$-dimensional column vectors and
$\mat{B}_{\nu}$ are $\eta\times \eta$ matrices which are strictly
lower-triangular (because the feedback is causal).
Also, as previously mentioned, we restrict the inner
encoders to produce sequences of inputs to the original
MAC $x_1,x_2\mapsto Y$ that satisfy the average block-power constraints
\eqref{eq:power} when the outer encoders feed them with zero-mean
sequences of unit average block-power. 
By \eqref{eq:encgenn} this is the case whenever the trace constraints
\eqref{eq:powergen} are satisfied.  Thus, in
the following we only allow for vectors $\vect{a}_1$ and $\vect{a}_2$
and for strictly lower-triangular matrices $\B_1$ and $\B_2$
satisfying \eqref{eq:powergen}.

To describe our linear choice of the inner decoder, we stack the
$\eta$ outputs $Y_1,\ldots, Y_\eta$, which the original MAC produces
for the pairs of inputs $(X_{1,1}, X_{2,1}),\ldots,(X_{1,\eta},X_{2,\eta})$, into the $\eta$-dimensional column vector
\begin{equation*}
\vect{Y}\eqdef \trans{(Y_1,\ldots, Y_\eta)}. 
\end{equation*}
We can then express the estimates produced by the outer decoder
by 
\begin{IEEEeqnarray}{rCl} \label{eq:newenc}
\begin{pmatrix} \hat{\Xi}_1\\\hat{\Xi}_2 \end{pmatrix} = \mat{D}
\vect{Y},  
\end{IEEEeqnarray}
for some matrix of our choice $\mat{D} \in \Reals^{2 \times \eta}$.

In the following we describe the MAC $\xi_1,\xi_2\mapsto
(\hat{\Xi}_1,\hat{\Xi}_2)$ as induced by $\eta, \vect{a}_1, \vect{a}_2$,
$\B_1,\B_2,$ and $\mat{D}$. Given
inputs $\xi_1,\xi_2\in\Reals$, it produces the vector of estimates
\begin{IEEEeqnarray}{rCl}\label{eq:newdec}
\begin{pmatrix} \hat{\Xi}_1 \\ \hat{\Xi}_2 \end{pmatrix} = \mat{A}
\begin{pmatrix} \xi_1 \\\xi_2 \end{pmatrix} + \vect{T},
\end{IEEEeqnarray} 
where the $2\times 2$-matrix $\mat{A}$ is given by
\begin{IEEEeqnarray}{rCl}\label{eq:chmat}
\mat{A}  & = & \mat{D}\left( \I_{\eta} - (\B_1+\B_2)
  \right)^{-1} \Ab,
\end{IEEEeqnarray}
where $\Ab$ is defined in \eqref{eq:Ab}, and where
the noise vector $\vect{T}$ is a zero-mean bivariate
Gaussian
\begin{IEEEeqnarray}{rCl}\label{eq:noisevec}
\vect{T} & = & \mat{D} (\I_\eta - (\B_1+\B_2))^{-1} (
\B_1\vect{W}_1+\B_2 \vect{W}_2+\vect{Z}), \IEEEeqnarraynumspace
\end{IEEEeqnarray}
for $\vect{W}_{1}\eqdef\trans{(W_{1,1},\ldots, W_{1,\eta})}$,
$\vect{W}_{2}\eqdef\trans{(W_{2,1},\ldots, W_{2,\eta})}$, and
$\vect{Z}\eqdef \trans{(Z_1,\ldots,Z_\eta)}$.
(Notice, that since $\B_1$ and $\mat{{B}}_2$ are strictly lower-triangular matrices,
the matrix $\left( \I_{\eta}- (\B_1+\B_2)\right)$ is
nonsingular and the inverse exists.)
Defining the $2\times \eta$ matrix 
\begin{equation}\label{eq:defCD}
\mat{C} \eqdef \mat{D}(\I_\eta -(\B_1+\B_2))^{-1},  
\end{equation}
we can express the matrix $\mat{A}$ in \eqref{eq:chmat} as
\begin{equation}\label{eq:chmatnew}
\mat{A} = \mat{C} \Ab,
\end{equation}
and the noise vector in \eqref{eq:noisevec} can be expressed as
\begin{equation}\label{eq:noisevecnew}
\vect{T} = \mat{C} \left( \B_1 \vect{W}_1 +\B_2 \vect{W}_2 +\vect{Z}\right).
\end{equation}
For fixed $\eta,\B_1,\B_2$, the mapping \eqref{eq:defCD} from
$\mat{D}$ to $\C$ is one-to-one, and thus we can parameterize our
concatenated scheme for noisy feedback by the parameters $\para$.

Note that by choosing $\eta=1, a_1=\sqrt{P_1}, a_2=\sqrt{P_2},$ and
$\mat{C}$ as the $2\times 1$ matrix with unit entries,
our scheme reduces to the capacity-achieving scheme for the original
MAC $x_1,x_2\mapsto Y$ without feedback subject to the power
constraints \eqref{eq:power}. 

\subsubsection{Choice of Parameters}
\label{sec:choicenoisy}
Given channel parameters $\ch$, determining for each
rate pair in $\Rate{\ch}$ a set of parameters $\para$ that achieves this
rate pair seems to be analytically intractable. Instead, we
 present guidelines on how to choose parameters and discuss the
two choices of parameters in Sections~\ref{sec:enc} and
\ref{sec:powerstar} that lead to Corollary~\ref{th:co1} and
Proposition~\ref{th:perfectfbeta}.

For the purpose of
describing our guidelines, throughout this section, we replace the
symbols $\xi_1$ and $\xi_2$ fed to the inner encoders by the
independent standard Gaussians $\Xi_1$ and $\Xi_2$.

We start with the matrix $\C$. Given parameters $\eta,
\vect{a}_1,\vect{a}_2,\B_1,$ and $\B_2$, the matrix $\C$ should be
chosen as $\C=\C_{\textnormal{LMMSE}}$, where
\begin{IEEEeqnarray}{rCl}\label{eq:LMMSEmat}
\mat{C}_{\textnormal{LMMSE}}\triangleq \trans{\Ab} \left( \Ab \trans{\Ab} + N\I_\eta +
  \Bb (\KW \otimes
  \I_\eta)\trans{\Bb}\right)^{-1}.\nonumber \\\IEEEeqnarraynumspace
\end{IEEEeqnarray}
By \eqref{eq:newenc}, \eqref{eq:chmatnew}, and
\eqref{eq:noisevecnew} this choice implies that
\begin{equation}\label{eq:lmsmat}
\begin{pmatrix} \hat{\Xi}_1 \\\hat{\Xi}_2\end{pmatrix} =\E{
  \begin{pmatrix} {\Xi}_1 \\{\Xi}_2\end{pmatrix}\Bigg| Y_1,\ldots,
  Y_\eta},
\end{equation}
and hence we call the matrix $\C_{\textnormal{LMMSE}}$ the
\emph{LMMSE-estimation matrix}.  The choice
$\C=\C_{\textnormal{LMMSE}}$ is optimal in the sense that the
corresponding region $\RateP{\nch}{\eta,
  \vect{a}_1,\vect{a}_2,\B_1,\B_2,\C_{\textnormal{LMMSE}}}$ contains
all  regions
$\RateP{\nch}{\para}$ that correspond to other choices of $\C$. The
optimality of the LMMSE-estimation matrix can be argued as follows. When 
\eqref{eq:lmsmat} holds,   then even additionally revealing $\vect{Y}$
(or any linear combinations thereof) to the outer decoder does not
increase the set of achievable rates in our scheme. 
Obviously, choosing
$\mat{C}= \mat{U}\mat{C}_{\textnormal{LMMSE}}$ for any non-singular
2-by-2 matrix $\mat{U}$ is also optimal. In particular, when $\eta=2$
every non-singular matrix is an optimal choice for $\mat{C}$.

We next consider the choice of parameters
$\vect{a}_1,\vect{a}_2,\B_1,\B_2$ and first focus on the special case
of perfect feedback. This special case
is in view of Ozarow's capacity result \cite{ozarow85} only of limited
interest, but it provides insight on how to choose the parameters for
other settings, e.g., the perfect partial-feedback setting (see Section~\ref{sec:choicepartial})
and the noisy feedback-setting with receiver side-information (Section~\ref{sec:choice_si}).  

For perfect feedback and a fixed $\eta$,
the parameters $\vect{a}_1,\vect{a}_2,\B_1,\B_2$ should be chosen such
that Inner Encoder~$\nu$, for $\nu\in\{1,2\}$, produces as its
$\ell$-th channel input a scaled version of the LMMSE-estimation error
of $\Xi_\nu$ when observing $(Y_1,\ldots, Y_{\ell-1})$, i.e.,
\begin{subequations}\label{eq:LMMSEX}
\begin{equation}\label{eq:LMMSEX1}
X_{1,\ell} = \pi_{1,\ell} \left( \Xi_1 - \E{\Xi_1| Y^{\ell-1}}\right), \qquad \ell\in\{1,\ldots, \eta\},
\end{equation}
and 
\begin{equation}\label{eq:LMMSEX2}
X_{2,\ell} = \pi_{2,\ell} \left( \Xi_2 - \E{\Xi_2| Y^{\ell-1}}\right), \qquad \ell\in\{1,\ldots, \eta\},
\end{equation}
\end{subequations}
for some real numbers $\pi_{1,1},\ldots, \pi_{1,\eta}$ and
$\pi_{2,1},\ldots, \pi_{2,\eta}$. In fact, every choice of parameters
not satisfying~\eqref{eq:LMMSEX} can be
strictly improved (with an appropriate choice of $\C$) so as to
achieve a larger region, see
Appendix~\ref{sec:LMMSEopt}.

For general noisy feedback, it is not optimal to choose
$\vect{a}_1,\vect{a}_2,\B_1,\B_2$ as in \eqref{eq:LMMSEX} when the channel outputs $Y_1,\ldots,Y_{\ell-1}$
are replaced by the feedback outputs $V_{1,1},\ldots, V_{1,\ell-1}$
and $V_{2,1},\ldots, V_{2,\ell-1}$. Intuitively, the reason is that
with such a choice the inner encoders introduce too much feedback
noise into the forward communication.

For the general setup it seems infeasible to derive the set of
optimal parameters $\vect{a}_1,\vect{a}_2,\B_1,\B_2$. 
However, it is easily proved that the parameters
$\vect{a}_1,\vect{a}_2,\B_1,\B_2$ have to be chosen so that they
satisfy both power constraints \eqref{eq:powergen2} and
\eqref{eq:powergen1} with equality, since otherwise there exists a choice of
parameters which corresponds to a larger achievable
region. This readily follows from the alternative formulation of
$\RateP{\nch}{\eta,\vect{a}_1,\vect{a}_2,\B_1,\B_2,\C}$ in
Section~\ref{sec:achaltnoisy}, because the RHSs of
~\eqref{eq:recexp} (which determine
$\RateP{\nch}{\eta,\vect{a}_1,\vect{a}_2,\B_1,\B_2,\C}$) can always be
increased by changing the last entry of $\vect{a}_1$, i.e.,
$a_{1,\eta}$, or the last entry of $\vect{a}_2$, i.e., $a_{2,\eta}$.

We finally consider the choice of $\eta$. If the goal is to maximize
the single rates, it is trivially optimal to choose $\eta=1$
irrespective of the channel parameters $\ch$. If in contrast the goal
is to maximize the sum-rate it seems infeasible to derive the optimal
$\eta$. However, numerical results indicate that the larger the
feedback-noise variances are, the smaller the parameter $\eta$ should
be chosen. It is easily proved that in the extreme case of no feedback
the sum-rate is maximized by choosing $\eta=1$. In contrast, in the
extreme case of perfect feedback we prove in
Section~\ref{sec:proofremperf} that with the choice of parameters
suggested in Section~\ref{sec:powerstar} the maximum sum-rate of our
concatenated scheme converges to the perfect-feedback sum-rate
capacity as the parameter $\eta$ tends to infinity.

In the remaining, we discuss the two specific choices of the
parameters $\vect{a}_1,\vect{a}_2,\B_1,\B_2,\C$ given $\eta\in\Nat$
presented in Sections~\ref{sec:enc} and \ref{sec:powerstar}.  For both
choices, the parameter $\C$ is the LMMSE-estimation matrix and the
parameters $\vect{a}_1,\vect{a}_2,\B_1,\B_2$ are such that when
specialized to perfect feedback they satisfy~\eqref{eq:LMMSEX}. In the
choice in Section~\ref{sec:enc}, each inner encoder allocates the same
power for all channel inputs. The achievable region corresponding to
this choice is presented in Corollary~\ref{cor:ach}, and includes as
special case the result on
the symmetric setup in Corollary~\ref{th:co1}. 
In the choice in Section~\ref{sec:powerstar}, the inner encoders use
the power-allocation strategy suggested by \cite{kramer02} for perfect
feedback. The corresponding achievable region is presented in
Corollary~\ref{cor:alte}, and
includes as special case the achievable region for perfect feedback in
Remark~\ref{def:remarkperf} used in the proof of
Propositions~\ref{prop:maxsumratesigma} and
\ref{prop:splitmaxsumratesigma} and Theorem~\ref{th:C_sym}.

 \subsection{Extensions of the Concatenated Scheme}\label{sec:ext}
 In the following three subsections we present three extensions of our
concatenated scheme by rate-splitting it with other schemes.  The
 idea of rate-splitting was introduced in \cite{carleial82} and
 \cite{ozarow85}. 

\subsubsection{Rate-Splitting with No-Feedback Scheme}\label{sec:entireregion}
In this first extension we combine our scheme with a no-feedback
scheme employing IID Gaussian codewords. This extension was inspired by the
rate-splitting scheme proposed by Ozarow for perfect feedback
\cite{ozarow85}.  Only one transmitter applies the rate-splitting. For
the description we assume it is Transmitter~1. Thus, Transmitter~1
splits Message $M_1$ of rate $R_1$ into two independent parts: Message
$M_{1,\textnormal{NF}}$ of rate $R_{1,\textnormal{NF}}$ and Message
$M_{1,\textnormal{CS}}$ of rate $R_{1,\textnormal{CS}}$, where
$R_{1,\textnormal{NF}}$ and $R_{1,\textnormal{CS}}$ sum up to
$R_1$. Here, NF stands for ``no-feedback'' and CS stands for
``concatenated scheme''.

We first present a rough overview of the scheme. We start with the
encodings. Transmitter~1 uses a fraction of its available power
$P_1'$, for some $0\leq P_1'\leq P_1$, to produce a sequence by
encoding Message $M_{1,\textnormal{NF}}$ using Gaussian
codewords\footnote{To satisfy the powers constraints the Gaussian
  codewords should be of variance slightly less than $P_1'$. However,
  this is a technicality which we ignore.}  (without using the
feedback). With the rest of the power $(P_1-P_1')$ it produces a
sequence of the same length by encoding Message
$M_{1,\textnormal{CS}}$ using our concatenated scheme and the outputs
of the feedback link. It sends
the sum of the two produced sequences over the channel.
If the concatenated scheme is of parameter $\eta$ and its outer code
is of blocklength $n$, then both sequences are of length $\eta n$.
Transmitter~2 produces a sequence of equal length by encoding Message
$M_{2}$ with power $P_2$ using the concatenated scheme and sends this
sequence.

We next present a rough overview of the decoding at the receiver. The
receiver first decodes the pair $(M_{1,\textnormal{CS}},M_2)$ by using
the inner and the outer decoder of our concatenated scheme and
treating the transmission of Message $M_{1,\textnormal{NF}}$ as
additional noise. From its guess of $(M_{1,\textnormal{NF}},M_2)$ the
receiver cannot recover the sequences produced by our concatenated
scheme because it is incognizant of the feedback noise. Nevertheless,
it can form an estimate of both produced sequences (pretending that
its guess of $(M_{1,\textnormal{CS}},M_2)$ is correct) and subtract
the sum of the estimates from the received signal.  Based on the
resulting difference the receiver finally decodes message
$M_{1,\textnormal{NF}}$, which concludes the decoding.

In the following we describe the scheme in more detail.  Given
$M_{1,\textnormal{NF}}=m_{1,\textnormal{NF}}$, Transmitter~1 picks the
codeword $\vect{u}(m_{1,\textnormal{NF}})\eqdef \trans{(u_1,\ldots, u_{\eta
  n})}$ corresponding to $m_{1,\textnormal{NF}}$ from its Gaussian
codebook. Given $M_{1,\textnormal{CS}}=m_{1,\textnormal{CS}}$,
Transmitter~1 feeds $m_{1,\textnormal{CS}}$ to Outer Encoder~1, which
picks the codeword $\boldsymbol{\xi}_1(m_{1,\textnormal{CS}}) \eqdef
\trans{(\xi_{1,1},\ldots, \xi_{1,n})}$ corresponding to
$m_{1,\textnormal{CS}}$ from its codebook and feeds it to
Inner Encoder~1. Similarly, given $M_{2}=m_2$, Transmitter~2 feeds
$m_2$ to Outer Encoder~2, which picks the codeword
$\boldsymbol{\xi}_2(m_2)\eqdef\trans{(\xi_{2,1},\ldots, \xi_{2,n})}$
corresponding to $m_2$ and feeds it to Inner Encoder~2.  Denoting the
parameters of the inner encoders by $\vect{a}_1,\vect{a}_2,\B_1,$ and
$\B_2$, respectively, Inner Encoder~1 forms the $\eta$-dimensional
vectors 
\begin{equation}\label{eq:IX1U1}
\vect{a}_1 \xi_{1,k} + \B_1
\vect{V}_{1,k}, \qquad k \in \{1,\ldots, n\},
\end{equation}
and Inner Encoder~2 forms the $\eta$-dimensional vectors
\begin{equation}\label{eq:IX2U}
\vect{a}_2 \xi_{2,k} + \B_2 \vect{V}_{2,k}, \qquad k \in \{1,\ldots, n\},
\end{equation}
where for $\nu\in\{1,2\}$:
\begin{IEEEeqnarray*}{rCl}
\vect{V}_{\nu,k}& \eqdef& \trans{ (V_{\nu,(k-1)\eta+1},\ldots, V_{\nu,k\eta})}.
\end{IEEEeqnarray*}
The signal transmitted by Transmitter~1 is the
sum of the vectors in \eqref{eq:IX1U1}
and the vectors
\begin{IEEEeqnarray*}{rCl}
\vect{u}_k &\eqdef &\trans{ (u_{(k-1)\eta+1},\ldots, u_{k\eta})},
\qquad k \in \{1,\ldots, n\},
\end{IEEEeqnarray*}
i.e.,
\begin{equation}\label{eq:X1U1}
\vect{X}_{1,k} = \vect{u}_k+\vect{a}_1 \xi_{1,k} + \B_1
\vect{V}_{1,k}, \qquad k\in \{1,\ldots, n\},
\end{equation}
where 
\begin{equation*}
\vect{X}_{1,k} \eqdef \trans{ (X_{1,(k-1)\eta+1},\ldots, X_{1,k\eta})},
\qquad k \in \{1,\ldots, n\}.
\end{equation*}
The signal transmitted by 
Transmitter~2 is described by the vectors in \eqref{eq:IX2U} as follows:
\begin{equation}\label{eq:X2U}
\vect{X}_{2,k} = \vect{a}_2 \xi_{2,k} + \B_2 \vect{V}_{2,k}, \qquad k \in \{1,\ldots, n\},
\end{equation}
where 
\begin{equation*}
\vect{X}_{2,k} \eqdef \trans{ (X_{2,(k-1)\eta+1},\ldots, X_{2,k\eta})}.
\end{equation*}
Notice that if $\vect{a}_1,\vect{a}_2,\B_1,\B_2$ 
satisfy \eqref{eq:powergen} for powers
$(P_1-P_1')$ and $P_2$, noise variance $(N+P_1')$ and feedback-noise
covariance matrix $\KW$ and if the outer code's codewords are zero-mean
and average block-power constrained to 1, then for sufficiently large
blocklength $n$ the input sequences \eqref{eq:X1U1} and \eqref{eq:X2U} satisfy the power constraint \eqref{eq:power} with
arbitrary high probability.

We next describe the decoding. The receiver first decodes the pair
$(M_{1,\textnormal{CS}},M_{2})$ based on the tuple $(Y_1,\ldots Y_{
  \eta n})$ by treating the codeword $\vect{U}(M_{1,\textnormal{NF}})$
as additional noise and by applying the inner and outer decoders of the
concatenated scheme. Let $\hat{M}_{\textnormal{ICS}}$ and $\hat{M}_2$
denote the receiver's guesses of the messages $M_{\textnormal{ICS}}$
and $M_{2}$, and let $\left(\hat{\Xi}_{1,1}^{(\textnormal{Rx})},
  \ldots, \hat{\Xi}_{1,n}^{(\textnormal{Rx})}\right)$ and
$\left(\hat{\Xi}_{2,1}^{(\textnormal{Rx})}, \ldots,
  \hat{\Xi}_{2,n}^{(\textnormal{Rx})}\right)$ denote the corresponding
codewords in the outer codes.  The receiver then attempts to estimate
and subtract the influence of the concatenated scheme (see
\eqref{eq:IX1U1} and \eqref{eq:IX2U}) by computing for each
$k\in\{1,\ldots,n\}$ the difference
\begin{IEEEeqnarray}{rCl}\label{eq:YtildeU}
  \tilde{\bfY}_k & \eqdef & \left( \I_\eta - \B_1-\B_2\right)\bfY_k - \vect{a}_1
  \hat{\Xi}_{1,k}^{(\textnormal{Rx})} - \vect{a}_2
  \hat{\Xi}_{2,k}^{(\textnormal{Rx})},\IEEEeqnarraynumspace
\end{IEEEeqnarray} 
where the $\eta$-dimensional vector $\vect{Y}_k$ is defined as
\begin{equation*}
\vect{Y}_k \eqdef \trans{(Y_{(k-1)\eta+1},\ldots, Y_{k \eta})}.
\end{equation*}
If the receiver decoded $M_{1,\textnormal{CS}}$ and $M_2$ correctly,
i.e., if $\hat{M}_{1,\textnormal{CS}}=M_{1,\textnormal{CS}}$ and
$\hat{M}_2=M_2$, then \eqref{eq:YtildeU} corresponds to
\begin{IEEEeqnarray*}{rCl}
  \vect{U}_k+\B_1 \vect{W}_{1,k} + \B_2 \vect{W}_{2,k}
  +\vect{Z}_k,\qquad k \in\{1,\ldots, n\}.
\end{IEEEeqnarray*}
Finally, the receiver decodes Message $M_{1,\textnormal{NF}}$ based on the
differences $\{\tilde{\bfY}_i\}_{i=1}^n$ using 
 an optimal decoder for a Gaussian $\eta$-input antenna/$\eta$-output antenna channel where the noise sequences are
white but correlated across antennas. Notice that because of the
correlation of the noise sequences across antennas, the scheme might
be improved if correlated Gaussian codewords are used to transmit
Message $M_{1,\textnormal{NF}}$.

\subsubsection{Rate-Splitting with Carleial's Scheme}\label{sec:extCarleial}
Our second extension is based on modifying Carleial's rate-splitting
scheme \cite{carleial82}. Carleial's scheme combines a variation of
the Cover-Leung scheme \cite{coverleung81} with a no-feedback scheme
by means of rate-splitting. Here, we propose to modify his scheme by
replacing the no-feedback scheme with our concatenated scheme. Since
for $\eta=1, a_1=\sqrt{P_1},$ and $a_2=\sqrt{P_2}$ our concatenated
scheme results in an optimal no-feedback scheme, our proposed extension
includes Carleial's scheme as a special case.
In the following we roughly sketch the idea behind our extended scheme.
For more details see
Section~\ref{sec:RSCL2}.

 Our scheme is a Block-Markov scheme of blocklength
 $n'$. Each block of $n'$ channel uses is divided into
$(B+1)$ blocks, each of length $\eta n $ for positive integers $\eta$
and $n$, i.e., we assume that $n'=(B+1)\eta n$. Each transmitter splits its message into two sequences of
independent submessages:  Transmitter~$\nu$, for $\nu\in\{1,2\}$ splits its
message $M_\nu$ into a sequence of independent submessages
$\left\{M_{\nu,\textnormal{CL},1},\ldots,
  M_{\nu,\textnormal{CL},B}\right\}$ of rates $R_{\nu, \textnormal{CL}}$
and into a sequence of independent submessages
$\left\{M_{\nu,\textnormal{CS},1},\ldots,M_{\nu,\textnormal{CS},B}\right\}$
of rates $R_{\nu,\textnormal{CS}}$. The rates $R_{\nu, \textnormal{CL}}$
and $R_{\nu,\textnormal{CS}}$ should be nonnegative and sum to $R_{\nu}
\frac{B+1}{B}$, but otherwise can be chosen arbitrary depending on the
parameters of the setting. Similarly, for Transmitter~2. (Here, the
subscript CL stands for ``Cover-Leung'' and the subscript CS stands for
``concatenated scheme''.) 

As in Carleial's scheme, after each block $b\in\{1,\ldots, B\}$
Transmitter~$1$ and Transmitter~2 decode the other transmitter's
submessage $M_{2,\textnormal{CL},b}$ and $M_{1,\textnormal{CL},b}$
based on their feedback outputs. The two transmitters can accomplish
the decodings in two different ways. Transmitter~1 either directly
decodes Message $M_{2,\textnormal{CL},b}$, or it first decodes
$M_{2,\textnormal{CS},b}$ before decoding the desired message
$M_{2,\textnormal{CL},b}$. Which
alternative is better depends on the specific parameters of the
setting.

The encoding is performed as follows.  To encode
  messages $\left\{M_{\nu,\textnormal{CL},b}\right\}_{b=1}^B$
  Transmitter~$\nu$, for $\nu\in\{1,2\}$, uses
    Carleial's variation of the Cover-Leung scheme and to encode
    messages $\left\{M_{\nu,\textnormal{CS},b}\right\}_{b=1}^B$ it uses our
      concatenated scheme. More specifically, before the transmission
      in Block $b\in\{1,\ldots,B\}$ starts, Transmitter~$\nu$ chooses the
      codewords for messages $M_{\nu,\textnormal{CL},b}$,
      $M_{1,\textnormal{CL},b-1}$, and $M_{2,\textnormal{CL},b-1}$
      from the corresponding Gaussian codebooks and produces an $\eta
      n$-length sequence of power $P_\nu'$, for some $0\leq P_\nu' \leq
      P_\nu$, by taking a linear combination of the chosen codewords.
      It also produces an $\eta n$-length sequence of power
      $(P_\nu-P_\nu')$ by encoding message $M_{\nu,\textnormal{CS},b}$ using
      the outer and inner encoders of our concatenated scheme where
      $\eta$ is the parameter of the inner code and $n$ is the
      blocklength of the outer code. It sends the sum
      of the two produced sequences in Block $b$. In Block $(B+1)$
      Transmitter~$\nu$ picks the codewords for messages
      $M_{1,\textnormal{CL},B}$ and $M_{2,\textnormal{CL},B}$ from the
      corresponding Gaussian codebooks and sends a linear combination
      of power $P_\nu'$ of these codewords. 

After each Block $b\in\{1,\ldots, B\}$ the
receiver decodes messages  $M_{1,\textnormal{CS},b}$,
$M_{2,\textnormal{CS},b}$$M_{1,\textnormal{CL},b-1}$, and
$M_{2,\textnormal{CL},b-1}$. It
first decodes messages $M_{1,\textnormal{CS},b}$ and
$M_{2,\textnormal{CS},b}$ using inner and outer decoder of our
concatenated scheme and treating the sequences produced by encoding
messages $M_{1,\textnormal{CL},b-1}$, $M_{2,\textnormal{CL},b-1}$,
$M_{1,\textnormal{CL},b}$ and $M_{2,\textnormal{CL},b}$ as additional
noise.  From its guess of
$(M_{1,\textnormal{CS},b},M_{2,\textnormal{CS},b})$ the receiver
cannot recover the sequences produced by our concatenated scheme
because it is incognizant of the feedback noise. Nevertheless, it can
form an estimate of both produced sequences (pretending that its guess
is correct) and subtract the sum of the estimates from the received
signal. Based on the resulting difference and based on similar
differences which resulted in the previous block, it then
decodes messages $(M_{1,\textnormal{CL},b-1},
M_{2,\textnormal{CL},b-1})$. After the last block $(B+1)$ the receiver decodes
the pair $(M_{1,\textnormal{CL},B},M_{2,\textnormal{CL},B})$. More
general decoding orders at the receiver could be considered, but for
simplicity, we restrict attention to this order.

\subsubsection[Interleaving \& Rate-Splitting]{Interleaving \& Rate-Splitting with Carleial's Cover-Leung Scheme}\label{sec:CarleialInter}
Our third extension is based on rate-splitting an interleaved version
of Carleial's Cover-Leung scheme with an interleaved version of our
concatenated scheme. We only describe here the general structure of
the scheme. For more details see
Appendix~\ref{sec:interleaving}.

Our scheme is a Block-Markov scheme of blocklenght $n'$.
Each block of $n'$ channel uses is divided into $(B+1)$ blocks, each
of length $\eta n $ and each such block is further divided into $\eta$
subblocks of length $n$. Thus, it is assumed that $B,\eta,$ and $n$
are positive integers such that $n'=(B+1)\eta n$.  Similarly,  each
transmitter splits its message into two sequences of independent
submessages: Transmitter~$\nu$, for $\nu\in\{1,2\}$, splits its message $M_\nu$ into a
sequence of independent submessages
$\left\{M_{\nu,\textnormal{ICL},1},
  \ldots,M_{\nu,\textnormal{ICL},\eta B} \right\}$ and into a sequence
of independent submessages
$\left\{M_{\nu,\textnormal{ICS},1},\ldots,M_{\nu,\textnormal{ICS},B}\right\}$.
Notice that the first sequence of submessages is of length $\eta B$,
and the second of length $B$.  Messages
$\{M_{\nu,\textnormal{ICL},(b-1)\eta+\ell}\}_{b=1}^B$ are of rate $R_{\nu,
  \textnormal{ICL}, \ell}$, and Messages $\{M_{\nu,\textnormal{ICS},
  b}\}_{b=1}^B$ of rate $R_{1,\textnormal{ICS}}$. The rates $R_{\nu,
  \textnormal{ICL}, 1}, \ldots, R_{\nu,\textnormal{ICL},\eta}$, and
$R_{\nu,\textnormal{ICS}}$ should be nonnegative and sum to
$R_{\nu} \frac{B+1}{B}$, but otherwise can be chosen arbitrary depending
on the parameters of the setting. (The
subscript ICL stands for ``interleaved Cover-Leung'' and the subscript
ICS stands for ``interleaved concatenated scheme''.)

Similar to the previous extension and similar to Carleial's scheme,
the transmitters decode part of the other transmitter's messages based on
their feedback outputs. Specifically in this scheme, after each
subblock $\tilde{b}\in\{1,\ldots, B\eta\}$, Trasmitter~1 and
Transmitter~2 decode the other transmitter's submessage
$M_{2,\textnormal{ICL},\tilde{b}}$ and
$M_{1,\textnormal{ICL},\tilde{b}}$. Following this decoding step, the
transmitters compute ``cleaned'' feedback outputs, i.e., they mitigate
the influence of the Cover-Leung messages
$M_{1,\textnormal{ICL},\tilde{b}},$
$M_{2,\textnormal{ICL},\tilde{b}},$
$M_{1,\textnormal{ICL},\tilde{b}-\eta},$ and
$M_{2,\textnormal{ICL},\tilde{b}-\eta}$ transmitted in this block on
the observed feedback outputs.  Transmitter~1 computes its ``cleaned''
feedback output more specifically as follows. It first reconstructs
the sequence that was produced by Transmitter~2 in this subblock
$\tilde{b}$ to encode messages $M_{2,\textnormal{ICL},\tilde{b}}$,
$M_{1,\textnormal{ICL},\tilde{b}-\eta}$, and $M_{2,\textnormal{ICL},
  \tilde{b}-\eta}$ (pretending that its guesses of
$M_{2,\textnormal{ICL},\tilde{b}}$ and
$M_{2,\textnormal{ICL},\tilde{b}-\eta}$ are correct). It then
subtracts this reconstructed sequence and the sequence it produced
itself in this subblock to encode $M_{1,\textnormal{ICL},\tilde{b}}$,
$M_{1,\textnormal{ICL},\tilde{b}-\eta}$, and
$M_{2,\textnormal{ICL},\tilde{b}-\eta}$ from its observed feedback
outputs.  Similarly for Transmitter~2.


The encoding is performed as follows. To encode Messages
$\{M_{\nu,\textnormal{ICL},k}\}_{k=1}^{\eta B}$, Transmitter~$\nu$,
for $\nu\in\{1,2\}$, uses an interleaved version of Carleial's
Cover-Leung scheme, and to encode Messages
$\{M_{\nu,\textnormal{ICS},b}\}_{b=1}^B$ it uses an interleaved
version of our concatenated scheme. We describe these encodings in
more detail. In a fixed block ${b} \in \{1,\ldots, B\}$,
Transmitter~$\nu$ sends the sum of two $\eta n$-length sequences. The
first sequence is of power $P_\nu'$, for some $0\leq P_\nu'\leq
P_\nu$, and consists of $\eta$ subblocks. The $\ell$-th subblock of
the sequence, for $\ell\in\{1,\ldots, \eta\}$, Transmitter~$\nu$
chooses the $n$-length codewords for Messages
$M_{\nu,\textnormal{ICL},(b-1)\eta+\ell},
M_{1,\textnormal{ICL},(b-2)\eta+\ell}$, and
$M_{2,\textnormal{ICL},(b-2)\eta+\ell}$ from the corresponding
Gaussian codebooks and takes a linear combination of these chosen
codewords. We notice that here each pair of messages
$(M_{1,\textnormal{ICL},\tilde{b}},M_{2,\textnormal{ICL},\tilde{b}})$,
for $\tilde{b}\in\{1,\ldots, B\eta\}$, is encoded into Subblocks
$\tilde{b}$ and $\tilde{b}+\eta$, and not---as in Carleial's original
scheme---into Subblocks $\tilde{b}$ and $\tilde{b}+1$.  The second
sequence is of power $(P_1-P_1')$ and produced as follows:
Transmitter~$\nu$ first applies its outer encoder to encode Message
$M_{\nu,\textnormal{ICS},b}$, and then feeds the outcome to a modified
version of its inner encoder. The inner encoder is modified as
described by the following two items. 1.)  Instead of the original
feedback the modified inner encoder uses the ``cleaned'' feedback
mentioned above, where the influence of the interleaved Cover-Leung
type scheme is mitigated. 2.)  Unlike the original inner encoder where the
$\ell$-th fed codeword symbol is encoded into $\eta$ subsequent
symbols at positions $(\ell-1)\eta+1$ to $\ell \eta$, the modified
inner encoder encodes the $\ell$-th fed codeword symbol into the
$\eta$ symbols at positions $\ell,n+\ell, \ldots, (\eta-1)n+\ell$, for
$\ell\in\{1,\ldots, \eta\}$.

 Notice that the chosen
interleaving of the modified inner encoders preserves the causality of
the feedback.  Moreover, it implies that in the interleaved sequence
the symbols in Subblock $\tilde{b}$, for $\tilde{b}\in \{(b-1)\eta+1,
\ldots, b\eta\}$, only depend on feedback outputs of previous
subblocks $1, \ldots, \tilde{b}-1$ and not on feedback outputs of the
current Subblock $\tilde{b}$. This is the reason why the modified
inner encoder can use the ``cleaned'' feedback instead of the original
feedback.

 The receiver first decodes Messages
$\{M_{1,\textnormal{ICL},\tilde{b}}\}_{\tilde{b}=1}^{\eta B}$ and
$\{M_{2,\textnormal{ICL},\tilde{b}}\}_{\tilde{b}=1}^{\eta B}$ and only
thereafter decodes Messages $\{M_{1,\textnormal{ICS},b}\}_{{b}=1}^{
  B}$ and $\{M_{2,\textnormal{ICS},b}\}_{b=1}^B$. More specifically,
the receiver first decodes Messages $\{(M_{1,\textnormal{ICL},
  (b-1)\eta+1}, M_{2,\textnormal{ICL},(b-1)\eta +1})\}_{b=1}^{B}$,
followed by Messages $\{(M_{1,\textnormal{ICL},(b-1)\eta+ 2},
M_{2,\textnormal{ICL},(b-1)\eta +2})\}_{b=1}^{B}$, etc. The receiver
then reconstructs the sequences produced to encode these messages
(pretending its guesses are correct) and subtracts them from the
received signal.  Based on the resulting difference, which we call the
``cleaned'' output signal, the receiver decodes Messages
$\{M_{1,\textnormal{ICS},b}\}_{b=1}^B$ and
$\{M_{2,\textnormal{ICS},b}\}_{b=1}^B$. To this end, it first reverses
the interleaving and then applies the inner and outer decoders of our
concatenated scheme.

Notice that in the presented scheme, Messages
$\{M_{1,\textnormal{ICS},b}\}_{b=1}^B$ and
$\{M_{2,\textnormal{ICS},b}\}_{b=1}^B$ are decoded based on the
``cleaned'' output signal and they are encoded using the ``cleaned''
feedbacks. The ``cleaned'' output signal and the ``cleaned'' feedbacks
correspond to the output signals and the feedbacks in a situation
where only the interleaved concatenated scheme is employed but not the
interleaved version of Carleial's Cover-Leung scheme. Therefore, in
the presented rate-splitting scheme there is no degradation in
performance of the interleaved concatenated scheme due to the
rate-splitting with Carleial's Cover-Leung scheme.

Further, notice that in a given Block $b\in\{1,\ldots,B\}$ 
the sum of the two sequences produced to encode Messages
$M_{1,\textnormal{ICS},b}$ and $M_{2,\textnormal{ICS},b}$ is of
different power in each of the $\eta$ subblocks. Thus, these sequences
introduce different noise levels on
the receiver's decoding of Messages $\{(M_{1,\textnormal{ICL}, (b-1)\eta+\ell},
M_{2,\textnormal{ICL},(b-1)\eta +\ell})\}_{\ell=1}^{\eta}$, and consequently
the rates $\{R_{1,\textnormal{ICL},\ell}\}_{\ell=1}^{\eta}$ and 
$\{R_{2,\textnormal{ICL},\ell}\}_{\ell=1}^{\eta}$ should be
chosen depending on $\ell$.

\subsection{Proofs}\label{sec:proofs}
\subsubsection{Proof of Proposition~\ref{eq:monotonicity}}\label{sec:monoto}
 We first prove Part~\ref{pa}). To this end, we show that
for every fixed $\eta\in \Nat$ and fixed $\eta$-dimensional vectors
$\vect{a}_1,\vect{a}_2$, $\eta\times\eta$-dimensional matrices
$\B_1,\B_2$, and $2\times\eta$-dimensional matrix $\C$, the following
two statements hold:
\begin{enumerate}
\item[i)] For all positive semidefinite matrices $\KW$ and  $\KW'$: 
\begin{IEEEeqnarray*}{lCl}\lefteqn{
\Big(\KW \succeq \KW'\Big)\Longrightarrow } \quad \\ 
&& \Big(\RateP{\nch}{\eta,\vect{a}_1,\vect{a}_2,\B_1,\B_2,\C} \\ & &\hspace{0.5cm}
 \subseteq
\RateP{N,\KW'}{\eta,\vect{a}_1,\vect{a}_2,\B_1,\B_2,\C}\Big).
\end{IEEEeqnarray*}
\item[ii)] If the choice of parameters $ \eta, \vect{a}_1,\vect{a}_2, \B_1,\B_2,\C$
  satisfies the power constraints \eqref{eq:powergen} for a covariance matrix $\KW$, then it
  also satisfies these power constraints for all covariance matrices
  $\KW'$ for which $\KW\succeq \KW'$.
\end{enumerate}
By Definition~\ref{def:R}, Statements i) and ii) imply that
\begin{IEEEeqnarray*}{rCl}\lefteqn{
\Big(\KW \succeq \KW'\Big)  \Longrightarrow} \quad\\ & & \Big(\Rate{P_1,P_2,N,\KW}
\subseteq \Rate{P_1,P_2,N,\KW'}\Big),
\end{IEEEeqnarray*}
and thus conclude the proof of Part~\ref{pa}).

We start by proving Statement i).  Fix a tuple
$(\eta,\vect{a}_1,\vect{a}_2, \B_1,\B_2,\C)$.  We only prove Statement
i) for the case where $\mat{C}\trans{\mat{C}}$ is nonsingular. For the
case where $\mat{C}\trans{\mat{C}}$ is singular but $\mat{C}\neq \mat{0}$ the proof is analogous
and therefore omitted; for $\mat{C}=\mat{0}$ the proof is trivial.  To establish Statement i) when
$\mat{C}\trans{\mat{C}}$ is nonsingular, it suffices to show that all
three RHSs of \eqref{eq:Rgen1} are
monotonically decreasing in $\KW$ with respect to the Loewner order.
We only prove the monotonicity of the RHS of
\eqref{eq:R1gen1}; the monotonicities of the RHSs of
\eqref{eq:R2gen1} and \eqref{eq:R12gen1} can be shown analogously.
Thus, in the following we fix two positive semidefinite $2\times 2$
matrices $\KW$ and $\KW'$ satisfying $\KW \succeq \KW'$ and we show
that:
\begin{IEEEeqnarray}{lCl}\label{eq:mon}
  \lefteqn{\frac{1}{2\eta}\log\left( \frac{ \left| \C (
          \vect{a}_1\trans{\vect{a}}_1 + N\I_\eta +
          \Bb (\KW \otimes \I_{\eta})
          \trans{\Bb} )\trans{\C}\right|}{\left| \C (
          N\I_\eta + \Bb (\KW \otimes
          \I_{\eta}) \trans{\Bb}
          )\trans{\C}\right|}\right)}\nonumber \\
   \geq \frac{1}{2\eta}\log\left(\!\frac{ \left| \C (
        \vect{a}_1\trans{\vect{a}}_1 + N\I_\eta +
        \Bb (\KW' \otimes \I_{\eta})
        \trans{\Bb} )\trans{\C}\right|}{\left| \C (
        N\I_\eta + \Bb (\KW' \otimes \I_{\eta})
        \trans{\Bb} ) \trans{\C}\right|}\!\right)\!.\nonumber \\
\end{IEEEeqnarray}

Before proving \eqref{eq:mon} we recall the following well-known properties of positive semidefinite
matrices. For all  positive semidefinite
$n \times n$ matrices $\mat{K},\mat{K}_1,\mat{K}_2$ satisfying
$\mat{K}_1\succeq \mat{K}_2$ and for all  $m\times n$
matrices $\mat{M}$ the following properties hold:
\begin{IEEEeqnarray}{rCl}
\mat{M}\mat{K}_1 \trans{\mat{M}} & \succeq & \mat{M}\mat{K}_2
\trans{\mat{M}}, \label{eq:p1}\\
\mat{K}+\mat{K}_1 & \succeq& \mat{K}+\mat{K}_2, \label{eq:p2}\\
\mat{K}\mat{K}_1 & \succeq & \mat{K}\mat{K}_2, \label{eq:p3}\\
\mat{K}_1^{-1} & \preceq & \mat{K}_2^{-1}, \label{eq:p4}
\end{IEEEeqnarray}
and 
\begin{IEEEeqnarray}{rCl}
 |\mat{K}_1| &\geq &  |\mat{K}_2|, \label{eq:p5}\\
\tr{\mat{K}_1} & \geq & \tr{\mat{K}_2}. \label{eq:p6}
\end{IEEEeqnarray}
Based on these properties and the definition $\mat{A}_{1} \eqdef
\vect{a}_1\trans{\vect{a}}_1$ the following sequence of
implications can be proved:
\begin{IEEEeqnarray}{rCl}
  \lefteqn{\Big(\KW \succeq \KW' \Big)}\nonumber \\ &\Longrightarrow& \Big(
  \left(\KW \otimes
    \I_{\eta} \right)\succeq\left( \KW' \otimes \I_{\eta}\right) \Big)\label{eq:ptt1}\\
  &  \Longrightarrow & \Big(\left( \Bb\left( \KW
      \otimes \I_\eta \right) \trans{\Bb} \right)
  \succeq \left( \Bb \left(\KW' \otimes \I_\eta
    \right) \trans{\Bb}\right)\Big)\nonumber \\\label{eq:ptt2}\\
  & \Longrightarrow & \Big( \left( N\I_\eta + \Bb
    (\KW \otimes \I_{\eta}) \trans{\Bb} \right) \nonumber \\ && \hspace{0.7cm}
  \succeq \left(
    N\I_\eta + \Bb (\KW' \otimes \I_{\eta})
    \trans{\Bb}\right)\Big) \label{eq:ptt3}
  \\
   & \Longrightarrow &\Big(\left( \C ( N\I_\eta +
       \Bb (\KW \otimes \I_{\eta})
       \trans{\Bb} )\trans{\C}\right) \nonumber \\ && \hspace{0.7cm}   \succeq
    \left(\C ( N\I_\eta + \Bb (\KW'
      \otimes \I_{\eta}) \trans{\Bb}
      )\trans{\C}
\right) \Big) \label{eq:ptt4}\\
  &\Longrightarrow & \Big( \left(\C ( N\I_\eta +
    \Bb (\KW \otimes \I_{\eta})
    \trans{\Bb} )\trans{\C}\right)^{-1} \nonumber \\ & & \hspace{0.7cm} \preceq
 \left(\C ( N\I_\eta + \Bb (\KW'
    \otimes \I_{\eta}) \trans{\Bb}
    )\trans{\C}\right)^{-1}\Big) \label{eq:ptt5}\\
  & \Longrightarrow & \nonumber \\
\lefteqn{ \Big(\left( \I_2+ \C
    \mat{A}_1\trans{\C}\left(\C ( N\I_\eta +
      \Bb (\KW \otimes \I_{\eta})
      \trans{\Bb}
      )\trans{\C}\right)^{-1} \right)}\nonumber \\
\lefteqn{  \preceq \left(\I_2 +\C
  \mat{A}_1\trans{\C} \left(\C ( N\I_\eta +
    \Bb (\KW' \otimes \I_{\eta})
    \trans{\Bb}
    )\trans{\C}\right)^{-1}\right)\Big) }\nonumber \\\label{eq:ptt6}
\end{IEEEeqnarray}
where \eqref{eq:ptt1} follows by the linearity of the Kronecker
product $\otimes$ and because for every positive semidefinite matrix
$\mat{K}$ also the Kronecker product $\mat{K}\otimes \I_\eta$ is
positive semidefinite\footnote{That $\mat{K}\succeq 0$ implies
  $(\mat{K}\otimes \I)\succeq 0$ can be seen as follows. For
  every ${2\eta}$-dimensional vector $\vect{x}\eqdef
  \trans{(x_1,\ldots, x_{2\eta})}$, where we define
  $\vect{x}_{i}\eqdef\trans{(x_{2i-1}, x_{2i})}$ for
  $i\in\{1,\ldots, \eta\}$, and every $2\times 2$ positive
  semidefinite matrix $\mat{K}$ the term
  $\trans{\vect{x}}\left(\mat{K} \otimes \I_{\eta}\right)\vect{x} $
  can be written as $\sum_{i=1}^\eta \trans{\vect{x}}_i \mat{K}\vect{x}_i$,
   which is 
  nonnegative since $\mat{K}$ is positive semidefinite.}; where
\eqref{eq:ptt1} follows by \eqref{eq:p1}; where \eqref{eq:ptt3}
follows by \eqref{eq:p2} and because $N\I_\eta\succeq 0$; where
\eqref{eq:ptt4} follows by \eqref{eq:p1}; where \eqref{eq:ptt5}
follows by \eqref{eq:p4}; where \eqref{eq:ptt6} follows by
\eqref{eq:p3} and \eqref{eq:p4} and because
$\mat{A}_1\succeq 0$, and thus, by \eqref{eq:p1}, also
$\C\mat{A}_1\trans{\C}\succeq
0$.

Inequality~\eqref{eq:mon} follows then from~\eqref{eq:ptt6}, from
\eqref{eq:p5}, from the monotonicity of the $\log$-function, and from
the fact that for every $2\times2$ positive semidefinite matrix $\mat{K}$,
for $\mat{A}_1$ as defined above, and when $\mat{C}\trans{\mat{C}}$ is nonsingular:
\begin{IEEEeqnarray*}{rCl}
  \lefteqn{\frac{1}{2\eta}\log\left( \frac{ \left| \C( \vect{a}_1\trans{\vect{a}}_1 + N\I_\eta +
          \Bb (\mat{K} \otimes \I_{\eta})
          \trans{\Bb} )\trans{\C}\right|}{\left| \C( N\I_\eta + \Bb
          (\mat{K} \otimes \I_{\eta}) \trans{\Bb}
          )\trans{\C}\right|}\right)}\nonumber \\
  & = & \frac{1}{2\eta}\log\Big( \Big| \I_2 +\C
      \mat{A}_1\trans{\C} \\ & & \hspace{1.5cm} \cdot\left(\C ( N\I_\eta +
        \Bb (\KW' \otimes \I_{\eta})
        \trans{\Bb} )\trans{\C}\right)^{-1}\Big|
  \Big)
\end{IEEEeqnarray*}
which holds because for all nonsingular square matrices $\mat{M}_1$ and $\mat{M}_2$ of
the same dimension $\frac{|\mat{M}_1|}{|\mat{M}_2|}= \left|\mat{M}_1
  \mat{M}_2^{-1}\right|$. This concludes the proof of Statement i).

We next prove Statement ii). It suffices to show that for fixed
parameters $\eta,\vect{a}_1,\vect{a}_2,\B_1,\B_2,\C$, the left-hand
sides of the power constraints \eqref{eq:powergen} are monotonically increasing in $\KW$ with
respect to the Loewner order.  Similarly to the proof of Statement i),
this can be shown by a sequence of implications based on
\eqref{eq:p1}, on \eqref{eq:p2}, on \eqref{eq:p6}, on the fact that
$\KW \succeq \KW'$ implies $\left(\KW \otimes \I_{\eta} \right)\succeq
\left(\KW'\otimes \I_{\eta}\right)$, and on the fact that the trace of
a sum equals the sum of the traces. The details are omitted.

We prove Part~\ref{pb}).  The inclusion of the LHS in the
RHS is trivial, because for every positive $\delta$ all
choices of parameters $\para$ satisfying the power constraints
\eqref{eq:powergen} for powers
$(P_1-\delta)$ and $(P_2-\delta)$ satisfy the power constraints also
for powers $P_1$ and $P_2$.

The inclusion of the RHS in the LHS is proved
as follows. We fix a rate pair $(R_1^\circ,R_2^\circ)$ in the interior
of $\Rate{\ch}$, i.e.,
\begin{equation}\label{eq:interior1}
(R_1^\circ,R_2^\circ)\in \IRate{\ch},
\end{equation}
and show that for all sufficiently small $\delta >0$ the rate pair can
also be achieved with powers $P_1-\delta$ and $P_2-\delta$, i.e.,
\begin{equation}\label{eq:deltaeq}
(R_1^{\circ},R_2^{\circ})\in \Rate{P_1-\delta,P_2-\delta,N,\KW}.
\end{equation}

We first choose parameters $\eta', \vect{a}_1', \vect{a}_2', \B_1',
\B_2', \C'$ so that the power constraints \eqref{eq:powergen} are satisfied for powers $P_1$ and $P_2$ and so
that
\begin{equation}\label{eq:interior2}
(R_1^\circ,R_2^\circ) \in \IRateP{\nch}{\eta', \vect{a}_1', \vect{a}_2', \B_1', \B_2', \C'}.
\end{equation}
By \eqref{eq:interior1}, such a choice always exists. Moreover, for
such a choice the matrix $\mat{C}'$ differs from the all-zero matrix
and both vectors $\vect{a}_1'$ and $\vect{a}_2'$ differ from the
all-zero vector. This can be argued as follows. It is
easily shown that if $\mat{C}'=\mat{0}$, $\vect{a}_1'=\vect{0}$, or
$\vect{a}_2'=\vect{0}$ then  the region $\RateP{\nch}{\eta',
  \vect{a}_1', \vect{a}_2', \B_1', \B_2', \C'}$ is degenerate, i.e.,
either $R_1=0$ for all points in the region or $R_2=0$ for all points
in the region. Consequently, the region $\RateP{\nch}{\eta',
  \vect{a}_1', \vect{a}_2', \B_1', \B_2', \C'}$ cannot contain any
interior points of $\Rate{\ch}$, thus contradicting
\eqref{eq:interior2}.

We next define for each $\delta>0$ the quantities $\kappa_1(\delta)$
and $\kappa_2(\delta)$ as in \eqref{eq:defk12} on top of the next page,
\begin{figure*}
\begin{subequations}\label{eq:defk12}
\begin{IEEEeqnarray}{rCl}
  \kappa_1(\delta)&\eqdef &\sqrt{1- \frac{\delta \eta} {  \tr{  \blkr{\I_\eta}{0}   \left(\I_{2\eta} - \Bbb    \right)^{-1}  \Ad\trans{\Ad}   \left(\I_{2\eta} - \Bbb   \right)^{-\T} \blkc{\I_\eta}{0} }}} \label{eq:defk1}\\
  \kappa_2(\delta)&\eqdef &\sqrt{1-\frac{\delta \eta}{ \tr{ \blkr{0}{\I_\eta} \left(\I_{2\eta} - \Bbb
      \right)^{-1} \Ad\trans{\Ad} \left(\I_{2\eta} - \Bbb
      \right)^{-\T} \blkc{0}{\I_\eta} }}} \label{eq:defk2}
\end{IEEEeqnarray}
\end{subequations}
\hrulefill
\end{figure*}
and we define
\begin{equation*}\label{eq:defk}
\kappa(\delta)=\min\{\kappa_1(\delta), \kappa_2(\delta)\}. 
\end{equation*}
Since $\vect{a}_1'$ and $\vect{a}_2'$ both differ from $\vect{0}$, the
denominators in \eqref{eq:defk1} and \eqref{eq:defk2} are non-zero and
the quantities $\kappa_{1}(\delta)$, $\kappa_2(\delta)$, and
$\kappa(\delta)$ are well defined. Moreover, $\kappa(\delta)$ tends to
$1$ as $\delta\downarrow 0$.

The desired inclusion \eqref{eq:deltaeq} is then established by showing that for all sufficiently small $\delta>0$ the following two statements hold.
 \begin{enumerate}
 \item[i)]\label{i1}  The
parameters $\eta',\kappa(\delta)\vect{a}_1',
\kappa(\delta)\vect{a}_2',\B_1',\B_2',\C'$ satisfy the power constraints
\eqref{eq:powergen}
  for powers
$(P_1-\delta)$ and $(P_2-\delta)$.
\item[ii)]\label{i2} The rate pair $(R_1^{\circ},R_2^{\circ})$ lies in the
  region $\RateP{N,\KW}{\eta',
\kappa(\delta)\vect{a}_1', \kappa(\delta)\vect{a}_2',\B_1',\B_2',\C'}.$
\end{enumerate}
Statement~i) is easily verified by substituting the parameters
$\eta', \kappa(\delta)\vect{a}_1',
\kappa(\delta)\vect{a}_2',\B_1',\B_2',\C'$ into the LHSs of the power constraints
\eqref{eq:powergen} and using the
fact that the parameters $\eta', \vect{a}_1',
\vect{a}_2',\B_1',\B_2',\C'$ satisfy these power constraints for powers
$P_1$ and $P_2$.  Statement~ii) follows because for given
parameters $\eta, \B_1,\B_2,$ and $\C$ the RHSs of
Constraints~\eqref{eq:Rgen1}---which define the
region $\RateP{\ch}{}$ when $\mat{C}\neq 0$---are continuous in the
entries of $\vect{a}_1$ and $\vect{a}_2$, and because $\kappa(\delta)$
tends to 1 as $\delta\downarrow 0$.

We finally prove Part~\ref{pc}), i.e., Equality~\eqref{eq:Part2}. The
inclusion of the LHS in the RHS is trivial,
because replacing the intersection on the LHS by the
specific choice $\mat{K}=\mat{0}$ can only increase the region, and
because the region $\Rate{P_1,P_2,N, \mat{0}}$ is closed. The
interesting inclusion is that the LHS contains the
RHS. To prove this inclusion, we first notice that
\begin{IEEEeqnarray*}{lCl}
\lefteqn{\cl{ \Ue \IK \Rate{P_1,P_2,N, \mat{K}}}}\\
 & \supseteq & \cl{ \Ue \IK  \Rate{P_1,P_2,N,\tr{\mat{K}}
 \I_2}}  \\
& = &  \cl{ \Ue   \Rate{P_1,P_2,N,
  \sigma^2 \I_2} },
\end{IEEEeqnarray*}
where the inclusion and the equality both follow by the monotonicity
proved in Part~\ref{pa}).  Thus, it remains to show that
\begin{IEEEeqnarray}{rCl}\label{eq:q}
  \cl{ \Ue   \Rate{P_1,P_2,N,
  \sigma^2 \I_2} } \supseteq \Rate{P_1,P_2,N,\mat{0}}. \IEEEeqnarraynumspace
\end{IEEEeqnarray}
  To prove~\eqref{eq:q}, we fix a rate pair
$(R_1^{\circ},R_2^{\circ})$ in the interior of
$\Rate{P_1,P_2,N,\mat{0}}$, and show that for all sufficiently
small $\sigma^2>0$ there exists a set of
parameters $\eta, \vect{a}_1, \vect{a}_2, \mat{B}_1, \mat{B}_2,
\mat{C}$ satisfying the following two statements.
\begin{enumerate}
\item[i)]\label{s1}The parameters $\para$ satisfy the power 
constraints \eqref{eq:powergen} for feedback-noise covariance matrix $\KW=\sigma^2\I_2$ and
powers $P_1$ and $P_2$.
\item[ii)]\label{s2} The rate pair
$(R_1^{\circ},R_2^{\circ})$ lies in the region $\RateP{N,\sigma^2
\I_2}{\eta,\vect{a}_1,\vect{a}_2,\B_1,\B_2,\C}$.
\end{enumerate}

We first notice that by Part~\ref{pb}),  for all sufficiently small
$\delta>0$ the pair $(R_1^{\circ},R_2^{\circ})$  lies in the
interior of $\Rate{P_1-\delta,P_2-\delta,N,\mat{0}}$, i.e.,
\begin{equation*}\label{eq:sdfak}
(R_1^{\circ},R_2^{\circ}) \in \IRate{P_1-\delta,P_2-\delta,N,\mat{0}}.
\end{equation*}
This implies that for all sufficiently small $\delta>0$
there exists a set of parameters $(\eta(\delta),\vect{a}_1(\delta),\vect{a}_2(\delta),\B_1(\delta),\B_2(\delta),\C(\delta))$
so that 
\begin{itemize}
\item the power constraints \eqref{eq:powergen} are satisfied for  feedback-noise covariance matrix $\KW=\mat{0}$ and  powers $(P_1-\delta)$
and $(P_2-\delta)$; and 
\item the rate pair $(R_1^{\circ},R_2^{\circ})$
satisfies 
\begin{IEEEeqnarray}{rCl}\label{eq:aaa}
\lefteqn{(R_1^{\circ},R_2^{\circ})} \nonumber \\ &\in&
\IRateP{N,\mat{0}}{\eta(\delta),\vect{a}_1(\delta),\vect{a}_2(\delta),\B_1(\delta),\B_2(\delta),\C(\delta)}.\nonumber \\
\end{IEEEeqnarray}
\end{itemize}
The proof is then established by fixing a sufficiently small
$\delta>0$, and showing that for all sufficiently small $\sigma^2>0$ the choice
of parameters
$(\eta(\delta),\vect{a}_1(\delta),\vect{a}_2(\delta),\B_1(\delta),\B_2(\delta),\C(\delta))$
satisfies the above Statements i) and ii). 

Statement~i) holds because for $\KW=\sigma^2\I_2$ the LHSs
of the power constraints \eqref{eq:powergen}
are continuous in $\sigma^2>0$, and because the parameters
$(\eta(\delta),\vect{a}_1(\delta),\vect{a}_2(\delta),\B_1(\delta),\B_2(\delta),\C(\delta))$
satisfy the power constraints for feedback-noise covariance matrix
$\KW=\mat{0}$ and powers $(P_1-\delta)$ and $(P_2-\delta)$.
Statement~ii) holds because for $\KW=\sigma^2\I_2$ the RHSs of
Constraints \eqref{eq:Rgen1}---which define the
region region $\RateP{\nch}{\para}$ when $\mat{C}\neq \mat{0}$---are
continuous in $\sigma^2$, and because of Inclusion~\eqref{eq:aaa}.

\subsubsection{Proof of Remark~\ref{th:perfectfbeta}}\label{sec:proofremperf}

Fix $P_1,P_2,N>0$. 
Specializing our concatenated scheme to the specific choice
of parameters in
Remark~\ref{def:remarkperf} obviously cannot
outperform our concatenated scheme for general parameters. Thus,
\begin{equation}\label{eq:easytoprove}
\cl{\bigcup_{\eta\in\Nat}
  \set{\tilde{R}}_{\eta}\left(P_1,P_2,N,\mat{0}\right)} \subseteq
  \Rate{P_1,P_2,N,\mat{0}}.
\end{equation}
We shall show in the following that 
\begin{equation}\label{eq:whattoprove}
  \set{R}_{\textnormal{Oz}}^{\rho^*}(P_1,P_2,N) \subseteq\cl{\bigcup_{\eta\in\Nat}
  \set{\tilde{R}}_{\eta}\left(P_1,P_2,N,\mat{0}\right)}, 
\end{equation}
which combined with~\eqref{eq:easytoprove} establishes the remark.

Recall that for fixed $\eta\in\Nat$ the region
$\tilde{\set{R}}_{\eta}(P_1,P_2,N,\mat{0})$ is
defined as the set of all rate pairs $(R_1,R_2)$ satisfying
Constraints~\eqref{eq:3constr} on top of the next page
\begin{figure*}
\begin{subequations}\label{eq:3constr}
\begin{IEEEeqnarray}{lCl}
R_1& \leq &\frac{1}{2\eta}\log \left( 1+ \frac{r P_1}{N}\right)
+
\sum_{\ell=
2}^{\eta}\frac{1}{2\eta}\log\left(1+  \frac{P_1(1-\rho_{\ell-1}^2) }{N } \right)\\
R_2 & \leq  &\frac{1}{2\eta }\log \left( 1+
    \frac{r P_2 }{N}\right)  +\sum_{\ell=2}^{\eta}\frac{1}{2\eta} \log \left(1+
  \frac{P_2(1-\rho_{\ell-1}^2) }{N }\right)\\
R_1+R_2 & \leq &
\frac{1}{2\eta}\log\left(1+\frac{rP_1+
      rP_2}{N}\right) 
 +\sum_{\ell=2}^{\eta}\frac{1}{2\eta}\log \left(1+
  \frac{P_1+P_2+2 \sqrt{P_1P_2}(-1)^{\ell-1}\rho_{\ell-1} }{N }
\right)
\end{IEEEeqnarray}
\end{subequations}
\hrulefill
\end{figure*}
where
\begin{subequations}\label{eq:rhosube}
\begin{equation}\label{eq:rho1f}
\rho_1=-\rho^*(P_1,P_2,N)
\end{equation}
and for $\ell\in\{2,\ldots, \eta-1\}$:
\begin{equation}\label{eq:rho_ell}
\rho_{\ell} = \frac{\rho_{\ell-1} N -
  (-1)^{\ell-1}\sqrt{P_1P_2}(1-\rho_{\ell-1}^2)
}{\sqrt{P_1(1-\rho_{\ell-1}^2)+N}\sqrt{P_2(1-\rho_{\ell-1}^2)+N} }, 
\end{equation} 
\end{subequations}
and where $r$ is the unique solution in $[0,1]$ to \eqref{eq:r1r2cor},
i.e., to 
\[\sqrt{ \frac{r^2P_1 P_2 }{(rP_1+N)(rP_2+N)}}=\rho^*(P_1,P_2,N).\]

We shall shortly prove that the solution to the
recursion~\eqref{eq:rhosube} is 
\begin{equation}\label{eq:condfix}
 \rho_{\ell}=(-1)^{\ell}\rho^*(P_1,P_2,N), \quad \ell\in\Nat.
\end{equation}
This implies that for all $\ell \in\Nat$ larger than 1:
\begin{IEEEeqnarray*}{rCl}
\rho_{\ell-1}^2&=& \rho^{*2},\\
(-1)^{\ell-1} \rho_{\ell-1} & = & \rho^*,
\end{IEEEeqnarray*}
and hence for fixed $\eta\in\Nat$ the region $\tilde{\set{R}}_{\eta}(P_1,P_2,N,\mat{0})$ contains
all rate pairs $(R_1,R_2)$ satisfying:
\begin{subequations}\label{eq:rhos}
\begin{IEEEeqnarray}{rCl}
R_1& \leq &\frac{1}{2\eta}\log \left( 1+ \frac{r P_1}{N}\right)
 \nonumber \\
 & & +\frac{\eta-1}{2\eta}\log\left(1+  \frac{P_1(1-\rho^{*2}) }{N } \right),\IEEEeqnarraynumspace\label{eq:rhos1}\\
R_2 & \leq  &\frac{1}{2\eta }\log \left( 1+
    \frac{r P_2 }{N}\right) \nonumber \\ 
& &+\frac{\eta-1}{2\eta} \log \left(1+
  \frac{P_2(1-\rho^{*2}) }{N }\right),\label{eq:rhos2} \\
R_1+R_2 & \leq &
\frac{1}{2\eta}\log\left(1+\frac{rP_1+
      rP_2}{N}\right) \nonumber \\ 
&& +\frac{\eta-1}{2\eta}\log \left(1+
  \frac{P_1+P_2+2 \sqrt{P_1P_2}\rho^* }{N }
\right).\nonumber \\\label{eq:rhos12} 
\end{IEEEeqnarray}
\end{subequations}
Notice that when $\eta$ tends to infinity, the RHSs of
\eqref{eq:rhos1}--\eqref{eq:rhos12} tend to the RHSs of
the three Constraints \eqref{eq:OZ1}--\eqref{eq:C_PerfectFB} evaluated
for $\rho=\rho^*$. 
Since Constraints~\eqref{eq:OZ1}--\eqref{eq:C_PerfectFB} evaluated for
$\rho=\rho^*$ determine the region
$\set{R}_{\textnormal{Oz}}^{\rho^*}(P_1,P_2,N)$,
Inclusion~\eqref{eq:whattoprove} follows immediately by
\eqref{eq:rhos} and by letting $\eta$ tend to
infinity.

  In the remaining, we prove~\eqref{eq:condfix} in two steps. In the
  first step we show that $\rho^*(P_1,P_2,N)$ is a fix point of the
  function $\L(\cdot)$ defined as
\begin{IEEEeqnarray*}{rCl}
\L:& & [0,1] \rightarrow \Reals, \nonumber \\
\L(\rho) &=&  \frac{\sqrt{P_1P_2}(1-\rho^2)-\rho N
}{\sqrt{P_1(1-\rho^2)+N}\sqrt{P_2(1-\rho^2)+N} }.
\end{IEEEeqnarray*}
Notice that $\L(\cdot)$ has at least one fix point in $[0,1]$ because
$\L(0)>0$ whereas $\L(1)<0$, and because $\L(\cdot)$ is continuous.
Further notice that every fix point of $\L(\cdot)$ must also be a
solution to 
\[1-\L(\rho)^2 = 1-\rho^2,\]
 i.e., a solution to
\begin{equation}\label{eq:Lxeq}
\frac{\N(\N+\pow_1+\pow_2+2 \sqrt{\pow_1 \pow_2}\rho)}{(\N+ \pow_1(1-\rho^2))(\N+\pow_2(1-\rho^2))}(1-\rho^2)  =  
(1-\rho^2).
\end{equation}
The solutions in $[0,1]$ to \eqref{eq:Lxeq} are given by $\rho=1$ and
by the solutions to
\begin{IEEEeqnarray}{rCl}\label{eq:eqxstar2}
\lefteqn{\N(\N+\pow_1+\pow_2+2 \sqrt{\pow_1 \pow_2}\rho) }\nonumber
\qquad \qquad \\& =&  
(\N+ \pow_1(1-\rho^2))(\N+\pow_2(1-\rho^2)). \IEEEeqnarraynumspace
\end{IEEEeqnarray}
Since $\rho=1$ is not a fix point of $\L(\cdot)$ and since
$\rho^*(P_1,P_2,N)$ is the unique solution in $[0,1]$ to
\eqref{eq:eqxstar2} (see Definition~\ref{def:rho}),
$\rho^*(P_1,P_2,N)$ must be a fix point of $\L(\cdot)$. This concludes
the first step.

In the second step we use the derived fix-point property of $\L(\cdot)$
to prove \eqref{eq:condfix}. The proof is lead by induction. For $\ell=1$ Condition
\eqref{eq:condfix} holds by definition. Assuming that
\eqref{eq:condfix} holds for some fixed $\ell\geq 1$, we have
\begin{IEEEeqnarray}{rCl}
\lefteqn{\rho_{\ell+1}}\nonumber \\ & = & 
  \frac{-(-1)^\ell\sqrt{P_1P_2}(1-|\rho_\ell|^2)+\rho_\ell N}{\sqrt{P_1(1-|\rho_{\ell}|^2)+N}\sqrt{P_2(1-|\rho_{\ell}|^2)+N} }
\label{1} \\
& = &(-1)^{\ell+1} \frac{ \sqrt{P_1P_2}(1-|\rho_{\ell}|^2)-|\rho_\ell|N
}{\sqrt{P_1(1-|\rho_{\ell}|^2)+N}\sqrt{P_2(1-|\rho_{\ell}|^2)+N}
}\nonumber \\\label{2} \\
& = & (-1)^{\ell+1}\L(|\rho_\ell|)\label{3} \\
& = & (-1)^{\ell+1}\rho^*(P_1,P_2,N),\label{4}\IEEEeqnarraynumspace
\end{IEEEeqnarray}
where \eqref{1} follows by the
definition of the sequence $\{\rho_{\ell}\}$ for $\ell>1$; \eqref{2}
follows because by the induction assumption
sign$(\rho_{\ell})=(-1)^{\ell}$; \eqref{3} follows by the definition
of the function $\L(\cdot)$; and finally \eqref{4} follows because by
the induction assumption $|\rho_\ell|=\rho^*$ and because $\rho^*$, as
shown in the first step, is
a fix point of $\L(\cdot)$.
Thus, \eqref{eq:condfix} holds also for $(\ell+1)$, which concludes
the induction step and the proof of the remark.

 \subsubsection{Proof of Lemma~\ref{rem:Cperfectentire} }\label{sec:Cperfectentire}

 We only prove Inclusion~\eqref{eq:Rrho1}; Inclusion~\eqref{eq:Rrho2}
 can be proved analogously.

Fix $\rho\in[0,\rho^*]$, and define $\alpha(\rho)$ as the
unique solution in $[0,1[$ to
\begin{equation}\label{eq:alp}
\frac{P_1+P_2+2\sqrt{P_1P_2}\rho+N}{P_1(1-\rho^2)+N} = 1+
\frac{P_2\left(1-\frac{\rho^2}{1-\alpha}\right)}{\alpha P_1+N}.
\end{equation}
That \eqref{eq:alp} has exactly one solution in $[0,1)$ follows by the
Intermediate Value Theorem and the following observations: The
RHS of \eqref{eq:alp} is continuous and strictly
decreasing in $\alpha$; for $\alpha=0$ the RHS of
\eqref{eq:alp} is larger or equal to the LHS because $0\leq
\rho\leq \rho^*$ and by Remark~\ref{rem:incdec}; and for $\alpha$
tending to $1$ the RHS tends to $-\infty$ and thus is
smaller than the LHS.

 Further, define 
 \begin{IEEEeqnarray*}{rCl}
P_1'&\eqdef &\alpha(\rho) P_1,\\
P_1''&\eqdef& (1-\alpha(\rho))P_1,\\
 N'&\eqdef& P_1'+N,\\
\rho'&\eqdef&\frac{\rho}{\sqrt{1-\alpha(\rho)}}
\end{IEEEeqnarray*}
and notice that by these definitions:
\begin{IEEEeqnarray}{rCl}\label{eq:rhobareq}
\lefteqn{
N'(N'+P_1''+P_2+2\sqrt{P_1''P_2}\rho')} \qquad \nonumber \\ &=&
(N'+P_1''(1-\rho'^2))(N'+P_2(1-\rho'^2)), \IEEEeqnarraynumspace
\end{IEEEeqnarray}
and hence
\begin{equation*}
\rho'=\rho^*(P_1'',P_2,N').
\end{equation*}

Also, define
$(R_{1,\textnormal{Oz}}^{\rho},R_{2,\textnormal{Oz}}^{\rho})$ as the dominant corner
point of the rectangle $\set{R}_{1,\textnormal{Oz}}^{\rho}(P_1,P_2,N)$. The
following two remarks on
$(R_{1,\textnormal{Oz}}^{\rho},R_{2,\textnormal{Oz}}^{\rho})$ are from
\cite{ozarow85}, and based on \eqref{eq:rhobareq}.
\begin{remark}\label{eq:prop_alpha}
  The rate point $(R_{1,\textnormal{Oz}}^{\rho},
  R_{2,\textnormal{Oz}}^{\rho})$ can be expressed as
\begin{IEEEeqnarray*}{rCl}
R_{1,\textnormal{Oz}}^{\rho}&=&R_{1,1,\textnormal{Oz}}^{\rho}+R_{1,2,\textnormal{Oz}}^{\rho},\\
R_{2,\textnormal{Oz}}^{\rho} & = & \frac{1}{2}\log\left(1+\frac{P_2(1-\rho'^2)}{N'} \right), 
\end{IEEEeqnarray*}
where
\begin{IEEEeqnarray*}{rCl}
R_{1,1,\textnormal{Oz}}^{\rho} & \eqdef& \frac{1}{2}
\log\left(1+\frac{ P_1'}{N} \right), \\
R_{1,2,\textnormal{Oz}}^{\rho} & \eqdef & \frac{1}{2} \log \left(1+ \frac{P_1''(1-\rho'^2) }{ N'} \right). 
\end{IEEEeqnarray*}
\end{remark}

\begin{remark}\label{rem:dom}
 The rate point
$(R_{1,2,\textnormal{Oz}}^{\rho},R_{2,\textnormal{Oz}}^{\rho})$
corresponds to the dominant corner point of the rectangle
$\set{R}_{\textnormal{Oz}}^{\rho'}(P_1'',P_2,N')$, where $\rho'=\rho^*(P_1'',P_2,N')$.
\end{remark}

We are now ready to prove Inclusion~\eqref{eq:Rrho1}. 
For $\KW=\mat{0}$ the RHS of \eqref{eq:77} equals
$\frac{1}{2}\log\left(1+\frac{P_1'}{N}\right)$, irrespective of the
parameters $\vect{a}_1,\vect{a}_2,\B_1,\B_2,\C$. Therefore, the region
$\Rateone{P_1',P_1'',P_2,N,\mat{0}}$ is given by the set of all rate
pairs $(R_1,R_2)$ which for some nonnegative
$R_{1,\textnormal{CS}},R_{1,\textnormal{NF}}$ summing to $R_1$ satisfy
\begin{subequations}\label{eq:GKK}
\begin{IEEEeqnarray}{rCl}
(R_{1,\textnormal{CS}},R_{2}) \; & \in & \; 
\Rate{P_1'',P_2,N',\mat{0}} , \label{eq:GK1}\\
R_{1,\textnormal{NF}} \;& \leq & \; \frac{1}{2} \log\left(1+
  \frac{P_1'}{N}\right).\label{eq:GK}\IEEEeqnarraynumspace
\end{IEEEeqnarray}
\end{subequations}
Since by  Remark~\ref{rem:dom} and Remark~\ref{th:perfectfbeta}:
\begin{equation*}
\left( R_{1,2,\textnormal{Oz}}^{\rho},
  R_{2,\textnormal{Oz}}^{\rho}\right) \in
  \Rate{P_1'', P_2,N',\mat{0}}
\end{equation*}
and  by Remark~\ref{eq:prop_alpha}: 
\begin{equation*}
R_{1,1,\textnormal{Oz}}^{\rho} \leq  \frac{1}{2} \log\left(1+
  \frac{P_1'}{N}\right),
\end{equation*}
the triple $\left(
 R_{1,1,\textnormal{Oz}}^{\rho}, R_{1,2,\textnormal{Oz}}^{\rho},
  R_{2,\textnormal{Oz}}^{\rho}\right)$  satisfies \eqref{eq:GKK}, and hence
\begin{equation}\label{eq:pointinc}
\left( R_{1,\textnormal{Oz}}^{\rho},
  R_{2,\textnormal{Oz}}^{\rho}\right) \in
  \Rateone{P_1',P_1'', P_2,N,\mat{0}}.
\end{equation}
Inclusion \eqref{eq:Rrho1} finally follows because
$\left( R_{1,\textnormal{Oz}}^{\rho},
  R_{2,\textnormal{Oz}}^{\rho}\right)$ is the dominant corner point of
the rectangle $\set{R}_{1,\textnormal{Oz}}^{\rho}(P_1,P_2,N)$, and therefore
\eqref{eq:pointinc} implies that the entire region
$\set{R}_{1,\textnormal{Oz}}^{\rho}(P_1,P_2,N)$ is contained in  $\Rateone{P_1',P_1'', P_2,N,\mat{0}}$.

 \subsubsection{Proof of Proposition~\ref{prop:splitmaxsumratesigma} }\label{sec:ProofProp2}

 We only prove Inclusion~\eqref{eq:largeR1};
 Inclusion~\eqref{eq:largeR2} can be proved analogously.

To this end, fix a $\rho\in[0,\rho^*(P_1,P_2,N)]$ and choose a power $P_1'\in[0,P_1]$ such
that
\begin{IEEEeqnarray}{rCl}\label{eq:importantinclusion}
\Rateone{P_1', (P_1-P_1'), P_2, N,
\mat{0}}\supseteq \set{R}_{1,\textnormal{Oz}}^{\rho}(P_1,P_2,N).\IEEEeqnarraynumspace
\end{IEEEeqnarray}
Notice that by Remark~\ref{rem:Cperfectentire} such a power $P_1'$
always exists. Inclusion~\eqref{eq:largeR1} follows then because by
Proposition~\ref{prop:monotonicity2}, Part 2.:
\begin{IEEEeqnarray*}{rCl}\lefteqn{
\cl{ \Ue \IKtwo 
\Rateone{P_1', (P_1-P_1'), P_2, N,
\mat{K}}}} \qquad \\ 
&=&
\Rateone{P_1', (P_1-P_1'), P_2, N,
\mat{0}}.\hspace{2cm} \label{eq:74}
\end{IEEEeqnarray*}

\subsubsection{Proof of Theorem~\ref{th:C_sym}}\label{sec:conv}

Fix $P_1,P_2,N>0$. The proof of the $\subseteq$-direction
follows trivially because replacing the intersection on the LHS by
the specific choice $\mat{K}=\mat{0}$ can only increase the region,
because $\capa_{\textnormal{NoisyFB}}(P_1,P_2,N,
\mat{0})=\capa_{\textnormal{PerfectFB}}(P_1,P_2,N)$, and because by
definition the region $\capa_{\textnormal{PerfectFB}}(P_1,P_2,N)$ is
closed.

The $\supseteq$-direction, i.e., 
\begin{IEEEeqnarray*}{rCl}
\lefteqn{\cl{\Ue \IKtwo
    \capa_{\textnormal{NoisyFB}}\left(P_1,P_2,N,\K\right)}}\qquad  \\ &
    \supseteq & \capa_{\textnormal{PerfectFB}}(P_1,P_2,N), \hspace{2cm}
\end{IEEEeqnarray*}
follows from the sequence of inclusions \eqref{eq:30}--\eqref{eq:34}
on top of the next page. 
\begin{figure*}[!t]
  \normalsize
\allowdisplaybreaks[4]
\begin{IEEEeqnarray}{rCl}
\lefteqn{\cl{\Ue \IK 
\capa_{\textnormal{NoisyFB}}\left( P_1,P_2,N,\K \right)}}\nonumber\\
& \supseteq  & \label{eq:30}
\textnormal{cl}\left( \Ue \IK  \left( \left( \bigcup_{\substack{
 P_1'\in[0,P_1]}}
\Rateone{P_1',
(P_1-P_1'),P_2,N,\mat{K}}\right)\right. \right. \nonumber \\ & &
 \left.\phantom{\textnormal{cl}\left(  \Ue \IK \right) } \quad \left.\cup \left(\bigcup_{\substack{
 P_2'\in[0,P_2]}}
 \Ratetwo{P_1,P_2',
 (P_2-P_2'),N,\mat{K}} \right) \right)\right) \label{eq:31} \\
& \supseteq  &
\cl{ \bigcup_{ P_1'\in[0,P_1]}\cl{   \Ue \IK
                                \Rateone{P_1',
(P_1-P_1'),P_2,N,\mat{K}} }} \nonumber \\  & & \cup \;\;\cl{ \bigcup_{ P_2'\in[0,P_2]} \cl{
      \Ue
     \IK 
 \Ratetwo{P_1,P_2',
 (P_2-P_2'),N,\mat{K}} } } \label{eq:32}\\
& \supseteq  &
 \cl{\bigcup_{\rho\in [0,\rho^*(P_1,P_2,N)]}
 \set{R}_{1,\textnormal{Oz}}^{\rho}(P_1,P_2,N)} \cup  \;\cl{\bigcup_{ \rho
   \in [0,\rho^*(P_1,P_2,N)]}
 \set{R}_{2,\textnormal{Oz}}^{\rho}(P_1,P_2,N) }\label{eq:33}\\
& = & \capa_{\textnormal{PerfectFB}}( P_1,P_2,N),\label{eq:34}
\end{IEEEeqnarray}
 \hrulefill
\end{figure*}
Inclusion \eqref{eq:30} follows from Proposition~\ref{prop:0};
\eqref{eq:32} follows by basic rules on sets; \eqref{eq:33} follows
from Proposition~\ref{prop:splitmaxsumratesigma}; and \eqref{eq:34}
follows by Remark~\ref{rem:alternPFBCap}.

\section{Partial Feedback}\label{sec:par}

We now focus on the setup with noisy or perfect partial feedback. For
this setup we again present new achievable regions, and based on these
new regions we derive new qualitative properties of the capacity
region (Section~\ref{sec:res_par}). We also present the coding schemes
corresponding
to these new achievable regions (Sections~\ref{sec:simple_par}--\ref{sec:ext_par}). They are obtained from
the noisy-feedback schemes in Sections~\ref{sec:simple}--\ref{sec:ext}
by restricting the set of parameters and in the case of the extended schemes by
additionally specializing Carleial's scheme to noisy partial feedback.


\subsection{Results}\label{sec:res_par}

We first present  results for noisy partial feedback
(Section~\ref{sec:ach_partial}) and then results that hold only for perfect
partial feedback (Section~\ref{sec:achppf}).

\subsubsection{Results for Noisy Feedback}\label{sec:ach_partial} 
Evaluating the rates achieved by our concatenated scheme with general
parameters in Section~\ref{sec:conc_par_des} ahead leads to the 
achievability result in Theorem~\ref{th:generalP}. Before stating the
result we define:
\begin{definition}\label{def:npf}  Let $\eta$ be a positive integer, let
  $\vect{a}_1,\vect{a}_2$ be $\eta$-dimensional vectors, let $\B_2$
  be a strictly lower-triangular $\eta\times \eta$ matrix, and let
  $\mat{C}_{\textnormal{P}}$ be a $2\times \eta$ matrix. Then,
  depending on the matrix $\mat{C}_{\textnormal{P}}$ the rate region
  $\RatePpar{N,\sigma_2^2}{\eta,\vect{a}_1,\vect{a}_2,\B_2,\C_{\textnormal{P}}}$
  is defined as follows:
\begin{itemize}
\item If the product $\mat{C}_{\textnormal{P}}\trans{\C}_{\textnormal{P}}$ is
  nonsingular,\footnote{Whenever
    $\eta \in \Nat$ is larger than 1, there is no loss in optimality
    in restricting attention to matrices $\C_{\textnormal{P}}$ so that $\C_{\textnormal{P}}\trans{\C}_{\textnormal{P}}$
    is nonsingular.} then
  $\RatePpar{N,\sigma_2^2}{\eta,\vect{a}_1,\vect{a}_2,\B_2,\C_{\textnormal{P}}}$
  is defined as the set of all rate pairs $(R_1,R_2)$ satisfying
\begin{subequations}\label{eq:Rgen1P}
\begin{IEEEeqnarray}{lCl}
  R_1 & \leq & \frac{1}{2\eta}\log \frac{\left| \mat{C}_{\textnormal{P}} \left( \vect{a}_1
        \trans{\vect{a}_1} +N \I_{\eta}+ \sigma_2^2\B_2 \trans{\B_2}
      \right)\trans{\mat{C}_{\textnormal{P}}} \right|}{
    \left| \mat{C}_{\textnormal{P}}
      \left( N \I_{\eta} +\sigma_2^2 \B_2\trans{\B_2}\right) \trans{\mat{C}_{\textnormal{P}}}
    \right| }, \nonumber \\\label{eq:R1gen1P}
  \\
  R_2 & \leq &\frac{1}{2\eta}\log \frac{\left| \mat{C}_{\textnormal{P}} \left( \vect{a}_2
        \trans{\vect{a}_2}
        +N\I_{\eta}+\sigma_2^2\B_2\trans{\B_2}\right)\trans{\mat{C}_{\textnormal{P}}} \right|}{
    \left| \mat{C}_{\textnormal{P}} \left(  N \I_{\eta}
  +\sigma_2^2\B_2\trans{\B_2}\right)\trans{\mat{C}_{\textnormal{P}}}
  \right| },\nonumber \\\label{eq:R2gen1P}
\\
\lefteqn{  R_1+R_2} \nonumber \\ & \leq & \frac{1}{2\eta}\log \frac{\left| \mat{C}_{\textnormal{P}} \left( \Ab
        \trans{\Ab} +N\I_{\eta}+\sigma_2^2\B_2 \trans{\B_2} \right)
  \trans{\mat{C}_{\textnormal{P}}}
    \right|}{ \left| \mat{C}_{\textnormal{P}} \left( N \I_{\eta} +
        \sigma_2^2 \B_2 \trans{\B_2}\right)\trans{\mat{C}_{\textnormal{P}}} \right| }
  ,\nonumber \\\label{eq:R12gen1P}
\end{IEEEeqnarray}
\end{subequations}
 where $\Ab$ is defined in \eqref{eq:Ab}.
 \item If the product $\C_{\textnormal{P}}\trans{\C}_{\textnormal{P}}$ is singular
  but $\C_{\textnormal{P}}\neq \mat{0}$, then
  $\RatePpar{N,\sigma_2^2}{\eta,\vect{a}_1,\vect{a}_2,\B_2,\C_{\textnormal{P}}}$
  is defined as the set of all rate pairs $(R_1,R_2)$ satisfying 
\eqref{eq:Rgen1P} when the $2\times \eta$ matrix
  $\C_{\textnormal{P}}$ is replaced by the $\eta$-dimensional
  row-vector obtained by choosing one of its non-zero rows.
\item If $\C_{\textnormal{P}}=\mat{0}$, then
  $\RatePpar{N,\sigma_2^2}{\eta,\vect{a}_1,\vect{a}_2,\B_2,\C_{\textnormal{P}}}$
  is defined as the set containing only the origin.
\end{itemize}
\end{definition}
An alternative formulation
of the region
$\RatePpar{N,\sigma_2^2}{\eta,\vect{a}_1,\vect{a}_2,\B_2,\C_{\textnormal{P}}}$ is
presented in Section~\ref{sec:achaltnoisypar}.
\begin{definition}\label{def:Rp}
Define
\begin{IEEEeqnarray}{rCl}
\lefteqn{\Ratepar{P_1,P_2,N,\sigma_2^2}} \nonumber \\ & \eqdef& \cl{ \bigcup_{\eta,\vect{a}_1,\vect{a}_2,\B_2,\C_{\textnormal{P}}}
  \RatePpar{N,\sigma_2^2}{\eta,\vect{a}_1,\vect{a}_2,\B_2,\C_{\textnormal{P}}}} ,\IEEEeqnarraynumspace\label{eq:parunion}
\end{IEEEeqnarray}
where the union is over all tuples
$(\eta,\vect{a}_1,\vect{a}_2,\B_2,\C_{\textnormal{P}})$ satisfying
  the
trace constraints 
\begin{subequations}\label{eq:powerpart}
\begin{IEEEeqnarray}{rCl}\label{eq:power1part}
\trans{\vect{a}}_1\vect{a}_1 &\leq& \eta P_1
\end{IEEEeqnarray}
and 
\begin{IEEEeqnarray}{rCl}\lefteqn{
\textnormal{tr}\bigg((\I_{\eta}-\B_2)^{-1} \Big({ \vect{a}}_2\trans{\vect{a}}_2 + 
      \B_2 \vect{a}_1\trans{\vect{a}}_1 \trans{\B}_2 } \nonumber \\ & & \hspace{1.2cm} + (N+\sigma_2^2) \B_2
      \trans{\B}_2\Big) (\I_{\eta}-\B_2)^{-\T}\bigg) \leq \eta
      P_2.\IEEEeqnarraynumspace \label{eq:power2part}
\end{IEEEeqnarray}
\end{subequations}

\end{definition}
\begin{theorem}[Noisy Partial Feedback]\label{th:generalP}
  The capacity region
  $\capa_{\textnormal{NoisyPartialFB}}(P_1,P_2,N,\sigma_2^2)$ of the
  two-user AWGN MAC with noisy partial feedback to Transmitter~2
  contains the rate region $\Ratepar{P_1,P_2,N,\sigma_2^2}$, i.e., 
\begin{equation*}
\capa_{\textnormal{NoisyPartialFB}}(P_1,P_2,N,\sigma_2^2) \supseteq
\Ratepar{P_1,P_2,N,\sigma_2^2}.
\end{equation*}
\end{theorem}

\begin{proof}
Follows from Theorem~\ref{th:general} by choosing $\B_1$ as the
all-zero matrix.
\end{proof}
  \begin{remark} Evaluating the achievable region
    $\Ratepar{P_1,P_2,N,\sigma_2^2}$ seems to be difficult even
    numerically.  More easily computable (but possibly smaller)
    achievable regions are obtained by taking the union on the RHS of
    \eqref{eq:parunion} only over a subset of the parameters
    $\eta,\vect{a}_1,\vect{a}_2,\B_2,\C_{\textnormal{P}}$ satisfying
    \eqref{eq:powerpart}. In
    Remark~\ref{rem:partialfb} we present such a subset of parameters.
    In Section~\ref{sec:choicepartial} we present general guidelines
    on how to choose the parameters
    $\eta,\vect{a}_1,\vect{a}_2,\B_2,\C_{\textnormal{P}}$.
\end{remark}

Specializing Theorem~\ref{th:generalP} to equal powers channels, i.e.,
$P_1=P_2=P$, and to $\eta=2$ and the choice of the parameters
presented in Section~\ref{sec:enc} yields the following Corollary~\ref{th:co2}. \begin{corollary}[Equal Powers and Noisy Partial Feedback]\label{th:co2}
  The capacity region
  $\capa_{\textnormal{NoisyPartialFB}}(P,P,N,\sigma_2^2)$ of the two-user
  AWGN MAC with noisy partial feedback
  to Transmitter~2 and equal powers $P_1=P_2=P$ contains all rate
  pairs $(R_1,R_2)$ satisfying
 Constraints~\ref{eq:eqpow} on top of the next page.
 \begin{figure*}[!t]
  \normalsize
 \begin{subequations}\label{eq:eqpow}
\begin{IEEEeqnarray}{rCl}
  R_1& \leq & \frac{1}{4}\log \left(1+\frac{2P}{N} \right)+ \frac{1}{4}\log \left( 1 - \frac{P}{2P+N}
    \frac{P\frac{P}{N}\sigma_2^2}{\left(2P+N+\sigma_2^2+\frac{P}{N}\sigma_2^2\right)\left(P+N+\sigma_2^2+\frac{P}{N}\sigma_2^2\right)} \right) \label{eq:eqpow1} \\
  R_2& \leq& \frac{1}{4}\log\left(1+\frac{2 P}{N} \right)
  +\frac{1}{4}\log\left(1+\frac{P}{2P+N}\cdot\frac{P}{P+N+\sigma_2^2+\frac{P}{N}\sigma_2^2}  \right)\label{eq:eqpow2}\\
 R_1 +R_2 & \leq& \frac{1}{2} \log \left( 1+ \frac{2
      P}{N}\right) \nonumber \\  & &+ \frac{1}{4} \log \left(
    1+\frac{2P^2}{(2P+N)^2}\left(\sqrt{
        1+\frac{P(P+N+\sigma_2^2)}{\left(P+N+\sigma_2^2+\frac{P}{N}\sigma_2^2\right)^2}}-1\right) \right.
   \nonumber  \\
& & \hspace{1.4cm} \left.+\left(\frac{P}{2P+N}\right)^2 
\frac{(P+N)(2P+N+\sigma_2^2)}{(2P+N+\sigma_2^2+\frac{P}{N}\sigma_2^2)(P+N+\sigma_2^2+\frac{P}{N}\sigma_2^2 
      )} \right) \label{eq:eqpow12}
\end{IEEEeqnarray}
\end{subequations}
\hrulefill
\end{figure*}
\end{corollary}
From Corollary~\ref{th:co2} it follows immediately that for
equal-powers channels noisy partial feedback increases the capacity,
no matter how large the noise variance $\sigma^2\geq 0$ is. The following
stronger result holds:
\begin{theorem}[Noisy Partial Feedback is Always Beneficial]\label{th:C_unsympar}
 For all $ N,P_1,P_2>0$ and $\sigma_2^2\geq 0$
\begin{IEEEeqnarray}{rCl}\label{eq:incnoisy}
\capaMAC(P_1,P_2,N) & \subset & 
    \capa_{\textnormal{NoisyPartialFB}}(P_1,P_2,N, \sigma_2^2),\IEEEeqnarraynumspace
  \end{IEEEeqnarray}
where the inclusion is strict.
\end{theorem}
\begin{proof}
 See Section~\ref{sec:useful}.
\end{proof}

\subsubsection{Results for Perfect Partial Feedback}\label{sec:achppf}

Specializing Theorem~\ref{th:generalP} to perfect partial feedback,
i.e., to $\sigma_2^2=0$ yields:
\begin{corollary}[Perfect Partial Feedback]\label{th:ppf}
The capacity region $\capa_{\textnormal{PerfectPartialFB}}(P_1,P_2,N)$
  of the two-user AWGN MAC with perfect partial feedback to
  Transmitter~2 contains the rate region
  $\Ratepar{P_1,P_2,N, 0}$, i.e.,
\begin{equation*}
\capa_{\textnormal{PerfectPartialFB}}(P_1,P_2,N)\supseteq
  \Ratepar{P_1,P_2,N, 0}.
\end{equation*}
\end{corollary}
Specializing Corollary~\ref{th:ppf} to $\eta=2$ and the choice of parameters in
Section~\ref{sec:enc}  yields:
\begin{corollary}\label{th:co2b}
  The capacity region $\capa_{\textnormal{PerfectPartialFB}}(P_1,P_2,N)$ of
  the two-user AWGN MAC with perfect
  partial feedback to Transmitter 2 contains all rate
  pairs $(R_1,R_2)$ satisfying
\begin{IEEEeqnarray*}{lCl}
 R_1 &\leq& \frac{1}{4}\log\left( 1+\frac{2P_1}{N}\right),\\
 R_2 &\leq &\frac{1}{4} \log \left( 1+
  \frac{P_2\left(2+\frac{P_2}{P_1+N}\right)}{N}\right),\\
 \lefteqn{R_1+R_2} \nonumber \\& \leq& \frac{1}{4}\log\left(1+\frac{P_1+P_2}{N}
\right) \nonumber \\ & &
 +\frac{1}{4}\log\left(1+\frac{P_1\frac{P_2+N}{P_1+P_2+N}+P_2}{N}
 \right. \\ & & \left.\hspace{1.7cm} +\frac{
 2\sqrt{P_1P_2\frac{P_1}{P_1+N}\frac{P_2}{P_1+P_2+N}}}{N} \right). \nonumber\\
\end{IEEEeqnarray*}
\end{corollary}
With this Corollary~\ref{th:co2b} at hand we can answer the question
by Van der Meulen in \cite{vandermeulen87} whether the Cover-Leung
region equals the capacity region of the MAC with perfect
partial feedback.
\begin{theorem}\label{th:partial_R_ach}
  Consider a two-user AWGN MAC with perfect partial feedback. For some
  powers $P_1, P_2$ and noise variance $N$ the
  inclusion
\begin{equation*}
 \mathcal{R}_{\textnormal{CL}}(P_1,P_2,N)\; \subset \;\capa_{\textnormal{PerfectPartialFB}}(P_1,P_2,N)
\end{equation*}
is strict.
\end{theorem}
\begin{proof}
  The inclusion is proved in Section~\ref{sec:proof_partial} by
  showing that for powers $\pow_1=1, \pow_2=5$ and noise variance
  $\N=5$ the region in Corollary~\ref{th:co2b} contains rate points
  that lie strictly outside the Cover-Leung region.
\end{proof}

The last two results are achieved by modifying the
rate-splitting schemes for noisy feedback in
Sections~\ref{sec:extCarleial} and \ref{sec:CarleialInter} so as to
apply also for perfect partial feedback. For details see
Section~\ref{sec:ext_par}. 
\begin{proposition}[Rate-Splitting for Perfect Partial Feedback I]\label{prop:CarleialRSsimple}
  The capacity region
  $\capa_{\textnormal{PerfectPartialFB}}(P_1,P_2,N)$ of the two-user
  AWGN MAC with perfect partial feedback to Transmitter~2 contains all
  rate pairs $(R_1,R_2)$ which for some nonnegative $R_{1,\textnormal{CL}},
  R_{1,\textnormal{CS}}$ summing to $R_1$, for some nonnegative
  $R_{2,\textnormal{CL}}, R_{2, \textnormal{CS}}$ summing to $R_2$, and
  for some choice of $\rho_1,\rho_2 \in [0,1]$ and $P_1'\in [0,P_1],
  P_2'\in[0,P_2]$ satisfy
\begin{IEEEeqnarray*}{rCl}
  (R_{1,\textnormal{CL}},R_{2,\textnormal{CL}})\; & \in & \; {\set{R}}_{\textnormal{CL}}^{(\rho_1,\rho_2)}(
  P_1', P_2',N),\\
  (R_{1,\textnormal{CS}},R_{2,\textnormal{CS}})\; & \in & \;\Ratepar{
   (P_1-P_1'), ( P_2-P_2'), N_{\textnormal{CS}}},
\end{IEEEeqnarray*}
where $N_{\textnormal{CS}}\eqdef (N + P_1' +P_2'
    +2 \sqrt{ P_1'P_2'}\rho_1 \rho_2)$.
\end{proposition}
\begin{proof}
  The rate region is achieved by modifying the rate-splitting
  scheme for noisy feedback in Section \ref{sec:extCarleial} as
  described in Section~\ref{sec:ext_par}. Here, the version of the
  scheme in Section~\ref{sec:extCarleial} is chosen where
  Transmitter~2 decodes the submessages encoded with the concatenated
  scheme before decoding the submessages encoded with Carleial's
  Cover-Leung scheme.  The analysis of the rate-splitting scheme is
  based on a genie-aided argument as in \cite{rimoldiurbanke96} and
  \cite{wozencraftjacobs65}. The details are omitted.
\end{proof}
\begin{proposition}[Rate-Splitting for Perfect Partial Feedback II]\label{prop:CarleialRS}
  The capacity region
  $\capa_{\textnormal{PerfectPartialFB}}(P_1,P_2,N)$ of the two-user
  AWGN MAC with perfect partial feedback to Transmitter~2 contains all
  rate pairs
  $(R_1,R_2)$
  which for nonnegative $(R_{1,\textnormal{ICL},1},
  R_{1,\textnormal{ICL},2},R_{1,\textnormal{ICS}})$ summing to $R_1$;
  nonnegative $(R_{2,\textnormal{ICL},1},
  R_{2,\textnormal{ICL},2},R_{2,\textnormal{ICS}})$ summing to $R_2$;
 and for some choice of $\rho_1, \rho_2\in [0,1]$ and $P_1'\in[0,P_1],
P_2'\in[0,P_2]$ satisfy all the 11 constraints~\eqref{eq:ICSICL} on
  top of the next page,  
\begin{figure*}
\begin{subequations}\label{eq:ICSICL}
\begin{IEEEeqnarray}{rCl}
  R_{1,\textnormal{ICS}} &\leq&
  \frac{1}{4}\log\left(1+\frac{2(\pow_1-P_1')}{\N}\right)\\
  R_{2,\textnormal{ICS}} & \leq & \frac{1}{4}\log\left(1+\frac{(\pow_2-P_2')
      \left(2+\frac{\pow_2-P_2'}{\pow_1-P_1'+\N}\right)}{\N}\right)\\
   R_{1,\textnormal{ICS}}+ R_{2,\textnormal{ICS}}& \leq &
  \frac{1}{4}\log\left(1+\frac{\pow_1-P_1'+\pow_2-P_2'}{\N}\right)
  +
  \frac{1}{4}\log\left(1+\frac{ \N_2 }{\N}\right)
  \\
  R_{1,\textnormal{ICL},1} & \leq & \frac{1}{4} \log\left(1+
    \frac{(1-\rho_1^2)
      \pow_1'}{\pow_1'+\N}\right) \\
  R_{1,\textnormal{ICL},1} & \leq & \frac{1}{4} \log\left(1+ \frac{
      (1-\rho_1^2) \pow_1'}{\N_1+\N} \right)  +
  \frac{1}{4}\log\left(1+\frac{\left(\sqrt{ \rho^2_1 \pow_1'}+\sqrt{
          \rho^2_2 \pow_2'}\right)^2}{\N_1 + \N + (1-\rho_1^2) \pow_1'
      + (1-\rho_2^2) \pow_2'}\right) 
\\ R_{2,\textnormal{ICL},1} & \leq &
  \frac{1}{4} \log\left(1+ \frac{ (1-\rho_2^2)
      \pow_2'}{\N_1+\N}\right)
  \\
R_{1,\textnormal{ICL},1} + R_{2,\textnormal{ICL},1} & \leq & \frac{1}{4}
  \log\left(1+ \frac{\pow_1'+\pow_2' +
      2\sqrt{ \pow_1'\pow_2' \rho^2_1
        \rho^2_2}}{\N_1+\N}\right)
  \\
  R_{1,\textnormal{ICL},2} & \leq & \frac{1}{4} \log\left(1+ \frac{
      (1-\rho_1^2)
      \pow_1'}{\pow_1'\frac{\N}{ \pow_1'+\N}+\N}\right) \\
  R_{1,\textnormal{ICL},2} & \leq & \frac{1}{4} \log\left(1+ \frac{
      (1-\rho_1^2) \pow_1'}{\N_2+\N} \right) +\frac{1}{4}\log\left(1+
    \frac{\left(\sqrt{ \rho^2_1 \pow_1'}+\sqrt{ \rho^2_2
          \pow_2'}\right)^2} {\N_2 + \N + (1-\rho_1^2) \pow_1' +
       (1-\rho_2^2) \pow_2'
    }\right)\\
  R_{2,\textnormal{ICL},2} &\leq & \frac{1}{4} \log\left(1+ \frac{
      (1-\rho_2^2) \pow_2'}{\N_2+\N}\right)\\
 R_{1,\textnormal{ICL},2} + R_{2,\textnormal{ICL},2} & \leq& \frac{1}{4}
  \log\left(1+ \frac{\pow_1'+ \pow_2' +
      2\sqrt{ \pow_1'\pow_2' \rho^2_1
        \rho^2_2}}{\N_2+\N}\right)
\end{IEEEeqnarray}
\end{subequations}
\hrulefill
\end{figure*}
where 
\begin{IEEEeqnarray*}{rCl}
\N_1&\triangleq&\pow_1-P_1'+ \pow_2-P_2', \\
\N_2&\triangleq&\frac{( \pow_1-P_1')(\pow_2-P_2') +\N}{(\pow_1-P_1'+
  \pow_2-P_2'+\N)}+ (P_2-P_2') \nonumber \\ & &+ 2
\sqrt{\frac{(\pow_1-P_1')^2}{(P_1-P_1'
    +\N)} \frac{ (\pow_2-P_2')^2}{(\pow_1-P_1'+
    \pow_2-P_2'+\N)}} .
\end{IEEEeqnarray*}  
\end{proposition}
\begin{proof}
  The rate region is achieved by modifying the rate-splitting scheme
  for noisy feedback in Section~\ref{sec:CarleialInter} so as to apply
  also for perfect partial feedback (see Section~\ref{sec:ext_par}),
  and by choosing the parameters of the concatenated scheme as
  $\eta=2$ and as described in Remark~\ref{rem:partialfb}. The proof
  follows by accordingly combining Corollary~\ref{th:co2b} and the
  rate constraints which arise from the decodings in Carleial's
  variation of the Cover-Leung scheme.  Again, a genie-aided argument
  is used in the analysis.  The details are omitted.
\end{proof}

\begin{remark}
  In the case of perfect partial feedback, for all channel parameters
  $P_1,P_2,N>0$, the achievable regions by Carleial \cite{carleial82}
  and Willems et al.  \cite{willemsvandermeulenschalkwijk83-2}
  (Appendices~\ref{sec:Carleial} and \ref{sec:Willems}) correspond to
  the Cover-Leung region $\set{R}_{\textnormal{CL}}(P_1,P_2,N)$ (see,
  e.g., the explanation in \cite[Section~II-C]{carleial82}).  Since
  irrespective of $P_1,P_2,N>0$, the Cover-Leung region is contained
  in the two achievable regions in
  Propositions~\ref{prop:CarleialRSsimple} and \ref{prop:CarleialRS},
  we conclude that Propositions~\ref{prop:CarleialRSsimple} and
  \ref{prop:CarleialRS} include also Carleial's and Willems et al.'s
  regions for perfect partial feedback.
\end{remark}

\subsection{Simple Scheme}\label{sec:simple_par}
If in the simple scheme for noisy feedback in Section~\ref{sec:simple}
the parameter $b_1$ is restricted to be 0, then the scheme applies also
to noisy partial feedback. In particular, in this case it
achieves all nonnegative rate pairs $(R_1,R_2)$ that satisfy
\begin{IEEEeqnarray*}{rCl}
  R_1 & \leq & \frac{1}{4}\log \left( 1+
  \frac{a_{1,1}^2}{N}+\frac{a_{2,2}^2}{b_2^2\sigma_2^2 +N}  \right) \\
  R_2 & \leq & \frac{1}{4}\log \left( 1+ \frac{a_{2,1}^2}{N} +
  \frac{a_{2,2}^2}{b_2^2 \sigma_2^2+N} \right)\\
  R_1+R_2 & \leq & \frac{1}{4}\log\Bigg(1+
    \frac{a_{1,1}^2+a_{2,1}^2}{N} + \frac{a_{1,2}^2+a_{2,2}^2}{b_2^2
  \sigma_2^2+ N}  \nonumber \\ & & \hspace{2.2cm}+\frac{(a_{1,1} a_{2,2}-a_{2,1}a_{1,2})^2}{N(b_2^2 \sigma_2^2+ N)}\Bigg),
\end{IEEEeqnarray*}
for some choice of parameters $a_{1,1},a_{1,2},a_{2,1},a_{2,2},b_2$
satisfying
\begin{IEEEeqnarray*}{rCl}
a_{1,1}^2+ a_{1,2}^2 \leq 2 P_1, 
\end{IEEEeqnarray*}
and 
\begin{IEEEeqnarray*}{rCl}
a_{2,1}^2+ (a_{2,2}- b_2 a_{2,1})^2 + b_2^2(a_{1,1}^2+ N+  \sigma_2^2) \leq 2 P_2.
\end{IEEEeqnarray*}

The simple scheme for noisy partial feedback is included as a special
case in the concatenated scheme for noisy partial feedback described
in the next-following section. However, the simple scheme suffices to
prove Corollaries~\ref{th:co2} and \ref{th:co2b} and
Theorem~\ref{th:partial_R_ach}.


\subsection{Concatenated Scheme}\label{sec:conc_par}

\subsubsection{Scheme}\label{sec:conc_par_des}
If in the concatenated scheme for noisy feedback in
Section~\ref{sec:gen} the parameter $\B_1$ is restricted to be the
all-zero matrix, then this scheme applies also to noisy partial
feedback. In this case, applying the inner encoders with parameters
$\eta, \vect{a}_1,\vect{a}_2, \B_1=\mat{0}, \B_2$, and $\mat{D}$
induces a ``new'' MAC $\xi_1,\xi_2\mapsto (\hat{\Xi}_1,\hat{\Xi}_2)$ of
channel law
\begin{IEEEeqnarray}{rCl}\label{eq:newdecpartial}
\begin{pmatrix} \hat{\Xi}_1 \\ \hat{\Xi}_2 \end{pmatrix} = \mat{A}_{\textnormal{P}}
\begin{pmatrix} \xi_1 \\\xi_2 \end{pmatrix} + \vect{T}_{\textnormal{P}},
\end{IEEEeqnarray} 
where the $2\times 2$ matrix $\mat{A}_{\textnormal{P}}$ is given by
\begin{IEEEeqnarray}{rCl}
\mat{A}_{\textnormal{P}}  & = & \mat{D}\left( \I_{\eta} -\B_2
  \right)^{-1} \Ab;\label{eq:chmatP}
\end{IEEEeqnarray}
where $\Ab$ is defined as in \eqref{eq:Ab}; and where the noise vector $\vect{T}_{\textnormal{P}}$ is a zero-mean
bivariate Gaussian 
\begin{IEEEeqnarray}{rCl}\label{eq:noisevecP}
\vect{T}_{\textnormal{P}} & = & \mat{D} \left( \I_{\eta} -\B_2
  \right)^{-1} \left( \B_2 \vect{W}_2 + \vect{Z}\right). \IEEEeqnarraynumspace
\end{IEEEeqnarray}
Defining the $2 \times \eta$ matrix
\begin{equation}\label{eq:DCP}
\mat{C}_{\textnormal{P}} \eqdef \mat{D}\left( \I_{\eta} -\B_2
  \right)^{-1},
\end{equation}
the channel matrix in \eqref{eq:chmatP} and the noise
vector in \eqref{eq:noisevecP} can be expressed as
\begin{IEEEeqnarray}{rCl}
\mat{A}_{\textnormal{P}} & = & \mat{C}_{\textnormal{P}}\Ab, \label{eq:chmatnewP}\\
\vect{T}_{\textnormal{P}} & = & \C_{\textnormal{P}} \left( \B_2
  \vect{W}_2 +\vect{Z}\right).\label{eq:noisevecnewP}
\end{IEEEeqnarray}
For fixed $\eta$ and $\B_2$ the mapping \eqref{eq:DCP} from $\mat{D}$ to
$\C_{\textnormal{P}}$ is one-to-one, and thus we can parameterize our
concatenated scheme for noisy partial feedback by the parameters $\eta, \vect{a}_1,\vect{a}_2,\B_2,\C_{\textnormal{P}}$.

Specializing also the power constraints~\eqref{eq:powergen} to the choice $\B_1=\mat{0}$ and to noisy
partial feedback we see that only parameters $\eta,\vect{a}_1,\vect{a}_2$, and
$\B_2$ satisfying \eqref{eq:powerpart} are allowed.

\subsubsection{Choice of Parameters}
\label{sec:choicepartial}

In the following we describe guidelines on how to choose the
parameters of the concatenated scheme for noisy partial feedback. The
guidelines parallel the guidelines presented in
Section~\ref{sec:choicenoisy} for noisy feedback. Similarly, the
proofs why some of these guidelines are optimal parallel those in
Section~\ref{sec:choicenoisy} and are omitted.

Let $P_1,P_2,N>0$,
$\sigma_2^2\geq 0$ be given, and for the purpose of description replace
the symbols $\xi_1$ and $\xi_2$ fed to the inner encoders by the
independent standard Gaussians $\Xi_1$ and $\Xi_2$.

We start with the matrix $\C_{\textnormal{P}}$. Given parameters $\eta,\vect{a}_1,\vect{a}_2,\B_2$ the matrix
$\C_{\textnormal{P}}$ should be chosen as
$\C_{\textnormal{P}}=\mat{C}_{\textnormal{P,LMMSE}}$, where
\begin{equation}\label{eq:LMMSEmatP}
\mat{C}_{\textnormal{P,LMMSE}} \triangleq \trans{\Ab} \left( \Ab \trans{\Ab} + N\I_\eta +
  \sigma_2^2\B_2\trans{\B}_2\right)^{-1}.
\end{equation}
 The matrix
$\mat{C}_{\textnormal{P,LMMSE}}$ in \eqref{eq:LMMSEmatP} is called the
LMMSE-estimation matrix, since by~\eqref{eq:newdecpartial},
\eqref{eq:chmatnewP}, and \eqref{eq:noisevecnewP}, choosing
$\C_{\textnormal{P}}=\mat{C}_{\textnormal{P,LMMSE}}$ implies:
\begin{equation*}
\begin{pmatrix} \hat{\Xi}_1 \\\hat{\Xi}_2\end{pmatrix} =\E{
  \begin{pmatrix} {\Xi}_1 \\{\Xi}_2\end{pmatrix}\bigg| Y_1,\ldots,
  Y_\eta}.
\end{equation*}
Choosing $\C_{\textnormal{P}}=\mat{C}_{\textnormal{P,LMMSE}}$ is
optimal in the sense that the corresponding region
$\RatePpar{\sigma_2^2,N}{\eta,\vect{a}_1,\vect{a}_2,\B_2,\mat{C}_{\textnormal{P,LMMSE}}}$
contains all regions
$\RatePpar{\sigma_2^2,N}{\vect{a}_1,\vect{a}_2,\B_2,\mat{C}_{\textnormal{P}}}$
corresponding to other choices of the parameter $\C_{\textnormal{P}}$.
Choosing
$\mat{C}_{\textnormal{P}}=\mat{U}\mat{C}_{\textnormal{P,LMMSE}}$ for
some non-singular 2-by-2 matrix $\mat{U}$ is also optimal, and for
$\eta=2$ choosing $\C_{\textnormal{P}}$ as any non-singular matrix is
optimal.

We next consider the choice of the parameters
$\vect{a}_1,\vect{a}_2,\B_2$, and first focus on the special case of perfect
partial feedback.  For perfect partial feedback the parameters
$\vect{a}_1,\vect{a}_2,\B_2$ should be chosen so that the inputs
produced by Inner Encoder~2 correspond to scaled versions of the
LMMSE-estimation errors of $\Xi_2$ when observing the past feedback
outputs. Thus, for $\ell\in\{1,\ldots ,n\}$, they should satisfy
\begin{equation}\label{eq:LMMSEX2f}
X_{2,\ell} = \pi_{2,\ell} \left( \Xi_2 - \E{\Xi_2| Y_1,\ldots,
    Y_{\ell-1}}\right),
\end{equation}
for some real numbers $\pi_{2,1},\ldots, \pi_{2,\eta}$. Otherwise
there exists a choice of parameters satisfying \eqref{eq:LMMSEX2f}
that---with an appropriate choice of the matrix
$\mat{C}_{\textnormal{P}}$---strictly improves on the original choice,
i.e., corresponds to a larger region than the original choice.

A similar choice for noisy partial feedback is not optimal, and it
seems analytically infeasible to determine the optimal choice of the
parameters $\vect{a}_1,\vect{a}_2,\B_1,\B_2$.  However, it is easily
seen that for noisy partial feedback the parameters
$\vect{a}_1,\vect{a}_2,\B_1,\B_2$ should be chosen so that both power
constraints \eqref{eq:power1part} and \eqref{eq:power2part} are
satisfied with equality; otherwise there exists a choice of 
parameters satisfying \eqref{eq:power1part} and \eqref{eq:power2part}
that strictly improves on the original choice.

In Remark~\ref{rem:partialfb}, we present for every $\eta\in\Nat$ a
specific (suboptimal) choice of the parameters
$\vect{a}_1,\vect{a}_2,\B_2,$ and $\C_{\textnormal{P}}$. For this
specific choice, the parameter $\C_{\textnormal{P}}$ is the
LMMSE-estimation matrix, the parameters $\eta,\vect{a}_1,\vect{a}_2,\B_2$ satisfy the
power constraints \eqref{eq:power1part} and \eqref{eq:power2part} with
equality, and when specialized to perfect partial feedback
$\vect{a}_1,\vect{a}_2,\B_2$ satisfy~\eqref{eq:LMMSEX2}.  We present
the corresponding achievable region for $\eta=2$ and equal powers,
i.e., $P_1=P_2=P$, in Corollary~\ref{th:co2} and for $\eta=2$ and
perfect partial feedback in Corollary~\ref{th:co2b}.

 \subsection{Extensions of Concatenated Scheme}\label{sec:ext_par}
The schemes in Sections~\ref{sec:ext} apply also to noisy
partial feedback, if the parameter $\B_1$ is restricted to be the
all-zero matrix, and
if Carleial's variation of the Cover-Leung scheme is specialized to
noisy partial feedback. For more details see
Sections~\ref{sec:appPerPar1} and \ref{sec:appPerPar2}.

\subsection{Proofs}

\subsubsection{Proof of Theorem~\ref{th:C_unsym}}\label{sec:useful}

To prove Part~2)~we distinguish between the case of equal powers and
of unequal powers.  In the case of equal powers, $P_1=P_2=P$, we
consider the achievable region in Corollary \ref{th:co2}, and notice
that, irrespective of $P, N>0$ and $\sigma_2^2\geq 0$, the RHS of the
sum-rate constraint \eqref{eq:eqpow12} is smaller than the sum of the
RHSs of the single-rate constraints \eqref{eq:eqpow1} and
\eqref{eq:eqpow2}. Thus, for equal powers the achievable region in
Corollary~\ref{th:co2} is a pentagon (and not a rectangle) and there
exist achievable pairs $(R_1,R_2)$ of sum-rate equal to the RHS of
\eqref{eq:eqpow12}, which is larger than
$\frac{1}{2}\log\left(1+\frac{2P}{N} \right)$. This concludes the
proof in the case of equal powers.

To prove Part~2) in the case of unequal powers, $P_1\neq P_2$, we use
the following rate-splitting/time-sharing strategy. We assume
$P_1>P_2$; the case $P_1<P_2$ can analogously be treated.
Transmitter~1 splits its message $M_1$ into two independent
submessages: submessage $M_{1,1}$ of rate $R_{1,1}$
and submessage $M_{1,2}$ of rate $R_{1,2}$. During a
fraction of time $\frac{P_1-P_2}{P_1+P_2}$ Transmitter~1 sends Message
$M_{1,2}$ using an optimal no-feedback scheme of power $(P_1+P_2)$
while Transmitter~2 is quiet. During the remaining fraction of time
$\frac{2P_2}{P_1+P_2}$ Transmitters 1 and 2 use equal powers
$\frac{P_1+P_2}{2}$ to send messages $M_{1,1}$ and $M_2$ with the
concatenated scheme in Section~\ref{sec:gen}. Choosing the parameters
of the concatenated scheme as proposed in
Remark~\ref{rem:partialfb}, by
Corollary~\ref{th:co2} (where we replace $P$ by $\frac{P_1+P_2}{2}$)
and by the capacity of a AWGN single-user channel, the
described rate-splitting/time-sharing scheme achieves the rate pair
$(R_1=R_{1,1}+R_{1,2},
R_2)$  where $R_{1,1}, R_{1,2}$, and $R_2$ are given by Equations
\eqref{eq:ratessplitted} on top of this page. 
\begin{figure*}[!t]
\begin{subequations}\label{eq:ratessplitted}
\begin{IEEEeqnarray}{rCl} 
R_{1,1} & = & \frac{P_1-P_2}{2(P_1+P_2)}\log
\left(1 +\frac{P_1+P_2}{N}\right),\\
R_2= R_{1,2}& =& \frac{P_2}{2(P_1+P_2)} \log \left( 1+  \frac{P_1+P_2}{N}\right)
 \nonumber \\& & + \frac{P_2}{2(P_1+P_2)}\log \Bigg( 1+{\left(\sqrt{
      1+\frac{\frac{P_1+P_2}{2}(\frac{P_1+P_2}{2}+N+\sigma_2^2)}{\left(\frac{P_1+P_2}{2}+N+\sigma_2^2+\frac{P_1+P_2}{2N}\sigma_2^2\right)^2}}-1\right)} \frac{\frac{1}{2}(P_1+P_2)^2}{(P_1+P_2+N)} \nonumber \\ & & \hspace{4.2cm}  
+\frac{\left(\frac{P_1+P_2}{2}\right)\left(\frac{P_1+P_2}+N\right)}{(P_1+P_2+N)^2\left(\frac{P_1+P_2}{2}+N+\sigma_2^2+\frac{P_1+P_2}{2N}\sigma_2^2  \right)}
\nonumber \\ & & \hspace{8.5cm} \cdot \frac{(P_1+P_2+N+\sigma_2^2)
}{\left(P_1+P_2+N+\sigma_2^2+\frac{P_1+P_2}{2N}\sigma_2^2\right) }
\Bigg)\nonumber \\
\end{IEEEeqnarray}
\end{subequations}
\end{figure*}
The proof of \eqref{eq:incnoisy} follows then by noting that for every $P_1,P_2,N>0$ and every
$\sigma_2^2 \geq 0$ the rate pair
$(R_1,R_2)$ has a sum-rate which is
strictly larger than
$\frac{1}{2}\log\left(1+\frac{P_1+P_2}{N}\right)$, and therefore lies
strictly outside the no-feedback capacity region
$\capaMAC(P_1,P_2,N)$.

\subsubsection{Proof of Theorem~\ref{th:partial_R_ach}}\label{sec:proof_partial}

We consider an AWGN MAC with powers $P_1=1, P_2=5$,
noise variance $N=5$, and with perfect partial feedback. We prove the theorem by showing that for this
channel the rate point $(\bar{R}_1,\bar{R}_2)$,
\begin{IEEEeqnarray*}{rCl}
\bar{R}_{1} & =&\frac{1}{4}\log \left(\frac{7}{5}
\right),%
 \\
\bar{R}_2&=&\frac{1}{4}\log\left( 3 + \frac{3}{7}
  + \frac{2}{7}\sqrt{\frac{11}{6}}\right), %
\end{IEEEeqnarray*}
---which by Corollary~\ref{th:co2b} is achievable---lies 
outside the Cover-Leung region $\set{R}_{\textnormal{CL}}(P_1,P_2,N)$.
This implies that the capacity region
$\capa_{\textnormal{PerfectPartialFB}}(P_1,P_2,N)$ is strictly
larger than the Cover-Leung region
$\set{R}_{\textnormal{CL}}(P_1,P_2,N)$ for $P_1=1$ and $P_2=N=5$ .

Before starting with the proof, we have a closer
look at the region $\set{R}_{\textnormal{CL}}(P_1,P_2,N)$ and show the
following lemma.
\begin{lemma}\label{prop:CL}
For $P_1,P_2,N>0$ and  for every $\rho_1\in[0,1)$ which satisfies
\begin{equation}\label{eq:condrho1}
\frac{P_2}{N}\geq \frac{\rho_1^2}{1-\rho_1^2},
\end{equation}
the rate point
$(R_1(\rho_1),R_2(\rho_1))$ given by 
\begin{IEEEeqnarray}{lCl}
R_1(\rho_1) & = & \frac{1}{2} \log\left( 1+
  \frac{P_1\left(1-\rho_1^2\right)}{N}\right), \label{eq:R1rho1}
\end{IEEEeqnarray}
and by Equation~\eqref{eq:R2rho1} on top of the next page.
\begin{figure*}
\begin{IEEEeqnarray}{rCl}
R_2(\rho_1)& =& \max_{\rho_2\in[0,1]} \Bigg\{\min\Bigg\{ \frac{1}{2}\log\Bigg( 1+
  \frac{P_2\left(1-\rho_2^2\right)}{N}\Bigg), 
    \frac{1}{2}\log\Bigg(
  \frac{ P_1+P_2+2\sqrt{P_1 P_2 }\rho_1
  \rho_2+N}{P_1\left(1-\rho_1^2\right)+N}\Bigg)
  \Bigg\}\Bigg\}, \label{eq:R2rho1}
\end{IEEEeqnarray}
\hrulefill
\end{figure*}
lies on the boundary of $\set{R}_{\textnormal{CL}}(P_1,P_2,N)$ in the sense
that for every $\epsilon>0$ 
\begin{IEEEeqnarray*}{rCl}%
\left(R_1(\rho_1),R_{2}(\rho_1)+\epsilon\right)\;\notin\;
\set{R}_{\textnormal{CL}}(P_1,P_2,N).
\end{IEEEeqnarray*}
\end{lemma}
\begin{proof} As a first step we examine Expression~\eqref{eq:R2rho1}
  and characterize $R_2(\rho_1)$ more explicitly. To this end, we
  consider a fixed $\rho_1\in[0,1]$ that satisfies
  \eqref{eq:condrho1}.  Then, we notice that in the minimization in
  \eqref{eq:R2rho1} the first term is strictly decreasing in
  $\rho_2\in[0,1]$ whereas the second term is strictly increasing in
  $\rho_2$.  Also, for $\rho_2=1$ the first term in the maximization
  in \eqref{eq:R2rho1} is smaller than the second term, whereas by
  Condition \eqref{eq:condrho1} for $\rho_2=0$ the second term is smaller.
  Thus, for fixed $\rho_1\in[0,1]$ satisfying \eqref{eq:condrho1} the maximum
  in \eqref{eq:R2rho1} is achieved when both terms are equal, i.e.,
  for $\bar{\rho}_2$ given by the unique solution in $[0,1]$ to
\begin{IEEEeqnarray*}{rCl}
\lefteqn{ \frac{1}{2}\log\left( 1+
  \frac{P_2\left(1-\rho_2^2\right)}{N}\right) } \qquad \\ &=&\frac{1}{2}\log\left(
  \frac{ P_1+P_2+2\sqrt{P_1 P_2 }\rho_1
  \rho_2+N}{P_1\left(1-\rho_1^2\right)+N}\right).
\end{IEEEeqnarray*}
This implies that the rate pair $(R_1(\rho_1),R_2(\rho_1))$ satisfies
all three rate constraints defining the rectangle
$\set{R}_{\textnormal{CL}}^{(\rho_1,\bar{\rho_2})}(P_1,P_2,N)$ with
equality, i.e.,
\begin{subequations}\label{eq:RCLeq}
\begin{IEEEeqnarray}{rCl}
R_1(\rho_1) & =  & \frac{1}{2} \log\left( 1+
  \frac{P_1\left(1-\rho_1^2\right)}{N}\right),\label{eq:R1CLeq} \\
R_2(\rho_1) & =& \frac{1}{2}\log\left( 1+
  \frac{P_2\left(1-\left(\bar{\rho}_2\right)^2\right)}{N}\right),\IEEEeqnarraynumspace\label{eq:R2CLeq}\end{IEEEeqnarray}
and 
\begin{IEEEeqnarray}{rCl}
\lefteqn{R_1(\rho_1)+R_2(\rho_1)} \nonumber\quad 
 \\ & = &  \frac{1}{2}\log\left(1+
  \frac{ P_1+P_2+2\sqrt{P_1 P_2 }\rho_1
  \bar{\rho}_2}{N}\right).\IEEEeqnarraynumspace\label{eq:R12CLeq}
\end{IEEEeqnarray}
\end{subequations}
Hence, $(R_1(\rho_1),R_2(\rho_1))$ is the dominant corner point of the
rectangle
$\set{R}_{\textnormal{CL}}^{(\rho_1,\bar{\rho}_2)}(P_1,P_2,N)$, and
for all $\epsilon>0$ the rate point
$(R_1(\rho_1),R_2(\rho_1)+\epsilon)$ lies outside the rate
region $\set{R}_{\textnormal{CL}}^{(\rho_1,\bar{\rho}_2)}(P_1,P_2,N)$.
In the remaining we show that the rate point
$(R_1(\rho_1),R_2(\rho_1))$ also lies outside the regions
$\set{R}_{\textnormal{CL}}^{(\rho_1',\rho_2')}(P_1,P_2,N)$ for all
$\rho_1',\rho_2'\in[0,1]$ not equal to the pair $
({\rho}_1,\bar{\rho}_2)$, and therefore also
$(R_1(\rho_1),R_2(\rho_1)+\epsilon)$ lies outside these regions for
every $\epsilon>0$. This will then conclude the proof of the lemma.
We distinguish the following three cases: 1) $\rho_1'>\rho_1$ and 
$\rho_2'$ arbitrary; 2) $\rho_1'\leq \rho_1$ and
$\rho_2'>\bar{\rho}_2$; and 3)
$\rho_{1}'\leq \rho_1$ and  $\rho_2'<\bar{\rho}_2$. In case 1) the rate
point $(R_1(\rho_1),R_2(\rho_1))$ lies outside the region
$\set{R}_{\textnormal{CL}}^{(\rho_1',\rho_2')}(P_1,P_2,N)$ because
$R_1(\rho_1)$ violates the single-rate constraint, see \eqref{eq:R1CL}
and \eqref{eq:R1CLeq}. Similarly, in case 2) the rate point lies
outside the region
$\set{R}_{\textnormal{CL}}^{(\rho_1',\rho_2')}(P_1,P_2,N)$ because in
this case $R_{2}(\rho_1)$ violates the single-rate constraint, see
\eqref{eq:R2CL} and \eqref{eq:R2CLeq}. Finally, in case 3) the rate
point lies outside the region
$\set{R}_{\textnormal{CL}}^{(\rho_1',\rho_2')}(P_1,P_2,N)$ because the
product $\rho_1' \cdot \rho_2'$ is strictly smaller than the product
$\rho_1\cdot \bar{\rho}_2$, and thus the sum
$R_{1}(\rho_1)+R_{2}(\rho_1)$ violates the sum-rate constraint, see
\eqref{eq:R12CL} and \eqref{eq:R12CLeq}. \end{proof}

We are now ready to prove that the achievable rate point
$(\bar{R}_1,\bar{R}_2)$ lies outside the Cover-Leung region
$\set{R}_{\textnormal{CL}}(P_1,P_2,N)$.  To this end, we choose
$\rho_1=\sqrt{6-\sqrt{35}}$ and notice that it satisfies Condition
\eqref{eq:condrho1} for $P_2=N=5$. Hence, Lemma~\ref{prop:CL} applies
and the rate point $(R_{1}^{\set{B}},R_{2}^{\set{B}})$,
\begin{IEEEeqnarray}{rCl}
R_1^{\set{B}} & \eqdef& \frac{1}{2} \log\left( \sqrt{\frac{7}{5}}\right),\nonumber
\\
R_2^{\set{B}} & \eqdef&  \max_{\rho_2\in[0,1]}\Bigg\{ \min\Bigg\{ \frac{1}{2}\log\left( 1+
  \left(1-\rho_2^2\right)\right), \nonumber \\ & & \hspace{1.55cm} 
    \frac{1}{2}\log\left( \frac{11
    +2\sqrt{5 (6-\sqrt{35}) } \rho_2}{ \sqrt{35}}\right)
\Bigg\}\Bigg\}\nonumber \\\label{eq:R2b}
\end{IEEEeqnarray}
lies on the boundary of the Cover-Leung region
$\set{R}_{\textnormal{CL}}(P_1,P_2,N)$, and in particular for every
$\epsilon>0$ the rate point
$(R_{1}^{\set{B}},R_{2}^{\set{B}}+\epsilon)$ lies strictly outside the
Cover-Leung region $\set{R}_{\textnormal{CL}}(P_1,P_2,N)$. Since
\begin{equation*}
R_{1}^{\set{B}}=\bar{R}_1, 
\end{equation*}
in order to show that the rate point $(\bar{R}_1,\bar{R}_2)$ lies
strictly outside $ \set{R}_{\textnormal{CL}}(P_1,P_2,N)$ it suffices to show that
\begin{equation}\label{eq:ineq}
R_{2}^{\set{B}}<\bar{R}_2. 
\end{equation}
To prove \eqref{eq:ineq} we could compute $\bar{\rho}_2$---the value of
$\rho_2$ which maximizes \eqref{eq:R2b}---and $R_{2}^{\set{B}}$ and then
check Condition \eqref{eq:ineq}. However, it is
easier---and sufficient---to show that for all $\rho_2\in[0,1]$ either
\begin{IEEEeqnarray}{rCl}\label{eq:casea}
 \frac{1}{2}\log\left(1+ \frac{P_2(1-\rho_2^2)}{N}\right)& = &
 \frac{1}{2}\log\left(2-\rho_2^2 \right)\nonumber \\ & <& \bar{R}_2
\end{IEEEeqnarray}
or 
\begin{IEEEeqnarray}{rCl}\label{eq:caseb}
\lefteqn{\frac{1}{2}\log\left( \frac{P_1+P_2+2\sqrt{P_1P_2}\rho_1
    \rho_2+N}{P_1(1-\rho_1^2)+N}\right)}\nonumber\qquad  \\  & = &\frac{1}{2}\log\left(
    \frac{11+2\sqrt{5(6-\sqrt{35})}\rho_2}{\sqrt{35}}\right)\nonumber
    \\  & <&\bar{R}_2.
\end{IEEEeqnarray}
To this end, note first that the LHS of \eqref{eq:casea} is
 decreasing in $\rho_2\in[0,1]$, and therefore for all
$\sqrt{\frac{1}{7}}\leq \rho_2\leq 1$ it follows that 
\begin{equation*}
  \frac{1}{2}\log\left(2-\rho_2^2 \right) \leq
  \frac{1}{4}\log\left(3+\frac{3}{7}+ \frac{1}{49}\right) < \bar{R}_2.
\end{equation*}
On the other hand, the LHS of \eqref{eq:caseb} is 
increasing in $\rho_2$, and thus for all $0\leq \rho_2 \leq
\sqrt{\frac{1}{7}}$ 
\begin{IEEEeqnarray*}{rCl}
\lefteqn{\frac{1}{2}\log\left(
  \frac{11+2\sqrt{5(6-\sqrt{35})}\rho_2}{\sqrt{35}}\right)} \\
&\leq &
\frac{1}{4}\log\Bigg( 3+\frac{3}{7}+\frac{1}{35}  +
  \frac{44\sqrt{\frac{5}{7}\left(6-\sqrt{35}\right)}}{35}\\ & & \hspace{4cm}+\frac{\frac{4}{7}\left(30-5\sqrt{35}\right)}{35} \Bigg)
\\
&= & \frac{1}{4}\log\Bigg( 3 + \frac{3}{7}+ 
 \frac{2}{7}\Bigg( \frac{1}{10}+
  \frac{22}{\sqrt{35}}\sqrt{(6-\sqrt{35})} \\ & &\hspace{4.3cm} +
  \frac{12}{7}-2\sqrt{\frac{5}{7}}\Bigg)\Bigg) \\ &<& \bar{R}_2,
\end{IEEEeqnarray*}
where the inequality follows because 
\begin{equation*}
 \frac{1}{10}+ \frac{22}{\sqrt{35}}\sqrt{(6-\sqrt{35})} +
 \frac{12}{7}-2\sqrt{\frac{5}{7}} < \sqrt{\frac{11}{6}}.
\end{equation*}
This concludes the proof of the theorem.

  \section{ Noisy Feedback with Receiver Side-Information}\label{sec:noisysi}

  For the setup with receiver side-information we present a new
  achievable region (Section~\ref{sec:ressi}) and a scheme that
  achieves this region (Section~\ref{sec:conc_si}). The proposed
  scheme is an extension of the concatenated scheme for noisy feedback
  in Section~\ref{sec:gense} and exploits the side-information at the
  receiver. The simple scheme (Section~\ref{sec:simple}) and the
  extended schemes (Sections~\ref{sec:ext}) can be analogously
  extended to this setup with receiver side-information. For brevity,
  we omit the description of these latter extensions.

\subsection{Results}\label{sec:ressi}

\begin{definition}
Let $\eta$ be a positive integer;$\vect{a}_1,\vect{a}_2$ be
$\eta$-dimensional vectors; $\B_1,\B_2$ be strictly lower-triangular
$\eta\times\eta$ matrices; and $\C_\textnormal{SI}$ be a $2\times\eta$
matrix. Depending on the matrix $\mat{C}_{\textnormal{SI}}$ the rate region
$\RatePsi{N,\KW}{\eta,\vect{a}_1,\vect{a}_2,\B_1,\B_2,\C_\textnormal{SI}}$
is defined as follows:
\begin{itemize}
\item If the product $\mat{C}_{\textnormal{SI}}\trans{\C}_{\textnormal{SI}}$ is
  nonsingular,\footnote{Whenever $\eta \in \Nat$ is larger than 1,
    there is no loss in optimality in restricting attention to
    matrices $\C_{\textnormal{SI}}$ so that
    $\C_{\textnormal{SI}}\trans{\C}_{\textnormal{SI}}$ is
    nonsingular.} then
  $\RatePsi{N,\sigma_2^2}{\eta,\vect{a}_1,\vect{a}_2,\B_2,\C_{\textnormal{SI}}}$ is defined as the set of all rate pairs $(R_1,R_2)$ satisfying
\begin{subequations}\label{eq:Rgen1SI}
\begin{IEEEeqnarray}{rCl}
  R_1 & \leq & \frac{1}{2\eta}\log  \frac{\left|\C_{\textnormal{SI}} \left( \vect{a}_1
        \trans{\vect{a}_1} +N\I_\eta\right)
        \trans{\C}_{\textnormal{SI}}\right|}{N \left| \C_{\textnormal{SI}} \trans{\C}_{\textnormal{SI}}  \right|} ,
  \label{eq:R1gen1SI} \IEEEeqnarraynumspace
  \\
  R_2 & \leq  &\frac{1}{2\eta}\log \frac{\left| \C_{\textnormal{SI}}
      \left(   \vect{a}_2 \trans{\vect{a}_2} + N\I_{\eta}\right)
      \trans{\C}_{\textnormal{SI}}   \right|}{ N\left|
      \C_{\textnormal{SI}}  \trans{\C}_{\textnormal{SI}}  \right|} ,\label{eq:R2gen1SI} \IEEEeqnarraynumspace
\\
R_1+R_2 & \leq  &
\frac{1}{2\eta}\log \frac{\left| \C_{\textnormal{SI}}  \left(  \Ab \trans{\Ab}+ N
      \I_{\eta}\right)  \trans{\C}_{\textnormal{SI}}   \right|}{ N\left|
    \C_{\textnormal{SI}}    \trans{\C}_{\textnormal{SI}}  \right| } , \label{eq:R12gen1SI} \IEEEeqnarraynumspace
\end{IEEEeqnarray}
\end{subequations}
  where $\Ab$ is defined in \eqref{eq:Ab}.
\item If $\C_{\textnormal{SI}}\trans{\C}_{\textnormal{SI}}$ is singular
  but $\C_{\textnormal{SI}}\neq \mat{0}$, then
  $\RatePsi{N,\sigma_2^2}{\eta,\vect{a}_1,\vect{a}_2,\B_2,\C_{\textnormal{SI}}}$
  is defined as the set of all rate pairs $(R_1,R_2)$ satisfying 
\eqref{eq:Rgen1SI} when the $2\times \eta$ matrix
  $\C_{\textnormal{SI}}$ is replaced by the $\eta$-dimensional
  row-vector obtained by choosing one of the non-zero rows of
  $\C_{\textnormal{SI}}$.
\item If $\C_{\textnormal{SI}}=\mat{0}$, then
  $\RatePsi{N,\sigma_2^2}{\eta,\vect{a}_1,\vect{a}_2,\B_2,\C_{\textnormal{SI}}}$
  is defined as the set containing only the origin.
\end{itemize}
\end{definition}
(An alternative formulation of the region
$\RatePsi{\nch}{\eta,\vect{a}_1,\vect{a}_2,\B_1,\B_2,\C_\textnormal{SI}}$ is
presented in Section~\ref{sec:achaltnoisysi}.)
\begin{definition}\label{def:RSI}
Define the region
\begin{IEEEeqnarray}{lCl}\label{eq:siunion}
\lefteqn{\Ratesi{P_1,P_2,N,\KW}} \quad \nonumber \\   \eqdef \cl{
  \bigcup 
  \RatePsi{N,\KW}{\eta,\vect{a}_1,\vect{a}_2,\B_1,\B_2,\C_\textnormal{SI}}},\IEEEeqnarraynumspace
\end{IEEEeqnarray}
where the union is over all tuples $(\eta,\vect{a}_1,\vect{a}_2, \B_1,
\B_2,\C_\textnormal{SI})$ satisfying the trace constraints
\eqref{eq:powergen}.
\end{definition}
\begin{theorem}[Noisy Feedback with Receiver Side-Information]\label{th:SIgen} 
  The capacity region $\capa_{\textnormal{NoisyFBSI}}(P_1,P_2,N,\KW)$ of the
  two-user AWGN MAC with noisy feedback where the receiver is
  cognizant of the realization of the feedback-noise sequences
  contains the rate region $\Ratesi{P_1,P_2,N,\KW}$, i.e., 
\begin{IEEEeqnarray*}{rCl}
\lefteqn{\capa_{\textnormal{NoisyFBSI}}(P_1,P_2,N,\KW)}\qquad \\ & \supseteq& \Ratesi{P_1,P_2,N,\KW}.
\end{IEEEeqnarray*}
\end{theorem}
\begin{proof} The achievability result is based on the concatenated scheme in Section~\ref{sec:conc_des_si}. It is obtained from Theorem~\ref{th:general} by setting
  $\sigma_1^2=\sigma_2^2=0$ in the rate expressions in 
  \eqref{eq:Rgen1} (but not in the power constraints
  \eqref{eq:powergen}). The reason why in
 \eqref{eq:Rgen1} we may set
  $\sigma_1^2=\sigma_2^2=0$ is because in the scheme in Section~\ref{sec:conc_des_si}, prior to the decoding, the
  receiver subtracts off the influence of the feedback-noise sequences
  $\{W_{1,t}\}$ and $\{W_{2,t}\}$. The details of the proof are
  omitted.
\end{proof}
 
\begin{remark}
  Evaluating the achievable region $\Ratesi{P_1,P_2,N,\KW}$ seems to
  be difficult even numerically.  More easily computable (but possibly
  smaller) achievable regions are obtained by taking the union on the
  RHS of \eqref{eq:siunion} only over a subset of the parameters
  $\para_{\textnormal{SI}}$ satisfying \eqref{eq:powergen}. In Section~\ref{sec:parametersSI} we present
  such a subset of parameters and its corresponding achievable region
  (Corollary~\ref{th:R_ach_SI}). In Section~\ref{sec:choice_si} ahead we
  present more general guidelines on how to choose the parameters
  $\para_{\textnormal{SI}}$ for noisy feedback with receiver
  side-information.
\end{remark}

\subsection{Concatenated Scheme}\label{sec:conc_si}

\subsubsection{Scheme}\label{sec:conc_des_si}

In this section we extend our concatenated scheme to noisy feedback
with receiver side-information.
We use the same outer code and the same inner encoders as in the
setting without side-information. The difference is only in the inner
decoder. Thus, when fed the pair of symbols $(\xi_1,\xi_2)$, the
inner encoders produce, as before, sequences of channel inputs
\begin{equation} \label{eq:encgennSI}
\vect{X}_{\nu} =\vect{a}_\nu \xi_\nu+ \mat{B}_{\nu} \vect{V}_{\nu}, \qquad \nu\in\{1,2\},
\end{equation}
where $\vect{X}_{\nu}\eqdef \trans{(X_{\nu,1},\ldots, X_{\nu,\eta})}$,
$\vect{V}_{\nu}\eqdef \trans{(V_{\nu,1},\ldots,V_{\nu,\eta-1})}$, and
where $\vect{a}_1,\vect{a}_2$ are $\eta$-dimensional vectors and
$\B_1,\B_2$ are strictly lower-triangular $\eta\times \eta$ matrices
satisfying the power constraints \eqref{eq:powergen}. But, we modify the structure of the inner decoder so
that it computes the estimates
$(\hat{\Xi}_1,\hat{\Xi}_2)$ not
only as a function of the output sequence but also of the
feedback-noise sequences. Again, we choose a linear mapping, i.e., for
$\vect{Y}\eqdef\trans{(Y_{1},\ldots, Y_\eta)}$,
$\vect{W}_{1}\eqdef\trans{(W_{1,1},\ldots, W_{1,\eta})}$, and
$\vect{W}_{2}\eqdef\trans{(W_{2,1},\ldots, W_{2,\eta})}$, the inner
decoder computes
\begin{equation*}
\begin{pmatrix} \hat{\Xi}_1\\ \hat{\Xi}_2 \end{pmatrix} =
\mat{D}_{0} \vect{Y} + \mat{D}_{1}
\vect{W}_{1}  + \mat{D}_{2} \vect{W}_{2},
\end{equation*}
for $2\times \eta$ matrices $\mat{D}_0,\mat{D}_1,\mat{D}_2$ of our choice. Given
$\vect{a}_1,\vect{a}_2,\B_1,\B_2,$ and $\mat{D}_0$ an optimal choice for
the matrices $\mat{D}_1$ and $\mat{D}_2$ subtracts off the contributions to
$\mat{D}_0\vect{Y}$ that come about from the feedback-noise
sequences, i.e., an optimal choice of $\mat{D}_1$ and $\mat{D}_2$ satisfies
\begin{subequations}\label{eq:optD}
\begin{IEEEeqnarray}{rCl}\label{eq:optD1} 
\mat{D}_1 &= & -\mat{D}_0 (\I_\eta - (\B_1+\B_2))^{-1} \B_1, \\
\mat{D}_2 & = &- \mat{D}_0 (\I_\eta - (\B_1+\B_2))^{-1} \B_2.
\label{eq:optD2}
\end{IEEEeqnarray}
\end{subequations}
Such a choice leads to the following description of the
``new'' MAC $\xi_1,\xi_2\mapsto (\hat{\Xi}_1, \hat{\Xi}_2)$:
\begin{IEEEeqnarray}{rCl}\label{eq:newdecSI}
\begin{pmatrix} \hat{\Xi}_1 \\ \hat{\Xi}_2 \end{pmatrix} = \mat{A}_{\textnormal{SI}}
\begin{pmatrix} \xi_1 \\\xi_2 \end{pmatrix} + \vect{T}_{\textnormal{SI}},
\end{IEEEeqnarray} 
where the $2\times 2$ matrix $\mat{A}$ is given by
\begin{IEEEeqnarray}{rCl}\label{eq:chmatSI}
\mat{A}_{\textnormal{SI}}  & =  & \mat{D}_0\left( \I_{\eta} - (\B_1+\B_2)\right)^{-1} \Ab,
\end{IEEEeqnarray}
where $\Ab$ is defined as in \eqref{eq:Ab}, and where the noise vector $\vect{T}$ is a zero-mean bivariate Gaussian
\begin{IEEEeqnarray}{rCl}\label{eq:noisevSI}
\vect{T}_{\textnormal{SI}} & = & \mat{D}_0 \left( \I_{\eta}-  (\B_1+\B_2)\right)^{-1}  \vect{Z}.
 \IEEEeqnarraynumspace
\end{IEEEeqnarray}
In the following we shall always assume that $\mat{D}_1$ and
$\mat{D}_2$ are optimally chosen so that the ``new'' MAC is given
by \eqref{eq:newdecSI}--\eqref{eq:noisevSI}.
We define the $2\times \eta$ matrix
\begin{equation}\label{eq:DCSI}
\mat{C}_\textnormal{SI}  \eqdef  \mat{D}_0 \left( \I_{\eta} - (\B_1+\B_2)\right)^{-1},
\end{equation}
and hence $\mat{A}_{\textnormal{SI}}$ in \eqref{eq:chmatSI} and the
noise vector $\vect{T}_{\textnormal{SI}}$ in
\eqref{eq:noisevSI} can be expressed as 
\begin{IEEEeqnarray}{rCl}
\mat{A}_{\textnormal{SI}}&=& \C_{\textnormal{SI}} \Ab, \label{eq:chmatnewSI}\\
\vect{T}_{\textnormal{SI}} & = & \C_{\textnormal{SI}} \vect{Z}.\label{eq:noisevecnewSI}
\end{IEEEeqnarray}
For fixed $\eta, \B_1, \B_2$ the mapping \eqref{eq:DCSI} from
$\mat{D}_0$, to $\C_{\textnormal{SI}}$ is one-to-one, and thus we can
parameterize our concatenated scheme for noisy feedback with receiver
side-information by
$\eta,\vect{a}_1,\vect{a}_2,\B_1,\B_2,\mat{C}_{\textnormal{SI}}$.

All parameters $\eta,\vect{a}_1,\vect{a}_2,\B_1,\B_2$ that satisfy the
power constraints \eqref{eq:powergen} are allowed.

\subsubsection{Choice of Parameters}\label{sec:choice_si}
As in the previously studied setups we present guidelines on how to
choose the parameters $\eta, \vect{a}_1,\vect{a}_2,\B_1,\B_2,
\C_{\textnormal{SI}}$ of the concatenated scheme. The guidelines
parallel the guidelines for noisy feedback and noisy partial feedback
in Sections~\ref{sec:choicenoisy} and~\ref{sec:choicepartial};
likewise, also the proofs of optimality parallel the proofs in
Section~\ref{sec:choicenoisy} and are omitted. 

Let $P_1,P_2,N>0$ and  $\KW \succeq 0$ be given, and for the purpose of
describing our guidelines replace the symbols $\xi_1,\xi_2$ fed to the
inner encoders by the independent standard Gaussians $\Xi_1,\Xi_2$.

We first present the optimal choice of the parameter
$\C_{\textnormal{SI}}$. Given $\eta,\vect{a}_1,\vect{a}_2,\B_1,\B_2$
the parameter $\C_\textnormal{SI}$ should be chosen as
$\C_\textnormal{SI}=\C_{\textnormal{SI,LMMSE}}$, where
\begin{equation}\label{eq:LMMSESIf}
\C_{\textnormal{SI,LMMSE}} = \trans{\Ab} \left( \Ab \trans{\Ab}
  + N \I_\eta\right)^{-1},
\end{equation}
since the corresponding achievable region contains all regions
corresponding to other choices of the matrix $\C_{\textnormal{SI}}$.
The matrix $\C_\textnormal{SI,LMMSE}$ is called the
\emph{LMMSE-estimation matrix with side-information}, since by
\eqref{eq:newdecSI}, \eqref{eq:chmatnewSI}, and
\eqref{eq:noisevecnewSI} the choice in \eqref{eq:LMMSESIf}---combined
with the optimal choices of $\mat{D}_1$ and $\mat{D}_2$ defined by
\eqref{eq:optD} and \eqref{eq:DCSI}---implies that
\begin{equation*}
\begin{pmatrix}\hat{\Xi}_1\\\hat{\Xi}_2\end{pmatrix} =
\E{\begin{pmatrix} \Xi_1\\\Xi_2\end{pmatrix} \bigg|
  Y_1^{\eta},W_{1}^{\eta},W_{2}^{\eta}}.
\end{equation*}
Obviously, also choosing $\mat{C}_{\textnormal{SI}}=\mat{U}
\C_{\textnormal{SI,LMMSE}}$ for some non-singular 2-by-2 matrix
$\mat{U}$ is optimal, and for $\eta=2$ choosing $\C_{\textnormal{SI}}$
as any non-singular matrix is optimal.

We next consider the choice of parameters
$\vect{a}_1,\vect{a}_2,\B_1,\B_2$ and focus on the following two special
cases: 
\begin{itemize}
\item[a)] $\eta\in \Nat$ is arbitrary and
$\varrho =1$, i.e., the feedback noises are perfectly correlated,
\item[b)] $\eta=2$ and $\varrho \in [-1,1)$ arbitrary.
\end{itemize}
In these cases, given parameter $\eta\in \Nat$,  the parameters $\vect{a}_1,\vect{a}_2,\B_1,\B_2$ should be
chosen so that the inner encoders produce
\begin{subequations}\label{eq:LMMSEXV}
\begin{equation}\label{eq:LMMSEXV1}
X_{1,\ell} = \pi_{1,\ell} \left( \Xi_1 - \E{\Xi_1| V_{1}^{\ell-1}
    }\right), \qquad \ell\in\{1,\ldots, \eta\},
\end{equation}
and 
\begin{equation}\label{eq:LMMSEXV2}
X_{2,\ell} = \pi_{2,\ell} \left( \Xi_2 - \E{\Xi_2| V_{2}^{\ell-1}}\right), \qquad \ell\in\{1,\ldots, \eta\},
\end{equation}
\end{subequations}
for some real numbers $\pi_{1,1},\ldots, \pi_{1,\eta}$ and
$\pi_{2,1},\ldots, \pi_{2,\eta}$. Otherwise, there exists a choice of parameters
$\eta, \vect{a}_1,\vect{a}_2,\B_1,\B_2,\C_{\textnormal{SI}}$ of the form \eqref{eq:LMMSEXV} that strictly improves on the original choice.

In general, it seems difficult to determine the optimal choice of the
parameters $\vect{a}_1,\vect{a}_2,\B_1,\B_2$. However, it is easily
proved that the parameters $\eta, \vect{a}_1,\vect{a}_2,\B_1,\B_2,
\C_{\textnormal{SI}}$ should be chosen so as to satisfy the power
constraints \eqref{eq:powergen2} and \eqref{eq:powergen1} with
equality; otherwise there exists a choice of parameters satisfying
\eqref{eq:powergen2} and \eqref{eq:powergen1} with equality that
strictly improves on the original choice.

In Section~\ref{app:si_nonsym}, we present a specific choice of the
parameters $\vect{a}_1,\vect{a}_2,\B_1,\B_2,\C_\textnormal{SI}$ that
guarantees that $\C_\textnormal{SI}$ is the LMMSE-estimation matrix with
side-information, the power constraints \eqref{eq:powergen} are
satisfied with equality, and $\eta, \vect{a}_1,\vect{a}_2,\B_1,\B_2$
satisfy \eqref{eq:LMMSEXV} for all $\eta\in\Nat$ and $\varrho \in
[-1,1]$. We present the corresponding achievable region in
Corollary~\ref{th:R_ach_SI}.

\section{Summary}\label{sec:summary}
We have studied four different kinds of two-user AWGN MACs with
imperfect feedback:
\begin{itemize}
\item \emph{noisy feedback}, where the feedback links to both
  transmitters are
  corrupted by AWGN;
\item  \emph{noisy partial feedback}, where one transmitter has
  noisy feedback and the other no feedback;
\item \emph{perfect partial feedback}, where one transmitter has
  noise-free feedback and the other no feedback; and
\item \emph{noisy feedback with receiver side-information}, where both
  transmitters have noisy feedback and the feedback-noise sequences
  are perfectly known to the receiver.
\end{itemize}
For each of these settings we have presented a coding scheme (called
concatenated scheme) with general parameters, and we have stated the
corresponding achievable regions (Theorem~\ref{th:general},
Theorem~\ref{th:generalP}, Corollary~\ref{th:ppf}, and
Theorem~\ref{th:SIgen}). We have  improved the concatenated
scheme by rate-splitting it either with a simple no-feedback scheme or
with Carleial's version of the Cover-Leung scheme. The achievable regions
corresponding to these improvements are stated in
Proposition~\ref{prop:0} (noisy feedback) and
Propositions~\ref{prop:CarleialRSsimple} and~\ref{prop:CarleialRS}
(perfect partial feedback).

The two achievable regions for noisy feedback in
Theorem~\ref{th:general} and Proposition~\ref{prop:0} exhibit the
following three properties: 1. They are monotonically decreasing in
the feedback-noise covariance matrix with respect to the Loewner order
(Propositions~\ref{eq:monotonicity} and \ref{prop:monotonicity2}). 2.
They are continuous in the transmit-powers
(Propositions~\ref{eq:monotonicity} and \ref{prop:monotonicity2}). 3.
They converge to Ozarow's perfect-feedback regions when the feedback
noise-variances tend to 0, irrespective of the feedback-noise
correlations (Propositions~\ref{prop:maxsumratesigma} and
\ref{prop:splitmaxsumratesigma}).

We have further presented guidelines for choosing the parameters of
our concatenated schemes (Sections~\ref{sec:choicenoisy},
\ref{sec:choicepartial}, and \ref{sec:choice_si}), and have suggested
(suboptimal) specific choices of the parameters (Sections
\ref{sec:enc}, \ref{sec:powerstar}, and \ref{sec:parametersSI}). The
achievable regions corresponding to these specific choices are
presented in Corollary~\ref{th:co1}, Corollary~\ref{th:co2},
Corollary~\ref{th:co2b}, Corollary~\ref{cor:ach},
Corollary~\ref{cor:alte}, Remark~\ref{def:remarkperf}, and
Corollary~\ref{th:R_ach_SI}.

These achievable regions---combined with the previously described properties
of the achievable regions for noisy feedback in
Theorem~\ref{th:general} and Proposition~\ref{prop:0}---allowed us to
infer:
\begin{itemize}
\item [1)] Feedback---no matter how noisy---is
strictly better than no feedback. I.e., irrespective of the
feedback-noise variances, the capacity region with one or two noisy
feedback links is strictly larger than the no-feedback capacity region
(Theorems~\ref{th:C_unsym} and~\ref{th:C_unsympar}). 
\item[2)]The noisy-feedback capacity region converges
  to the perfect-feedback capacity region as the feedback-noise
  variances on both links tend to 0---irrespective of the
  feedback-noise correlations (Theorem~\ref{th:C_sym}). 
\item [3)] The Cover-Leung region in general does
not equal capacity for perfect partial feedback channels
(Theorem~\ref{th:partial_R_ach}).  This answers in the negative a
question posed by van der Meulen in \cite{vandermeulen87}.
\end{itemize}

\section*{Acknowledgment}
We gratefully acknowledge helpful discussions with Michael Gastpar and Gerhard Kramer.

\begin{appendix}

\subsection{Carleial's region}\label{sec:Carleial}
Carleial proved  the achievability result for the AWGN MAC
  with noisy feedback in Theorem~\ref{th:car} ahead \cite{carleial82}.

\begin{definition}\label{eq:defCl}
  Define the rate region $\set{R}_{\textnormal{Car}}(P_1,P_2,N,\sigma_1^2,\sigma_2^2)$ as
  the set of all rate pairs $(R_1,R_2)$ which for some nonnegative
  numbers $R_{1,0},R_{1,1}$ summing to $R_1$, for some nonnegative
  numbers $R_{2,0},R_{2,2}$ summing to $R_2$, and for some choice of
  parameters $\alpha_1,\alpha_2,\beta_1,\beta_2,\upsilon \in [0,1]$
  satisfy the 13 conditions \eqref{eq:Car} shown on top of the next
  page. 
\begin{figure*}
\begin{subequations}\label{eq:Car}
\begin{IEEEeqnarray}{lCl}
R_{1,0} & \leq &  \frac{1}{2}\log\left(1+\frac{\alpha_1 \bar{\beta}_1
    P_1}{\alpha_1 \beta_1 P_1 + N + \sigma_2^2} \right)\label{eq:Car1} \\
R_{2,0} & \leq &  \frac{1}{2}\log\left(1+\frac{\alpha_2 \bar{\beta}_2
    P_2}{\alpha_2 \beta_2 P_2 + N + \sigma_1^2} \right) \label{eq:Car2}\\
R_{1,1} & \leq & \frac{1}{2}\log\left(1+\frac{\alpha_1 \beta_1 P_1}{N}
    \right), \label{eq:Car3}\\ 
R_{2,2} & \leq & \frac{1}{2} \log \left(1+ \frac{\alpha_2 \beta_2
    P_2}{N} \right)\label{eq:Car4}\\
{R_{1,0}+R_{20}} & \leq  & \frac{1}{2}\log \left(1+
    \frac{\alpha_1\bar{\beta}_1 P_1 + \alpha_2 \bar{\beta}_2 P_2}{N}
    \right)+\l{\bar{\alpha}_1P_1+ \bar{\alpha}_2 P_2 + 2
    \sqrt{\bar{\alpha}_1 \bar{\alpha}_2 P_1 P_2 } }{N + \alpha_1 P_1
    +\alpha_2 P_2}\label{eq:Car5}\\
{R_{1,0}+R_{2,2}} & \leq & \l{\alpha_1 \bar{\beta}_1 P_1 + \alpha_2
    \beta_2 P_2 }{N} + \l{\upsilon\left(\bar{\alpha}_1P_1+ \bar{\alpha}_2 P_2 + 2
    \sqrt{\bar{\alpha}_1 \bar{\alpha}_2 P_1 P_2 }\right) }{N + \alpha_1 P_1
    +\alpha_2 P_2}\label{eq:Car7}\\
{R_{2,0}+R_{1,1}} & \leq & \l{\alpha_1 \beta_1 P_1+\alpha_2
    \bar{\beta}_2 P_2 }{N}+ \l{\bar{\upsilon}\left(\bar{\alpha}_1P_1+ \bar{\alpha}_2 P_2 + 2
    \sqrt{\bar{\alpha}_1 \bar{\alpha}_2 P_1 P_2 }\right) }{N + \alpha_1 P_1
    +\alpha_2 P_2}\label{eq:Car8}\\
{R_{1,1}+R_{2,2}}& \leq & \l{\alpha_1 \beta_1 P_1+\alpha_2 \beta_2
    P_2}{N} \label{eq:Car10}\\
{R_{1}+R_{2,0}} & \leq & \l{\alpha_1 P_1 +\alpha_2
    \bar{\beta}_2 P_2 }{N}+\l{\bar{\alpha}_1P_1+ \bar{\alpha}_2 P_2 + 2
    \sqrt{\bar{\alpha}_1 \bar{\alpha}_2 P_1 P_2 } }{N + \alpha_1 P_1
    +\alpha_2 P_2}\label{eq:Car11}\\
{R_{1,0}+R_{2}} & \leq & \l{\alpha_1\bar{\beta}_1 P_1 +\alpha_2
     P_2 }{N}+\l{\bar{\alpha}_1P_1+ \bar{\alpha}_2 P_2 + 2
    \sqrt{\bar{\alpha}_1 \bar{\alpha}_2 P_1 P_2 } }{N + \alpha_1 P_1
    +\alpha_2 P_2}\label{eq:Car12}\\
{R_{1}+R_{2,2}}& \leq & \l{\alpha_1 P_1 +\alpha_2
    \beta_2 P_2 }{N} +\l{\upsilon\left(\bar{\alpha}_1P_1+ \bar{\alpha}_2 P_2 + 2    \sqrt{\bar{\alpha}_1 \bar{\alpha}_2 P_1 P_2 } \right)}{N + \alpha_1 P_1
    +\alpha_2 P_2}\label{eq:Car13}\\
{R_{1,1}+R_{2}} & \leq & \l{\alpha_1 \beta_1 P_1 +\alpha_2
     P_2 }{N}  +\l{\bar{\upsilon}\left(\bar{\alpha}_1P_1+ \bar{\alpha}_2 P_2 + 2
    \sqrt{\bar{\alpha}_1 \bar{\alpha}_2 P_1 P_2 }\right) }{N + \alpha_1 P_1
    +\alpha_2 P_2}\label{eq:Car14}\\
{R_{1}+R_{2}} & \leq &
    \l{P_1+P_2+2\sqrt{\bar{\alpha}_1 \bar{\alpha}_2 P_1 P_2}}{N}\label{eq:Car15}\IEEEeqnarraynumspace
\end{IEEEeqnarray}
\end{subequations}
\hrulefill
\end{figure*}
where for $x\in[0,1]$ we define $\bar{x}\eqdef (1-x)$.
\end{definition}  
\begin{theorem}[Carleial \cite{carleial82}]\label{th:car}
Consider an AWGN MAC with noisy feedback of transmit powers $P_1,P_2$,
  noise variance $N$, and feedback-noise covariance matrix $\KW =\begin{pmatrix}\sigma_1^2 &\varrho
  \sigma_1 \sigma_2 \\ \varrho \sigma_1 \sigma_2 & \sigma_2^2
  \end{pmatrix}$. Irrespective of the noise correlation $\varrho\in[-1,1]$, the region 
  $\set{R}_{\textnormal{Ach}}(P_1,P_2,N, \sigma_1^2,\sigma_2^2)$ is
  achievable for this channel,
  i.e., 
\begin{equation*}
\set{R}_{\textnormal{Ach}}(P_1,P_2,N, \sigma_1^2,\sigma_2^2) \subseteq
\capa_{\textnormal{NoisyFB}}(P_1,P_2,N, \KW).
\end{equation*}
\end{theorem}

\begin{lemma}\label{th:usecar}
The rate region
$\set{R}_{\textnormal{Car}}(P_1,P_2,N,\sigma_1^2, \sigma_2^2)$ collapses to the
no-feedback capacity region $\capaMAC(P_1,P_2,N)$ when the
feedback-noise variances $\sigma_1^2,\sigma_2^2$ exceed a
certain threshold depending on the parameters $P_1,P_2,$ and $N$. 
In particular, 
\begin{equation*}
\set{R}_{\textnormal{Car}}(P_1,P_2,N,\sigma_1^2, \sigma_2^2)
=\capaMAC(P_1,P_2,N) , \end{equation*}
for $ \sigma_1^2 \geq
P_1\left(\frac{3}{2}+\frac{P_2}{N}\right)$ and $ 
\sigma_2^2 \geq
P_2\left(\frac{3}{2}+\frac{P_1}{N}\right)$.
\end{lemma}
\begin{proof}
  For all values of $\sigma_1^2,\sigma_2^2$ the region
  $\set{R}_{\textnormal{Car}}(P_1,P_2,N,\sigma_1^2,\sigma_2^2)$
  trivially contains the no-feedback capacity region
  $\capaMAC(P_1,P_2,N)$ because the region obtained by substituting
  $\alpha_1=\alpha_2=\beta_1=\beta_2=1$ into
  \eqref{eq:Car} coincides with
  $\capaMAC(P_1,P_2,N)$. Thus, it remains to prove that
  $\set{R}_{\textnormal{Car}}(P_1,P_2,N,\sigma_1^2,\sigma_2^2)$ is included in
  $\capaMAC(P_1,P_2,N)$ for all $\sigma_1^2,\sigma_2^2$ exceeding some
  threshold depending on $P_1,P_2$, and $N$.
  
  To this end, we choose $\sigma_1^2,\sigma_2^2>0$ and we fix a rate
  pair $(R_1,R_2)$ in
  $\set{R}_{\textnormal{Car}}(P_1,P_2,N,\sigma_1^2,\sigma_2^2)$. We then fix parameters
  $\alpha_1,\alpha_2,\beta_1,\beta_2,\upsilon\in [0,1]$, two nonnegative
  numbers $R_{1,0}$ and $R_{1,1}$ summing to $R_1$, and two nonnegative
    numbers $R_{2,0}$ and $R_{2,2}$ summing to $R_2$ so that
    Constraints \eqref{eq:Car} are satisfied. We show in the
  following that if $\sigma_1^2,\sigma_2^2>0$ are sufficiently large,
  then $(R_1,R_2)$ lies in $\capaMAC(P_1,P_2,N)$. 

By \eqref{eq:Car1} and \eqref{eq:Car3} the rate
  $R_1$ satisfies
\begin{IEEEeqnarray}{rCl}\label{eq:quad1}
R_{1} & \leq & \l{\alpha_1 \bar{\beta}_1 P_1}{\alpha_1
    \beta_1 P_1 +N+ \sigma_2^2 } \nonumber \\ & & +\l{\alpha_1 \beta_1 P_1 }{N}\nonumber\\
& \leq & \l{P_1}{N},
\end{IEEEeqnarray}
and by \eqref{eq:Car2} and \eqref{eq:Car4} the rate $R_2$ satisfies
\begin{IEEEeqnarray}{rCl}\label{eq:quad2}
R_{2} & \leq & \l{\alpha_2 \bar{\beta}_2 P_2}{\alpha_2
    \beta_2 P_2 +N+ \sigma_1^2 } \nonumber \\ & & +\l{\alpha_2 \beta_2P_2 }{N}\nonumber \\
& \leq & \l{P_1}{N}, 
\end{IEEEeqnarray}
Furthermore, by \eqref{eq:Car1}, \eqref{eq:Car2}, and \eqref{eq:Car10}
the sum of the rates $R_1+R_2$ satisfies Inequality
\eqref{eq:Rsumth} on the next page.
\begin{figure*}
\begin{IEEEeqnarray}{rCl}\label{eq:Rsumth}
  R_1+R_2  & \leq &
\frac{1}{2}\log\left(1+\frac{\alpha_1 \bar{\beta}_1    P_1}{\alpha_1
    \beta_1 P_1 + N + \sigma_2^2} \right) +
\frac{1}{2}\log\left(1+\frac{\alpha_2 \bar{\beta}_2 P_2}{\alpha_2
    \beta_2 P_2 + N + \sigma_1^2} \right)\nonumber \\ & &  +
\l{\alpha_1 \beta_1 P_1 +\alpha_2 \beta_2 P_2}{N}\nonumber \\
  &\leq &\frac{1}{2} \log \Bigg(1+\frac{\alpha_1 \beta_1 P_1+\alpha_2
      \beta_2 P_2}{N}   \nonumber \\
& & \phantom{ \frac{1}{2} \log \Bigg(1)} +\frac{\alpha_1 \bar{\beta}_1
     P_1}{N}\Bigg(\frac{N+\alpha_1 \beta_1 P_1 +\alpha_2 \beta_2 P_2}{\alpha_1 \beta_1 P_1
            +N+\sigma_2^2}+ \frac{(\alpha_1 \beta_1 P_1 +\frac{1}{2}N)\frac{\alpha_2
         \bar{\beta}_2 P_2}{\alpha_2 \beta_2 P_2+
            N+\sigma_1^2}}{\alpha_1 \beta_1 P_1
            +N+\sigma_2^2}\Bigg) \nonumber \\
& & \phantom{ \frac{1}{2} \log \Bigg(1)} +\frac{\alpha_2 \bar{\beta}_2
     P_2}{N}\Bigg(\frac{N+\alpha_1 \beta_1 P_1 +\alpha_2 \beta_2 P_2}{\alpha_2 \beta_2 P_2
            +N+\sigma_1^2}
+ \frac{(\alpha_2 \beta_2 P_2 +\frac{1}{2}N)\frac{\alpha_1
         \bar{\beta}_1 P_1}{\alpha_1 \beta_1 P_1+
            N+\sigma_2^2}}{\alpha_2 \beta_2 P_2
            +N+\sigma_1^2}\Bigg) \Bigg)
\end{IEEEeqnarray}
\end{figure*}
Notice that for $\sigma_1^2,\sigma_2^2$ larger than some threshold
depending on $P_1,P_2,N$---and in particular for $\sigma_1^2>
P_1\left( \frac{3}{2}+\frac{P_2}{N}\right)$ and $\sigma_2^2 >
P_2\left(\frac{3}{2}+\frac{P_1}{N}\right)$---irrespective of the chosen
parameters $\alpha_1,\alpha_2,\beta_1,\beta_2,\upsilon$:
\begin{IEEEeqnarray}{rCl}
\lefteqn{\frac{N+\alpha_1 \beta_1 P_1 +\alpha_2 \beta_2 P_2}{\alpha_1 \beta_1 P_1
            +N+\sigma_2^2}} \;\;\nonumber \\ & &  +\frac{
       (\alpha_1 \beta_1 P_1 +\frac{1}{2}N)(\alpha_2
         \bar{\beta}_2 P_2)}{(\alpha_2 \beta_2 P_2+
            N+\sigma_1^2)(\alpha_1 \beta_1 P_1
            +N+\sigma_2^2)}<1\IEEEeqnarraynumspace\label{eq:neq1}
\end{IEEEeqnarray}
and 
\begin{IEEEeqnarray}{rCl}
\lefteqn{\frac{N+\alpha_1 \beta_1 P_1+ \alpha_2 \beta_2 P_2}{\alpha_2 \beta_2 P_2
            +N+\sigma_1^2} } \nonumber\;\; \\ & & +\frac{
       (\alpha_2 \beta_2 P_2 +\frac{1}{2}N)(\alpha_1
         \bar{\beta}_1 P_1)}{(\alpha_1 \beta_1 P_1+
            N+\sigma_2^2)(\alpha_2 \beta_2 P_2
            +N+\sigma_1^2)}<1.\IEEEeqnarraynumspace\label{eq:neq2}
\end{IEEEeqnarray}
Thus, when $\sigma_1^2,\sigma_2^2$ exceed a certain threshold
depending on $P_1,P_2$, and $N$, the RHS of
\eqref{eq:Rsumth} is upper bounded by $\frac{1}{2}\log\left(1+\frac{P_1+P_2}{N}\right)$.
We conclude that when $\sigma_1^2,\sigma_2^2$ are sufficiently large,
then by~\eqref{eq:quad1}--\eqref{eq:neq2} the rate pair
$(R_1,R_2)$ satisfies
\eqref{eq:C_NOFB} and hence lies in the
no-feedback capacity region 
$\capaMAC(P_1,P_2,N)$. This concludes the proof.
\end{proof}

 \subsection{Willems et al.'s region}\label{sec:Willems}
Willems et al.  proved  an achievability result for
the discrete memoryless MAC with imperfect feedback \cite{willemsvandermeulenschalkwijk83-2}. The result
can easily be extended to the two-user AWGN MAC with noisy
feedback (Theorem~\ref{th:wil} ahead). 
\begin{definition}\label{def:Wl}
 Define the rate region
$\set{R}_{\textnormal{Wil}}(P_1,P_2,N,\sigma_1^2,\sigma_2^2)$ as the set
of all  rate pairs
$(R_1,R_2)$ which for some nonnegative numbers $R_{1,1}$ and $R_{1,2}$
summing to $R_1$, for some nonnegative numbers $R_{2,1}$ and $R_{2,2}$
summing to $R_2$, and for some parameters
$\delta_1,\delta_2, \rho_1,\rho_2\in[0,1]$ satisfy the following five constraints:
\begin{IEEEeqnarray*}{lCl}
R_{1,1} & \leq & \frac{1}{2} \log \left( 1+\frac{\delta_1
    P_1}{N}\right) \\
R_{1,0} & \leq & \frac{1}{2} \log \left(1+\frac{\bar{\delta}_1
    P_1(1-\rho_1^2)}{\delta_1 P_1+N+\sigma_2^2} \right)\\
R_{2,0} & \leq & \frac{1}{2}
    \log\left(1+\frac{\bar{\delta}_2P_2(1-\rho_2^2)}{\delta_2 P_2 +N +
    \sigma_1^2}\right)\\
R_{2,2} & \leq & \frac{1}{2}\log\left(1+\frac{\delta_2 P_2}{N} \right)
    \\ 
\lefteqn{R_{1,1}+R_{2,2}}\nonumber \\ &\leq & \frac{1}{2} \log \left( 1+ \frac{\delta_1
    P_1+\delta_2 P_2}{N}\right)\\
\lefteqn{R_1+R_2 }\\ &\leq & \frac{1}{2}
    \log\left(1+\frac{P_1+P_2+2\sqrt{\bar{\delta}_1\bar{\delta}_2
    P_1P_2}\rho_1 \rho_2}{N} \right).
\end{IEEEeqnarray*}
\end{definition}
\begin{theorem}[Willems et al. \cite{willemsvandermeulenschalkwijk83-2}]\label{th:wil}
Consider an AWGN MAC with noisy feedback of transmit powers $P_1,P_2$,
noise variance $N$, and 
feedback-noise covariance matrix  $\KW=\begin{pmatrix}\sigma_1^2 &\varrho
  \sigma_1 \sigma_2 \\ \varrho \sigma_1 \sigma_2 & \sigma_2^2
  \end{pmatrix}$. Irrespective of the feedback-noise
  correlation $\varrho\in[-1,1]$, the region
$\set{R}_{\textnormal{Wil}}(P_1,P_2,N,\sigma_1^2,\sigma_2^2)$ is 
achievable for this channel, i.e., 
\begin{equation*}
\set{R}_{\textnormal{Wil}}(P_1,P_2,N, \sigma_1^2,\sigma_2^2) \subseteq
\capa_{\textnormal{NoisyFB}}(P_1,P_2,N, \KW).
\end{equation*}
\end{theorem}
\begin{lemma}\label{eq:prop_useful}
The rate region
$\set{R}_{\textnormal{Wil}}(P_1,P_2,N,\sigma_1^2,\sigma_2^2)$ collapses to the
no-feedback capacity region $\capaMAC(P_1,P_2,N)$ when the
feedback-noise variances $\sigma_1^2,\sigma_2^2$ exceed a
certain threshold depending on the parameters $P_1,P_2,$ and $N$. 
In particular, 
\begin{equation*}
\set{R}_{\textnormal{Wil}}(P_1,P_2,N,\sigma_1^2,\sigma_2^2 )
=\capaMAC(P_1,P_2,N) , 
\end{equation*}
for $ \sigma_1^2 \geq
P_1\left(\frac{3}{2}+\frac{P_2}{N}\right)$ and $ 
\sigma_2^2 \geq
P_2\left(\frac{3}{2}+\frac{P_1}{N}\right).$
\end{lemma}
\begin{proof}
Follows along similar lines as the proof of Lemma~\ref{th:usecar} in
the previous appendix, and
is omitted. \end{proof}

\subsection{Optimality of LMMSE-Estimation Error
  Parameters for Perfect Feedback }\label{sec:LMMSEopt} We show that
in our concatenated scheme for perfect feedback it is optimal to
choose the parameters $\eta,\vect{a}_1,\vect{a}_2,\B_1,\B_2$ so that
the two inner encoders produce as their $\ell$-th channel inputs
scaled versions of the LMMSE-estimation errors when estimating the fed
symbols $\Xi_1$ and $\Xi_2$ based on the previous outputs $Y_1,\ldots,
Y_{\ell-1}$, see \eqref{eq:LMMSEX}.

\begin{proposition}\label{eq:propapp}
Assume that $\KW=\mat{0}$, i.e., 
perfect  feedback. If the parameters
$\eta,\vect{a}_1,\vect{a}_2,\B_1,\B_2,\C$ satisfy the power
constraints \eqref{eq:powergen} but not Conditions
\eqref{eq:LMMSEX}, then there exist
parameters
$\eta,{\vect{a}}^*_1,{\vect{a}}^*_2,{\B}^*_1,{\B}^*_2,{\C}^*$
satisfying both \eqref{eq:powergen} and
\eqref{eq:LMMSEX}, and 
\begin{IEEEeqnarray*}{rCl}
\lefteqn{\RateP{N, \mat{0}}{\para}} \qquad \\ & \subset& \RateP{N, \mat{0}}{\eta,{\vect{a}}_1^*,{\vect{a}}^*_2,{\B}^*_1,{\B}^*_2,{\C}^*}
\end{IEEEeqnarray*}
with the inclusion being strict.
\end{proposition}
The proof is given after the following lemma. 

\begin{lemma}\label{lem:app}
 Assume that $\KW=\mat{0}$, i.e.,
  assume perfect feedback. If the parameters
  $\eta, \vect{a}_1',\vect{a}_2',\B_1',\B_2' ,\C'$ satisfy
  \eqref{eq:powergen} but violate
  \eqref{eq:LMMSEX} then there exist parameters 
$\eta,\hat{\vect{a}}_1,\hat{\vect{a}}_2,\hat{\B}_1,\hat{\B}_2,\hat{\C}$ 
satisfying the following three conditions:
\begin{enumerate} 
\item\label{item:p1} the parameters 
  $\eta,\hat{\vect{a}}_1,\hat{\vect{a}}_2,\hat{\B}_1,\hat{\B}_2,\hat{\C}$
  satisfy 
  \eqref{eq:LMMSEX};
\item \label{item:p2} $\RateP{N,\mat{0}}{\eta,
    \vect{a}_1',\vect{a}_2',\B_1',\B_2' ,\C'}$\\ $= \RateP{N,\mat{0}}{\eta,
    \hat{\vect{a}}_1,\hat{\vect{a}}_2,\hat{\B}_1,\hat{\B}_2 ,\hat{\C}}$;
\item \label{item:p3}the parameters ${\eta,
    \hat{\vect{a}}_1,\hat{\vect{a}}_2,\hat{\B}_1,\hat{\B}_2
    ,\hat{\C}}$ satisfy  
    \eqref{eq:powergen2} and \eqref{eq:powergen1}, and at least one of them with strict inequality.
\end{enumerate}
\end{lemma}
\begin{proof}
Fix parameters $\eta, \vect{a}_1',\vect{a}_2',\B_1',\B_2' ,\C'$
satisfying \eqref{eq:powergen} but violating
\eqref{eq:LMMSEX}. Define the following new parameters.
\begin{itemize}
\item Let $\hat{\vect{a}}_1=\vect{a}_1'$ and $\hat{\vect{a}}_2=\vect{a}_2'$.
\item Let $\hat{\B}_1$ and $\hat{\B}_2$ be so that
  $\hat{\vect{a}}_1,\hat{\vect{a}}_2,\hat{\B}_1,\hat{\B}_2$ satisfy
  \eqref{eq:LMMSEX}. (Notice that given the 
  parameters $\hat{\vect{a}}_1$ and $\hat{\vect{a}}_2$ there exists
  exactly one choice of the parameters $\hat{\B}_1$ and $\hat{\B}_2$
  satisfying \eqref{eq:LMMSEX}. I.e., the
  scaling coefficients $\{\pi_{1,\ell}\}_{\ell=1}^\eta$ and
  $\{\pi_{2,\ell}\}_{\ell=1}^\eta$ in \eqref{eq:LMMSEX} are determined by
  $\hat{\vect{a}}_1$ and $\hat{\vect{a}}_2$.)
\item Let
$\hat{\C}=\C'$.
\end{itemize}
By construction, our choice ${
    \hat{\vect{a}}_1,\hat{\vect{a}}_2,\hat{\B}_1,\hat{\B}_2
    ,\hat{\C}}$ trivially satisfies
Condition~\ref{item:p1} in the lemma. Moreover, since for $\KW=\mat{0}$ the region
$\RateP{N,\mat{0}}{\para}$ 
depends  only on $\vect{a}_1,\vect{a}_2$, and $\C$ but not on $\B_1$
and $\B_2$, see Definition~\ref{def:Reta}, 
 the regions $\RateP{N,\mat{0}}{\eta,
  \vect{a}_1',\vect{a}_2',\B_1',\B_2' ,\C'}$ and $
\RateP{N,\mat{0}}{\eta,
  \hat{\vect{a}}_1,\hat{\vect{a}}_2,\hat{\B}_1,\hat{\B}_2 ,\hat{\C}}$
coincide. Thus also Condition~\ref{item:p2} is satisfied.

We are left with proving that the parameters $\hat{\vect{a}}_1,
\hat{\vect{a}}_2, \hat{\B}_1, \hat{\B}_2, \hat{\C}$ satisfy
also Condition~\ref{item:p3}. Before doing so, 
we introduce some helpful assumptions and
notation. Assume in the following that Inner Encoder~1 and Inner
Encoder~2 are fed the independent standard Gaussians $\Xi_1$ and
$\Xi_2$, respectively. Let $Y_1', \ldots, Y_\eta'$ denote the
$\eta$ channel outputs of the original MAC $x_1,x_2\mapsto Y$ when the
inner encoders use the parameters
$\eta,\vect{a}_1',\vect{a}_2',\B_1',\B_2',\C'$, and similarly, let
$\hat{Y}_1, \ldots, \hat{Y}_\eta$ denote the $\eta$ channel outputs of
the original MAC $x_1,x_2\mapsto Y$ when the inner encoders use
the parameters
$\eta,\hat{\vect{a}}_1,\hat{\vect{a}}_2,\hat{\B}_1,\hat{\B}_2,\hat{\C}$.
Also, let $a_{1,\ell}', a_{2,\ell}', \hat{a}_{1,\ell}$, and $\hat{a}_{2,\ell}$
denote the $\ell$-th entry of the vectors $\vect{a}_1',
\vect{a}_2',\hat{\vect{a}}_{1},$ and $\hat{\vect{a}}_2$, respectively, and let
$b_{1,\ell,j}', b_{2,\ell,j}', \hat{b}_{1,\ell,j},$ and $\hat{b}_{2,\ell,j}$
denote the row-$\ell$ column-$j$ entry of the matrices
$\B_1',\B_2',\hat{\B}_1,$ and $\hat{\B}_2$, respectively, for
$j,\ell\in\{1,\ldots, \eta\}$ and $\nu\in\{1,2\}$.

 Fix an $\ell \in \{1,\ldots,
\eta\}$. By the definition of LMMSE-estimation errors, for all 
$\nu\in\{1,2\}$ and all real
numbers $\{b_{\nu,\ell,j}\}_{j=1}^{\ell-1}$
\begin{IEEEeqnarray}{rCl}\label{eq:opt}
\lefteqn{\Var{\hat{a}_{\nu,\ell} \Xi_\nu - \sum_{j=1}^{\ell-1} b_{\nu,\ell,j}
  \hat{Y}_j}} \nonumber \qquad \\ & \geq& \Var{\hat{a}_{\nu,\ell} \Xi_\nu- \sum_{j=1}^{\ell-1} \hat{b}_{\nu,\ell,j}
  \hat{Y}_j},
\end{IEEEeqnarray}
with equality if, and only if, $b_{\nu,\ell,j}=\hat{b}_{\nu,\ell,j}$
for all $j\in\{1,\ldots, \ell-1\}$.  We would like to prove a similar
inequality to \eqref{eq:opt} but where in the RHS of \eqref{eq:opt}
the outputs $\{\hat{Y}_1,\ldots, \hat{Y}_{\ell-1}\}$ are replaced by
the outputs $\{Y_1',\ldots, Y_{\ell-1}'\}$. To this end, we notice
that since $\hat{\vect{{a}}}_1=\vect{a}_1'$ and
$\hat{\vect{a}}_2=\vect{a}_2'$ there exist real numbers
$\{b_{1,\ell,j}\}_{j=1}^{\ell-1}$ and
$\{b_{2,\ell,j}\}_{j=1}^{\ell-1}$ such that
\begin{equation*}
\left(\hat{a}_{\nu,\ell}\Xi_\nu- \sum_{j=1}^{\ell-1} {b}_{\nu,\ell, j}
\hat{Y}_{j}\right) = \left(a_{\nu,\ell}'\Xi_\nu- \sum_{j=1}^{\ell-1}
b_{\nu,\ell, j}'Y_{j}'\right)
\end{equation*}
with probability 1. 
Combining this observation with
Inequality~\eqref{eq:opt} the desired inequality follows:
\begin{IEEEeqnarray}{rCl}\label{eq:ineq1}
\lefteqn{
\Var{a_{\nu,\ell}'\Xi_\nu   - \sum_{j=1}^{\ell-1} b_{\nu,\ell,j}'
  Y_j'}} \nonumber \qquad \\  &
\geq & \Var{\hat{a}_{\nu,\ell}\Xi_\nu - \sum_{j=1}^{\ell-1} \hat{b}_{\nu,\ell,j}
  \hat{Y}_j},
\end{IEEEeqnarray}
with equality if, and only if,
\begin{equation}\label{eq:1l}
\left(a_{\nu,\ell}' \Xi_\nu - \sum_{j=1}^{\ell-1} b_{\nu,\ell,j}' Y_j'\right)= \left(\hat{a}_{\nu,\ell} \Xi_1- \sum_{j=1}^{\ell-1} \hat{b}_{\nu,\ell,j}
  \hat{Y}_j\right)
\end{equation}
with probability 1.
By~\eqref{eq:ineq1} and since the parameters $\eta,
\vect{a}_1',\vect{a}_2',\B_1',\B_2'$ satisfy the power constraints
\eqref{eq:powergen}, it further  follows that
also the parameters
$\eta,\hat{\vect{a}}_1,\hat{\vect{a}}_2,\hat{\B}_1,\hat{\B}_2$ satisfy
Constraints \eqref{eq:powergen}. Moreover,
since the pairs $(\B_1',\B_2')$ and $(\hat{\B}_1,\hat{\B}_2)$ differ,
not for all $\ell\in\{1,\ldots, \eta\}$ and all $\nu\in\{1,2\}$
equality \eqref{eq:1l} can hold and thus the parameters
$\eta,\hat{\vect{a}}_1,\hat{\vect{a}}_2,\hat{\B}_1,\hat{\B}_2$ satisfy
either \eqref{eq:powergen2} or \eqref{eq:powergen1} with strict
inequality. This concludes the proof of the lemma.
\end{proof}
\begin{proof}[Proof of Proposition~\ref{eq:propapp}] The proof uses Lemma~\ref{lem:app} twice.
  Fix parameters $\para$ satisfying \eqref{eq:powergen} but violating
  \eqref{eq:LMMSEX}. By Lemma~\ref{lem:app} there exist
  parameters
  $\eta,\hat{\vect{a}}_1,\hat{\vect{a}}_2,\hat{\B}_1,\hat{\B}_2,\hat{\C}$
  that satisfy \eqref{eq:LMMSEX} and
\begin{IEEEeqnarray}{rCl}\label{eq:thisisfirst}
\lefteqn{\RateP{\nch}{\para}}\quad \nonumber \\ & =& \RateP{\nch}{\eta,\hat{\vect{a}}_1,\hat{\vect{a}}_2,\hat{\B}_1,\hat{\B}_2,\hat{\C}},
\end{IEEEeqnarray}
and that satisfy \eqref{eq:powergen2} and \eqref{eq:powergen1},
whereby one of them with strict inequality.  Further, since the
parameters
$\eta,\hat{\vect{a}}_1,\hat{\vect{a}}_2,\hat{\B}_1,\hat{\B}_2,\hat{\C}$
satisfy either \eqref{eq:powergen2} or \eqref{eq:powergen1} with
strict inequality, there exist parameters $\eta,
\check{\vect{a}}_1,\check{\vect{a}}_2,\check{\B}_1,\check{\B}_2,\check{\C}$
that satisfy both \eqref{eq:powergen2} and \eqref{eq:powergen1} with
equality (but not necessarily \eqref{eq:LMMSEX}) and that correspond
to a strictly larger region (see Section~\ref{sec:choicenoisy}).
Thus, by \eqref{eq:thisisfirst}
\begin{IEEEeqnarray}{rCl}\label{eq:thisisit}
\lefteqn{\RateP{\nch}{\para} }\quad \nonumber \\ &\subset&  \RateP{\nch}{\eta,
\check{\vect{a}}_1,\check{\vect{a}}_2,\check{\B}_1,\check{\B}_2,\check{\C}}
\end{IEEEeqnarray}
with the inclusion being strict.

Applying Lemma~\ref{lem:app} again, this time to parameters $\eta,
\check{\vect{a}}_1,\check{\vect{a}}_2,\check{\B}_1,\check{\B}_2,\check{\C}$,
we conclude that there exists a choice of parameters
$\eta,\vect{a}_1^*,\vect{a}_2^*,\B_1^*,\B_2^*,\C^*$ satisfying both
\eqref{eq:powergen} and \eqref{eq:LMMSEX} and 
\begin{IEEEeqnarray*}{rCl}
\lefteqn{\RateP{\nch}{\eta,
\check{\vect{a}}_1,\check{\vect{a}}_2,\check{\B}_1,\check{\B}_2,\check{\C}}
} \quad \nonumber \\ &= &\RateP{\nch}{\eta,\vect{a}_1^*,\vect{a}_2^*,\B_1^*,\B_2^*,\C^*}.
\end{IEEEeqnarray*}
By \eqref{eq:thisisit} this implies
\begin{IEEEeqnarray*}{rCl}
\lefteqn{\RateP{\nch}{\para} } \quad \nonumber \\ &\subset& \RateP{\nch}{\eta,\vect{a}_1^*,\vect{a}_2^*,\B_1^*,\B_2^*,\C^*}
\end{IEEEeqnarray*}
with the inclusion being strict, which concludes the proof.
\end{proof}


\subsection{Alternative Formulation of Achievable Regions}

We derive an alternative formulation of the region achieved by our
concatenated scheme
$\RateP{\nch}{\eta,\vect{a}_1,\vect{a}_2,\B_1,\B_2,\C}$ when $\C$ is
chosen as the LMMSE-estimation matrix $\C_{\textnormal{LMMSE}}$ (as
defined in \eqref{eq:LMMSEmat}) and
$\eta,\vect{a}_1,\vect{a}_2,\B_1,\B_2$ are arbitrary.  Recall that
there is no loss in optimality in restricting attention to the choice
$\C=\C_{\textnormal{LMMSE}}$, see Section~\ref{sec:choicenoisy}.
Similarly, we derive an alternative formulation for the achievable region
$\RatePpar{N,\sigma_2^2}{\eta,\vect{a}_1,\vect{a}_2,\B_2,\C_{\textnormal{P}}}$
when $\C_{\textnormal{P}}=\C_{\textnormal{P,LMMSE}}$ (as defined
in~\eqref{eq:LMMSEmatP}), and an alternative formulation for the
achievable region
$\RatePsi{\nch}{\eta,\vect{a}_1,\vect{a}_2,\B_1,\B_2,\C_{\textnormal{SI}}}$
when $\C_{\textnormal{SI}}=\C_{\textnormal{SI,LMMSE}}$ (as defined
in~\eqref{eq:LMMSESIf}). These alternative formulations simplify the
description of the achievable regions corresponding to our specific
choices of parameters suggested in Appendices~\ref{sec:achproof},
\ref{sec:alterna}, and \ref{app:si_nonsym}. In particular, for perfect
feedback the alternative formulation is useful to describe the
achievable region corresponding to the choice of parameters in
Appendix~\ref{sec:alterna}, see Remark~\ref{def:remarkperf}. The region
in Remark~\ref{def:remarkperf} is used in
Section~\ref{sec:proofremperf} to prove that our concatenated scheme
for perfect feedback achieves all points in the interior of Ozarow's
region $\set{R}_{\textnormal{Oz}}^{\rho^*}(P_1,P_2,N)$
(Proposition~\ref{prop:maxsumratesigma}).

\subsubsection{Noisy Feedback}\label{sec:achaltnoisy}
 Given parameters
$\eta,\vect{a}_1,\vect{a}_2,\B_1,\B_2$ and
$\C=\C_{\textnormal{LMMSE}}$, we derive an alternative formulation for
the region achieved by our concatenated scheme $\RateP{\nch}{\para}$.

To simplify notation, in this section we assume that Inner Encoder~1
and Inner Encoder~2 are fed independent zero-mean unit-variance
Gaussian random variables and therefore we denote them by $\Xi_1$ and
$\Xi_2$ instead of $\xi_1$ and $\xi_2$.  The region achieved by our
concatenated scheme can then be expressed as the set of all
nonnegative rate pairs $(R_1,R_2)$ satisfying
\begin{subequations}\label{eq:Rgen}
\begin{IEEEeqnarray}{rCl}
  R_1 & \leq& \frac{1}{\eta} I(\Xi_1 ;\hat{\Xi}_1, \hat{\Xi}_2| \Xi_2),
\label{eq:R1gen}\\
 R_2 & \leq& \frac{1}{\eta} I(\Xi_2; \hat{\Xi}_1, \hat{\Xi}_2| \Xi_1),\label{eq:R2gen}\\
R_1+ R_2 & \leq& \frac{1}{\eta} I(\Xi_1, \Xi_2;\hat{\Xi}_1,
\hat{\Xi}_2), \label{eq:R12gen}
\end{IEEEeqnarray} 
\end{subequations}
where the conditional law of $(\hat{\Xi}_1,\hat{\Xi}_2)$ given
$\Xi_1=\xi_1$ and $\Xi_2=\xi_2$ is determined by the channel law
$\xi_1,\xi_2\mapsto (\hat{\Xi}_1,\hat{\Xi}_2)$ in
Equation~\eqref{eq:newdec} (Section~\ref{sec:gen}). Notice that since
$\C$ is the LMMSE-estimation matrix $\C_{\textnormal{LMMSE}}$
in~\eqref{eq:LMMSEmat} (Section~\ref{sec:choicenoisy}), by the
Gaussianity of the involved random variables the rate constraints in
\eqref{eq:Rgen} are equivalent to:
\begin{subequations}\label{eq:Rgennew}
\begin{IEEEeqnarray}{rCl}
  R_1 & \leq& \frac{1}{\eta} I(\Xi_1 ;Y_1,\ldots,Y_\eta| \Xi_2),
\label{eq:R1gennew}\\
R_2 & \leq& \frac{1}{\eta} I(\Xi_2; Y_1,\ldots, Y_\eta| \Xi_1),\label{eq:R2gennew}\\
R_1+ R_2 & \leq& \frac{1}{\eta} I(\Xi_1, \Xi_2;Y_1,\ldots, Y_\eta),
\label{eq:R12gennew}
\end{IEEEeqnarray} 
\end{subequations}
where $Y_1,\ldots, Y_\eta$ are the $\eta$ channel outputs produced by
the original channel $x_1,x_2\mapsto Y$ when the inner encoders are
fed the independent standard Gaussians $\Xi_1$ and $\Xi_2$.

Denote for each channel use $\ell\in
\{1,\ldots,\eta\}$ the receiver's innovation
by $I_\ell$, i.e., 
\begin{IEEEeqnarray}{rCl}\label{eq:inno1}
I_{\ell} &\triangleq & Y_{\ell}
-\E{Y_\ell|Y^{\ell-1}},
\end{IEEEeqnarray}
and the receiver's LMMSE-estimation errors about the symbols $\Xi_1$
and $\Xi_2$ by $E_{1,\ell}$ and $E_{2,\ell}$, i.e., 
\begin{subequations}\label{eq:epsi}
\begin{IEEEeqnarray}{rCl}
\Er_{1,\ell} & \triangleq  & \Xi_{1} - \E{\Xi_1|Y^{\ell}},
\label{eq:epsi1} \\
\Er_{2,\ell} & \triangleq  &
\Xi_{2} - \E{\Xi_2| Y^{\ell}}.\label{eq:epsi2} 
\end{IEEEeqnarray}
\end{subequations}
Then, notice that there exists a bijective mapping between the innovations
$I_1,\ldots, I_\eta$ and the channel outputs $Y_1,\ldots,
Y_\eta$, and by the Gaussianity of the involved random variables, for
each $\ell\in\{1,\ldots, \eta\}$, the tuple
$(E_{1,\ell},E_{2,\ell},I_{\ell})$ is independent of the previous
outputs and innovations $(Y_{1},\ldots, Y_{\ell-1}, I_{1},\ldots,
I_{\ell-1})$. By the chain rule of mutual information we
can therefore rewrite Constraints
\eqref{eq:Rgennew} as
\begin{subequations}\label{eq:Rinno} 
\begin{IEEEeqnarray}{rCl}
  R_1 & \leq& \frac{1}{\eta} \sum_{\ell=1}^{\eta} I(\Xi_1 ;I_\ell| \Xi_2),
\label{eq:R1inno}\\
R_2 & \leq& \frac{1}{\eta} \sum_{\ell=1}^{\eta}I(\Xi_2; I_\ell | \Xi_1),\label{eq:R2inno}\\
R_1+ R_2 & \leq& \frac{1}{\eta}\sum_{\ell=1}^{\eta} I(\Xi_1, \Xi_2;I_\ell).
\label{eq:R12inno}
\end{IEEEeqnarray} 
\end{subequations}
In the following we give a more explicit description of the
innovations $\{I_{\ell}\}_{\ell=1}^{\eta}$ in terms of the entries of
  the parameters $\vect{a}_1,\vect{a}_2,\B_1, \B_2$. For each
  $\ell\in\{1,\ldots, \eta\}$, let $a_{\nu,\ell}$ denote the $\ell$-th
  entry of the vector $\vect{a}_\nu$ and $b_{\nu,\ell,j}$ denote the
  row-$\ell$ column-$j$ entry of the matrix $\B_\nu$, for
  $j,\ell\in\{1,\ldots, \eta\}$ and $\nu\in\{1,2\}$. Also, let
  $\alpha_{1,\ell} \triangleq \Var{\Er_{1,\ell}}$ and $\alpha_{2,\ell}
  \triangleq \Var{\Er_{2,\ell}}$ denote the variances of $E_{1,\ell}$
  and $E_{2,\ell}$, and $\rho_{\ell} \triangleq
  \frac{\Cov{\Er_{1,\ell}}{\Er_{2,\ell}}}{\sqrt{\alpha_{1,\ell}\alpha_{2,\ell}}}$
  their correlation coefficient. 
We can then write the innovations as 
\begin{subequations}\label{eq:innooo}
\begin{equation}\label{eq:II1}
I_1=Y_1 =a_{1,1} \Xi_1+a_{2,1} \Xi_2+Z_1;
\end{equation}
and for  $\ell\in\{2,\ldots, \eta\}$ as
\begin{IEEEeqnarray}{rCl}
I_\ell &=& a_{1,\ell} E_{1,\ell-1}+a_{2,\ell}E_{2,\ell-1} \nonumber \\
& & +
(W_{\ell-1}-\E{W_{\ell-1}|Y^{\ell-1}}) +Z_\ell, \nonumber \\
 & = &  \kappa_{1,\ell-1}  E_{1,\ell-1} +
 \kappa_{2,\ell-1}E_{2,\ell-1} \nonumber \\ && +
 W_{\perp,\ell-1}+Z_\ell, \label{eq:inno2}
\end{IEEEeqnarray}
\end{subequations}
where 
\begin{IEEEeqnarray}{rCl}
W_{\ell-1} & \eqdef & \sum_{j=1}^{\ell-1} b_{1,\ell,j} W_{1,j}
  + \sum_{j=1}^{\ell-1} b_{2,\ell,j}  W_{2,j},\label{eq:Wsum} \\
 W_{\perp,\ell-1} & \eqdef&  W_{\ell-1} - \E{
   W_{\ell-1}|\Er_{1,\ell-1}, \Er_{2,\ell-1}, Y^{\ell-1}}\nonumber \\\label{eq:WperpY} \\
 & =& W_{\ell-1} - \E{
   W_{\ell-1}|\Er_{1,\ell-1}, \Er_{2,\ell-1},  I^{\ell-1}},\nonumber \\\label{eq:Wperp}
 \end{IEEEeqnarray}
and 
\begin{IEEEeqnarray}{rCl}
  \kappa_{1,\ell-1} & \eqdef & a_{1,\ell}+ \frac{ \alpha_{2,\ell-1} \Cov{
      \Er_{1,\ell-1}}{ W_{\ell-1}
    } }{(1-\rho_{\ell-1}^2)  \alpha_{1,\ell-1} \alpha_{2,\ell-1}}
      \nonumber \\ & & -\frac{ \rho_{\ell-1} \sqrt{\alpha_{1,\ell-1} \alpha_{2,\ell-1}}
    \Cov{\Er_{2,\ell-1}}{ W_{\ell-1}}
  }{(1-\rho_{\ell-1}^2)  \alpha_{1,\ell-1}
      \alpha_{2,\ell-1}},\nonumber \\\label{eq:k1}
\\
  \kappa_{2,\ell-1} & \eqdef &a_{2,\ell}+ \frac{ \alpha_{1,\ell-1} \Cov{
      \Er_{2,\ell-1}}{ W_{\ell-1}
    }}{ (1-\rho_{\ell-1}^2)  \alpha_{1,\ell-1} \alpha_{2,\ell-1}}
      \nonumber \\ & & - \frac{\rho_{\ell-1} \sqrt{\alpha_{1,\ell-1} \alpha_{2,\ell-1}}
    \Cov{\Er_{1,\ell-1}}{W_{\ell-1} }} {(1-\rho_{\ell-1}^2)
    \alpha_{1,\ell-1} \alpha_{2,\ell-1}}.\nonumber \\\label{eq:k2}
\end{IEEEeqnarray}
Evaluating the mutual information expressions in
~\eqref{eq:Rinno} for the innovations
in~\eqref{eq:innooo}, we conclude that our
concatenated scheme for noisy feedback with parameters
$\eta,\vect{a}_1,\vect{a}_2,\B_1,\B_2$ and
$\C=\C_{\textnormal{LMMSE}}$ achieves all rate pairs $(R_1,R_2)$
satisfying Constraints~\eqref{eq:recexp} on the top of next page, where
we defined  $\alpha_{1,0}\eqdef 1$, $\alpha_{2,0}\eqdef 1$,
$\rho_0\eqdef  0$, $\kappa_{1,0}\eqdef a_{1,1}$, $\kappa_{2,0}\eqdef a_{2,1}$,
$W_{\perp,0}\eqdef 0$. 
\begin{figure*}[!t]
\begin{subequations}\label{eq:recexp}
\begin{IEEEeqnarray}{rCll}
R_1 & \leq & \frac{1}{\eta} \sum_{\ell=1}^{\eta}
\frac{1}{2}\log\Bigg(1 &+
\frac{\kappa_{1,\ell-1}^2 \alpha_{1,\ell-1}\left(1-\rho_{\ell-1}^2\right)}{\Var{W_{\perp,\ell-1}}+N} \Bigg) \label{eq:recexp1}\\
R_2 & \leq & \frac{1}{\eta} \sum_{\ell=1}^{\eta}
\frac{1}{2}\log\Bigg(1&+
\frac{\kappa_{2,\ell-1}^2 \alpha_{2,\ell-1}\left(1-\rho_{\ell-1}^2\right)}{\Var{W_{\perp,\ell-1}}+N}\Bigg) \label{eq:recexp2}\\
 R_1 + R_2 & \leq & \frac{1}{\eta} \sum_{\ell=1}^{\eta}
\frac{1}{2}\log\Bigg( 1&+\frac{ \kappa_{1,\ell-1}^2\alpha_{1,\ell-1} +
 \kappa_{2,\ell-1}^2 \alpha_{2,\ell-1} + 2 
 \kappa_{1,\ell-1}\kappa_{2,\ell-1} \sqrt{ \alpha_{1,\ell-1}
    \alpha_{2,\ell-1}}\rho_{\ell-1}}{\Var{W_{\perp,\ell-1}}+N}\Bigg)\IEEEeqnarraynumspace \label{eq:recexp12}
\end{IEEEeqnarray}
\end{subequations} 
\end{figure*}

We conclude this section with a recursive characterization of
the variances $\{\alpha_{1,\ell}\}_{\ell=1}^{\eta}$ and
  $\{\alpha_{2,\ell}\}_{\ell=1}^{\eta}$, and the correlation
    coefficients $\{\rho_{\ell}\}_{\ell=1}^{\eta}$. 
Defining
$E_{1,0} \triangleq \Xi_1$, $E_{2,0}  \triangleq  \Xi_2$, we find 
 for $\ell\in\{1,\ldots,\eta\}$:
\begin{subequations}\label{eq:El}
\begin{IEEEeqnarray}{rCl}\label{eq:E1l}
\Er_{1,\ell} & = & \Er_{1,\ell-1}-
\frac{\Cov{\Er_{1,\ell-1} }{I_\ell }}{\Var{I_\ell }}I_{\ell}, \\
\Er_{2,\ell}  & = & \Er_{2,\ell-1} - \frac{\Cov{\Er_{2,\ell-1}
  }{I_\ell }}{\Var{I_\ell }}I_{\ell},\label{eq:E2l}
\end{IEEEeqnarray}
\end{subequations}
and consequently, by \eqref{eq:innooo} the recursive  expressions
\eqref{eq:alpharec11}--\eqref{eq:rho33} on top of the next page.
\begin{figure*}
\begin{IEEEeqnarray}{rCll}
\alpha_{1,\ell} &=& \alpha_{1,\ell-1} \Bigg(
  \frac{\kappa_{1,\ell-1}^2
\alpha_{1,\ell-1} + \kappa_{2,\ell-1}^2
\alpha_{2,\ell-1} + 2\kappa_{1,\ell-1}\kappa_{2,\ell-1} \sqrt{\alpha_{1,\ell-1}\alpha_{2,\ell-1}}\rho_{\ell-1}+  \Var{W_{\perp,\ell-1}}+N  }{\kappa_{2,\ell-1}^2 \alpha_{2,\ell-1}
  (1-\rho_{\ell-1}^2)+\Var{W_{\perp,\ell-1}}+N}  
\label{eq:alpharec11}
  \\
\alpha_{2,\ell} &=& \alpha_{2,\ell-1} \bigg( 
  \frac{\kappa_{1,\ell-1}^2
\alpha_{1,\ell-1}+\kappa_{2,\ell-1}^2
\alpha_{2,\ell-1}+ 2\kappa_{1,\ell-1}\kappa_{2,\ell-1} \sqrt{\alpha_{1,\ell-1}\alpha_{2,\ell-1}}\rho_{\ell-1} + \Var{W_{\perp,\ell-1}}+N   }{\kappa_{1,\ell-1}^2 \alpha_{1,\ell-1}
  (1-\rho_{\ell-1}^2)+\Var{W_{\perp,\ell-1}}+N}
\label{eq:alpharec22}\\
\rho_{\ell} & = & \frac{- \kappa_{1,\ell-1}\kappa_{2,\ell-1} \sqrt{\alpha_{1,\ell-1}
  \alpha_{2,\ell-1}}(1-\rho_{\ell-1}^2) +\rho_{\ell-1} (
  \Var{W_{\perp,\ell-1}}+N)}{\sqrt{\kappa_{1,\ell-1}^2 \alpha_{1,\ell-1}
  (1-\rho_{\ell-1}^2) +\Var{W_{\perp,\ell-1}}+N} \sqrt{\kappa_{2,\ell-1}^2 \alpha_{2,\ell-1}
  (1-\rho_{\ell-1}^2) +\Var{W_{\perp,\ell-1}}+N } } \label{eq:rho33}
\end{IEEEeqnarray} 
\hrulefill
\end{figure*}

This alternative formulation is used in Corollaries~\ref{cor:ach}
and~\ref{cor:alte} in Appendices~\ref{sec:achproof} and
\ref{sec:alterna} ahead to describe the regions achieved by our
concatenated scheme for noisy feedback with the specific choices of
parameters described in Sections~\ref{sec:enc} and
\ref{sec:powerstar}. In particular, it is used to describe the region
achieved in the special case of perfect feedback when the parameters
are chosen as in Section~\ref{sec:powerstar}, see
Remark~\ref{def:remarkperf}.

\subsubsection{Noisy Partial Feedback}~\label{sec:achaltnoisypar}
The desired alternative formulation of 
$\RatePpar{N,\sigma_2^2}{\eta,\vect{a}_1,\vect{a}_2,\B_1,\B_2,\C_{\textnormal{P}}}$
can be derived along the  lines shown in the previous
subsection~\ref{sec:achaltnoisy}. We omit the details and only present
the result.

 Fix a choice of
parameters $\eta,\vect{a}_1,\vect{a}_2,\B_1,\B_2$ and $\C_{\textnormal{P}}=\C_{\textnormal{P,LMMSE}}$.
Denote the $\ell$-th entry of
the vector $\vect{a}_\nu$ by $a_{\nu,\ell}$ and
denote the row-$\ell$ column-$j$ entry of the matrix $\B_2$ by
$b_{2,\ell,j}$, for $j,\ell\in\{1,\ldots,\eta\}$ and $\nu\in\{1,2\}$.
Our concatenated scheme for noisy partial feedback and
parameters $\eta,\vect{a}_1,\vect{a}_2,\B_2, \C_{\textnormal{P}}=\C_{\textnormal{P,LMMSE}}$ achieves all rate
pairs $(R_1,R_2)$ satisfying Constraints~\eqref{eq:recp} on top of the
next page, where recall that $\alpha_{1,0}= 1$, $\alpha_{2,0}= 1$, $\rho_0
=0$, $\kappa_{1,0}= a_{1,1}$, $\kappa_{2,0}= a_{2,1}$,
$W_{\perp,0}=0$, and where
$\{\alpha_{1,\ell}\}_{\ell=1}^{\eta-1}$,
$\{\alpha_{2,\ell}\}_{\ell=1}^{\eta-1}$,
$\{\rho_{\ell}\}_{\ell=1}^{\eta-1}$,
$\{\kappa_{1,\ell}\}_{\ell=1}^{\eta-1}$,
$\{\kappa_{2,\ell}\}_{\ell=1}^{\eta-1}$, and
$\{W_{\perp,\ell}\}_{\ell=1}^{\eta-1}$ are defined by $E_{1,0}=
\Xi_1$, $E_{2,0}= \Xi_2$, and
Equations~\eqref{eq:innooo}--\eqref{eq:rho33}
(Subsection~\ref{sec:achaltnoisy}) except that \eqref{eq:Wsum} should
be replaced by
\begin{IEEEeqnarray*}{rCl}
W_{\ell-1} & = & \sum_{j=1}^{\ell-1}
b_{2,\ell,j} W_{2,j}.
\end{IEEEeqnarray*}
\begin{figure*}
\begin{subequations}\label{eq:recp}
\begin{IEEEeqnarray}{rCll}
R_1 & \leq & \frac{1}{\eta} \sum_{\ell=1}^{\eta}
\frac{1}{2}\log\Bigg(1&+
\frac{\kappa_{1,\ell-1}^2
  \alpha_{1,\ell-1}\left(1-\rho_{\ell-1}^2\right)}{\Var{W_{\perp,\ell-1}}+N} \Bigg)
\label{eq:recp1}\\
R_2 & \leq & \frac{1}{\eta} \sum_{\ell=1}^{\eta}
\frac{1}{2}\log\Bigg(1&+
\frac{
 \kappa_{2,\ell-1}^2 \alpha_{2,\ell-1}\left(1-\rho_{\ell-1}^2\right)}{\Var{W_{\perp,\ell-1}}+N}\Bigg)
\label{eq:recp2} \\
 R_1 + R_2  & \leq & \frac{1}{\eta} \sum_{\ell=1}^{\eta}
\frac{1}{2}\log\Bigg( 1&+\frac{ \kappa_{1,\ell-1}^2\alpha_{1,\ell-1} +
  \kappa_{2,\ell-1}^2\alpha_{2,\ell-1} +  2 
  \kappa_{1,\ell-1}\kappa_{2,\ell-1}\sqrt{ \alpha_{1,\ell-1} \alpha_{2,\ell-1}}
 \rho_{\ell-1} }{\Var{W_{\perp,\ell-1}}+N}\Bigg)\;\;\IEEEeqnarraynumspace\label{eq:recp12}
\end{IEEEeqnarray}
\end{subequations}
\end{figure*}

\subsubsection{Noisy Feedback with Receiver Side-Information}~\label{sec:achaltnoisysi}

We derive an alternative formulation of the rate region achieved by our
concatenated scheme
$\RatePsi{N,\KW}{\eta,\vect{a}_1,\vect{a}_2,\B_1,\B_2,\C_{\textnormal{SI}}}$
for a fixed choice of parameters $\eta,\vect{a}_1,\vect{a}_2,\B_1,\B_2$
and $\C_{\textnormal{SI}}=\C_{\textnormal{SI,LMMSE}}$. Denote the $\ell$-th entry of the vector
$\vect{a}_\nu$ by $a_{\nu,\ell}$ and denote the row-$\ell$ column-$j$
entry of the matrix $\B_\nu$ by $b_{\nu,\ell,j}$, for
$j,\ell\in\{1,\ldots, \eta\}$ and $\nu\in\{1,2\}$. The 
desired alternative formulation of
$\RatePsi{\nch}{\eta,\vect{a}_1,\vect{a}_2,\B_1,\B_2,\C_{\textnormal{SI}}}$ can
be derived along the lines described in Subsection~\ref{sec:achaltnoisy}
but with the following two modifications. Instead of being defined as in
\eqref{eq:epsi}, the LMMSE-estimation errors
$\Er_{1,\ell}$ and $\Er_{2,\ell}$, for $\ell \in \{1,\ldots, \eta\}$,
are defined as
\begin{subequations}\label{eq:epsSI}
\begin{IEEEeqnarray}{rCl}\label{eq:eps1SI}
\Er_{1,\ell} & \triangleq  &
\Xi_{1}-\E{\Xi_1| Y^\ell, W_{1}^{\ell-1},
  W_{2}^{\ell-1}},  \IEEEeqnarraynumspace
\\
\Er_{2,\ell} & \triangleq  &
\Xi_{2}-\E{\Xi_2| Y^\ell, W_{1}^{\ell-1},
  W_{2}^{\ell-1}},
\label{eq:eps2SI}
\end{IEEEeqnarray}
\end{subequations}
and instead of being defined as in \eqref{eq:inno1}, the innovation
$I_\ell$, for $\ell\in\{1,\ldots,\eta\}$, is defined as
\begin{IEEEeqnarray}{rCl}\label{eq:innoSI}
I_{\ell} &\triangleq &
Y_{\ell}-\E{Y_\ell| Y^{\ell-1},W_{1}^{\ell-1}, W_{2}^{\ell-1}}.\IEEEeqnarraynumspace
\end{IEEEeqnarray}
Notice that by \eqref{eq:epsSI}  and \eqref{eq:innoSI}:
\begin{IEEEeqnarray*}{rCl}
I_{\ell}& = &
a_{1,\ell}\Er_{1,\ell-1}+a_{2,\ell}\Er_{2,\ell-1}+Z_{\ell},
\qquad \ell \in \{1,\ldots, \eta\}. 
\end{IEEEeqnarray*}

We omit the details of the derivation and only state the resulting
alternative formulation of the region
$\RatePsi{\nch}{\eta,\vect{a}_1,\vect{a}_2,\B_1,\B_2,\C_{\textnormal{SI}}}$.
Our concatenated scheme for noisy feedback with receiver
side-information and parameters $\eta,\vect{a}_1,\vect{a}_2,\B_1,\B_2,
\C_{\textnormal{SI}}$ achieves all rate pairs $(R_1,R_2)$ satisfying
Constraints \eqref{eq:recexpsi} on top of the next page, where recall
that  $\alpha_{1,0}=1$, $\alpha_{2,0}=1$, $\rho_{0}=0$, and
where $\{\alpha_{1,\ell}\}_{\ell=1}^{\eta}$,
$\{\alpha_{2,\ell}\}_{\ell=1}^{\eta}$ and
$\{\rho_{\ell}\}_{\ell=1}^{\eta}$ are defined through
Recursions~\eqref{eq:alpha1re}--\eqref{eq:rhore}, also on top of the
next page.
\begin{figure*}
\begin{subequations}\label{eq:recexpsi}
\begin{IEEEeqnarray}{rCll}
R_1 & \leq & \frac{1}{\eta} \sum_{\ell=1}^{\eta}
\frac{1}{2}\log\Bigg(1&+
\frac{ a_{1,\ell}^2\alpha_{1,\ell-1}\left(1-\rho_{\ell-1}^2\right)}{N} \Bigg) \label{eq:recexpsi1}\\
R_2 & \leq & \frac{1}{\eta} \sum_{\ell=1}^{\eta}
\frac{1}{2}\log\Bigg(1&+
\frac{ a_{2,\ell}^2\alpha_{2,\ell-1}\left(1-\rho_{\ell-1}^2\right)}{N}\Bigg) \label{eq:recexpsi2} \\
  R_1 + R_2 & \leq & \frac{1}{\eta} \sum_{\ell=1}^{\eta}
\frac{1}{2}\log\Bigg( 1& +\frac{
    a_{1,\ell}^2\alpha_{1,\ell-1}+a_{2,\ell}^2  \alpha_{2,\ell-1}+2
    a_{1,\ell}a_{2,\ell}\rho_{\ell-1}\sqrt{\alpha_{1,\ell-1}\alpha_{2,\ell-1}}}{N}\Bigg) \label{eq:recexpsi12}\IEEEeqnarraynumspace
\end{IEEEeqnarray}
\end{subequations}
\begin{IEEEeqnarray}{rCll}
  \alpha_{1,\ell} &=& \alpha_{1,\ell-1} \bigg( \frac{a_{1,\ell}^2
    \alpha_{1,\ell-1} + a_{2,\ell}^2 \alpha_{2,\ell-1} + 2 a_{1,\ell}
    a_{2,\ell}
    \sqrt{\alpha_{1,\ell-1}\alpha_{2,\ell-1}}\rho_{\ell-1}+N }{a_{2,\ell}^2
    \alpha_{2,\ell-1} (1-\rho_{\ell-1}^2)+N}\bigg)^{-1}
  \label{eq:alpha1re}\\
  \alpha_{2,\ell}&=& \alpha_{2,\ell-1} \bigg( \frac{a_{1,\ell}^2
    \alpha_{1,\ell-1}+a_{2,\ell}^2 \alpha_{2,\ell-1} + 2a_{1,\ell}
    a_{2,\ell}
    \sqrt{\alpha_{1,\ell-1}\alpha_{2,\ell-1}}\rho_{\ell-1}+N }{a_{2,\ell}^2
    \alpha_{2,\ell-1}
    (1-\rho_{\ell-1}^2)+N} \bigg)^{-1}\label{eq:alpha2re}\\
  \rho_{\ell} & = & \frac{- a_{1,\ell} a_{2,\ell}
  \sqrt{\alpha_{1,\ell-1}
    \alpha_{2,\ell-1}}(1-\rho_{\ell-1}^2)+\rho_{\ell-1} N}{ \sqrt{a_{1,\ell}^2 \alpha_{1,\ell-1}
  (1-\rho_{\ell-1}^2) +N} 
 \sqrt{a_{2,\ell}^2 \alpha_{2,\ell-1}
  (1-\rho_{\ell-1}^2) +N } } \label{eq:rhore}
\end{IEEEeqnarray} 
\hrulefill
\end{figure*}

\subsection{Choice of Parameters I} \label{sec:achproof}

In Section~\ref{sec:enc}, we present a specific choice of the
parameters ${\vect{a}}_1, {\vect{a}}_2, {\B}_1,{\B}_2, {\C}$ for given
$\eta\in\Nat$. We treat the noisy-feedback setting and
the noisy or perfect partial-feedback setting. We denote our choice
for noisy feedback by $\bar{\vect{a}}_1, \bar{\vect{a}}_2,
\bar{\B}_1,\bar{\B}_2, \bar{\C}$ and our choice for partial feedback
by $\bar{\vect{a}}_{1,\textnormal{P}}, \bar{\vect{a}}_{2,\textnormal{P}}, \bar{\B}_{2,\textnormal{P}}, \bar{\C}_\textnormal{P}$.

As we shall see, our choices are such that $\bar{\C}$ and
$\bar{\C}_{\textnormal{P}}$ are LMMSE-estimation matrices. Thus, the
region achieved by our concatenated scheme for noisy feedback with
parameters $\eta, \bar{\vect{a}}_1, \bar{\vect{a}}_2,
\bar{\B}_1,\bar{\B}_2, \bar{\C}$ is obtained by substituting the
parameters into the RHSs of \eqref{eq:recexp} in
Section~\ref{sec:achaltnoisy}. The resulting achievable region is
presented in Corollary~\ref{cor:ach} ahead. Similarly, the region
achieved by our concatenated scheme for partial feedback with
parameters $\eta, \bar{\vect{a}}_{1,\textnormal{P}},
\bar{\vect{a}}_{2,\textnormal{P}}, \bar{\B}_{2,\textnormal{P}},
\bar{\C}_{\textnormal{P}}$ is obtained by substituting the parameters into
the RHSs of~\eqref{eq:recp1}--\eqref{eq:recp12} in
Section~\ref{sec:achaltnoisypar}. For brevity we do not
present this latter achievable region.

\subsubsection{Description of Parameters}\label{sec:enc}
Let a positive integer $\eta \in \Nat$ be given. We
first consider the noisy-feedback setting; the partial-feedback
setting is treated only shortly in Remark~\ref{rem:partialfb} at the
end of this section.

Instead of describing our choice
$\bar{\vect{a}}_1,\bar{\vect{a}}_2, \bar{\B}_1,\bar{\B}_2,$ and
$\bar{\mat{C}}$ directly, we will describe how Inner Encoder~1 and Inner
Encoder~2 map the fed symbols to the sequences of channel inputs
$X_{1,1},\ldots,X_{1,\eta}$ and $X_{2,1},\ldots,X_{2,\eta}$.
This then determines $\bar{\vect{a}}_1,\bar{\vect{a}}_2,
\bar{\B}_1,\bar{\B}_2$. The matrix $\bar{\C}$ is chosen as the
LMMSE-estimation matrix. For the purpose of describing
our choice we replace the pair of input symbols $\xi_1$ and $\xi_2$ by
the independent standard Gaussians $\Xi_1$ and $\Xi_2$.

The inner encoders are chosen so as to produce
\begin{IEEEeqnarray}{rCl}\label{eq:X1a}
X_{1,1} & = & \sqrt{P_1}\Xi_1, \\
X_{2,1} & = & \sqrt{P_2}\Xi_2,\label{eq:X2a}
\end{IEEEeqnarray}
and for $\ell\in \{2,\ldots, \eta\}$: 
\begin{IEEEeqnarray}{rCl}
X_{1,\ell} &=&\sqrt{\frac{\pow_1}{{\D}_{1,\ell-1}}} \left( \Xi_1 -
\trans{ {\boldsymbol{\gamma}}_{1,\ell-1}} V_{1}^{\ell-1}
\right),\label{eq:X1b} \\
X_{2,\ell}& = & (-1)^{\ell-1} \sqrt{\frac{\pow_2}{{\D}_{2,\ell-1}}}\left( \Xi_2 -
  \trans{{\boldsymbol{\gamma}}_{2,\ell-1}} \mat{M}_{\ell-1} V_{2}^{\ell-1}\right),
\nonumber \\\label{eq:X2b}
\end{IEEEeqnarray}
where for $\ell\in\{1,\ldots, \eta-1\}$
\begin{IEEEeqnarray}{rCl}
\label{eq:M}
\mat{M}_{\ell}&\triangleq&\diag{1,
  -1, 1, \ldots, (-1)^{\ell-1} },\\
{\D}_{1,\ell} &\triangleq & \Var{\Xi_{1} - \trans{{\boldsymbol{\gamma}}_{1,\ell}}
  V_{1}^{\ell}}, \nonumber \\\label{eq:Dell1} \\
{\D}_{2,\ell}& \triangleq &\Var{\Xi_{2} - \trans{{\boldsymbol{\gamma}}_{2,\ell}} \mat{M}_{\ell}
  V_{2}^{\ell}},\nonumber \\ \label{eq:Dell2}\\
\boldsymbol{\gamma}_{1,\ell}&  \triangleq &\left(
  (\sigma_1^2+\sigma_2^2 - 2 \varrho \sigma_1 \sigma_2)\frac{P_1}{N}
  \mat{I}_{\ell}+
  \mat{K}_{V_{1}^{\ell}}\right)^{-1}
  \mat{K}_{V_{1}^{\ell}, \Xi_1},\nonumber \\  \label{eq:coeff_choice1}\\
\boldsymbol{\gamma}_{2,\ell}&  \triangleq &\left(
  (\sigma_1^2+\sigma_2^2 - 2 \varrho \sigma_1 \sigma_2) \frac{P_2}{N} \mat{I}_\ell+
  \mat{K}_{V_{2}^{\ell}}\right)^{-1}
  \mat{K}_{V_{2}^{\ell}, \Xi_2}.\nonumber \\  \label{eq:coeff_choice2}\end{IEEEeqnarray}

Notice that Inner Encoder~2 modulates its inputs with an alternating
sequence of $+1$ or $-1$ (which is inspired by the Fourier-MEC scheme
in \cite{kramer02}), and it multiplies the noisy feedback vectors by
the matrix $\mat{M}_{\ell-1}$ before further processing it (which
accounts for the modulation of past inputs). The presented choice of
the inner encoders ensures that the input sequences to the original
MAC $x_1,x_2\mapsto Y$ satisfy the average block-power constraints
\eqref{eq:power}. In particular, with the presented choice all input
symbols $X_{1,1},\ldots, X_{1,\eta}$ have the same expected power
$P_1$, and
all input symbols $X_{2,1},\ldots, X_{2,\eta}$ have the same expected
power $P_2$.

This encoding scheme corresponds to the following parameters of the
concatenated scheme:
\begin{IEEEeqnarray*}{rCl}
\bar{\vect{a}}_1 & \eqdef & \trans{\begin{pmatrix} \sqrt{P_1}&
    \sqrt{\frac{P_1}{{\beta}_{1,1}}} & \ldots&
    \sqrt{\frac{P_1}{{\beta}_{1,\eta-1}}} \end{pmatrix}}, \\ 
\bar{\vect{a}}_2 & \eqdef & \trans{\begin{pmatrix} \sqrt{P_2}&
   - \sqrt{\frac{P_2}{{\beta}_{2,1}}} & \ldots& (-1)^{\eta-1}
    \sqrt{\frac{P_2}{{\beta}_{2,\eta-1}}} \end{pmatrix}},
\end{IEEEeqnarray*} 
and 
\begin{IEEEeqnarray*}{rCl}
\bar{\B}_1 & \eqdef &\trans{ \begin{pmatrix}\vect{0} &
    - \sqrt{\frac{P_1}{{\beta}_{1,1}}} \boldsymbol{{\gamma}}_{1,1}^{(0)} & \ldots &- \sqrt{\frac{P_1}{\beta_{1,\eta-1}}}  \boldsymbol{\gamma}_{1,\eta-1}^{(0)}
    \end{pmatrix}},\\ 
\bar{\B}_2 & \eqdef &\trans{ \begin{pmatrix}\vect{0} &
     \sqrt{\frac{P_2}{{\beta}_{2,1}}}  \boldsymbol{\gamma}_{2,1}^{(0)} & \ldots & (-1)^{\eta}
    \sqrt{\frac{P_2}{{\beta}_{2,\eta-1}}} \boldsymbol{\gamma}_{2,\eta-1}^{(0)} \end{pmatrix}},
\end{IEEEeqnarray*}
where the vectors $\left\{\boldsymbol{\gamma}_{1,
    \ell}^{(0)}\right\}_{\ell=1}^{\eta-1}$ and
$\left\{\boldsymbol{\gamma}_{2,\ell}^{(0)}\right\}_{\ell=1}^{\eta-1}$ are
defined as the $\eta$-dimensional vectors obtained by stacking the
$\ell$-dimensional column-vector $\boldsymbol{\gamma}_{\nu,\ell}$ on top of an
$(\eta-\ell)$-dimensional column-vector with all zero entries, i.e.,
\begin{equation}\label{eq:vectb0}
\boldsymbol{\gamma}_{\nu, \ell}^{(0)} \eqdef \begin{pmatrix}\boldsymbol{\gamma}_{\nu,\ell} \\
  \vect{0} \end{pmatrix}. 
\end{equation}

The parameter $\bar{\C}$ is chosen as the LMMSE-estimation matrix $\C_{\textnormal{LMMSE}}$,
where recall
\begin{IEEEeqnarray}{rCl}
\mat{C}_{\textnormal{LMMSE}} =\trans{\bar{\mat{A}}_{\textnormal{r}}} \left(
  {\bar{\mat{A}}_{\textnormal{r}}} \trans{\bar{\mat{A}}}_{\textnormal{r}} +N
  \I_{\eta} + \bar{\B}_{\textnormal{r}} (\KW \otimes
  \I_\eta)\trans{\bar{\B}}_{\textnormal{r}}\right)^{-1}, \nonumber \\\label{eq:choiceC}
\end{IEEEeqnarray}
where $\bar{\mat{A}}_{\textnormal{r}}\eqdef \begin{pmatrix} \bar{\vect{a}}_1 &
\bar{\vect{a}}_2\end{pmatrix}$ and $\bar{\mat{B}}_{\textnormal{r}} \eqdef
\begin{pmatrix}\bar{\B}_1 & \bar{\B}_2 \end{pmatrix}$.

\begin{remark}\label{rem:partialfb}
  A similar choice of the parameters can also be made in the case of
  partial feedback. In this case, we choose the parameters
  corresponding to the inner
  encoders and the inner decoder as in \eqref{eq:X1a}--\eqref{eq:choiceC} except for replacing \eqref{eq:X1b} by
\begin{equation*}
X_{1,\ell} = \sqrt{P_1}\Xi_1, \qquad \ell \in \{2,\ldots, \eta\},
\end{equation*}
and replacing \eqref{eq:coeff_choice2} by
\begin{equation*}
\boldsymbol{\gamma}_{2,\ell}  \triangleq \left(
  \sigma_2^2  \frac{P_2}{N} \mat{I}_\ell+
  \mat{K}_{V_{2}^{\ell}}\right)^{-1}
  \mat{K}_{V_{2}^{\ell}, \Xi_2}. 
\end{equation*}
We denote the parameters of the concatenated scheme corresponding to
this choice by
$\bar{\vect{a}}_{1,\textnormal{P}},\bar{\vect{a}}_{2,\textnormal{P}},
\bar{\B}_{2,\textnormal{P}},$  and $\bar{\C}_\textnormal{P}$.  
\end{remark}

\subsubsection{Achievable Region}\label{sec:achreg}
We present the achievable region corresponding to our concatenated
scheme for noisy feedback with parameters as presented in the previous section.
\begin{definition}\label{def:Retabar}
  For a positive integer $\eta$, define
  $\bar{\set{R}}_{\eta}(P_1,P_2,N,\KW)$ as the set of all rate-pairs
  $(R_1,R_2)$ satisfying the three rate constraints~\eqref{eq:R_ach}
  on top of the next page,
\begin{figure*}
\begin{subequations}\label{eq:R_ach}
\begin{IEEEeqnarray}{rCl}
R_1& \leq& \frac{1}{2\eta}\sum_{\ell=1}^{\eta} \log\left( 1+
  \frac{P_1\bar{\kappa}_{1,\ell-1}^2 \frac{\alpha_{1,\ell-1}}{\D_{1,\ell-1}}  \left(1-\rho_{\ell-1}^2\right)
  }{\Var{W_{\perp,\ell-1}}+N}
\right)
\label{eq:R1ach} \\
R_2& \leq& \frac{1}{2\eta}\sum_{\ell=1}^{\eta}
\log\left( 1+ \frac{P_2  \bar{\kappa}_{2,\ell-1}^2\frac{
\alpha_{2,\ell-1}}{\D_{2,\ell-1}}    \left(1-\rho_{\ell-1}^2\right)
  }{\Var{W_{\perp,\ell-1}}+N}    \right)\label{eq:R2ach}\\
 R_1+R_2 &\leq&
  \frac{1}{2\eta}\sum_{\ell=1}^{\eta} \log\Bigg( 1 + \frac{P_1\bar{\kappa}_{1,\ell-1}^2
    \frac{\alpha_{1,\ell-1}}{\D_{1,\ell-1}}
    +\pow_2  \bar{\kappa}_{2,\ell-1}^2\frac{ \alpha_{2,\ell-1}}{\D_{2,\ell-1}}}{\Var{W_{\perp,\ell-1}}+N}
 +\frac{ 2\sqrt{P_1P_2}  \bar{\kappa}_{1,\ell-1}\bar{\kappa}_{2,\ell-1}\sqrt{\frac{
        \alpha_{1,\ell-1}}{\D_{1,\ell-1}}} \sqrt{\frac{
        \alpha_{2,\ell-1}}{\D_{2,\ell-1}}}
  \rho_{\ell-1} }{\Var{W_{\perp,\ell-1}}+N}\Bigg)\nonumber \\
  \label{eq:R12ach}
\end{IEEEeqnarray}
\end{subequations}
\begin{IEEEeqnarray}{rCl}
    \alpha_{1,\ell}&=&\alpha_{1,\ell-1} \Bigg( \frac{P_1\bar{\kappa}_{1,\ell-1}^2\frac{\alpha_{1,\ell-1}}{\D_{1,\ell-1}}+P_2\bar{\kappa}_{2,\ell-1}^2\frac{\alpha_{2,\ell-1}}{\D_{2,\ell-1}} }{P_2
    \bar{\kappa}_{2,\ell-1}^2\frac{\alpha_{2,\ell-1}}{\D_{2,\ell-1}}(1-\rho_{\ell-1}^2)+\Var{W_{\perp,\ell-1}}+N}\nonumber
    \\ & & \hspace{2.6cm}  + \frac{2\sqrt{P_1P_2}\bar{\kappa}_{1,\ell-1}\bar{\kappa}_{2,\ell-1}\sqrt{\frac{
          \alpha_{1,\ell-1}
          \alpha_{2,\ell-1}}{\D_{1,\ell-1}\D_{2,\ell-1}}}
      (-1)^{\ell}\rho_{\ell-1}+\Var{W_{\perp,\ell-1}}+N}{P_2\bar{\kappa}_{2,\ell-1}^2
    \frac{\alpha_{2,\ell-1}}{\D_{2,\ell-1}}(1-\rho_{\ell-1}^2)+\Var{W_{\perp,\ell-1}}+N}\Bigg)^{-1}   \label{eq:alphanu1}
  \\
 \alpha_{2,\ell}&=&\alpha_{2,\ell-1} \Bigg( \frac{P_1\bar{\kappa}_{1,\ell-1}^2\frac{\alpha_{1,\ell-1}}{\D_{1,\ell-1}}+P_2\bar{\kappa}_{2,\ell-1}^2\frac{\alpha_{2,\ell-1}}{\D_{2,\ell-1}} }{P_1\bar{\kappa}_{1,\ell-1}^2
    \frac{\alpha_{1,\ell-1}}{\D_{1,\ell-1}}(1-\rho_{\ell-1}^2)+\Var{W_{\perp,\ell-1}}+N}\nonumber
    \\ & & \hspace{2.6cm} + \frac{2\sqrt{P_1P_2}\bar{\kappa}_{1,\ell-1}\bar{\kappa}_{2,\ell-1}\sqrt{\frac{
          \alpha_{1,\ell-1}
          \alpha_{2,\ell-1}}{\D_{1,\ell-1}\D_{2,\ell-1}}}
      (-1)^{\ell}\rho_{\ell-1}+\Var{W_{\perp,\ell-1}}+N}{P_1\bar{\kappa}_{1,\ell-1}^2
    \frac{\alpha_{1,\ell-1}}{\D_{1,\ell-1}}(1-\rho_{\ell-1}^2)+\Var{W_{\perp,\ell-1}}+N}\Bigg)^{-1}
     \label{eq:alphanu2}\\ 
  \rho_{\ell}&= & \rho_{\ell-1} \frac{ -
  \sqrt{P_1P_2}\bar{\kappa}_{1,\ell-1}\bar{\kappa}_{2,\ell-1}
     \sqrt{ \frac{\alpha_{1,\ell-1}}{\D_{1,\ell-1}}}\sqrt{
      \frac{\alpha_{2,\ell-1}}{\D_{2,\ell-1}}}(1-\rho_{\ell-1}^2) +\rho_{\ell-1}(\Var{W_{\perp,\ell-1}}+N)}{\sqrt{
     P_1\bar{\kappa}_{1,\ell-1}^2 \frac{\alpha_{1,\ell-1}}{\D_{1,\ell-1}}(1-\rho_{\ell-1}^2)
      +\Var{W_{\perp,\ell-1}} +N}\sqrt{
      P_2\bar{\kappa}_{2,\ell-1}^2\frac{\alpha_{2,\ell-1}}{\D_{2,\ell-1}}(1-\rho_{\ell-1}^2)
      +\Var{W_{\perp,\ell-1}} +N}}\nonumber \\
\label{eq:rhonu}
\end{IEEEeqnarray}
\hrulefill
\end{figure*}
 where recall that $\alpha_{1,0}= 1$, $\alpha_{2,0}= 1$,
$\beta_{1,0}= 1$, $\beta_{2,0}= 1$,
$\rho_0= 0$, 
 $W_{\perp,0}=0$, where $\{\alpha_{1,\ell}\}_{\ell=1}^{\eta-1}$,
$\{\alpha_{2,\ell}\}_{\ell=1}^{\eta-1}$,
$\{\rho_\ell\}_{\ell=1}^{\eta-1}$ are defined by
Recursions~\eqref{eq:alphanu1}--\eqref{eq:rhonu} also on top of the next
page, and where $\bar{\kappa}_{1,0}\eqdef 1$,
$\bar{\kappa}_{2,0}\eqdef 1$, and 
 $\{\bar{\kappa}_{1,\ell}\}_{\ell=1}^{\eta-1}$ and
$\{\bar{\kappa}_{2,\ell}\}_{\ell=1}^{\eta-1}$ are defined
 by\footnote{Notice that for each $\nu\in\{1,2\}$ and each
 $\ell\in\{1,\ldots, \eta-1\}$ we have $\bar{\kappa}_{\nu,\ell}=
 \sqrt{\frac{\beta_{\nu,\ell}}{P_\nu}}\kappa_{\nu,\ell}$, when
 $\kappa_{\nu,\ell}$ is defined as in \eqref{eq:k1} or \eqref{eq:k2}
 in Section~\ref{sec:achaltnoisy}.}
\begin{IEEEeqnarray}{rCl}
 \bar{\kappa}_{1,\ell} & \triangleq&1+\sqrt{ \frac{
      \D_{1,\ell}}{P_1}}  \Bigg(\frac{ \alpha_{2,\ell} \Cov{
      \Er_{1,\ell}}{ W_{\ell}
    } }{(1-\rho_{\ell}^2)  \alpha_{1,\ell} \alpha_{2,\ell}}\nonumber
      \\ && \hspace{2cm} -\frac{ \rho_{\ell} \sqrt{\alpha_{1,\ell} \alpha_{2,\ell}}
    \Cov{\Er_{2,\ell}}{ W_{\ell}}
  }{(1-\rho_{\ell}^2)  \alpha_{1,\ell} \alpha_{2,\ell}}\Bigg),
     \nonumber \\ \label{eq:bkappa1}\\
 \bar{ \kappa}_{2,\ell} & \eqdef &1+ \sqrt{ \frac{
      \D_{2,\ell}}{P_2}}\Bigg(\frac{ \alpha_{1,\ell} \Cov{
      \Er_{2,\ell}}{ W_{\ell}
    }}{ (1-\rho_{\ell}^2)  \alpha_{1,\ell} \alpha_{2,\ell}} \nonumber
      \\ & &\hspace{2cm}- \frac{\rho_{\ell} \sqrt{\alpha_{1,\ell} \alpha_{2,\ell}}
    \Cov{\Er_{1,\ell}}{W_{\ell} }} {(1-\rho_{\ell}^2)
    \alpha_{1,\ell} \alpha_{2,\ell}}\Bigg),
\nonumber \\\label{eq:bkappa2}
\end{IEEEeqnarray}
and  $\{\D_{1,\ell}\}_{\ell=1}^{\eta-1}$,
$\{\D_{2,\ell}\}_{\ell=1}^{\eta-1}$,
$\{E_{1,\ell}\}_{\ell=1}^{\eta-1}$, $\{E_{2,\ell}\}_{\ell=1}^{\eta-1}$,
$\{W_{\ell}\}_{\ell=1}^{\eta-1}$, and
$\{W_{\perp,\ell}\}_{\ell=1}^{\eta}$ are defined by Equations
\eqref{eq:X1a}--\eqref{eq:coeff_choice2} and
by Equations \eqref{eq:epsi},~\eqref{eq:Wsum}, and \eqref{eq:WperpY}.
\end{definition}

\begin{corollary}[Noisy Feedback]\label{cor:ach}
 The capacity region of the
two-user AWGN MAC with noisy feedback contains all rate regions
$\bar{\set{R}}_{\eta}(P_1,P_2,N,\KW)$ for positive integers $\eta$, i.e., 
\begin{IEEEeqnarray*}{rCl}
\lefteqn{\capa_{\textnormal{NoisyFB}}(P_1,P_2,N, \KW) }\qquad
  \nonumber \\ &\supseteq &\cl{ \bigcup_{\eta \in\Nat}
  \bar{\set{R}}_{\eta}(P_1,P_2,N,\KW)}.
\end{IEEEeqnarray*}
\end{corollary}

\subsection{Choice of Parameters II}\label{sec:alterna}
In Section~\ref{sec:powerstar} we present a second choice of the
parameters ${\vect{a}}_1,{\vect{a}}_2, {\B}_1, {\B}_2$, and ${\C}$
given $\eta\in\Nat$. We only treat the noisy-feedback setting. The
choice we propose is based on extending the choice of parameters in
Section~\ref{sec:enc} with a form of power allocation as suggested in
\cite{kramer02}. We denote this choice by
$\tilde{\vect{a}}_1,\tilde{\vect{a}}_2,\tilde{\B}_1,\tilde{\B}_2,
\tilde{\C}$.

As we shall see, for our choice $\tilde{\C}$ is the LMMSE-estimation
matrix. Thus, the achievable region of our concatenated scheme with
parameters $\eta,
\tilde{\vect{a}}_1,\tilde{\vect{a}}_2,\tilde{\B}_1,\tilde{\B}_2,
\tilde{\C}$ is obtained by substituting the parameters
$\tilde{\vect{a}}_1,\tilde{\vect{a}}_2,\tilde{\B}_1,\tilde{\B}_2$ into
the RHSs of \eqref{eq:recexp} in
Section~\ref{sec:achaltnoisy}. The resulting achievable region is
presented in Corollary~\ref{cor:alte} ahead.

\subsubsection{Description of Parameters} \label{sec:powerstar}

We only consider the noisy feedback setting. An analogous choice of
the parameters for the partial feedback setting is obtained by similar
modifications as in Remark~\ref{rem:partialfb} in the previous appendix.

We first describe how Inner Encoder~1 and Inner Encoder~2 map the fed
symbols to the sequences of channel inputs $X_{1,1},\ldots,
X_{1,\eta}$ and $X_{2,1},\ldots, X_{2,\eta}$. This, then determines
$\tilde{\vect{a}}_1,\tilde{\vect{a}}_2,\tilde{\B}_1,\tilde{\B}_2$. The
matrix $\tilde{\C}$ is chosen as the LMMSE-estimation matrix. 

The inner encoders use the same linear strategies as in
Section~\ref{sec:enc}, with the only difference that here for every
fed symbol, Inner Encoder~1 scales the first produced symbol by a
constant $\sqrt{r}$, and similarly Inner Encoder~2 scales the first
produced symbol by the same constant $\sqrt{r}$, where $r\in[0,1]$ is
defined as the solution to
\begin{equation}\label{eq:r1r2}
\sqrt{\frac{ r^2 P_1P_2}{(rP_1+N)(rP_2+N)}}=\rho^*(P_1,P_2,N).
\end{equation}
Equation \eqref{eq:r1r2} has a unique solution in $[0,1]$ because 
\eqref{eq:r1r2} is strictly  increasing in
$r\in[0,1]$ and by
\begin{equation}\label{eq:inequ}
0<\rho^*(P_1,P_2,N)<\sqrt{\frac{P_1P_2}{(P_1+N)(P_2+N)}}.
\end{equation}
Here, Equation \eqref{eq:inequ} holds by the continuity of the expressions in
\eqref{eq:rhostar}, and because for $\rho=0$ the RHS of
\eqref{eq:rhostar} is strictly larger than its LHS, whereas
for $\rho=\sqrt{ \frac{P_1P_2}{(P_1+N)(P_2+N)}}$ the LHS of
\eqref{eq:rhostar} is strictly larger than its RHS.

 The reason for scaling the first produced symbols by $\sqrt{r}<1$
is to ensure that the correlation coefficient $\rho_1$ satisfies
$\rho_1=-\rho^*(P_1,P_2,N)$. This property is used in the proof of
Remark~\ref{th:perfectfbeta} in Section~\ref{sec:proofremperf}, where
we show that for perfect feedback and with the choice of parameters
presented in this section our concatenated scheme achieves the
sum-rate capacity.

The trick of reducing the powers of
certain channel inputs $X_{1,t}$ and $X_{2,t}$ in order to control the
next correlation coefficient $\rho_{t}$ was introduced in Kramer's
perfect-feedback scheme \cite{kramer02}. Ozarow uses a different trick
in his scheme \cite{ozarow85}. He assumes that the two transmitters
share a common randomness, which allows them to vary a
specific correlation coefficient $\rho_t$ by adding a scaled version
of the common randomness to their channel inputs $X_{1,t}$ and
$X_{2,t}$.

For the detailed description of the inner encoders we again replace
the fed symbols $\xi_1,\xi_2$ by the independent standard Gaussians
$\Xi_1$ and $\Xi_2$. Then, Inner Encoder~1 produces
\begin{IEEEeqnarray}{rCl}%
X_{1,1} & = & \sqrt{r P_1} \Xi_1, \label{eq:1A} \\
X_{1,\ell} &= &\sqrt{\frac{\pow_1}{\D_{1,\ell-1}}} \left( \Xi_1 -
\trans{ \boldsymbol{\gamma}_{1,\ell-1}} V_{1}^{\ell-1}
\right), \;\; \ell \in \{2,\ldots, \eta\}, \nonumber
\end{IEEEeqnarray}
and Inner Encoder~2 produces
\begin{IEEEeqnarray}{rCl}%
X_{2,1} & = & \sqrt{ rP_2} \Xi_2, \label{eq:2A}\\
X_{2,\ell}& = & (-1)^{\ell-1} \sqrt{\frac{\pow_2}{\D_{2,\ell-1}}}\left( \Xi_2 -
  \trans{\boldsymbol{\gamma}}_{2,\ell-1} \mat{M}_{\ell-1}
  V_{2,1}^{\ell-1}\right), \nonumber \\ && \hspace{4.6cm}\ell \in \{2,\ldots, \eta\},\nonumber
\end{IEEEeqnarray}
where $\{\mat{M}_{\ell}\}_{\ell=1}^{\eta-1}$,
$\{\D_{1,\ell}\}_{\ell=1}^{\eta-1}$,
$\{\D_{2,\ell}\}_{\ell=1}^{\eta-1}$,
$\{\boldsymbol{\gamma}_{1,\ell}\}_{\ell=1}^{\eta-1}$,
$\{\boldsymbol{\gamma}_{2,\ell}\}_{\ell=1}^{\eta-1}$ are defined as in
the previous appendix when the channel inputs $X_{1,1}$ and $X_{2,1}$
rather than being defined by~\eqref{eq:X1a} and \eqref{eq:X2a} are now
defined by~\eqref{eq:1A} and \eqref{eq:2A}, and where $r$ is defined
by \eqref{eq:r1r2}.

The described encodings correspond to the following parameters in the
concatenated scheme:
\begin{IEEEeqnarray*}{rCl}
\tilde{\vect{a}}_1 & \eqdef & \trans{\begin{pmatrix} \sqrt{rP_1}&
    \sqrt{\frac{P_1}{\beta_{1,1}}} & \ldots&
    \sqrt{\frac{P_1}{\beta_{1,\eta-1}}} \end{pmatrix}} ,\\ 
\tilde{\vect{a}}_2 & \eqdef & \trans{\begin{pmatrix} \sqrt{rP_2}&
   - \sqrt{\frac{P_2}{\beta_{2,1}}} & \ldots& (-1)^{\eta-1}
    \sqrt{\frac{P_2}{\beta_{2,\eta-1}}} \end{pmatrix}},
\end{IEEEeqnarray*} 
and 
\begin{IEEEeqnarray*}{rCl}
\tilde{\B}_1 & \eqdef &\trans{ \begin{pmatrix}\vect{0} &
  - \sqrt{\frac{P_1}{\beta_{1,1}}} \boldsymbol{\gamma}_{1,1}^{(0)} & \ldots &   -\sqrt{\frac{P_1}{\beta_{1,\eta-1}}}\boldsymbol{\gamma}_{1,\eta-1}^{(0)}
    \end{pmatrix}},\\ 
\tilde{\B}_2 & \eqdef &\trans{ \begin{pmatrix}\vect{0} &
   \sqrt{\frac{P_2}{\beta_{2,1}}}  \boldsymbol{\gamma}_{2,1}^{(0)} & \ldots &  (-1)^{\eta}
    \sqrt{\frac{P_2}{\beta_{2,\eta-1}}} \boldsymbol{\gamma}_{2,\eta-1}^{(0)} \end{pmatrix}},
\end{IEEEeqnarray*}
where $\boldsymbol{0}$ denotes the all-zero column-vector
 and where
$\left\{\boldsymbol{\gamma}_{1,\ell}^{(0)}\right\}_{\ell=1}^{\eta-1}$
and
$\left\{\boldsymbol{\gamma}_{2,\ell}^{(0)}\right\}_{\ell=1}^{\eta-1}$
are defined as in the previous appendix.

The matrix $\tilde{\C}$ is chosen as the LMMSE-estimation matrix
$\C_{\textnormal{LMMSE}}$, where recall that
\begin{IEEEeqnarray*}{rCl}%
\mat{C}_{\textnormal{LMMSE}}  =\trans{\tilde{\mat{A}}}_{\textnormal{r}} \left(
  \tilde{\mat{A}}_{\textnormal{r}} \trans{\tilde{\mat{A}}}_{\textnormal{r}} + N
  \I_{\eta} + \tilde{\mat{B}}_{\textnormal{r}} (\KW \otimes
  \I_\eta)\trans{\tilde{\mat{B}}}_{\textnormal{r}}\right)^{-1}, 
\end{IEEEeqnarray*}
where $\tilde{\mat{A}}_{\textnormal{r}}\eqdef \begin{pmatrix} \tilde{\vect{a}}_1 &
\tilde{\vect{a}}_2\end{pmatrix}$ and $\tilde{\B}_{\textnormal{r}} \eqdef
\begin{pmatrix}\tilde{\B}_1 & \tilde{\B}_2 \end{pmatrix}$. 

\subsubsection{Achievable Region}

\begin{definition}\label{def:Rtilde}
 For each $\eta\in \Nat$ define the rate region $\tilde{\set{R}}_{\eta}(P_1,P_2,N,\KW)$ as
  the set of all rate-pairs $(R_1,R_2)$ satisfying
  Constraints~\eqref{eq:R_achalt} on top of the next page, 
\begin{figure*}
\begin{subequations}\label{eq:R_achalt}
\begin{IEEEeqnarray}{rCl}
R_1 & \leq& \frac{1}{2\eta} \log\left(1+ \frac{rP_1}{N}\right)  +\frac{1}{2\eta}\sum_{\ell=2}^{\eta} \log\left( 1+ \frac{\pow_1\bar{\kappa}_{1,\ell-1}^2  \frac{ \alpha_{1,\ell-1}}{\D_{1,\ell-1}}\left(1-\rho_{\ell-1}^2\right)
  }{\Var{W_{\perp,\ell-1}}+N}
\right)\\
R_2 & \leq& \frac{1}{2\eta} \log\left(1+ \frac{rP_2}{N}\right)+\frac{1}{2\eta}\sum_{\ell=2}^{\eta}
\log\left( 1+ \frac{\pow_2 \bar{\kappa}_{2,\ell-1}^2   \frac{
  \alpha_{2,\ell-1}}{\D_{2,\ell-1}}   \left(1-\rho_{\ell-1}^2\right)
  }{\Var{W_{\perp,\ell-1}}+N}    \right) 
\\
R_1+R_2 &\leq& \frac{1}{2\eta} \log\left(1+
  \frac{rP_1+rP_2}{N}\right)\nonumber \\ && +\frac{1}{2\eta}\sum_{\ell=2}^{\eta}
\log\Bigg( 1 + \frac{\pow_1   \bar{\kappa}_{1,\ell-1}^2\frac{
  \alpha_{1,\ell-1}}{\D_{1,\ell-1}}
  +\pow_2  \bar{\kappa}_{2,\ell-1}^2 \frac{ \alpha_{2,\ell-1}}{\D_{2,\ell-1}}
 + 2\sqrt{\pow_1\pow_2 }\bar{\kappa}_{1,\ell-1}\bar{\kappa}_{2,\ell-1} \sqrt{\frac{
  \alpha_{1,\ell-1}}{\D_{1,\ell-1}}
 \frac{ \alpha_{2,\ell-1}}{\D_{2,\ell-1}}}
\rho_{\ell-1}
  }{\Var{W_{\perp,\ell-1}}+N}\Bigg)\nonumber \\
\end{IEEEeqnarray}
\end{subequations}
\hrulefill
\end{figure*}
where now (unlike in the previous appendix)
 \begin{IEEEeqnarray}{rCl}
\alpha_{1,1} & = & r P_1\frac{r P_2 +N}{r P_1+ rP_2
  +N},
\\
\alpha_{2,1} & = & r P_2\frac{r P_1 +N}{rP_1+r P_2 +N},
\\
\rho_{1} &= & -\rho^*(P_1,P_2,N),\label{eq:rho1}
\end{IEEEeqnarray}
where $r$ is the unique solution in $[0,1]$ to
\begin{equation}\label{eq:r1r2cor}
\sqrt{ \frac{r^2P_1 P_2 }{(rP_1+N)(rP_2+N)}}=\rho^*(P_1,P_2,N),
\end{equation}
and where the parameters
 $\{\alpha_{1,\ell}\}_{\ell=2}^{\eta-1}$, $\{\alpha_{2,\ell}\}_{\ell=2}^{\eta-1}$, $\{\rho_{\ell}\}_{\ell=2}^{\eta-1}$, $\{\beta_{1,\ell}\}_{\ell=1}^{\eta-1}$,
 $\{\beta_{2,\ell}\}_{\ell=1}^{\eta-1}$,
 $\{\bar{\kappa}_{1,\ell}\}_{\ell=1}^{\eta-1}$,
 $\{\bar{\kappa}_{2,\ell}\}_{\ell=1}^{\eta-1}$,
 $\{W_{\perp,\ell}\}_{\ell=1}^{\eta-1}$ are defined
 as in the previous appendix, if the input symbols $X_{1,1}$ and $X_{2,1}$ rather than being defined by \eqref{eq:X1a} and
  \eqref{eq:X2a} are now defined by~\eqref{eq:1A} and \eqref{eq:2A}.
\end{definition}

\begin{corollary}\label{cor:alte}
 For the
two-user AWGN MAC with noisy feedback our concatenated scheme with the
parameters described in Section~\ref{sec:powerstar} achieves all rate pairs in the regions
$\set{\tilde{R}}_{\eta}(P_1,P_2,N,\KW)$ for positive integers $\eta$,
i.e., 
\begin{IEEEeqnarray*}{rCl}
\lefteqn{\capa_{\textnormal{NoisyFB}}(P_1,P_2,N,\KW)}\qquad \\ &  \supseteq &
\cl{\bigcup_{\eta\in\Nat} \set{\tilde{R}}_{\eta}(P_1,P_2,N,\KW)}.
\end{IEEEeqnarray*}
\end{corollary}

\begin{remark} \label{def:remarkperf}
   Specializing the region in
  Definition~\ref{def:Rtilde} to perfect feedback, i.e., to
  $\KW=\mat{0}$, results in the region
  $\set{\tilde{R}}_{\eta}\left(P_1,P_2,N, \mat{0}\right)$, which is
  defined as the set of all rate pairs $(R_1,R_2)$ satisfying
\begin{subequations}\label{eq:Perfmod}
\begin{IEEEeqnarray}{rCl}
R_1& \leq &\frac{1}{2\eta}\log \left( 1+ \frac{r P_1}{N}\right)
\nonumber \\ & &+ 
\sum_{\ell=
2}^{\eta}\frac{1}{2\eta}\log\left(1+  \frac{P_1(1-\rho_{\ell-1}^2) }{N } \right),\;\;\label{eq:Perf1mod}\\
R_2 & \leq  &\frac{1}{2\eta }\log \left( 1+
    \frac{r P_2 }{N}\right)\nonumber \\ &&+\sum_{\ell=2}^{\eta}\frac{1}{2\eta} \log \left(1+
  \frac{P_2(1-\rho_{\ell-1}^2) }{N }\right),\;\;\label{eq:Perf2mod} \\
R_1+R_2 & \leq &
\frac{1}{2\eta}\log\left(1+\frac{rP_1+
      rP_2}{N}\right) 
 \nonumber \\& & +\sum_{\ell=2}^{\eta}\frac{1}{2\eta}\log \left(1+
  \frac{P_1+P_2+2 }{N} \right.\nonumber \\ & & \left. \hspace{2.4cm}+\frac{\sqrt{P_1P_2}(-1)^{\ell-1}\rho_{\ell-1} }{N }
\right)\nonumber \\\label{eq:Perf12mod}
\end{IEEEeqnarray}
\end{subequations}
where the sequence
$\{\rho_{\ell}\}_{\ell=1}^{\eta-1}$ is recursively defined by
$\rho_1=-\rho^*(P_1,P_2,N)$ and for $\ell \in \{2,\ldots, \eta\}$ by
\begin{equation}\label{eq:a3}
\rho_{\ell} = \frac{\rho_{\ell-1} N -
  (-1)^{\ell-1}\sqrt{P_1P_2}(1-\rho_{\ell-1}^2)
}{\sqrt{P_1(1-\rho_{\ell-1}^2)+N}\sqrt{P_2(1-\rho_{\ell-1}^2)+N} }, \qquad 
\end{equation} 
and where $r$ is the unique solution in $[0,1]$ to \eqref{eq:r1r2cor}.
 \end{remark}
\begin{proof}
Notice that if $\KW=\mat{0}$, then trivially $W_{\perp,\ell}=0$, for
$\ell\in\{1,\ldots, \eta-1\}$, and   
Definitions \eqref{eq:Dell1}--\eqref{eq:coeff_choice2},
 \eqref{eq:bkappa1}, and \eqref{eq:bkappa2} result in
\begin{IEEEeqnarray}{rCl}
\boldsymbol{\gamma}_{\nu,\ell} &=& \mat{K}_{Y^{\ell}}^{-1}\mat{K}_{Y^{\ell},\Xi_{\nu}}, 
\\
{\beta}_{\nu,\ell} & = & \Var{\Xi_{\nu} -  \trans{\mat{K}}_{Y^{\ell},
    \Xi_{\nu}}\mat{K}_{Y^{\ell}}^{-1} Y^{\ell}}=\alpha_{\nu,\ell},\\
\bar{\kappa}_{\nu,\ell}&=&1. \label{eq:aa4}
\end{IEEEeqnarray}
Thus, for perfect feedback the parameters suggested in
Section~\ref{sec:powerstar} are LMMSE-estimation error parameters, which
are optimal for perfect feedback in the sense discussed in
Section~\ref{sec:choicenoisy}. The rate
expressions in~\eqref{eq:R_ach} then result in
 Expressions~\eqref{eq:Perfmod}, and 
Recursion \eqref{eq:rhonu} results in \eqref{eq:a3}.
This concludes the proof of the remark.
\end{proof}

\subsection{Choice of Parameters III}\label{app:si_nonsym}
In this section we consider the noisy-feedback setup with receiver
side-information, and  we present for
each $\eta \in\Nat$ a specific choice of the
parameters $\vect{a}_1,\vect{a}_2,\B_1,\B_2,\C_{\textnormal{SI}}$,
which we call
$\breve{\vect{a}}_1,\breve{\vect{a}}_2,\breve{\B}_1,\breve{\B}_2,\breve{\C}_{\textnormal{SI}}$.
As we shall see, the matrix $\breve{\C}_{\textnormal{SI}}$ is chosen
as the LMMSE-estimation matrix. Thus, the achievable region of our
concatenated scheme with parameters
$\eta,\breve{\vect{a}}_1,\breve{\vect{a}}_2,\breve{\B}_1,\breve{\B}_2,\breve{\C}_{\textnormal{SI}}$
is obtained by substituting the parameters
$\breve{\vect{a}}_1,\breve{\vect{a}}_2,\breve{\B}_1,\breve{\B}_2$ into
\eqref{eq:recexpsi}. The resulting achievable
region is presented in Corollary~\ref{th:R_ach_SI} ahead.

\subsubsection{Description of Parameters}\label{sec:parametersSI}
Let a positive integer $\eta\in\Nat$ be given.  We
first describe how Inner Encoder~1 and Inner Encoder~2 map the fed
symbols to the channel inputs. This then determines
$\breve{\vect{a}}_1,\breve{\vect{a}}_2,\breve{\B}_1,\breve{\B}_2,\breve{\C}$.
To simplify the description we replace the symbols $\xi_1$ and $\xi_2$
fed to the inner encoders by the independent standard Gaussians
$\Xi_1$ and $\Xi_2$. We choose the inner encoders to produce
\begin{IEEEeqnarray}{rCl}\label{eq:X1SIa}
X_{1,1} &=& \sqrt{P_1}\Xi_1,\\
X_{2,1} & = & \sqrt{P_2}\Xi_2,\label{eq:X2SIa}
\end{IEEEeqnarray}
and  for $\ell \in
\{2,\ldots, \eta\}$:
\begin{IEEEeqnarray}{rCl}
X_{1,\ell} &=&\sqrt{\frac{\pow_1}{\breve{\beta}_{1,\ell-1}}} \left( \Xi_1 -
\breve{\boldsymbol{\gamma}}_{1,\ell-1}^{\T} V_{1}^{\ell-1}
\right),  \label{eq:X1SIb} \\
X_{2,\ell}& = & (-1)^{\ell-1} \sqrt{\frac{\pow_2}{\breve{\beta}_{2,\ell-1}}}\left( \Xi_2 -
  \breve{\boldsymbol{\gamma}}_{2,\ell-1}^{\T} \mat{M}_{\ell-1}
  V_{2}^{\ell-1}\right),\nonumber \\ \label{eq:X2SIb}
\end{IEEEeqnarray}
where for $\ell\in\{1,\ldots, \eta-1\}$ the matrix $\mat{M}_{\ell}$ is
defined as in \eqref{eq:M} and
\begin{IEEEeqnarray}{rCl}
\label{eq:Dell1SI}\breve{\beta}_{1,\ell} &\triangleq & \Var{\Xi_{1} - \breve{\boldsymbol{\gamma}}_{1,\ell}^{\T}
  V_{1}^{\ell}}, \\
\breve{\beta}_{2,\ell}& \triangleq &\Var{\Xi_{2} - {\breve{\boldsymbol{\gamma}}_{2,\ell}^{\T}} \mat{M}_{\ell}
  V_{2}^{\ell}},\label{eq:Dell2SI}\\
\label{eq:coeff_choiceSI1}
\breve{\boldsymbol{\gamma}}_{1,\ell}&  = &  \mat{K}_{V_{1}^{\ell}}^{-1}
  \mat{K}_{V_{1}^{\ell}, \Xi_1},\\
\label{eq:coeff_choiceSI2}
\breve{\boldsymbol{\gamma}}_{2,\ell}&  = &  \mat{K}_{V_{2}^{\ell}}^{-1}
  \mat{K}_{V_{2}^{\ell}, \Xi_2}.
\end{IEEEeqnarray}
Notice that this choice
 implies that the $\ell$-th channel input
produced by Inner Encoder~1 is a scaled version of
 the LMMSE-estimation error of $\Xi_1$ based
 on the past feedback outputs $V_{1,1}, \ldots, V_{1,\ell-1}$.
Similarly, for Inner Encoder~2. 

The described encodings correspond to the following parameters of the
concatenated scheme:
\begin{IEEEeqnarray*}{rCl}
\breve{\vect{a}}_1 & \eqdef & \trans{\begin{pmatrix} \sqrt{P_1}&
    \sqrt{\frac{P_1}{\breve{\beta}_{1,1}}} & \ldots&
    \sqrt{\frac{P_1}{\breve{\beta}_{1,\eta-1}}} \end{pmatrix}},\\ 
\breve{\vect{a}}_2 & \eqdef & \trans{\begin{pmatrix} \sqrt{P_2}&
   - \sqrt{\frac{P_2}{\breve{\beta}_{2,1}}} & \ldots& (-1)^{\eta-1}
    \sqrt{\frac{P_2}{\breve{\beta}_{2,\eta-1}}} \end{pmatrix}},
\end{IEEEeqnarray*} 
and 
\begin{IEEEeqnarray*}{rCl}
\breve{\B}_1 & \eqdef &\trans{ \begin{pmatrix}\vect{0} &
    - \sqrt{\frac{P_1}{\breve{\boldsymbol{\gamma}}_{1,1}}} \breve{\boldsymbol{\gamma}}_{1,1}^{(0)} & \ldots &- \sqrt{\frac{P_1}{\breve{\beta}_{1,\eta-1}}}  \breve{\boldsymbol{\gamma}}_{1,\eta-1}^{(0)}
    \end{pmatrix}},\\ 
\breve{\B}_2 & \eqdef &\trans{ \begin{pmatrix}\vect{0} &
     \sqrt{\frac{P_2}{\breve{\beta}_{2,1}}}  \breve{\boldsymbol{\gamma}}_{2,1}^{(0)} & \ldots & (-1)^{\eta}
    \sqrt{\frac{P_2}{\breve{\beta}_{2,\eta-1}}} \breve{\boldsymbol{\gamma}}_{2,\eta-1}^{(0)} \end{pmatrix}},
\end{IEEEeqnarray*}
where the
vectors $\left\{\breve{\boldsymbol{\gamma}}_{1, \ell}^{(0)}\right\}_{\ell=1}^{\eta-1}$ and
$\left\{\breve{\boldsymbol{\gamma}}_{2,\ell}^{(0)}\right\}_{\ell=1}^{\eta-1}$ are defined as the $\eta$-dimensional vector
obtained by stacking the $\ell$-dimensional column-vector
$\breve{\boldsymbol{\gamma}}_{\nu,\ell}$ on top of an $(\eta-\ell)$-dimensional
column-vector with all zero entries, i.e., 
\begin{equation*}
\breve{\boldsymbol{\gamma}}_{\nu, \ell}^{(0)} \eqdef \begin{pmatrix}\breve{\boldsymbol{\gamma}}_{\nu,\ell} \\
  \vect{0} \end{pmatrix}, \qquad \ell \in \{1,\ldots, \eta-1\}, \quad \nu\in\{1,2\}.
\end{equation*}

The matrix $\breve{\C}_{\textnormal{SI}}$ is chosen as the LMMSE-estimation matrix
with side-information, i.e., 
\begin{IEEEeqnarray*}{rCl}
\breve{\mat{\C}}_{\textnormal{SI}} = \trans{\breve{\mat{A}}}_{\textnormal{r}} (
\breve{\mat{A}}_{\textnormal{r}}\trans{\breve{\mat{A}}}_{\textnormal{r}} + N \I_\eta)^{-1},
\end{IEEEeqnarray*}
where $\breve{\mat{A}}_{\textnormal{r}}\eqdef \begin{pmatrix} \breve{\vect{a}}_1 & \breve{\vect{a}}_2\end{pmatrix} $.

\subsubsection{Achievable Region}\label{sec:Achsiss}

\begin{definition}
For each $\eta\in\Nat$ define the region
$\breve{\set{R}}_{\eta}(P_1,P_2,N,\KW)$ as the set of
all rate pairs $(R_1,R_2)$ satisfying 
\begin{IEEEeqnarray*}{rCl}
 R_1&  \leq & \frac{1}{2 \eta} \sum_{\ell=1}^{\eta} \log\left(1+
 \frac{ \pow_1\frac{\alpha_{1,\ell-1}}{\breve{\beta}_{1,\ell-1}}
    \left(1-
    \rho_{\ell-1}^2\right)}{\N} 
  \right), \\
 R_2 & \leq & \frac{1}{2 \eta} \sum_{\ell=1}^{\eta} \log\left(1+ \frac{ \pow_2 \frac{\alpha_{2,\ell-1}}{\breve{\beta}_{2,\ell-1}}
    \left(1-\rho_{\ell-1}^2\right)}{\N}
  \right),\\
\lefteqn{ R_1+R_2}\nonumber \\ & \leq &  \frac{1}{2 \eta}  \sum_{\ell=1}^{\eta} \log \left(
 1 + \frac{
 \pow_1\frac{\alpha_{1,\ell-1}}{\breve{\beta}_{1,\ell-1}}+\pow_2\frac{\alpha_{2,\ell-1}}{\breve{\beta}_{2,\ell-1}}}{N} \right.
  \nonumber \\ & & \left. +\frac{2\sqrt{\pow_1\pow_2}
    \sqrt{\frac{\alpha_{1,\ell-1}\alpha_{2,\ell-1}}{\breve{\beta}_{1,\ell-1}\breve{\beta}_{2,\ell-1}}} \left(1+ (-1)^{\ell-1}
    \rho_{\ell-1}\right)}{\N}\right),%
\end{IEEEeqnarray*}
where  $\alpha_{1,0} =1$,  $\alpha_{2,0}
= 1$, $\rho_0= 0$, and $\{\alpha_{1,\ell}\}_{\ell=1}^{\eta}$, $\{\alpha_{2,\ell}\}_{\ell=1}^{\eta},$ and
$\{\rho_{\ell}\}_{\ell=1}^{\eta}$ are recursively given by Recursions
\eqref{eq:1alpha}--\eqref{eq:0rho} displayed on top of the next page.
\begin{figure*}
\begin{IEEEeqnarray}{rCl}
\alpha_{1,\ell}
&=& 
\alpha_{1,\ell-1}\left( 1+\frac{\pow_1\frac{\alpha_{1,\ell-1}}{\breve{\beta}_{1,\ell-1}}+ 2 \sqrt{\pow_1\pow_2}
    \left( 1 +(-1)^{\ell-1} \rho_{\ell-1}\right)
    \sqrt{\frac{ \alpha_{1,\ell-1}
        \alpha_{2,\ell-1}  }{ \breve{\beta}_{1,\ell-1}
        \breve{\beta}_{2,\ell-1}  }} }{\frac{\alpha_{2,\ell-1}}{\breve{\beta}_{2,\ell-1}} \pow_2
    \left( 1 - \rho_{\ell-1}^2\right) +
    \N} \right)^{-1}\label{eq:1alpha} \\
\alpha_{2,\ell} & =& 
\alpha_{2,\ell-1}\left(1+\frac
{\pow_2 \frac{\alpha_{1,\ell-1}}{\breve{\beta}_{2,\ell-1}}+ 2 \sqrt{\pow_1\pow_2}
    \left( 1 +(-1)^{\ell-1} \rho_{\ell-1}\right)
    \sqrt{\frac{ \alpha_{1,\ell-1}
        \alpha_{2,\ell-1}  }{ \breve{\beta}_{1,\ell-1}
        \breve{\beta}_{2,\ell-1}  }} }
{\frac{\alpha_{1,\ell-1}}{\breve{\beta}_{1,\ell-1}} \pow_1
    \left( 1 - \rho_{\ell-1}^2\right) +
    \N} \right)^{-1}\label{eq:2alpha}\\
\rho_{\ell} & = & \frac{ -\sqrt{P_1P_2}\sqrt{\frac{
  \alpha_{1,\ell-1}  \alpha_{2,\ell-1}
}{\breve{\beta}_{1,\ell-1} \breve{\beta}_{2,\ell-1} }}
  (1-\rho_{\ell-1}^2)+\rho_{\ell-1} N  }{\sqrt{ P_1  \frac{\alpha_{1,\ell-1}
  }{\breve{\beta}_{1,\ell-1} } (1-\rho_{\ell-1}^2)+N}  \sqrt{ P_2  \frac{\alpha_{2,\ell-1}
  }{\breve{\beta}_{2,\ell-1} } (1-\rho_{\ell-1}^2)+N}  }\label{eq:0rho}
\end{IEEEeqnarray}
\hrulefill
\end{figure*}
and where $\breve{\beta}_{1,0} = 1,
\breve{\beta}_{2,0} = 1 $, and
$\{\breve{\beta}_{1,\ell}\}_{\ell=1}^{\eta-1}$ and
$\{\breve{\beta}_{2,\ell}\}_{\ell=1}^{\eta-1}$ are described by
\eqref{eq:X1SIa}--\eqref{eq:coeff_choiceSI2}.
\end{definition}
\begin{corollary}\label{th:R_ach_SI}
  The capacity region $\capa_{\textnormal{NoisyFBSI}}(P_1,P_2,N,\KW)$ of the
  two-user Gaussian MAC with noisy feedback and receiver side-information
  contains the rate regions
  $\breve{\set{R}}_{\eta}(P_1,P_2,N,\KW)$ for
  positive integers $\eta$, i.e., 
\begin{IEEEeqnarray*}{rCl}
 \lefteqn{\capa_{\textnormal{NoisyFBSI}}(P_1,P_2,N,\KW)}\qquad \\& \supseteq&
\cl{ \bigcup_{\eta \in \Nat} \breve{\set{R}}_{\eta}(P_1,P_2,N,\KW)}.
\end{IEEEeqnarray*}
\end{corollary}

\subsection{Rate-Splitting with Carleial's Cover-Leung Scheme}\label{sec:RSCL2}

In this section we describe the rate-splitting scheme in
Section~\ref{sec:extCarleial} in more detail. We consider the version
of the scheme where after each Block $b\in\{1,\ldots, B\}$
Transmitter~1 first decodes Message $M_{2,\textnormal{CS},b}$ before
decoding $M_{2,\textnormal{CL},b}$. Similarly, for Transmitter~2.

We first describe the encodings. We start with the encodings in
Block~$b$, for a fixed $b\in\{1,\ldots, B\}$, where we assume that
from decoding steps in the previous block $(b-1)$ both transmitters
are cognizant of the pair $(M_{1,\textnormal{CL},b-1},
M_{2,\textnormal{CL},b-1})$.  Given
$M_{\nu,\textnormal{CL},b}=m_{\nu,\textnormal{CL},b}$,
$M_{1,\textnormal{CL},b-1}=m_{1,\textnormal{CL},b-1}$, and
$M_{2,\textnormal{CL},b-1}=m_{2,\textnormal{CL},b-1}$,
Transmitter~$\nu$, for $\nu\in\{1,2\}$, picks the codewords
$\vect{u}_{\nu,b}(m_{\nu,\textnormal{CL},b})\eqdef
(u_{\nu,b,1},\ldots, u_{\nu,b,\eta n})$,
$\boldsymbol{\omega}_{1,b}(m_{1,\textnormal{CL},b-1}) \eqdef
(\omega_{1,b,1},\ldots, \omega_{1,b,\eta n} )$, and
$\boldsymbol{\omega}_{2,b}(m_{2,\textnormal{CL},b-1})\eqdef(\omega_{2,b,1},\ldots,
\omega_{2,b,\eta n} )$ from the corresponding codebooks, which have
independently been generated by randomly drawing each entry according
to an IID zero-mean unit-variance Gaussian distribution\footnote{To
  satisfy the power constraints the Gaussian distribution should be of
  variance slightly less than 1. However, this is a technicality which
  we ignore.}. Fix correlation coefficients $\rho_{1},
\rho_2\in[0,1]$, which are constant over all blocks $b\in \{1,\ldots
B\}$. Transmitter~$\nu$  computes the following linear
combinations for $k\in\{1,\ldots,n\}$ and $\nu\in\{1,2\}$:
\begin{IEEEeqnarray}{rCl}\label{eq:t1}
  \sqrt{(1-\rho_{\nu}^2)P_\nu'} \vect{u}_{\nu,b,k}+ \sqrt{\frac{1}{2}
    \rho_\nu^2 P_\nu'} \left(\boldsymbol{\omega}_{1,b,k}+
    \boldsymbol{\omega}_{2,b,k}\right), \IEEEeqnarraynumspace
\end{IEEEeqnarray}
where
\begin{IEEEeqnarray*}{rCl}
  \vect{u}_{\nu,b,k} \eqdef \trans{ (u_{\nu, b,(k-1)\eta+1},\ldots,
    u_{\nu, b, k\eta})},
   \\
  \boldsymbol{\omega}_{\nu,b,k} \eqdef \trans{ (\omega_{\nu,
      b,(k-1)\eta+1},\ldots, \omega_{\nu, b, k\eta})}.  
\end{IEEEeqnarray*}
Moreover, Transmitter~$\nu$ uses our concatenated code to encode
Message $M_{\nu,\textnormal{CS},b}$. Specifically, given
$M_{\nu,\textnormal{CS},b}=m_{\nu,\textnormal{CS},b}$,
Transmitter~$\nu$ feeds $m_{\nu,\textnormal{CS},b}$ to Outer
Encoder~$\nu$, which picks the codeword
$\boldsymbol{\xi}_\nu(m_{\nu,\textnormal{CS},b}) \eqdef
\trans{(\xi_{\nu,b,1},\ldots, \xi_{\nu,b,n})}$ corresponding to
$m_{\nu,\textnormal{CS},b}$ and feeds it to Inner Encoder~$\nu$.
Denoting the parameters of Inner Encoder~$\nu$ by $\vect{a}_\nu$ and
$\B_\nu$, Inner Encoder~$\nu$ produces the $\eta$-dimensional vectors
\begin{IEEEeqnarray}{rCl}\label{eq:s1}
\vect{a}_\nu \xi_{\nu,b,k} + \B_\nu \vect{V}_{\nu,b,k}, \qquad k \in
\{k,\ldots, n\},\IEEEeqnarraynumspace
\end{IEEEeqnarray}
where 
\begin{IEEEeqnarray}{rCl}
\vect{V}_{\nu,b,k}& \eqdef& \trans{ (V_{\nu,(b-1) \eta
    n+(k-1)\eta+1},\ldots, V_{\nu,(b-1)\eta n+k\eta})}.\nonumber \IEEEeqnarraynumspace
\end{IEEEeqnarray}

The signal transmitted by Transmitter~$\nu$ is then described by the sum of the vectors in
\eqref{eq:t1} and \eqref{eq:s1} as follows. For $k \in \{1,\ldots,
n\}$ and $\nu \in\{1,2\}$
\begin{IEEEeqnarray}{rCl}\label{eq:X1u}
\vect{X}_{\nu,b,k} & =& \sqrt{(1-\rho_\nu^2)P_\nu'}
\vect{u}_{\nu,b,k} \nonumber \\ & & +
\sqrt{\frac{1}{2} \rho_\nu^2 P_\nu'}\left( \boldsymbol{\omega}_{1,b,k}
 + \boldsymbol{\omega}_{2,b,k}\right)   \nonumber \\ &&
+\vect{a}_\nu \xi_{\nu,b,k} + \B_\nu
\vect{V}_{\nu,b,k},
\end{IEEEeqnarray}
where
\begin{IEEEeqnarray}{rCl}
\vect{X}_{\nu,b,k}& \eqdef& \trans{ (X_{\nu, (b-1)\eta n+
    (k-1)\eta+1},\ldots, X_{\nu,(b-1)\eta n+ k\eta})}.
\nonumber
    \IEEEeqnarraynumspace
\end{IEEEeqnarray}

Notice that if $\vect{a}_1,\vect{a}_2,\B_1$, and $\B_2$  satisfy the
power constraints  \eqref{eq:powergen}
 for powers $(P_1-P_1')$ and $(P_2-P_2')$,
noise variance $(N+P_1'+ P_2' + 2 \sqrt{P_1'P_2'} \rho_1 \rho_2 )$,
and feedback-noise covariance matrix $\KW$ and if the outer code's
codewords $\{\boldsymbol{\Xi}_{1}(M_{1,\textnormal{CS},b})\}$ and
$\{\boldsymbol{\Xi}_{2}(M_{2,\textnormal{CS},b})\}$ are zero-mean and average block-power constrained to 1, then the channel input sequences satisfy the power constraints
with arbitrary high probability.

In Block $(B+1)$ the two transmitters only send information
about the pair $(M_{1,\textnormal{CL},B}, M_{2,\textnormal{CL},B})$. 
Given $M_{1,\textnormal{CL},B}=m_{1,\textnormal{CL},B}$ and
$M_{2,\textnormal{CL},B}=m_{2,\textnormal{CL},B}$, both transmitters
pick the  codewords
$\boldsymbol{\omega}_{1,B+1}(m_{1,\textnormal{CL,B}})\eqdef\trans{(\omega_{1, B+1, 1}, \ldots, \omega_{1,B+1, \eta n})}$ and 
$\boldsymbol{\omega}_{2,B+1}(m_{2,\textnormal{CL,B}})\eqdef
  \trans{(\omega_{2, B+1, 1}, \ldots, \omega_{2,B+1, \eta n})}$ from
  the corresponding codebooks and form a linear combination  of
  power $P_\nu'$. Thus, defining
\begin{IEEEeqnarray*}{rCl}
\vect{X}_{\nu,B+1}&\triangleq &\trans{(X_{\nu,B\eta n+ 1
    },\ldots, X_{\nu,(B+1)\eta n  })}, \\
\boldsymbol{\omega}_{\nu,B+1} & \triangleq & \trans{(\omega_{\nu, B+1, 1}, \ldots, \omega _{\nu,B+1, \eta n})},
\end{IEEEeqnarray*}
the signal transmitted by Transmitter~$\nu$ can be described as
\begin{IEEEeqnarray}{rCl}\label{eq:s1b}
  \vect{X}_{\nu,B+1} & = & \sqrt{ \frac{1}{2} \rho_\nu^2
    P_\nu'}\left(\boldsymbol{\omega}_{1,B+1}
  + \boldsymbol{\omega}_{2,B+1}\right).\IEEEeqnarraynumspace
\end{IEEEeqnarray}

Next, we describe the decodings. We start with the decoding at
Transmitter~2; the decoding at Transmitter~1 is performed similarly
and therefore omitted; and the decodings at the receiver are described
later on.

Recall that after a fixed block $b$, for $b\in \{1,\ldots, B\}$,
Transmitter~2 first decodes Message $M_{1,\textnormal{CS},b}$,
followed by  Message $M_{1, \textnormal{CL},b}$. After Block
$b$, Transmitter~2 observed
$\{\vect{V}_{2,b,1}, \ldots,\vect{V}_{2,b,n}\}$, and additionally is
cognizant of the realizations of $\{\vect{U}_{2,b,1}, \ldots,
\vect{U}_{2,b,n}\}$, $\{\boldsymbol{\Omega}_{1,b,1}, \ldots,
\boldsymbol{\Omega}_{1,b,n}\}$, $\{\boldsymbol{\Omega}_{2,b,1}, \ldots,
\boldsymbol{\Omega}_{2,b,n}\}$, and $\{\Xi_{2,b,1}, \ldots, \Xi_{2,b,n}\}$.
It can thus compute for $k\in\{1,\ldots,n\}$:
\begin{IEEEeqnarray*}{rCl}
  \vect{\tilde{V}}_{2,b,k} & \triangleq & 
(\I-(\B_1+\B_2))  \vect{V}_{2,b,k} - \sqrt{(1-\rho_2^2)P_2'} \vect{U}_{2,b,k} \nonumber
  \\ && - \left(
    \sqrt{\frac{1}{2} \rho_1^2 P_1'} + \sqrt{\frac{1}{2} \rho_2^2
      P_2'}\right) \left( \boldsymbol{\Omega}_{1,b,k} +
    \boldsymbol{\Omega}_{2,b,k}\right)\nonumber \\
  & &
  -\vect{a}_2 \Xi_{2,b,k} \nonumber\\
  & =& \vect{a}_1 \cdot \Xi_{1,b,k} +
  \sqrt{(1-\rho_1^2)P_1'}\vect{U}_{1,b,k} +
  \vect{Z}_{b,k} \nonumber \\ &&+ \vect{W}_{2,b,k}
  +  \B_1
  \left(\vect{W}_{1,b,k}-\vect{W}_{2,b,k}\right), 
  \IEEEeqnarraynumspace
\end{IEEEeqnarray*}
where 
\begin{IEEEeqnarray*}{rCl}
\vect{Z}_{b,k} & \eqdef & \trans{(Z_{ (b-1)\eta n + (k-1) n + 1},
  \ldots, Z_{(b-1)\eta n + kn})},  \\
\vect{W}_{\nu,b,k} & \eqdef & \trans{(W_{\nu, (b-1)\eta n + (k-1) n + 1},
  \ldots, W_{\nu,(b-1)\eta n + kn})}.
\end{IEEEeqnarray*}
Since the sequence $\left\{\vect{\tilde{V}}_{2,b,1}, \ldots,
  \vect{\tilde{V}}_{2,b,n}\right\}$ is independent of the additional
information $\{\vect{U}_{2,b,1}, \ldots, \vect{U}_{2,b,n}\}$,
$\{\boldsymbol{\Omega}_{1,b,1}, \ldots,
\boldsymbol{\Omega}_{1,b,n}\}$, $\{\boldsymbol{\Omega}_{2,b,1},
\ldots, \boldsymbol{\Omega}_{2,b,n}\}$, and $\{\Xi_{2,b,1}, \ldots,
\Xi_{2,b,n}\}$, Transmitter~2 can optimally decode Message
$M_{1,\textnormal{CS},b}$ based on $\left\{\vect{\tilde{V}}_{2,b,1},
  \ldots, \vect{\tilde{V}}_{2,b,n}\right\}$ only. To this end, it does
not apply the inner and outer decoder of the concatenated scheme, but
directly applies an optimal decoder for a Gaussian single-input
antenna/$\eta$-output antenna channel with temporally-white noise sequences which
are correlated across antennas. 
Let $\hat{M}_{1,\textnormal{CS}}^{(\textnormal{Tx}2)}$ denote Transmitter~2's guess of Message
$M_{1,\textnormal{CS}}$ 
 and let 
$\trans{\left(\hat{\Xi}_{1,b,1}^{(\textnormal{Tx}2)},\ldots,\hat{\Xi}_{1,b,n}^{(\textnormal{Tx}2)}\right)}$
be the corresponding codeword of the outer code. 

Transmitter~2 then decodes Message $M_{1,\textnormal{CL}, b}$ as
follows. It first attempts to
subtract the influence of the sequence produced by encoding 
$M_{1,\textnormal{CS},b}$ and to this end computes
\begin{IEEEeqnarray*}{rCl}
  \vect{\tilde{V}}^{(2)}_{2,b,k} & \triangleq &
  \vect{\tilde{V}}_{2,b,k}- \vect{a}_1\hat{\Xi}_{1,b,k}^{(\textnormal{Tx}2)} , \qquad
  k\in\{1,\ldots,n\},
\end{IEEEeqnarray*}
which, if Transmitter~2 successfully decoded $M_{1,\textnormal{CS},b}$, equals 
\begin{IEEEeqnarray*}{rCl}
   \lefteqn{\sqrt{(1-\rho_1^2)P_1'}\vect{U}_{1,b,k} + \vect{Z}_{b,k}+\vect{W}_{2,b,k}}
 \nonumber \qquad \qquad \\ && + \B_1
  \left(\vect{W}_{1,b,k}-\vect{W}_{2,b,k}\right), \qquad 
  k\in\{1,\ldots,n\}.
\end{IEEEeqnarray*}
 Transmitter~2 then decodes Message
$M_{1,\textnormal{CL},b}$
based on the sequences $\left\{\vect{\tilde{V}}^{(2)}_{2,b,1}, \ldots,
\vect{\tilde{V}}^{(2)}_{2,b,n}\right\}$
using an optimal decoder for a Gaussian $\eta$-input antenna/$\eta$-output
antenna channel with temporally-white noise sequences correlated across antennas.

As a last element, we describe the decodings at the receiver. After
each block $b\in\{1,\ldots,B\}$ the receiver performs two decoding
steps. In the first step it decodes Messages
$(M_{1,\textnormal{CS},b},M_{2,\textnormal{CS},b})$ while treating the
sequences produced to encode ${M}_{1,\textnormal{CL},b-1},$
${M}_{2,\textnormal{CL},b-2},$ ${M}_{1,\textnormal{CL},b},$ and
${M}_{2,\textnormal{CL},b}$ as additional noise. For this decoding
step the receiver uses inner and outer decoders of the concatenated
scheme. Let
$\left(\hat{M}_{1,\textnormal{CS},b},\hat{M}_{2,\textnormal{CS},b}\right)$
denote the receiver's guess of the pair
$\left({M}_{1,\textnormal{CS},b},{M}_{2,\textnormal{CS},b}\right)$
produced in this first step, and
let $\trans{\left(\hat{\Xi}_{1,b,1}^{(\textnormal{Rx})}
  \ldots,\hat{\Xi}_{1,b,n}^{(\textnormal{Rx})}\right)}$ and
$\trans{\left(\hat{\Xi}_{2,b,1}^{(\textnormal{Rx})},\ldots,\hat{\Xi}_{2,b,n}^{(\textnormal{Rx})}\right)}
$ be the corresponding codewords of the outer code.  

In the second decoding step, the receiver decodes Messages
${M}_{1,\textnormal{CL},b-1}$ and ${M}_{2,\textnormal{CL},b-1}$. To
this end, it first pre-processes the outputs observed in blocks $b$ and
$b-1$ to mitigate the influence of the sequences produced to encode
Messages $(M_{1,\textnormal{CS},b},M_{2,\textnormal{CS},b})$. The
outputs in block $b$ are processed as follows: For each
$k\in\{1,\ldots, n\}$ the receiver computes
\begin{IEEEeqnarray}{rCl}\label{eq:Ytildu}
  \tilde{\bfY}_{b,k} & \eqdef & \bfY_{b,k} - \vect{a}_1\hat{\Xi}_{1,b,k}^{(\textnormal{Rx})}
 - \vect{a}_2\hat{\Xi}_{2,b,k}^{(\textnormal{Rx})}
  \nonumber \\ &&- \B_1 \bfY_{b,k}
  -\B_2 \bfY_{b,k} ,
\end{IEEEeqnarray}
where 
 \begin{IEEEeqnarray*}{l}
\vect{Y}_{b,k} \eqdef \trans{(Y_{(b-1)\eta n+ (k-1)\eta+1},\ldots,
  Y_{(b-1)\eta n+ k \eta})},
\end{IEEEeqnarray*}
Notice that in case the first decoding
step was successful, i.e,  in case that  $\hat{\Xi}_{1,b,k}^{(\textnormal{Rx})} =
\Xi_{1,b,k}$  and $\hat{\Xi}_{2,b,k}^{(\textnormal{Rx})} =
\Xi_{2,b,k}$ holds for all $k\in\{1,\ldots,n\}$, \eqref{eq:Ytildu} corresponds to
\begin{IEEEeqnarray*}{rCl}
\lefteqn{ \sqrt{(1-\rho_1^2) P_1'}\vect{U}_{1,b,k}+ \sqrt{(1-\rho_2^2)P_2'}\vect{U}_{2,b,k}}\quad  \nonumber \\
& &+ \left( \sqrt{\frac{1}{2} \rho_1^2
  P_1'}+\sqrt{\frac{1}{2}\rho_2^2
  P_2'}\right)\left(\boldsymbol{\Omega}_{1,b,k}+
  \boldsymbol{\Omega}_{2,b,k}\right) \nonumber \\ &&+ \B_1 \vect{W}_{1,b,k} + \B_2
  \vect{W}_{2,b,k} +\vect{Z}_{b,k}.
\end{IEEEeqnarray*}

Before describing how the receiver processes the outputs in block $b-1$, we
notice  that the receiver already decoded Messages
$M_{1,\textnormal{CL},b-2},$ $M_{2,\textnormal{CL},b-2}$,
$M_{1,\textnormal{CS},b-1},$ and $M_{2,\textnormal{CS},b-1}$ in
previous decoding steps. 
Let 
$\hat{M}_{1,\textnormal{CL},b-2}^{(\textnormal{Rx})},$ $\hat{M}_{2,\textnormal{CL},b-2}^{(\textnormal{Rx})}$,
$\hat{M}_{1,\textnormal{CS},b-1}^{(\textnormal{Rx})},$ and
$\hat{M}_{2,\textnormal{CS},b-1}^{(\textnormal{Rx})}$ denote the receiver's guess of these
messages. Also, for each $k\in\{1,\ldots,n\}$ 
let
$\boldsymbol{\hat{\Omega}}_{1,b-1,k}^{(\textnormal{Rx})}$ and
$\boldsymbol{\hat{\Omega}}_{2,b-1,k}^{(\textnormal{Rx})}$ denote the codewords
 that in the codebooks used in the $k$-th subblock of block $b-1$
correspond to the guesses $\hat{M}_{1,\textnormal{CL},b-2}^{(\textnormal{Rx})},$
$\hat{M}_{2,\textnormal{CL},b-2}^{(\textnormal{Rx})}$, and let $\trans{\left(\hat{\Xi}_{1,b-1,1}^{(\textnormal{Rx})}
  \ldots,\hat{\Xi}_{1,b-1,n}^{(\textnormal{Rx})}\right)}$ and
$\trans{\left(\hat{\Xi}_{2,b-1,1}^{(\textnormal{Rx})},\ldots,\hat{\Xi}_{2,b-1,n}^{(\textnormal{Rx})}\right)}$
denote the codewords that in the outer code used in block $b-1$ 
correspond to the guesses $\hat{M}_{1,\textnormal{CS},b-1}^{(\textnormal{Rx})},$ and
$\hat{M}_{2,\textnormal{CS},b-1}^{(\textnormal{Rx})}$. 
The receiver 
processes the outputs observed in the block
 $(b-1)$ by computing for $k\in\{1,\ldots, n\}$:
\begin{IEEEeqnarray*}{rCl}
\lefteqn{\vect{\tilde{Y}}^{(2)}_{b-1,k}}\\ & \eqdef & \bfY_{b-1,k} \nonumber \\ & & -\left( \sqrt{\frac{1}{2} \rho_1^2
  P_1'}+ \sqrt{\frac{1}{2}\rho_2^2
  P_2'}\right) \left(\boldsymbol{\hat{\Omega}}_{1,b-1,k}+
  \boldsymbol{\hat{\Omega}}_{2,b-1,k}\right)\nonumber \\
& & - \vect{a}_1{\Xi}_{1,b-1,k}
   - \vect{a}_2{\Xi}_{2,b-1,k}
   - \B_1 \bfY_{b,k}
  -\B_2 \bfY_{b,k}, \\
&= & \sqrt{(1-\rho_1^2)
   P_1'}\vect{U}_{1,b-1,k}+
 \sqrt{(1-\rho_2^2)P_2'}\vect{U}_{2,b-1,k} \nonumber \\ &&+ \B_1 \vect{W}_{1,b-1,k}
  + \B_2
  \vect{W}_{2,b-1,k} +\vect{Z}_{b-1,k}.
\end{IEEEeqnarray*}
Equipped with the sequences $\left\{\left(\vect{\tilde{Y}}_{b,i},
    \vect{\tilde{Y}}_{b-1,i}^{(2)}\right)\right\}_{i=1}^n$  the
    receiver finally decodes Messages $(M_{1,\textnormal{CL},b-1},M_{2,\textnormal{CL},b-1})$
using an
optimal decoder for a $2\eta$-input antenna/$2\eta$-output antenna
Gaussian MAC with temporally-white noise that is correlated across antennas.

After Block $(B+1)$ the receiver decodes Messages
$(M_{1,\textnormal{CL},B},M_{2,\textnormal{CL},B})$ based on
$\vect{\tilde{Y}}^{(2)}_{B,1},\ldots, \vect{\tilde{Y}}_{B,n}^{(2)}$
and based on the sequence $(Y_{B \eta n+1}, \ldots, Y_{(B+1)\eta n})$.
To this end, it again uses an optimal decoder for a $2\eta$-input
antenna/$2\eta$-output antenna Gaussian MAC with temporally-white
noise that is correlated across antennas.

\subsubsection{Noisy and Perfect Partial Feedback}\label{sec:appPerPar1}
The proposed extension applies also to settings with noisy or perfect partial
feedback to Transmitter~2, if $\B_1$ is set to the all-zero matrix and
if Carleial's scheme for partial feedback is
applied. Thus, our scheme should be modified so that there are no
decodings taking place at Transmitter~1 and so that in
\eqref{eq:X1u} and \eqref{eq:s1b} the term $\sqrt{\frac{1}{2}\rho_{\nu}^2 P_{\nu}'}
(\boldsymbol{\omega}_{1,b,i} + \boldsymbol{\omega}_{2,b,i})$ is
replaced by
$\sqrt{ \rho_{\nu}^2 P_{\nu}'} \boldsymbol{\omega}_{1,b,i}$.

Notice that in a setting with perfect partial feedback to
Transmitter~2 the components of the noise vectors corrupting
$\{\vect{\tilde{V}}_{2,b,i}\}$ are uncorrelated, similarly for
$\left\{\left(\vect{\tilde{V}}^{(2)}_{2,b,i}-\sqrt{(1-\rho_{1}^2
      P_1')} \vect{u}_{1,b-1,i}\right)\right\}$ and for
$\vect{\tilde{Y}}_{b,i}$ and $\vect{\tilde{Y}}_{b,i}^{(2)}$. Thus,
optimal decoders for Gaussian multi-input antenna/multi-output
antenna channels with uncorrelated white noise sequences can be used
to decode $M_{1,\textnormal{CL},b}$ at Transmitter~2 and to decode
$(M_{1,\textnormal{CL},b}, M_{2,\textnormal{CL},b})$ at the receiver.
Moreover, the observation
$\{\vect{Y}_{b,i}\}$ at the receiver is a degraded version of the
observation $\{\vect{\tilde{V}}_{1,b,i}\}$ at Transmitter~2. Thus, since the
receiver decodes $(M_{1,\textnormal{CS},b},M_{2,\textnormal{CS},b})$
based on $\{\vect{Y}_{b,i}\}$, in settings with perfect partial
feedback there is no loss in optimality in the presented
rate-splitting scheme if based on $\{\vect{\tilde{V}}_{1,b,i}\}$
Transmitter~2 first decodes message $M_{1,\textnormal{CS},b}$ before
decoding $M_{1,\textnormal{CL},b}$. In particular, the set of
achievable rates of the concatenated scheme is solely constrained by
the decoding at the receiver.

\subsection[Interleaving and Rate-Splitting]{Interleaving and Rate-Splitting with Carleial's Cover-Leung Scheme}\label{sec:interleaving}

We describe the scheme in Section~\ref{sec:CarleialInter} in more
detail. We start with the encodings and first consider the encodings
in the $\ell$-th subblock of Block $b$, for a fixed $b \in
\{1,\ldots, B\}$ and $\ell\in\{1,\ldots, \eta\}$. Define
$\tilde{b}=(b-1)\eta+\ell$. We assume that from decoding steps after
previous subblocks $((b-2)\eta+1),\ldots, (\tilde{b}-1)$, both
transmitters are cognizant of
$\{(M_{1,\textnormal{ICL},(b-2)\eta+1},M_{2,\textnormal{ICL},(b-2)\eta+1})$,
$\ldots,$ $(M_{1,\textnormal{ICL},\tilde{b}-1}$,$M_{2,\textnormal{ICL},\tilde{b}-1})\}$.

The encodings in Subblock $\tilde{b}$ consist of four steps. In the
first step Transmitter~1 produces an $n$-length vector to encode
messages $M_{1,\textnormal{ICL},\tilde{b}}$,
$M_{1,\textnormal{ICL},\tilde{b}-\eta}$, and
$M_{2,\textnormal{ICL},\tilde{b}-\eta}$ as follows.
Given $M_{1,\textnormal{ICL},\tilde{b}}=
m_{1,\textnormal{ICL},\tilde{b}}$,
$M_{1,\textnormal{ICL},\tilde{b}-\eta}=m_{1,\textnormal{ICL},\tilde{b}-\eta}$,
and
$M_{2,\textnormal{ICL},\tilde{b}-\eta}=m_{2,\textnormal{ICL},\tilde{b}-\eta}$,
Transmitter~$1$ first picks codewords
$\vect{u}_{1,\tilde{b}}(M_{1,\textnormal{ICL}, \tilde{b}})$,
$\boldsymbol{\omega}_{1,\tilde{b}}(M_{1,\textnormal{ICL},\tilde{b}-\eta})$,
and
$\boldsymbol{\omega}_{2,\tilde{b}}(M_{2,\textnormal{ICL},\tilde{b}-\eta})$
from the corresponding codebooks, which have independently been
generated by randomly drawing each entry according to an IID zero-mean
unit-variance Gaussian distribution\footnote{To satisfy the power
  constraints the Gaussian distribution should be of variance slightly
  less than 1. However, this is a technicality which we
  ignore.}. Transmitter~$1$ then completes the first step by computing the following linear
combination
 \begin{IEEEeqnarray}{rCl}\label{eq:t2}
\sqrt{(1-\rho_1^2)P_1'} \vect{u}_{1,\tilde{b}} + \sqrt{\frac{1}{2}
  \rho_{1}^2 P_{1}'} \left( \boldsymbol{\omega}_{1,\tilde{b}}+
  \boldsymbol{\omega}_{2,\tilde{b}} \right),\IEEEeqnarraynumspace
\end{IEEEeqnarray}
where $\rho_1\in[0,1]$ is a fixed chosen parameter of the scheme, which does
not depend on $\tilde{b}$.
Similarly, for Transmitter~2.

In the second step, Transmitter~$1$ computes
the ``cleaned'' feedback vectors $\vect{\bar{V}}_{\nu,(b-1)\eta+1},
\ldots, \vect{\bar{V}}_{\nu,(b-1)\eta+\ell-1}$, where
$\vect{\bar{V}}_{\nu,\tilde{b}'}$ for 
 $\tilde{b}'\in\{(b-1)\eta+1,\ldots,
(b-1)\eta+\ell-1\}$ is defined as:
\begin{IEEEeqnarray}{rCl}
\vect{\bar{V}}_{1,\tilde{b}'} &\triangleq& \vect{V}_{1, \tilde{b}'} -
\sqrt{(1-\rho_1^2) P_1'}\vect{U}_{1, \tilde{b}'} \nonumber \\ && -\sqrt{
  (1-\rho_2^2)P_2'} \vect{U}_{2, \tilde{b}'}%
\nonumber \\ && -\left( \sqrt{\frac{1}{2}\rho_1^2 P_1'} + \sqrt{\frac{1}{2} \rho_2^2
    P_2'}\right)
    \left(\boldsymbol{\Omega}_{1,\tilde{b}'}+\boldsymbol{\Omega}_{2,\tilde{b}'}\right),\nonumber \\
\label{eq:Vbar}
\end{IEEEeqnarray}
where $\vect{V}_{\nu,\tilde{b}'} \triangleq \trans{(V_{\nu,
    (\tilde{b}'-1)n +1}, \ldots, V_{\nu,\tilde{b}'n})}$.  Similarly,
for Transmitter~2.  Notice that for $\tilde{b}'
\in\{(b-1)\eta+1,\ldots, (b-1)\eta+\ell-1\}$ the ``cleaned'' feedback vectors
satisfy
\begin{equation*}
\vect{\bar{V}}_{1,\tilde{b}'} -  \vect{W}_{1,\tilde{b}'}=
\vect{\bar{V}}_{2,\tilde{b}'}-\vect{W}_{2,\tilde{b}'}, 
\end{equation*} 
where for $\tilde{b}'\in \{(b-1)\eta,\ldots, (b-1)\eta+\ell-1\}$ and 
 $\nu \in \{1,2\}$:
\begin{IEEEeqnarray*}{rCl}
  \vect{W}_{\nu,\tilde{b}'} & \triangleq & \trans{(W_{\nu,
      (\tilde{b}'-1)n +1}, \ldots, W_{\nu,\tilde{b}'n})}.
\end{IEEEeqnarray*}
Thus, they correspond to the feedback
vectors of a ``cleaned'' channel where the channel outputs are
described by the vectors  $\left\{(\vect{\bar{V}}_{1,\tilde{b}'} -
\vect{W}_{1,\tilde{b}'})\right\}$.

In the third step, Transmitter~1
produces an $n$-length vector to encode
Message $M_{1,\textnormal{ICS},b}$ using the ``cleaned'' feedback
vectors in \eqref{eq:Vbar} as explained shortly. Assume that at the
beginning of Block $b$ Transmitter~1 fed Message
$M_{1,\textnormal{ICS},b}$ to its outer encoder and that the outer
encoder produced the codeword
$\boldsymbol{\xi}_{1,b}$. Let
\begin{IEEEeqnarray*}{rCl}
\vect{a}_{1}&\triangleq&
\trans{(a_{1,1},\ldots, a_{1,\eta})},\\
\B_1 & \triangleq & \begin{pmatrix} b_{1, 1,1}& & \ldots &
  &b_{1,1,\eta}\\ \\ \ldots &&&& \ldots \\ \\   b_{1, \eta,1}& & \ldots &
  &b_{1,\eta,\eta} \end{pmatrix},
\end{IEEEeqnarray*}
denote the parameters of Transmitter~1's modified inner encoder. 
The modified inner encoder then  produces the $n$-length vector
\begin{IEEEeqnarray}{rCl}\label{eq:s2}
 a_{1,\ell} \boldsymbol{\xi}_{1,b} + \sum_{j=1}^{\ell-1}
  b_{1,\ell, j} \vect{\bar{V}}_{1,(b-1)\eta+j},
\IEEEeqnarraynumspace
\end{IEEEeqnarray}
which is also the $n$-length vector that Transmitter~1 produces in this
third step. Similarly, for Transmitter~2.

In the forth and last step, Transmitter~1 sums the $n$-length vectors
in~\eqref{eq:t2} and \eqref{eq:s2}, and sends the resulting symbols
over the channel. Similarly, for Transmitter~2.
 
Thus, the signal transmitted by Transmitter~$\nu$ in Subblock $\tilde{b}$
can be described as follows:
\begin{IEEEeqnarray}{rCl}\label{eq:x1}
  \vect{X}_{\nu,\tilde{b}} & = & \sqrt{(1-\rho_\nu^2)
    P_\nu'}\vect{u}_{\nu, \tilde{b}} +\sqrt{\frac{1}{2}\rho_\nu^2
    P_\nu'}\left(\boldsymbol{\omega}_{1,\tilde{b}}+\boldsymbol{\omega}_{2,\tilde{b}}\right)\nonumber
    \\ & &+
  \vect{a}_{\nu,\ell} \boldsymbol{\xi}_{\nu,b} +\sum_{j=1}^{\ell-1}
  b_{\nu,\ell, j} \vect{\bar{V}}_{\nu,(b-1)\eta+j},
    \IEEEeqnarraynumspace
\end{IEEEeqnarray} 
where
 $ \vect{X}_{\nu,\tilde{b}}  \triangleq  \trans{(X_{\nu,(\tilde{b}-1)n+1
 }, \ldots, X_{\nu,\tilde{b}n})}.$

Notice that if the parameters $(\vect{a}_1,\vect{a}_2,\mat{B}_1,\mat{B}_2)$
satisfy the power constraints \eqref{eq:powergen} for transmit powers $(P_1-P_1')$ and
$(P_2-P_2')$, noise variance $N$, and feedback-noise covariance matrix
$\KW$, then the input sequences satisfy the power constraints
\eqref{eq:power} with arbitrary high probability.

We next consider the encodings in the last Block $(B+1)$, where
 the two transmitters 
send information about the pairs of messages $\{(M_{1,\textnormal{CL},(B-1)
  \eta+1}, M_{2,\textnormal{CL}, (B-1)\eta+1}),$ $
\ldots,$ $(M_{1,\textnormal{CL},B\eta}, M_{2,\textnormal{CL},
 B\eta})\}$. We consider a fixed subblock $\tilde{b}\in\{B\eta+1,
 \ldots, (B+1)\eta\}$. The transmitters send  their channel inputs
 in this last block $(B+1)$ as follows. Given
$M_{1,\textnormal{CL}, \tilde{b}-\eta}=m_{1,\textnormal{CL},\tilde{b}-\eta}$ and
$ M_{2,\textnormal{CL}, \tilde{b}-\eta}=m_{2,\textnormal{CL},\tilde{b}-\eta}$,
both transmitters choose the codewords 
$\boldsymbol{\omega}_{1,\tilde{b}}(M_{1,\textnormal{CL},\tilde{b}-\eta})$, and
  $\boldsymbol{\omega}_{2,\tilde{b}}(M_{2,\textnormal{CL},\tilde{b}-\eta})$
  from the corresponding codebooks and send a linear combination of
 the chosen codewords over the channel. 
Thus, the signal transmitted by Transmitter~$\nu$ in Subblock $\tilde{b}$ can be described as 
\begin{IEEEeqnarray}{rCl}\label{eq:x1b}
  \vect{X}_{\nu,\tilde{b}} & = & \sqrt{\frac{1}{2}\rho_\nu^2
    P_\nu'}\left(\boldsymbol{\omega}_{1,\tilde{b}}+\boldsymbol{\omega}_{2,\tilde{b}}\right),\IEEEeqnarraynumspace
\end{IEEEeqnarray}
where
\begin{IEEEeqnarray*}{rCl}
  \vect{X}_{\nu,\tilde{b}} & \triangleq & \trans{(X_{\nu,(\tilde{b}-1)n+1
 }, \ldots, X_{\nu,\tilde{b}n})}.
\end{IEEEeqnarray*}

We next describe the decoding at Transmitter~2; the decoding at
Transmitter~1 is performed similarly and therefore omitted; and the
decoding at the receiver will be described later on. 

After each subblock $\tilde{b} \in \{1,\ldots, B\eta\}$
Transmitter~2 decodes Message $M_{1,\textnormal{ICL},\tilde{b}}$.  We
consider a fixed Subblock $\tilde{b}\in \{1,\ldots, B\eta\}$ and
define $b\in\{1,\ldots, B\}$ and $\ell\in\{1,\ldots, \eta\}$ so that
$\tilde{b}=(b-1)\eta+\ell$. Before describing the decoding of
Message $M_{1,\textnormal{ICL},\tilde{b}}$ at the end of this paragraph, we
notice the following. After Subblock $\tilde{b}$, Transmitter~2
observed the feedback vectors $\vect{V}_{2, (b-1)\eta+1}, \ldots,
\vect{V}_{2,(b-1)\eta+\ell}$ and is additionally cognizant of Messages
$M_{2,\textnormal{ICS},b}$, $\{M_{2,\textnormal{ICL},(b-1)\eta+1},
\ldots, M_{2,\textnormal{ICL}, (b-1)\eta+\ell}\}$, and (assuming its
previous decoding steps were successful) of Messages
$\{M_{1,\textnormal{ICL}, (b-1)\eta+1}, \ldots, M_{1,\textnormal{ICL},
  (b-1)\eta+\ell-1}\}$. It can therefore reconstruct the sequences
produced to encode these messages. Moreover, Transmitter~2 can estimate Transmitter~1's
feedback outputs $\vect{V}_{1, (b-1)\eta+1}, \ldots,
\vect{V}_{1,(b-1)\eta+\ell}$, (even though it cannot reconstruct them
because it is incognizant of the feedback noises). By 
subtracting the reconstructed sequences and the estimated sequence from
its feedback outputs Transmitter~2 can thus compute the $n$-dimensional vectors
$\vect{\tilde{N}}_{2,(b-1)\eta+1}, \ldots,
\vect{\tilde{N}}_{2,(b-1)\eta+\ell-1}$ and
$\vect{\tilde{V}}_{2,(b-1)\eta+\ell}$, which are defined as:
\begin{IEEEeqnarray*}{rCl}
\lefteqn{
  \vect{\tilde{V}}_{2,(b-1)\eta+\ell}}\nonumber\quad \\ & \triangleq &
  \vect{V}_{2,(b-1)\eta+\ell}- \sqrt{ (1-\rho_2^2) P_2'}
  \vect{U}_{2,(b-1)\eta+\ell} \nonumber \\
  & & - \left(\sqrt{\frac{1}{2} \rho_1^2 P_1'}+\sqrt{\frac{1}{2}
      \rho_2^2 P_2'}   \right)\nonumber \\ &&\qquad \cdot
(\boldsymbol{\Omega}_{1,(b-1)\eta+\ell}+\boldsymbol{\Omega}_{2,(b-1)\eta+\ell})
   \nonumber \\ &&
  - a_{2,\ell} \boldsymbol{\Xi}_{2,b} -
  \sum_{j=1}^{\ell-1} \left(b_{1,\ell, j}+ b_{2,\ell, j}\right)
  \vect{\bar{V}}_{2,(b-1)\eta+j}
  \\
  & = & \sqrt{ (1-\rho_1^2) P_1'} \vect{U}_{1,(b-1)\eta+\ell} +
  a_{1,\ell}\boldsymbol{\Xi}_{1,b}\nonumber \\ & & + \sum_{j=1}^{\ell-1} b_{1,\ell,j}
  \left( \vect{W}_{1, (b-1)\eta+j} -\vect{W}_{2,(b-1)\eta+j}\right)
  \nonumber \\ &&
  + \vect{Z}_{(b-1)\eta+\ell} + \vect{W}_{2,(b-1)\eta+\ell},
  \IEEEeqnarraynumspace
\end{IEEEeqnarray*}
where where we define the vector $\vect{Z}_{(b-1)\eta+\ell} \eqdef \trans{(Z_{((b-1)\eta+\ell-1)n+1},
  \ldots, Z_{((b-1)\eta+\ell)n})}$; and for $\tilde{b}'=(b-1)+\ell'$ and
$\ell'\in\{1,\ldots,\ell-1\}$:
\begin{IEEEeqnarray*}{rCl}
  \vect{\tilde{N}}_{2,\tilde{b}'} & \triangleq &
  \vect{V}_{2,\tilde{b}'}- \sqrt{ (1-\rho_1^2) P_1'}
  \vect{U}_{1,\tilde{b}'}\nonumber \\ & &- \sqrt{ (1-\rho_2^2) P_2'}
  \vect{U}_{2,\tilde{b}'}  \nonumber \\ & & - \left(\sqrt{\frac{1}{2} \rho_1^2
  P_1'}+\sqrt{\frac{1}{2} \rho_2^2 P_2'}
  \right)(\boldsymbol{\Omega}_{1,\tilde{b}'}+\boldsymbol{\Omega}_{2,\tilde{b}'})
 \nonumber \\ &&  - \sum_{j=1}^{\ell'-1}
  \left(b_{1,\ell', j}+ b_{2,\ell', j}\right) \vect{\bar{V}}_{2,(b-1)\eta+j} \nonumber \\ &&-  a_{2,\ell'} \boldsymbol{\Xi}_{2,b}
\\
  & = &  a_{1,\ell'}\boldsymbol{\Xi}_{1,b}
  \nonumber \\ && +
  \sum_{j=1}^{\ell'-1} b_{1,\ell',j} \left( \vect{W}_{1, (b-1)\eta+
      j} -\vect{W}_{2,(b-1)\eta+j}\right) \nonumber \\ &&+ \vect{Z}_{\tilde{b}'} +
  \vect{W}_{2,\tilde{b}'}, 
\end{IEEEeqnarray*}
where $\vect{Z}_{\tilde{b}'} \eqdef \trans{(Z_{(\tilde{b}'-1)n+1},
  \ldots, Z_{\tilde{b}'n})}$.  Transmitter~2 finally decodes Message
$M_{1,\textnormal{CL},\tilde{b}}$ based on $
\vect{\tilde{N}}_{2,(b-1)\eta+1},$ $\ldots,
\vect{\tilde{N}}_{2,(b-1)\eta+\ell-1}$, and
$\vect{\tilde{V}}_{2,(b-1)\eta+\ell}$ using an optimal decoder for a
single-input antenna/multi-output antenna Gaussian channel with
correlated but temporally-white noise sequences.

We next describe the decoding at the receiver. We first consider the
decoding of the pair $(M_{1,\textnormal{CL}, \tilde{b}},
M_{2,\textnormal{CL},\tilde{b}})$ after a fixed subblock $\tilde{b}
\in \{\eta+1,\ldots, (B+1)\eta\}$. Define $b\in\{2,\ldots, B+1\}$ and
$\ell \in\{1,\ldots, \eta\}$ so that $\tilde{b}=(b-1)\eta+\ell$.
Before describing the decoding of the pair $(M_{1,\textnormal{CL},
  \tilde{b}}, M_{2,\textnormal{CL},\tilde{b}})$ at the end of this
paragraph, we notice the following. In decoding steps after previous
subblocks the receiver has already decoded Messages
$\{(M_{1,\textnormal{ICL},(b-3)\eta +\ell'},
M_{2,\textnormal{ICL},(b-3)\eta +\ell'})\}_{\ell'=1}^{\ell}$,
$\{(M_{1,\textnormal{ICL},(b-2)\eta +\ell'},
M_{2,\textnormal{ICL},(b-2)\eta +\ell'})\}_{\ell'=1}^{\ell}$, and
$\{(M_{1,\textnormal{ICL},(b-1)\eta +\ell'},
M_{2,\textnormal{ICL},(b-1)\eta
  +\ell'})\}_{\ell'=1}^{\ell-1}$. Therefore, (assuming that these
decodings were successful)
the receiver can reconstruct the sequences produced to encode these
messages and subtract them from the output signal. Thus, the receiver
can compute for $b' \in\{b-1,b\}$ and $\ell'\in\{1,\ldots,\ell-1\}$
the ``cleaned'' output vector
\begin{IEEEeqnarray*}{rCl}
\lefteqn{\vect{\bar{Y}}_{(b'-1)\eta+\ell'} }\nonumber \\ &\triangleq&
\vect{Y}_{(b'-1)\eta+\ell'}  -
\sqrt{(1-\rho_1^2) P_1'}\vect{U}_{1, (b'-1)\eta+\ell'}  \nonumber \\&&l-\sqrt{
  (1-\rho_2^2)P_2'} \vect{U}_{2, (b'-1)\eta+\ell'}  \nonumber \\
&&-
\left( \sqrt{\frac{1}{2}\rho_1^2 P_1'} + \sqrt{\frac{1}{2} \rho_2^2
    P_2'}\right) \nonumber\\ & & \quad \cdot
\left(\boldsymbol{\Omega}_{1,(b'-1)\eta+\ell'}+\boldsymbol{\Omega}_{2,(b'-1)\eta+\ell'}\right), 
\nonumber \\
&=&  a_{1,\ell}\boldsymbol{\Xi}_{1,b'} +
a_{2,\ell'}\boldsymbol{\Xi}_{2,b'}
 \nonumber \\ &&+ \sum_{j=1}^{\ell'-1}\left(
b_{1,\ell',j}\vect{\bar{V}}_{1,(b'-1)\eta+j} +
 b_{2,\ell',j}\vect{\bar{V}}_{2,(b'-1)\eta+j}
 \right)
 \nonumber \\ & & + \vect{Z}_{(b'-1)\eta+\ell'},
\end{IEEEeqnarray*}
where \[\vect{Y}_{(b'-1)\eta+\ell'} \eqdef
\trans{(Y_{((b'-1)\eta+\ell'-1)n +1}, \ldots,
  Y_{((b'-1)\eta+\ell')n})},\] and it can compute
\begin{IEEEeqnarray*}{rCl}
\lefteqn{\vect{\bar{Y}}^{(2)}_{(b-2)\eta+\ell}}\nonumber \\ & \eqdef &
\vect{Y}_{(b-2)\eta+\ell} \nonumber \\ && -
\left(\sqrt{\frac{1}{2} \rho_1^2 P_1'} +
  \sqrt{\frac{1}{2} \rho_2^2 P_2'} \right)\nonumber \\ 
&& \cdot 
(\boldsymbol{\Omega}_{1,(b-2)\eta+\ell} +
\boldsymbol{\Omega}_{2,(b-2)\eta+\ell})\nonumber \\
& = & \sqrt{(1-\rho_1^2)P_1'} \vect{U}_{1,(b-2)\eta+\ell} \nonumber \\
&&+
\sqrt{(1-\rho_2^2)P_2'} \vect{U}_{2,(b-2)\eta+\ell} + a_{1,\ell} \boldsymbol{\Xi}_{1,b}  +
a_{2,\ell}\boldsymbol{\Xi}_{2,b} \nonumber \\
& &+ \sum_{j=1}^{\ell-1}\left(
b_{1,\ell,j}\vect{\bar{V}}_{1,(b-2)\eta+j} + b_{2,\ell,j} \vect{\bar{V}}_{2,(b-2)\eta+j}\right) \nonumber \\
& &+ \vect{Z}_{(b-2)\eta+\ell},
\end{IEEEeqnarray*}
where $\vect{Y}_{(b-2)\eta+\ell} \eqdef \trans{(Y_{((b-2)\eta+\ell-1)n
    +1}, \ldots, Y_{((b-2)\eta+\ell)n})}$.  Notice that the
``cleaned'' output vector $\vect{\bar{Y}}_{(b'-1)\eta+\ell'}$ equals
the difference
$\left(\vect{\bar{V}}_{1,(b'-1)\eta+\ell'}-\vect{W}_{1,(b'-1)\eta+\ell'}\right)$.
Notice further, that even though the ``cleaned'' outputs
$\vect{\bar{Y}}_{(b-2)\eta+1},\ldots,
\vect{\bar{Y}}_{(b-2)\eta+\ell-1}$ and
$\vect{\bar{Y}}_{(b-1)\eta+1},\ldots,
\vect{\bar{Y}}_{(b-1)\eta+\ell-1}$ do not depend on the pair
$(M_{1,\textnormal{CL}, \tilde{b}}, M_{2,\textnormal{CL},\tilde{b}})$,
they are correlated with the noise sequences corrupting
$\vect{\bar{Y}}^{(2)}_{(b-2)\eta+\ell}$ and
$\vect{Y}_{(b-1)\eta+\ell}$ and should be taken into account by the
receiver when decoding $(M_{1,\textnormal{CL}, \tilde{b}},
M_{2,\textnormal{CL},\tilde{b}})$. Thus, the receiver should decode
the pair $(M_{1,\textnormal{CL}, \tilde{b}},
M_{2,\textnormal{CL},\tilde{b}})$ based on the vectors
$\vect{\bar{Y}}_{(b-2)\eta+1},\ldots,
\vect{\bar{Y}}_{(b-2)\eta+\ell-1}$,
$\vect{\bar{Y}}_{(b-1)\eta+1},\ldots,
\vect{\bar{Y}}_{(b-1)\eta+\ell-1}$,
$\vect{\bar{Y}}^{(2)}_{(b-2)\eta+\ell}$, and
$\vect{\bar{Y}}_{(b-1)\eta+\ell}$. To this end, the receiver first
partly ``decorrelates'' the vectors by
computing
\begin{IEEEeqnarray*}{rCl}\lefteqn{
\vect{\tilde{Y}}_{(b-1)\eta+\ell} }\; \nonumber \\& \eqdef & \vect{Y}_{(b-1)\eta+\ell} -
\sum_{j=1}^{\ell-1} (b_{1,\ell, j}+b_{2,\ell,j})\vect{\bar{Y}}_{(b-1)\eta+j}
, \nonumber \\
 & = & \sqrt{(1-\rho_1^2)P_1'} \vect{U}_{1,(b-1)\eta+\ell}\nonumber \\
 & & +
\sqrt{(1-\rho_2^2)P_2'} \vect{U}_{2,(b-1)\eta+\ell}  \nonumber \\
& & + \left(  \sqrt{
    \frac{1}{2} \rho_1^2 P_1'}+ \sqrt{\frac{1}{2} \rho_2^2
    P_2'}\right) \nonumber \\ & & \qquad \cdot (
\boldsymbol{\Omega}_{1,(b-1)\eta+\ell}+\boldsymbol{\Omega}_{2,(b-1)\eta+\ell})
\nonumber \\ & & 
+ a_{1,\ell}\boldsymbol{\Xi}_{1,b} +
a_{2,\ell}\boldsymbol{\Xi}_{2,b}\nonumber\\
& &+ \sum_{j=1}^{\ell-1}\left(
b_{1,\ell,j}\vect{W}_{1,(b-1)\eta+j} +
 b_{2,\ell,j}\vect{W}_{2,(b-1)\eta+j}
 \right)\nonumber \\&&+ \vect{Z}_{(b-1)\eta+\ell},
\end{IEEEeqnarray*}
\begin{IEEEeqnarray*}{rCl}
\lefteqn{  \vect{\tilde{Y}}^{(2)}_{(b-2)\eta+\ell}}\nonumber \\ & \eqdef &
  \vect{\bar{Y}}^{(2)}_{(b-2)\eta+\ell} - \sum_{j=1}^{\ell-1}
  (b_{1,\ell, j}+b_{2,\ell,j}) \vect{\bar{Y}}_{(b-2)\eta+j}
  \IEEEeqnarraynumspace\\
  & = & \sqrt{(1-\rho_1^2)P_1'} \vect{U}_{1,(b-2)\eta+\ell} \nonumber
  \\ & & +
  \sqrt{(1-\rho_2^2)P_2'} \vect{U}_{2,(b-2)\eta+\ell}%
  \nonumber \\  & &+ a_{1,\ell}
  \boldsymbol{\Xi}_{1,b} +
  a_{2,\ell}\boldsymbol{\Xi}_{2,b} \nonumber \\
  & & + \sum_{j=1}^{\ell-1}\left(
    b_{1,\ell,j}\vect{\bar{W}}_{1,(b-2)\eta+j} + b_{2,\ell,j}
    \vect{\bar{W}}_{2,(b-2)\eta+j}\right)\nonumber \\ &&  + \vect{Z}_{(b-2)\eta+\ell},
\end{IEEEeqnarray*}
and for $b'\in\{b-1,b\},\; \ell'\in\{1,\ldots,\ell-1\}$:
\begin{IEEEeqnarray*}{rCl}
\lefteqn{\vect{\tilde{Y}}^{(3)}_{(b'-1)\eta + \ell'} }\nonumber \\ & \eqdef &\vect{\bar{Y}}_{(b'-1)\eta+\ell'}
 \nonumber \\  & & -
\sum_{j=1}^{\ell'-1} (b_{1,\ell', j}+b_{2,\ell',j}) \vect{\bar{Y}}_{(b'-1)\eta+j}
\nonumber \\
& = & a_{1,\ell'} \boldsymbol{\Xi}_{1,b'} +
a_{2,\ell'}\boldsymbol{\Xi}_{2,b'} 
\nonumber \\  & &  +\sum_{j=1}^{\ell'-1}\left(b_{1,\ell',j}\vect{W}_{1,(b'-1)\eta+j} + b_{2,\ell',j}
  \vect{W}_{2,(b'-1)\eta+j} \right)\nonumber \\ & & 
+ \vect{Z}_{(b'-1)\eta+\ell'}.
\end{IEEEeqnarray*} 
The receiver then decodes the pair of messages $(M_{1,\textnormal{ICL},
  \tilde{b}}, M_{2,\textnormal{ICL}, \tilde{b}})$ based on
$\vect{\tilde{Y}}_{(b-2)\eta+1}^{(3)}, \ldots,
\vect{\tilde{Y}}_{(b-2)\eta+\ell-1}^{(3)},
\vect{\tilde{Y}}_{(b-2)\eta+\ell}^{(2)}$ and
$\vect{\tilde{Y}}_{(b-1)\eta+1}^{(3)}, \ldots,
\vect{\tilde{Y}}_{(b-1)\eta+\ell-1}^{(3)}, \vect{\tilde{Y}}_{(b-1)\eta+\ell}$
using an optimal decoder for a $2$-input/$2\ell$-output antenna
Gaussian MAC with temporally-white noise sequences correlated across antennas.
  
After  decoding Messages
$\{(M_{1,\textnormal{ICL},\tilde{b}},M_{2,\textnormal{ICL},\tilde{b}})\}_{\tilde{b}=1}^{B\eta}$
the receiver decodes Messages
$\{(M_{1,\textnormal{ICS},b},M_{2,\textnormal{ICS},b})\}_{b=1}^{B}$.
 To this end, it  first reverses the interleaving introduced by the
 modified inner encoders on
the ``cleaned'' output vectors $\vect{\bar{Y}}_{1}, \ldots,
\vect{\bar{Y}}_{B \eta}$. That is, for $b\in\{1, \ldots, B\}$, it
constructs the $\eta n$-dimensional vector
\begin{IEEEeqnarray*}{rCl}
\vect{Y}_{\textnormal{DeInt},b} & \eqdef & 
  (\bar{Y}_{(b-1)\eta +1,1},
  \ldots, \bar{Y}_{b\eta,1},\\ & &   \bar{Y}_{(b-1)\eta+1,2}, \ldots,
  \bar{Y}_{b\eta, 2},\\ & &  \bar{Y}_{(b-1)\eta,n}, \ldots, \bar{Y}_{b,\eta n})^{\T},\nonumber
\end{IEEEeqnarray*}
where $\bar{Y}_{\tilde{b},i}$ denotes the $i$-th entry of vector
$\vect{\bar{Y}}_{\tilde{b}}$.  It then decodes Messages
$(M_{1,\textnormal{ICS},b},M_{2,\textnormal{ICS},b})$ applying inner
and outer decoder of the concatenated scheme to the vector
$\vect{Y}_{\textnormal{DeInt},b}$.

\subsubsection{Noisy and Perfect Partial Feedback}\label{sec:appPerPar2}
The proposed extension can also be applied in settings with noisy or perfect
partial feedback, if $\B_1$ is set to the all-zero
matrix and if Carleial's scheme for noisy or perfect partial feedback is
applied. Accordingly, our scheme should be modified so that there is no 
decoding taking place at Transmitter~1. Therefore,
in \eqref{eq:x1} and \eqref{eq:x1b} the term
$\sqrt{\frac{1}{2}\rho_{\nu}^2 P_{\nu}'}
(\boldsymbol{\omega}_{1,\tilde{b}} + \boldsymbol{\omega}_{2,\tilde{b}})$
should be replaced by $\sqrt{ \rho_{\nu}^2 P_{\nu}'}
\boldsymbol{\omega}_{1,\tilde{b}}$, for $\nu\in\{1,2\}$.

Notice that---as in the second extension---for perfect partial
feedback the various vectors computed for the
decodings at Transmitter~2 and for the decodings at the receiver have
uncorrelated noise components. Therefore, without loss in optimality,
Transmitter~2 and the receiver can use optimal decoders for Gaussian
multi-input antenna/multi-output antenna channels with independent
white noise sequences.

\end{appendix}

\bibliographystyle{plain}

\bibliography{header_short,mybiblio}

\begin{thebibliography}{10}

\bibitem{brosslapidothwigger08}
S.~I. Bross, A.~Lapidoth, and M.~A. Wigger.
\newblock The {G}aussian {MAC} with conferencing encoders.
\newblock In {\em Proc. IEEE Int. Symposium on Inf. Theory}, Toronto, Canada,
  July 6--11 2008.

\bibitem{carleial82}
A.~B. Carleial.
\newblock Multiple-access channels with different generalized feedback signals.
\newblock {\em IEEE Trans. Inform. Theory}, 28(6):841--850, Nov. 1982.

\bibitem{cover75}
T.~M. Cover.
\newblock Some advances in broadcast channels.
\newblock In A.~Viterbi, editor, {\em Advances in Communication Systems},
  volume~4. San Francisco: Academic Press, 1975.

\bibitem{coverleung81}
T.~M. Cover and C.~S.~K. Leung.
\newblock An achievable rate region for the multiple-access channel with
  feedback.
\newblock {\em IEEE Trans. Inform. Theory}, 27(3):292--298, May 1981.

\bibitem{gaarderwolf75}
N.~Th. Gaarder and J.~K. Wolf.
\newblock The capacity region of a multiple-access discrete memoryless channel
  can increase with feedback.
\newblock {\em IEEE Trans. Inform. Theory}, 21(1):100--102, Jan. 1975.

\bibitem{gastpar05}
M.~Gastpar.
\newblock On noisy feedback in {G}aussian networks.
\newblock In {\em Proc. 43rd Allerton Conf. Comm., Contr. and Comp.}, Allerton
  H., Monticello, Il, 2005.

\bibitem{gastparkramer06_1}
M.~Gastpar and G.~Kramer.
\newblock On cooperation via noisy feedback.
\newblock In {\em Proc. IZS}, pages 146--149, Feb. 22--24, 2006.

\bibitem{hekstrawillems89}
A.~P. Hekstra and F.~M.~J. Willems.
\newblock Dependence balance bounds for single-output two-way channels.
\newblock {\em IEEE Trans. Inform. Theory}, 35(1):44--53, Jan. 1989.

\bibitem{kimlapidothweissman06}
Y.-H. Kim, A.~Lapidoth, and T.~Weissman.
\newblock Bounds on the error exponent of the {AWGN} channel with
  {AWGN}-corrupted feedback.
\newblock In {\em Proc. of 24th IEEE Conv. of Electrical \& Electronics Eng. in
  Israle (IEEEI'06)}, pages 184--188, Eilat, Israel, Nov. 15--17 2006.

\bibitem{kramer98}
G.~Kramer.
\newblock {\em Directed information for channels with feedback}.
\newblock {Ph.D.} dissertation, ETH Zurich, Switzerland, Switzerland, 1998.

\bibitem{kramer02}
G.~Kramer.
\newblock Feedback strategies for white {Gaussian} interference networks.
\newblock {\em IEEE Trans. Inform. Theory}, 48(6):1423--1438, June 2002.

\bibitem{kramer03}
G.~Kramer.
\newblock Capacity results for the discrete memoryless network.
\newblock {\em IEEE Trans. Inform. Theory}, 49(1):4--21, Jan. 2003.

\bibitem{ozarow85}
L.~H. Ozarow.
\newblock The capacity of the white {Gaussian} multiple-access channel with
  feedback.
\newblock {\em IEEE Trans. Inform. Theory}, 30(4):623--629, July 1984.

\bibitem{rimoldiurbanke96}
B.~Rimoldi and R.~Urbanke.
\newblock A rate-splitting approach to the {G}aussian multiple-access channel.
\newblock {\em IEEE Trans. Inform. Theory}, 42:364--375, 1996.

\bibitem{schalkwijkkailath66}
J.~P.~M. Schalkwijk and T.~Kailath.
\newblock A coding scheme for additive noise channels with feedback-i:no
  bandwidth constraint.
\newblock {\em IEEE Trans. Inform. Theory}, 12:172--182, Apr. 1966.

\bibitem{shavivsteinberg08}
D.~Shaviv and Y.~Steinberg.
\newblock On the multiple access channel with common rate-limited feedback.
\newblock In {\em Proc. IZS}, pages 108--111, Mar. 12--14, 2008.

\bibitem{tandonulukus08}
R.~Tandon and S.~Ulukus.
\newblock Dependence balance based outer bounds for {Gaussian} networks with
  cooperation and feedback.
\newblock Submitted to \emph{IEEE Trans. Inform. Theory, December 2008}.
  http://arxiv.org/pdf/0812.1857.

\bibitem{vandermeulen87}
E.~C. van~der Meulen.
\newblock Capacity theorems for multiple-access channels with feedback.
\newblock In {\em ISICT'87}, Campinas, Brazil, July 27--August 01 1987.

\bibitem{vinodh07}
V.~Venkatesan.
\newblock Optimality of {Gaussian} inputs for a multi-access achievable region.
\newblock Semester Project, Signal and Inform. Proc. Lab., ETH Zurich,
  Switzerland, Supervised by A. Lapidoth and M. A. Wigger, June 2007.

\bibitem{willems82}
F.~M.~J. Willems.
\newblock The feedback capacity region of a class of discrete memoryless
  multiple-access channels.
\newblock {\em IEEE Trans. Inform. Theory}, 28(1):93--95, Jan. 1982.

\bibitem{willemsmeulen83}
F.~M.~J. Willems and E.~C. {van der Meulen}.
\newblock Partial feedback for the discrete memoryless multiple access channel.
\newblock {\em IEEE Trans. Inform. Theory}, 29(2):287--290, Mar. 1983.

\bibitem{willemsvandermeulenschalkwijk83}
F.~M.~J. Willems, E.~C. van~der Meulen, and J.~P.~M. Schalkwijk.
\newblock A coding scheme for the additive white {G}aussian noise multiple
  access channel with semi-feedback.
\newblock {\em Tijdschrift van het Nederlands Elektronica-en Radiogenootschap},
  48(3):103--107, 1983.

\bibitem{willemsvandermeulenschalkwijk83-2}
F.~M.~J. Willems, E.~C. van~der Meulen, and J.~P.~M. Schalkwijk.
\newblock Generalized feedback for the discrete memoryless multiple access
  channel.
\newblock In {\em Proc. 21th Allerton Conf. Comm., Contr. and Comp.}, pages
  284--292, Allerton H., Monticello, Il, Oct. 5--7, 1983.

\bibitem{wozencraftjacobs65}
J.~M. Wozencraft and I.~M. Jacobs.
\newblock {\em Principles of Communication Engineering}.
\newblock John Wiley \& Sons, 1965.

\bibitem{wyner74}
A.~D. Wyner.
\newblock Recent results in the {S}hannon theory.
\newblock {\em IEEE Trans. Inform. Theory}, 20:2--10, Jan. 1974.

\end{thebibliography}

\end{document}